\documentclass[]{aa}
\usepackage{graphicx,natbib}
\usepackage{floatrow}
\usepackage{subfigure}
\usepackage{amsmath,amssymb}
\usepackage{mathrsfs}
\usepackage{mathabx}

\usepackage{txfonts}
\usepackage{pifont}

\usepackage{hyperref}

\hypersetup{
   colorlinks=true,
   citecolor=blue,
   linkcolor=blue,
   urlcolor=black
   }

\usepackage{natbib}

\newcommand\crm{\cr\noalign{\medskip}}

\newcommand\be{\begin{equation}}
\newcommand\ee{\end{equation}}
\def\m@th{\mathsurround=0pt}
\newcommand\EQM[1]{\vcenter{\normalbaselines\m@th
    \ialign{${\displaystyle ##}$\hfil&&\ ${\displaystyle ##}$\hfil\crcr
    \mathstrut\crcr\noalign{\kern-\baselineskip}
    \noalign{\smallskip}
    #1\crcr\mathstrut\crcr\noalign{\kern-\baselineskip}}}}

\newcommand\A{A^{s}}

\newcommand{\mybf}[1]{#1}
\newcommand{\scor}[1]{#1}
\newcommand{\pcor}[1]{#1}
\newcommand{\jcor}[1]{#1}

\newsavebox\ltmcbox

\pdfpageattr{/Group <</S /Transparency /I true /CS /DeviceRGB>>}

\begin{document}

\title{Atmospheric tides in Earth-like planets}

\author{
P. Auclair-Desrotour\inst{1,2}
\and J. Laskar\inst{1}
\and S. Mathis\inst{2,3}
}

\institute{
IMCCE, Observatoire de Paris, CNRS UMR 8028, PSL, 77 Avenue Denfert-Rochereau, 75014 Paris, France\\
\email{pierre.auclair-desrotour@obspm.fr, jacques.laskar@obspm.fr}
\and Laboratoire AIM Paris-Saclay, CEA/DSM - CNRS - Universit\'e Paris Diderot, IRFU/SAp Centre de Saclay, F-91191 Gif-sur-Yvette, France
\and LESIA, Observatoire de Paris, PSL Research University, CNRS, Sorbonne Universit\'es, UPMC Univ. Paris 06, Univ. Paris Diderot, Sorbonne Paris Cit\'e, 5 place Jules Janssen, F-92195 Meudon, France\\
\email{stephane.mathis@cea.fr}
}

\date{Received ... / accepted ...}

\abstract
{Atmospheric tides can strongly affect the rotational dynamics of planets. In the family of Earth-like planets, such as Venus, this physical mechanism coupled with solid tides makes the angular velocity evolve over long timescales and determines the equilibrium configurations of their spin. }
{Contrary to the solid core, the atmosphere of a planet is submitted to both tidal gravitational potential and insolation flux coming from the star. The complex response of the gas is intrinsically linked to its physical properties. This dependence has to be characterized and quantified for an application to the large variety of extrasolar planetary systems. }
{We develop a theoretical global model where radiative losses, which are predominant in slowly rotating atmospheres, are taken into account. We analytically compute the perturbation of pressure, density, temperature and velocity field caused by a thermo-gravitational tidal perturbation. From these quantities, we deduce the expressions of atmospheric Love numbers and tidal torque exerted on the fluid shell by the star. The equations are written for the general case of a thick envelope and the simplified one of a thin isothermal atmosphere. }
{The dynamics of atmospheric tides depends on the frequency regime of the tidal perturbation: the thermal regime near synchronization and the dynamical regime characterizing fast-rotating planets. Gravitational and thermal perturbations imply different responses of the fluid, i.e. gravitational tides and thermal tides, which are clearly identified. The dependence of the torque on the tidal frequency is quantified using the analytic expressions of the model for Earth-like and Venus-like exoplanets and is in good agreement with the results given by Global Climate Models (GCM) simulations. Introducing dissipative processes such as radiation regularizes the tidal response of the atmosphere, otherwise it is singular at synchronization.}
{We demonstrate the important role played by the physical and dynamical properties of a super-Earth atmosphere (e.g. Coriolis, stratification, basic pressure, density, temperature, radiative emission) in the response of this latter to a tidal perturbation. We point out the key parameters defining tidal regimes (e.g. inertia, Brunt-V\"ais\"al\"a, radiative frequencies, tidal frequency) and characterize the behaviour of the fluid shell in the dissipative regime, which could not be studied without considering the radiative losses.}


\titlerunning{Atmospheric tides in Earth-like planets}
\authorrunning{Auclair-Desrotour, Laskar, Mathis}

\maketitle


\section{Introduction}

Since the pioneering work of \cite{Kelvin1863} about the gravitationally forced elongation of a solid spherical shell, tides have been recognized as a phenomenon of major importance in celestial mechanics. They are actually one of the key mechanisms involved in the evolution of planetary systems over long timescales. Because they affect both stars and planets, tides introduce a complex dynamical coupling between all the bodies that compose a system. This coupling results from the mutual interactions of bodies and is tightly bound to their internal physics. Solid tides have been studied for long \citep[see the reference works by][]{Love1911,GS1966}. The corresponding tidal dissipation, which is caused by viscous friction, can have a strong impact on the orbital elements of a planet \citep[e.g.][]{Mignard1979,Mignard1980,Hut1980,Hut1981,NSL1997,Henningetal2009,RMZL2012}. This evolution depends \jcor{smoothly} on the tidal frequency \citep[\jcor{e.g.}][]{EL2007,Greenberg2009}. Fluid tides cause internal dissipation as well as solid tides, and the energy dissipated is even more dependent on the tidal frequency than the latter because of a highly resonant behaviour \citep[][]{OL2004,Ogilvie2014,ADMLP2015}. As a consequence, in both cases, the phase shifted elongation of the shell induces a tidal torque which modifies the rotational dynamics of the body. This torque drives the evolution of the spin of planets and determines its possible states of equilibrium \citep[][]{GS1969,DI1980,CL01,Correia2003,CL2003,AS2010,Correia2014}. \\

The advent of exoplanets over the past two decades and the rapidly increasing number of discovered planetary systems \citep[][]{MQ1995,Perryman2011} have brought to light a huge diversity of existing orbital structures that can strongly differ from the one of the Solar system \citep[][]{Fabryckyb2012}. Therefore, tidal effects now need to be studied in a general context to characterize these new systems, understand their orbital and rotational dynamics, and constrain the physics of exoplanets with the informations provided by observational measurements. \\

 In this work, we focus on Earth-like exoplanets, which are interesting for their complex internal structure and possible habitability. Indeed, a planet such as the Earth or Venus, is composed of several solid and fluid layers with different physical properties, which implies that the tidal response of a layer cannot be simply correlated to its size. For instance, the Earth's ocean, in spite of its very small depth compared to the Earth's radius, is responsible for most of the tidal dissipation of the planet \citep[][]{Webb1980,ER2000,Ray2001}. The atmospheric envelope of a terrestrial exoplanet sometimes represents a non-negligible fraction of its mass (about $ 20 \ \%$ in the case of small Neptune-like bodies). It is also a layer of great complexity because it is submitted to both tidal gravitational potentials due to other bodies and thermal forcing of the host star. Therefore, it is necessary to develop theoretical models of atmospheric tides with a reduced number of physical parameters if possible. These models will be useful tools to explore the domain of parameters, and to quantify the tidal torque exerted on the spin axis of a planet and quantities used in the orbital evolution of planetary systems. One of these quantities is the second-order Love number ($ k_2 $) introduced by \cite{Love1911}, which measures the quadrupolar hydrostatic elongation. The adiabatic Love number $ k_2 $ has been estimated in the Solar System \citep[e.g.][]{KY1996,Lainey2007,Williamsetal2014}. \\

According to \cite{Wilkes1949}, the first theoretical global models of atmospheric tides have been developed at the end of the eighteenth century by Laplace, who published them in \textit{M\'ecanique c\'eleste} \jcor{\citep{Laplace1798}}. Laplace was interested in the case of an homogeneous atmosphere characterized by a constant pressure scale height and affected by tidal gravitational perturbations. He thus founded the classical theory of atmospheric tides. At the end of the nineteenth century, Lord Kelvin pointed out the unexpected importance of the Earth semidiurnal pressure oscillations in spite of the predominating diurnal perturbation \citep[][see also the measures of the daily variation of temperature and pressure, on Fig.~\ref{fig:mesures_pression}]{Kelvin1882,Hagan2003,Covey2009}. This question motivated many further studies \citep[e.g.][]{Lamb1911,Lamb1932,Taylor1929,Taylor1936,Pekeris1937}, which contributed to enrich the classical theory of tides. In the sixties, theoretical models have been generalized to study other modes, like diurnal oscillations \citep[see][]{Haurwitz1965,Kato1966,Lindzen1966,Lindzen1967a,Lindzen1967b,Lindzen1968}, but still remained focused on the case of the Earth. They revealed the impact of thermal tides due to Solar insolation by showing that these laters drive the Earth diurnal and semi-diurnal waves. Indeed, in the Earth atmosphere, the contribution of the tidal gravitational potential, which causes the elongation of solid layers, is negligible compared to the effect of the thermal forcing. A very detailed review of the general theory for thermal tides and of its history is given by \cite{ChapmanLindzen1970}, denoted CL70 in the following \citep[see also][]{Wilkes1949,Siebert1961}. \\

Leaning on this early work, we generalize here the existing theory to Earth-like planets and exoplanets. We introduce the mechanisms of radiative losses and thermal diffusion, which are usually not taken into account in geophysical models of atmospheric tides owing to their negligible effects on the Earth tidal response \citep[note that radiation was however introduced with a Newtonian cooling by][from which the present work is inspired]{LM1967,Dickinson1968}. These mechanisms cannot be ignored because they play a crucial role in the case of planets like Venus, that are very close to spin-orbit synchronization. Models based on perfect fluids hydrodynamics, such as the one detailed by CL70, fail to describe the atmospheric tidal response of such planets because they are singular at synchronization. In these models, the amplitude of perturbed quantities like pressure, density and temperature diverges at this point. The case of Venus was studied before by \cite{CL01}, who proposed for the tidal torque an empirical bi-parameter model describing two possible regimes: the regime of an atmosphere where dissipative processes can be ignored, far from synchronization, and its supposed regime at the vicinity of synchronization, where the perturbation is expected to annihilate. This behaviour was recently retrieved with GCM simulations by \cite{Leconte2015}, who took into account dissipative mechanisms, thus showing that these mechanisms drive the response of planetary atmospheres in the neighborhood of synchronization. \\

Hence, our objective in this work is to propose a new theoretical global model controlled by a small number of physical parameters, which could be used to characterize the response of any exoplanet atmosphere to a general tidal perturbation. We aim at answering the key following questions:
\begin{itemize}
   \item[$\bullet$] How does the amplitude of pressure, density, temperature, and velocity oscillations depend on the parameters of the system (i.e. rotation, stratification, radiative losses) ? 
   \item[$\bullet$] What are the possible regimes for atmospheric tides ? 
   \item[$\bullet$] How do Love numbers and tidal torques vary with tidal frequency ?
\end{itemize} 

First, in Sect. 2, we derive the general theory for thick atmospheres, which are characterized by a depth comparable to the radius of the planet. In this framework, we expand the perturbed quantities in Fourier series in time and longitude and in series of Hough functions \citep{Hough1898} in latitude. This allows us to compute the equations giving the vertical structure of the atmospheric response to the tidal perturbation. Then, we derive from the hydrodynamics the analytic expressions of the variations of pressure, density, temperature, velocity field and displacement, for any mode. At the end of the section, the density oscillations obtained are used to compute the atmospheric tidal gravitational potential, the Love numbers and the tidal torque exerted on the spin axis of the planet. In Section~3, we apply the equations of Section~2 to the case of a thin isothermal \scor{stably-stratified} atmosphere. In this simplified case, the terms associated with the curvature of the planet disappear and most of the parameters depending on the altitude, such as the pressure scale height of the atmosphere, become constant. \scor{In Section~\ref{sec:convective_atm}, we treat the case of slowly rotating convective atmospheres, such as near the ground in Venus \citep[][]{Marov1973,Seiff1980}.} In the next section, we derive the terms of thermal forcing from the Beer-Lambert law \citep{Bouguer1729,K1760,Beer1852} and the theory of temperature oscillations at the planetary surface. In \scor{Section~\ref{sec:application_Earth_Venus}}, the model is applied to the Earth and Venus. The corresponding torque and atmospheric Love numbers are computed from analytical equations. We end the study with a discussion and by giving our conclusions and prospects. To facilitate the reading, several technical issues have been deferred to the appendix where an index of notations is also provided. 

\section{Dynamics of a thick atmosphere forced thermally and gravitationally}
\label{sec:thick_atmos}

The reference book CL70 has set the bases of analytical approaches for atmospheric tides. This pioneering work focused on fast rotating telluric planets covered by a thin atmosphere, such as the Earth. \\

The goal of the present section is to generalize this work. Hence, we establish the equations that govern the dynamics of tides in a thick fluid shell in the whole range of  tidal frequencies, from synchronization to fast rotation.  We consider a spherical telluric planet of radius $ R $ covered by a stratified atmospheric layer of typical thickness $ H_{\rm atm} $. The atmosphere rotates uniformly with the rocky part at the angular velocity $ \Omega $ (the spin vector of the planet being denoted $ \boldsymbol{\Omega} $). Therefore, the dynamics will be written in the natural equatorial reference frame rotating with the planet, $ \mathscr{R}_{\rm E;T}:\left\{ O , \textbf{X}_{\rm E} , \textbf{Y}_{\rm E} , \textbf{Z}_{\rm E}  \right\} $, where $ O $ is the center of the planet, $ \textbf{X}_{\rm E} $ and $ \textbf{Y}_{\rm E} $ define the equatorial plane and $ \textbf{Z}_{\rm E} = \boldsymbol{\Omega} / \left| \boldsymbol{\Omega} \right| $. We use the spherical basis $ \left( \textbf{e}_r , \textbf{e}_\theta , \textbf{e}_\varphi \right) $ and coordinates $ \left( r , \theta , \varphi \right) $, where $ r $ is the radial coordinate, $ \theta $ the colatitude and $ \varphi $ the longitude. As usual, $ t $ stands for the time. \\

\begin{figure}[htb]
\centering
{\includegraphics[width=0.95\textwidth]
{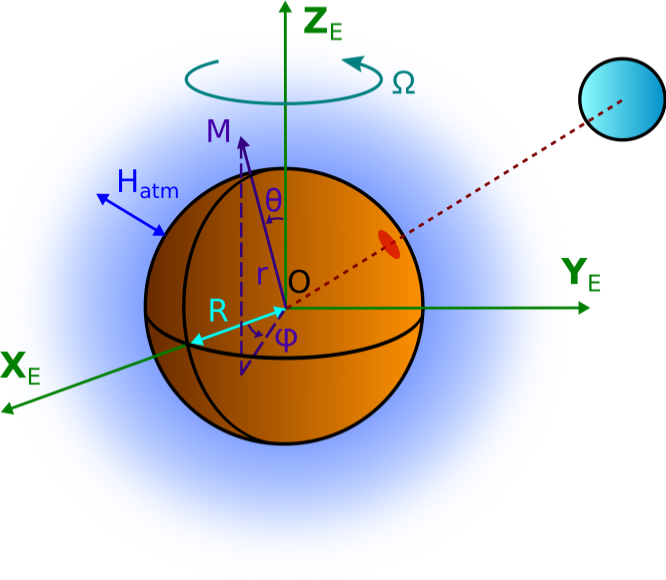}
\textsf{\caption{\label{fig:repere} Spherical coordinates system associated with the equatorial reference frame of the planet, and the geometrical parameters of the system.} }}
\end{figure}

The atmosphere is assumed to be a perfect gas homogeneous in composition, of molar mass $ M $, and stratified radially. Its pressure, density and temperature are denoted $ p $, $ \rho $ and $ T $ respectively. For the sake of simplicity, all dissipative mechanisms, such as viscous friction and heat diffusion, are ignored except the radiative losses of the gas, which play an important role in the vicinity of synchronization. Tides are considered as a small perturbation around a global static equilibrium state. The basic pressure $ p_0 $, density $ \rho_0 $, and temperature $ T_0 $, are supposed to vary with the radial coordinate only. The tidal response will be characterized by a combination of inertial waves, of typical frequency $ 2 \Omega $, due to Coriolis acceleration, and gravity waves which are restored by the stable stratification. The typical frequency of these waves, the Brunt-V\"ais\"al\"a frequency \citep[e.g.][]{Lighthill1978,GZ2008}, is given by

\begin{equation}
N^2 = g \left[ \frac{1}{\Gamma_1 p_0} \dfrac{d p_0}{dr} - \frac{1}{\rho_0} \dfrac{d \rho_0}{dr}  \right],
\label{sigmaBV}
\end{equation}

where $ \Gamma_1 = \left( \partial  \ln p_0 / \partial \ln \rho_0 \right)_S $ is the adiabatic exponent (the index $S$ being the specific macroscopic entropy) and $ g $ the gravity. We assume that the fluid follows the perfect gas law,

\begin{equation}
p = \mathscr{R}_s \rho  T,
\label{gaz_parfait}
\end{equation}

where $ \mathscr{R}_s = \mathscr{R}_{\rm GP} / M $ is the so-called specific gas constant of the atmosphere, $ \mathscr{R}_{\rm GP} = 8.3144621 \ {\rm J.mol^{-1}.K^{-1}} $ \citep[][]{codata2010} being the molar gas constant. In the following, the notations $ \Re $, $ \Im $ and $ ^* $ will be used to designate the real part, imaginary part and conjugate of a complex number. The subscript symbols $ _\Earth $ and $ _\Venus $ stand for the Earth and Venus respectively, which are taken as examples to illustrate the theory. All functions and fields displayed on figures are computed with the algebraic manipulator TRIP \citep{GL13} and plotted with gnuplot.

\subsection{Forced dynamics equations}

The atmosphere is affected by the tidal gravitational potential $ U $ and the heat power per mass unit $ J $. In this approach, the equations of dynamics are linearized around the equilibrium state. The small variations of this state induced by the tidal perturbation are the velocity field $ \textbf{V} = V_{\theta} \textbf{e}_\theta + V_{\varphi} \textbf{e}_\varphi  + V_r \textbf{e}_r  $, the corresponding displacement field, $ \boldsymbol{\xi} = \xi_\theta \textbf{e}_\theta + \xi_\varphi \textbf{e}_\varphi  + \xi_r  \textbf{e}_r $, being defined as

\begin{equation}
\textbf{V} = \dfrac{\partial \boldsymbol{\xi}}{\partial t},
\end{equation}

and the pressure ($ \delta p $), density ($ \delta \rho $) and temperature ($ \delta T $) fluctuations where

\begin{equation}
\begin{array}{ccc}
  p = p_0 + \delta p, & \rho = \rho_0 + \delta \rho, & T = T_0 + \delta T.
\end{array}
\end{equation} 

This represents six unknown quantities to compute ($ V_{r} $, $ V_{\theta} $, $ V_{\varphi} $, $ \delta p $, $ \delta \rho $, $ \delta T $), and we will therefore solve a six-equation system. We first introduce the Navier-Stokes equation. We assume the Cowling approximation in which the self gravitational potential fluctuations induced by tides are neglected \citep[][]{Cowling1941}. It is well adapted to waves characterized by a high vertical wavenumber as in the vicinity of synchronization as detailed further. \jcor{In} the equatorial reference frame $  \mathscr{R}_{\rm E;T} $, the linearized Navier-Stokes equation is \jcor{thus} written

\begin{equation}
\dfrac{\partial \textbf{V}}{\partial t} + 2  \boldsymbol{\Omega} \wedge \textbf{V} = -  \frac{1}{\rho_0} \boldsymbol{\nabla} p - \frac{g}{\rho_0} \delta \rho \, \textbf{e}_r -  \boldsymbol{\nabla} U,
\label{NS}
\end{equation}

which, projected on $ \textbf{e}_r $,  $ \textbf{e}_\theta $ and $ \textbf{e}_\varphi  $, gives

\begin{equation}
    \dfrac{\partial V_\theta}{\partial t} - 2 \Omega \cos \theta V_\varphi  = -  \displaystyle \frac{1}{r} \frac{\partial}{\partial \theta} \left( \frac{\delta p}{\rho_0} + U \right),
\label{NSbrut_1}
\end{equation}

\begin{equation}
    \dfrac{\partial V_\varphi}{\partial t} + 2 \Omega \cos \theta V_\theta  + 2 \Omega \sin \theta V_r = - \displaystyle \frac{1}{r \sin \theta} \dfrac{\partial }{ \partial \varphi} \left( \frac{\delta p}{\rho_0} + U \right),
    \label{NSbrut_2}
\end{equation}

\begin{equation}
    \dfrac{\partial V_r}{\partial t} - 2 \Omega \sin \theta V_\varphi = - \displaystyle \frac{1}{\rho_0} \dfrac{\partial \delta p}{\partial r} - g \frac{\delta \rho}{\rho_0} - \dfrac{\partial U}{ \partial r}.
    \label{NSbrut_3}
\end{equation}

Eqs.~(\ref{NSbrut_1}), (\ref{NSbrut_2}) and (\ref{NSbrut_3}) are coupled by the Coriolis terms (characterized by the factor $ 2 \Omega $) in the left-hand side. To simplify them, we assume the traditional approximation, that is to neglect the terms $ 2 \Omega \sin \theta V_r $ and $ 2 \Omega \sin \theta V_\varphi $ in Eqs.~(\ref{NSbrut_2}) and (\ref{NSbrut_3}) respectively. This hypothesis, commonly used in literature dealing with planetary atmospheres and stars hydrodynamics \citep[e.g.][]{Eckart1960,Mathis2008}, can be applied to stratified fluids where the tidal flow satisfies two conditions: (i) the buoyancy (given by $ g \delta \rho / \rho_0 \textbf{e}_r $ in Eq.~\ref{NS}) is strong compared to the Coriolis term ($ 2  \boldsymbol{\Omega} \wedge \textbf{V} $) and (ii) the tidal frequency (introduced below) is greater than the inertia frequency ($ 2 \Omega $) and smaller than the Brunt-V\"ais\"al\"a frequency. These conditions are satisfied by the Earth's atmosphere and fast rotating planets. The quantitative precision that can be expected with the traditional approximation \scor{may be decreased} in the vicinity of synchronization, where \pcor{the tidal frequency tends to zero ; moreover, we shall note that it leads to issues with momentum and energy conservation in the case of deep atmospheres} \citep[for a discussion about the limitations of the traditional approximation, see][]{GS2005,White2005,M09}. \\ 

Hence, we obtain:

\begin{equation}
    \dfrac{\partial V_\theta}{\partial t} - 2 \Omega \cos \theta V_\varphi  = -  \displaystyle \frac{1}{r} \frac{\partial}{\partial \theta} \left( \frac{\delta p}{\rho_0} + U \right),
    \label{NS1}
  \end{equation}

\begin{equation}
    \dfrac{\partial V_\varphi}{\partial t} + 2 \Omega \cos \theta V_\theta   = - \displaystyle \frac{1}{r \sin \theta} \dfrac{\partial }{ \partial \varphi} \left( \frac{\delta p}{\rho_0} + U \right),
    \label{NS2}
\end{equation}

\begin{equation}
 \rho_0 \dfrac{\partial V_r}{\partial t} = - \dfrac{\partial \delta p}{\partial r} - g \delta \rho - \rho_0 \dfrac{\partial U}{\partial r}.
 \label{NS3}
\end{equation}

In strongly stratified fluids, the left-hand side of Eq.~(\ref{NS3}) is usually ignored because it is very small compared to the terms in $ \delta p  $ and $ \delta \rho $ in typical waves regimes. The radial acceleration will only play a role in regimes where the tidal frequency is close to or exceeds the Brunt-V\"ais\"al\"a frequency, which corresponds to fast rotators. \\

The second equation of our system is the conservation of mass,

\begin{equation}
\dfrac{\partial \rho}{\partial t} + \boldsymbol{\nabla}. \left( \rho \textbf{V} \right) = 0, 
\end{equation}

which, in spherical coordinates, writes

\begin{equation}
\dfrac{\partial \delta \rho}{\partial t } + \frac{1}{r^2} \dfrac{\partial }{\partial r} \left( r^2 \rho_0 V_r  \right) + \frac{\rho_0}{r \sin \theta} \left[ \dfrac{\partial }{\partial \theta} \left( \sin \theta V_\theta \right) + \dfrac{\partial V_\varphi}{\partial \varphi} \right] = 0
\label{conservation_masse}
\end{equation}

The thermal forcing ($ J $) appears in the right-hand side of the linearized heat transport equation \citep[see CL70,][]{GZ2008}, given by

\begin{equation}
\frac{1}{\Gamma_1 p_0} \dfrac{\partial \delta p}{\partial t}  - \frac{1}{\rho_0} \dfrac{\partial \delta \rho}{\delta t} + \frac{N^2}{g} V_r = \kappa \frac{\rho_0}{p_0} \left[  J - J_{\rm rad} \right],
\label{transport_chaleur_1}
\end{equation}

where $ \kappa = \left( \Gamma_1 - 1 \right) / \Gamma_1 $ and $ J_{\rm rad} $ is the power per mass unit radiated by the atmosphere, supposed to behave as a grey body. We \mybf{consider} that $ J_{\rm rad} \ \propto \ \delta T $. This hypothesis is known as ``Newtonian cooling'' and was used by \cite{LM1967} to introduce radiation analytically in the classical theory of atmospheric tides \mybf{\citep[see also][]{Dickinson1968}}. \mybf{Physically, it corresponds to \scor{the case of} an optically thin atmosphere in which the flux emitted by a layer propagates upward or downward without being absorbed by the other layers. In optically thick atmospheres, such as Venus' one \citep[][]{Lacis1975}, this physical condition is not verified. Indeed, because of a stronger absorption, the power emitted by a layer is almost totally transmitted to the neighborhood. Therefore, this significant thermal coupling within the atmospheric shell should be ideally taken into account in a rigorous way, which would leads to great mathematical difficulties \scor{(e.g. complex radiative transfers, Laplacian operators)} in \scor{our} analytical approach.} \scor{However, recent numerical simulations of thermal tides in optically thick atmospheres by Leconte et al. (2015) showed behaviour of the flow in good agreement with a modeling with a radiative cooling. Therefore, we assume in this work \jcor{the Newtonian cooling} as a first modeling of the action of radiation on atmospheric tides.} The Newtonian cooling brings a new characteristic frequency, denoted $ \sigma_0 $, that we call radiative frequency and which depends on the thermal capacity of the atmosphere. The radiative power per unit mass is thus written

\begin{equation}
J_{\rm rad} = \frac{p_0 \sigma_0}{\kappa \rho_0 T_0} \delta T.
\label{phirad}
\end{equation}

Like the basic fields $ p_0 $, $ \rho_0 $ and $ T_0 $, the radiative frequency varies with $ r $ and defines the transition between the dynamical regime, where the radiative losses can be ignored, and the radiative regime, where they predominate in the heat transport equation. Assuming that the radiative emission of the gas is proportional to the local molar concentration $ C_0  = \rho_0 / M $, one may express $ J_{\rm rad} $ and $ \sigma_0 $ as functions of the physical parameters of the fluid (cf. Appendix~\ref{app:newtonian_cooling}),

\begin{equation}
J_{\rm rad} = \frac{8 \epsilon_a}{M} \mathscr{S} T_0^3 \delta T,
\label{phirad2}
\end{equation}

and


\begin{equation}
\sigma_0 \left( r \right) = \frac{8 \kappa \epsilon_a \mathscr{S} }{\mathscr{R}_{\rm GP} } T_0^3,
\label{sigma0}
\end{equation}

the parameter $ \epsilon_{\rm a} $ being an effective molar emissivity coefficient of the gas and $ \mathscr{S} = 5.670373 \times 10^{-8} \ {\rm W.m^{-2}.K^{-4} } $ the Stefan-Boltzmann constant \citep[][]{codata2010}. The substitution of Eq.~(\ref{phirad}) in Eq.~(\ref{transport_chaleur_1}) yields
 
 \begin{equation}
 \frac{1}{\Gamma_1 P_0} \dfrac{\partial \delta p}{\partial t}  - \frac{1}{\rho_0} \dfrac{\partial \delta \rho}{\delta t} + \frac{N^2}{g} V_r = \kappa \frac{\rho_0}{p_0} J - \sigma_0 \frac{\delta T}{T_0}.
 \label{transport_chaleur_2}
 \end{equation}

Finally, the system is closed by the perfect gas law

\begin{equation}
\frac{\delta p}{p_0}  = \frac{\delta T}{T_0} + \frac{\delta \rho}{\rho_0}.
\label{GPlaw}
\end{equation}

Substituting Eq.~(\ref{GPlaw}) in Eq.~(\ref{transport_chaleur_2}), we eliminate the unknown $ \delta T $, and obtain 

\begin{equation}
\frac{1}{\Gamma_1 p_0} \left(  \dfrac{\partial \delta p}{\partial t} + \Gamma_1 \sigma_0 \delta p \right) + \frac{N^2}{g} \dfrac{\partial \xi_r}{\partial t} = \frac{ \kappa \rho_0}{p_0} J + \frac{1}{\rho_0} \left( \dfrac{\partial \delta \rho}{\partial t} + \sigma_0 \delta \rho  \right).
\label{transport_chaleur_3}
\end{equation}

Because of the rotating motion of the perturber in the equatorial frame ($ \mathscr{R}_{\rm E;T} $), a tidal perturbation is supposed to be periodic in time ($ t $) and longitude ($ \varphi $). So, any perturbed quantity $ f $ of our model can be expanded in Fourier series of $ t $ and $ \varphi $

\begin{equation}
f = \sum_{m,\sigma}  f^{m,\sigma} \left( \theta , r \right) e^{i \left( \sigma t + m \varphi \right)},
\label{perturbation}
\end{equation}

the parameter $ \sigma $ being the tidal frequency of a Fourier component and $ m $ its longitudinal degree\footnote{In a binary star-planet system where the planet orbits circularly in the equatorial plane defined by $\textbf{X}_{\rm E}$ and $\textbf{Y}_{\rm E}$ at the orbital frequency $ n_{\rm orb} $, the semidiurnal tide corresponds to $ m = 2 $, $ \sigma = 2 \left( \Omega - n_{\rm orb} \right) $ and $ \nu = \Omega / \left( \Omega - n_{\rm orb} \right) $.}. We also introduce the spin parameter

\begin{equation}
\nu \left( \sigma \right) = \frac{2 \Omega}{\sigma},
\label{nu}
\end{equation}

which defines the possible regimes of tidal gravito-inertial waves: 
\begin{itemize}
\item[$ \bullet $] $ \left| \nu \right| \leq 1 $ corresponds to super-inertial waves, for which the tidal frequency is greater than the inertia frequency;
\item[$ \bullet $] $ \left| \nu \right| > 1 $ corresponds to sub-inertial waves, for which the tidal frequency is lower than the inertia frequency.
\end{itemize}

The parameters $ m $ and $ \nu $ determine the horizontal operators 

\begin{equation}
\mathcal{L}_\theta^{m,\nu} = \frac{1}{\left( 1 - \nu^2 \cos^2 \theta \right)} \left[ \dfrac{\partial}{\partial \theta} + m \nu \cot \theta  \right], 
\label{Ltheta}
\end{equation}

\begin{equation}
\mathcal{L}_\varphi^{m,\nu} = \frac{1}{\left( 1 - \nu^2 \cos^2 \theta \right)} \left[  \nu \cos \theta \dfrac{\partial}{ \partial \theta} + \frac{m}{\sin \theta}  \right] ;
\end{equation}

the operator \jcor{$ \mathcal{L}_{\theta}^{m,\nu} $ is associated with $ V_\theta^{m,\sigma} $ and $ \xi_\theta^{m,\sigma} $, while $ \mathcal{L}_\varphi^{m,\nu} $ is associated with $ V_\varphi^{m,\sigma} $ and $ \xi_\varphi^{m,\sigma} $} \citep[][]{LS1997,Mathis2008}. Hence, the traditional approximation makes us able to write the horizontal component of the velocity field as a function of the variations of the pressure and tidal gravitational potential only:

\begin{equation}
V_{\theta}^{m,\sigma} = \frac{ i   }{\sigma r  } \mathcal{L}_\theta^{m,\nu}  \left( y^{m,\sigma} + U^{m,\sigma}  \right),
\label{Vtheta}
\end{equation}

\begin{equation}
V_\varphi^{m,\sigma} = - \frac{1}{\sigma r  } \mathcal{L}_\varphi^{m,\nu} \left( y^{m,\sigma} + U^{m,\sigma} \right).
\label{Vphi}
\end{equation}

where $ y^{m,\sigma} = \delta p^{m,\sigma} / \rho_0 $. Since \jcor{$ \textbf{V}^{m,\sigma} = \partial_t \boldsymbol{\xi}^{m,\sigma} $}, we also get the corresponding horizontal component \jcor{$ \boldsymbol{\xi}_{\rm H}^{m,\sigma} = \xi_\theta^{m,\sigma} \textbf{e}_\theta + \xi_\varphi^{m,\sigma} \textbf{e}_\varphi $} of the displacement vector from Eq.~(\ref{Vtheta}) and (\ref{Vphi})

\begin{equation}
\xi_\theta^{m,\sigma} = \frac{1}{\sigma^2 r } \mathcal{L}_\theta^{m,\nu} \left( y^{m,\sigma} + U^{m,\sigma}  \right),
\label{xitheta}
\end{equation}

\begin{equation}
\xi_\varphi^{m,\sigma} = \frac{i}{\sigma^2 r  } \mathcal{L}_\varphi^{m,\nu} \left( y^{m,\sigma} + U^{m,\sigma} \right).
\label{xiphi}
\end{equation}

\begin{figure*}[htb]
 \centering
  \includegraphics[width=0.48\textwidth,clip]{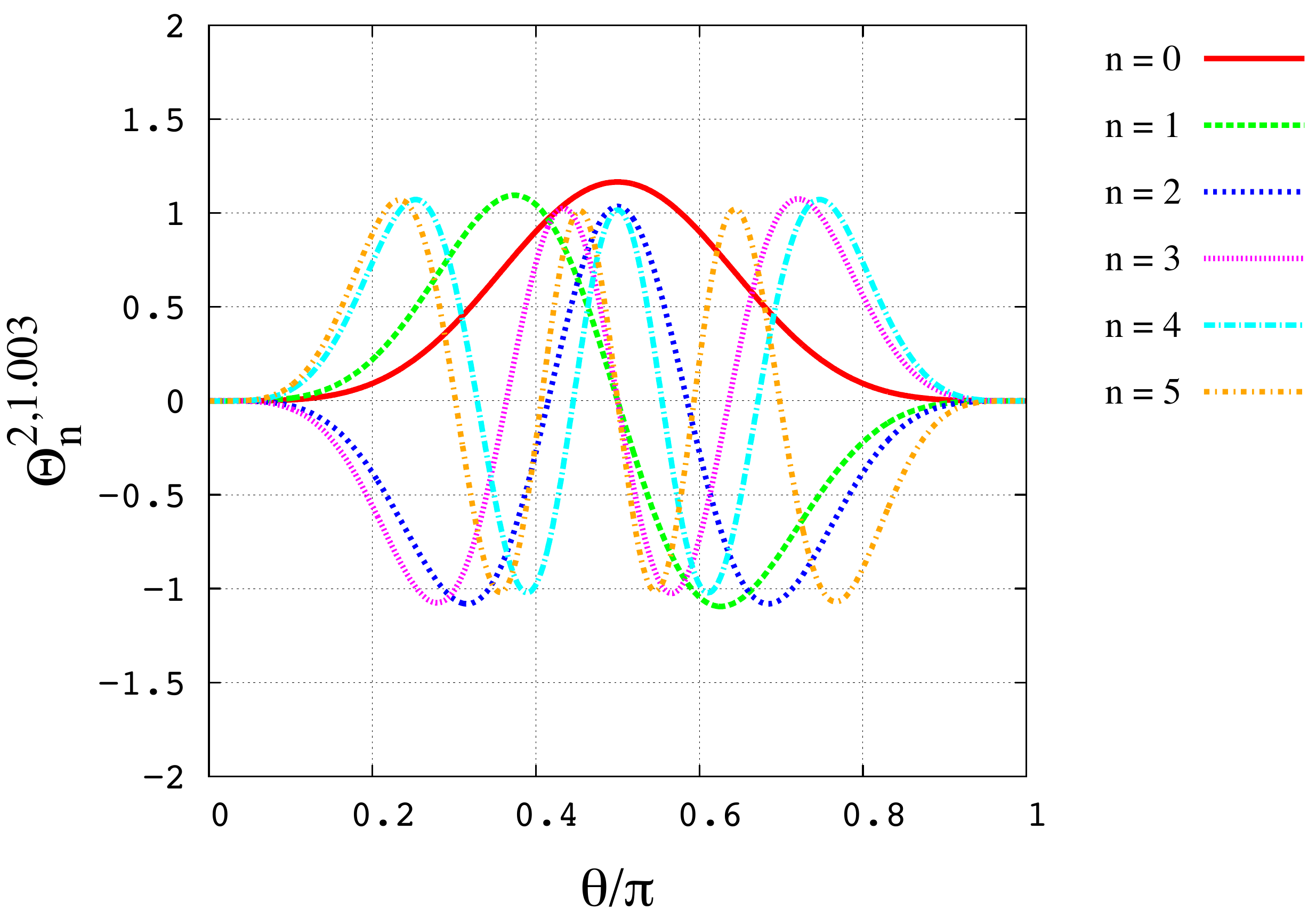} \hspace{2mm} 
 \includegraphics[width=0.48\textwidth,clip]{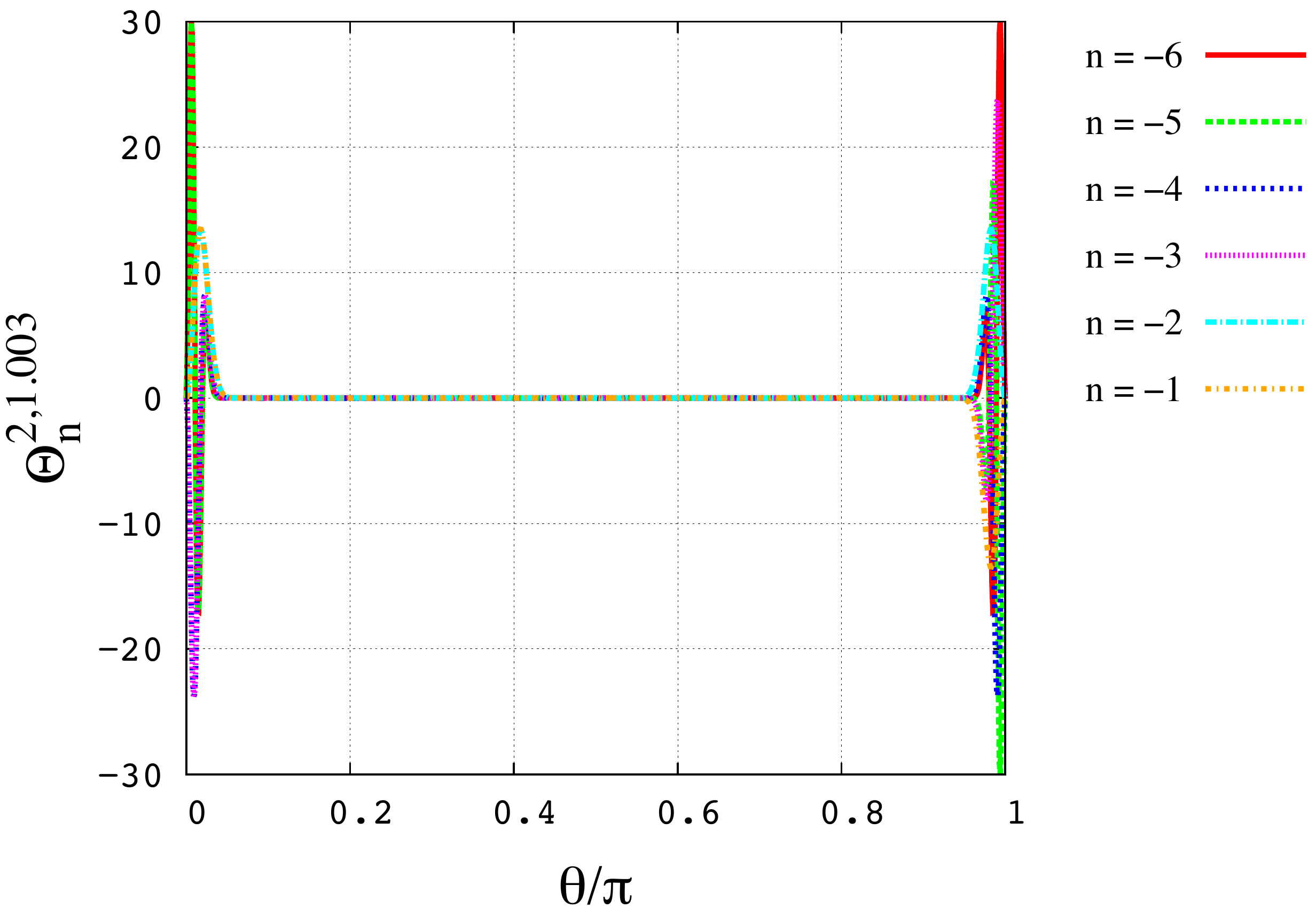}\\      
 \includegraphics[width=0.48\textwidth,clip]{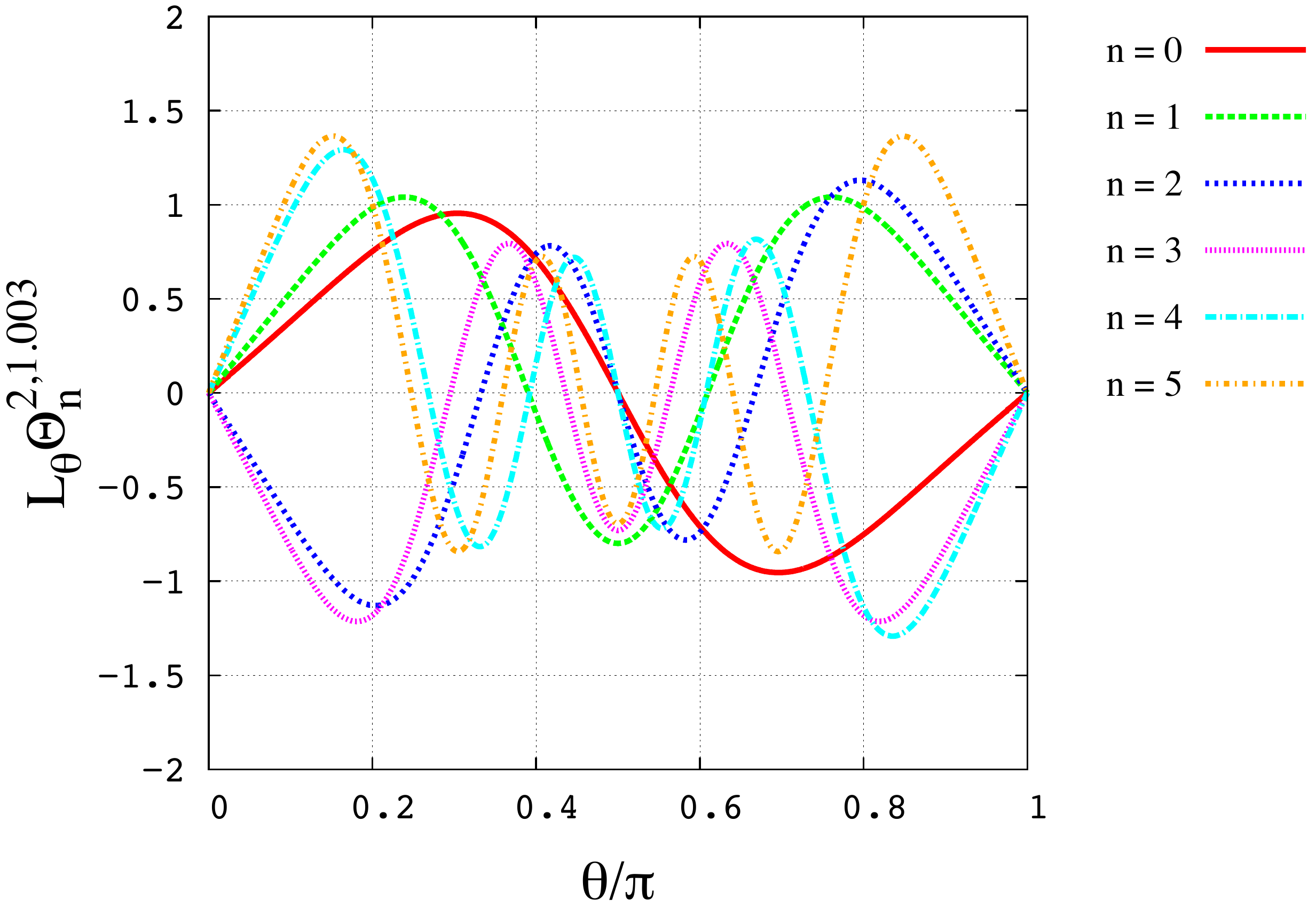} \hspace{2mm} 
  \includegraphics[width=0.48\textwidth,clip]{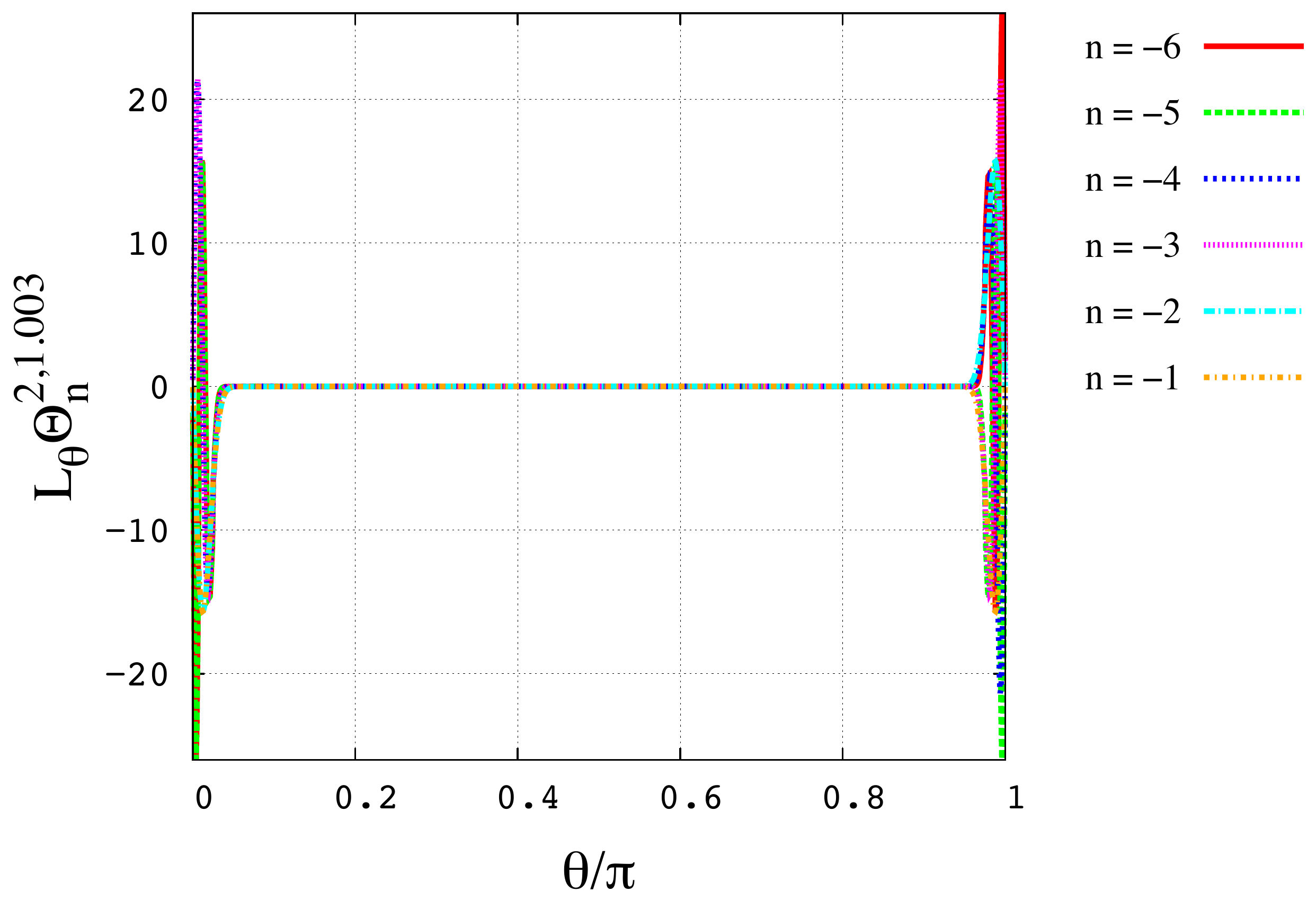}\\ 
   \includegraphics[width=0.48\textwidth,clip]{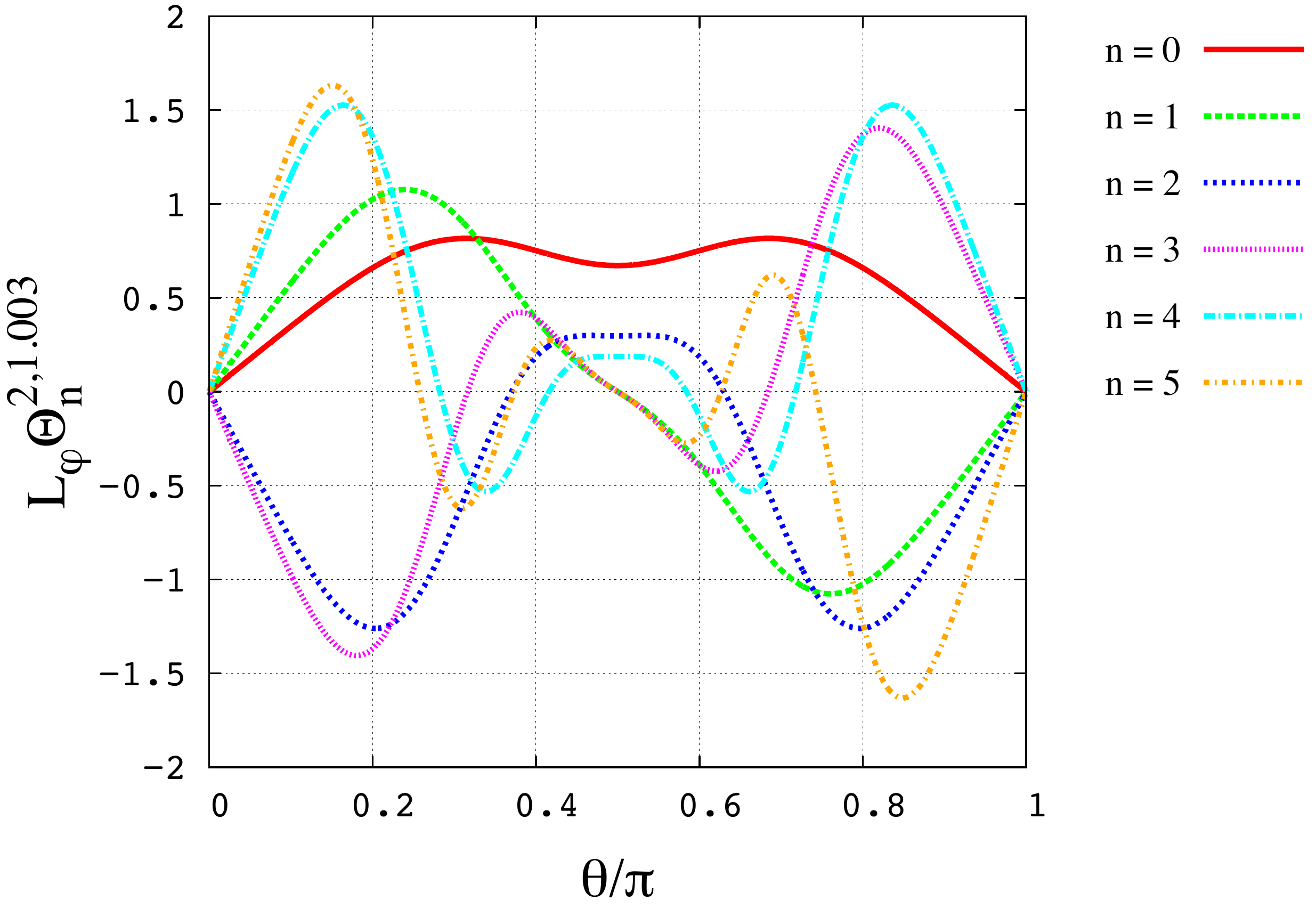} \hspace{2mm} 
  \includegraphics[width=0.48\textwidth,clip]{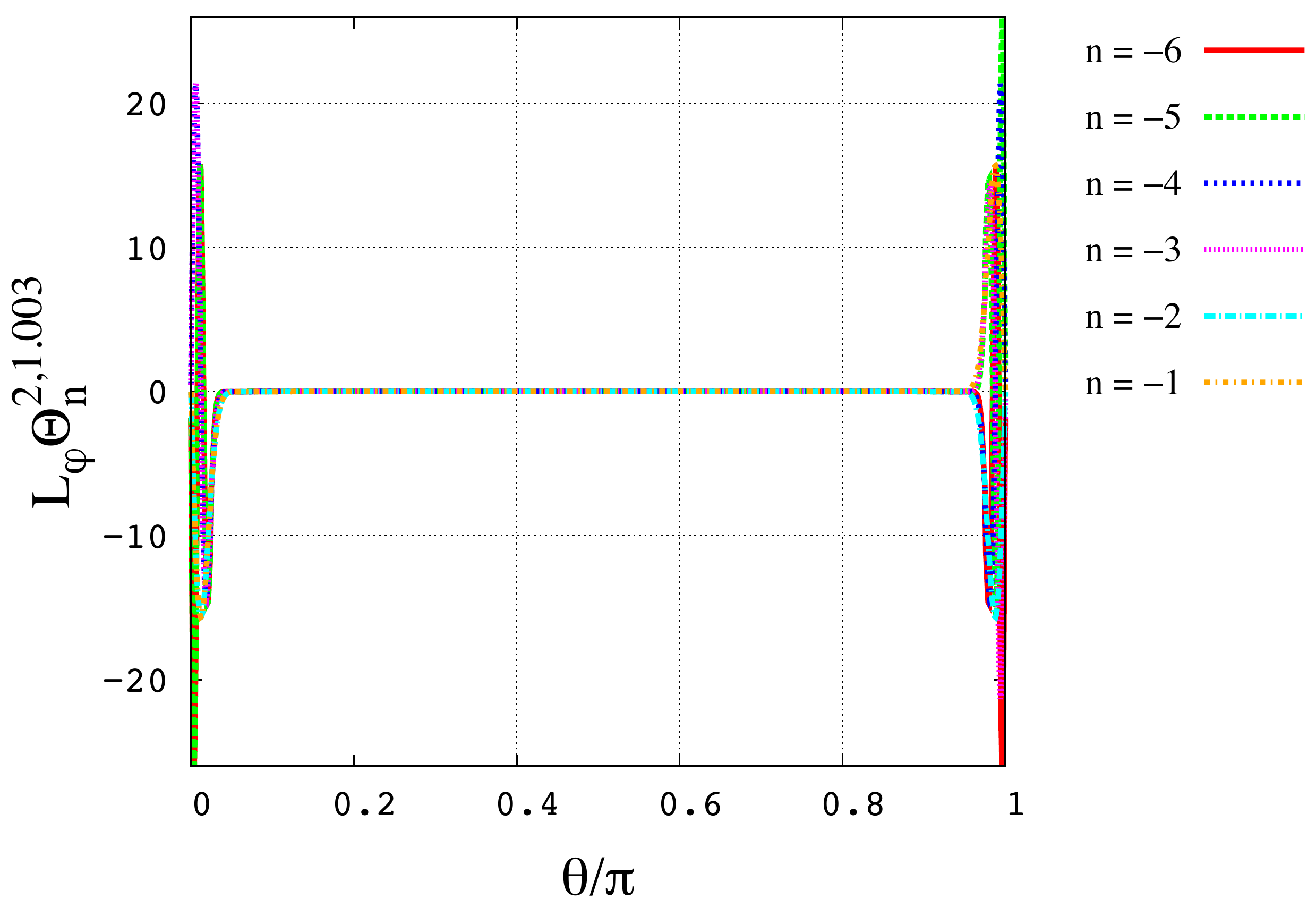}         
  \textsf{ \caption{\label{fig:fonctions_Hough_Terre} Normalized Hough functions $ \Theta_{n}^{m,\nu} $, $ \mathcal{L}_\theta^{m,\nu}   \Theta_{n}^{m,\nu} $ and $ \mathcal{L}_\varphi^{m,\nu}   \Theta_{n}^{m,\nu}  $ corresponding to the Earth's semidiurnal tide ($ m = 2 $ and $ \nu = 1.0030 $). \jcor{We recall that $ \nu = 2 \Omega / \sigma $.} {\bf Left:} First gravity modes ; $ 0 \leq n \leq 5 $. {\bf Right:} First Rossby modes: $ -6 \leq n \leq -1 $. In this case, $ \left| \nu \right| $ is very close to $ 1 $ and Rossby modes are tapped at the poles. The Earth semidiurnal tide is mainly described by gravity modes. For more details about the behaviour of Hough functions, see \mybf{Fig.~\ref{fig:fonctions_Hough} in Appendix~\ref{app:HF}}. }}
\end{figure*}

The substitution of Eqs.~(\ref{Vtheta}) and (\ref{Vphi}) in Eq.~(\ref{conservation_masse}) yields

\begin{equation}
\delta \rho^{m,\sigma} = - \frac{1}{r^2} \dfrac{\partial }{\partial r} \left( r^2 \rho_0 \xi_r^{m,\sigma} \right) - \frac{\rho_0}{\sigma^2 r^2} \mathcal{L}^{m,\nu} \left( y^{m,\sigma} + U^{m,\sigma} \right),
\label{divergent_2}
\end{equation}

where \jcor{$ \mathcal{L}^{m,\nu} $} is the Laplace's operator, parametrized by $ \nu $ and $ m $ and is expressed by

\begin{equation}
\begin{array}{rcl}
\mathcal{L}^{m,\nu} & = & \displaystyle \frac{1}{\sin \theta} \dfrac{\partial }{\partial \theta} \left( \frac{  \sin \theta}{1 - \nu^2 \cos^2 \theta} \dfrac{\partial}{\partial \theta} \right)  \\
\vspace{0.1mm}\\
& & \displaystyle - \frac{1}{1 - \nu^2 \cos^2 \theta} \left( m \nu \frac{1 + \nu^2 \cos^2 \theta}{1 - \nu^2 \cos^2 \theta} + \frac{m^2}{\sin^2 \theta} \right)
\end{array}
\label{Laplace}
\end{equation}

\jcor{with $ \nu = 2 \Omega / \sigma $ (Eq.~\ref{nu})}. When $ \nu = 0 $ (i.e. $ \Omega = 0 $), \jcor{$ \mathcal{L}^{m,\nu} $} is reduced to the horizontal Laplacian. Since the horizontal velocity field is not coupled with the other variables, only three unknowns remain. So, the new system is written

\begin{equation}
 \sigma^2 \rho_0 \xi_r^{m,\sigma} =  \dfrac{\partial \delta p^{m,\sigma}}{\partial r} + g \delta \rho^{m,\sigma} + \rho_0 \dfrac{\partial U^{m,\sigma}}{\partial r},
\label{eq_1}
\end{equation}

\begin{equation}
\delta \rho^{m,\sigma} = - \frac{1}{r^2} \dfrac{\partial }{\partial r} \left( r^2 \rho_0 \xi_r^{m,\sigma} \right) - \frac{\rho_0}{\sigma^2 r^2} \mathcal{L}^{m,\nu} \left( y^{m,\sigma} + U^{m,\sigma} \right),
\label{eq_2}
\end{equation}

\begin{equation}
\left(  i \sigma + \Gamma_1 \sigma_0 \right) \frac{\delta p^{m,\sigma}}{\Gamma_1 p_0} + i \sigma \frac{N^2}{g} \xi_r^{m,\sigma} = \frac{\kappa \rho_0}{p_0} J^{m,\sigma} + \left( i \sigma + \sigma_0 \right) \frac{\delta \rho^{m,\sigma}}{\rho_0},
\label{eq_3}
\end{equation}

\begin{figure}[htb]
 \centering
  \includegraphics[width=0.96\textwidth,clip]{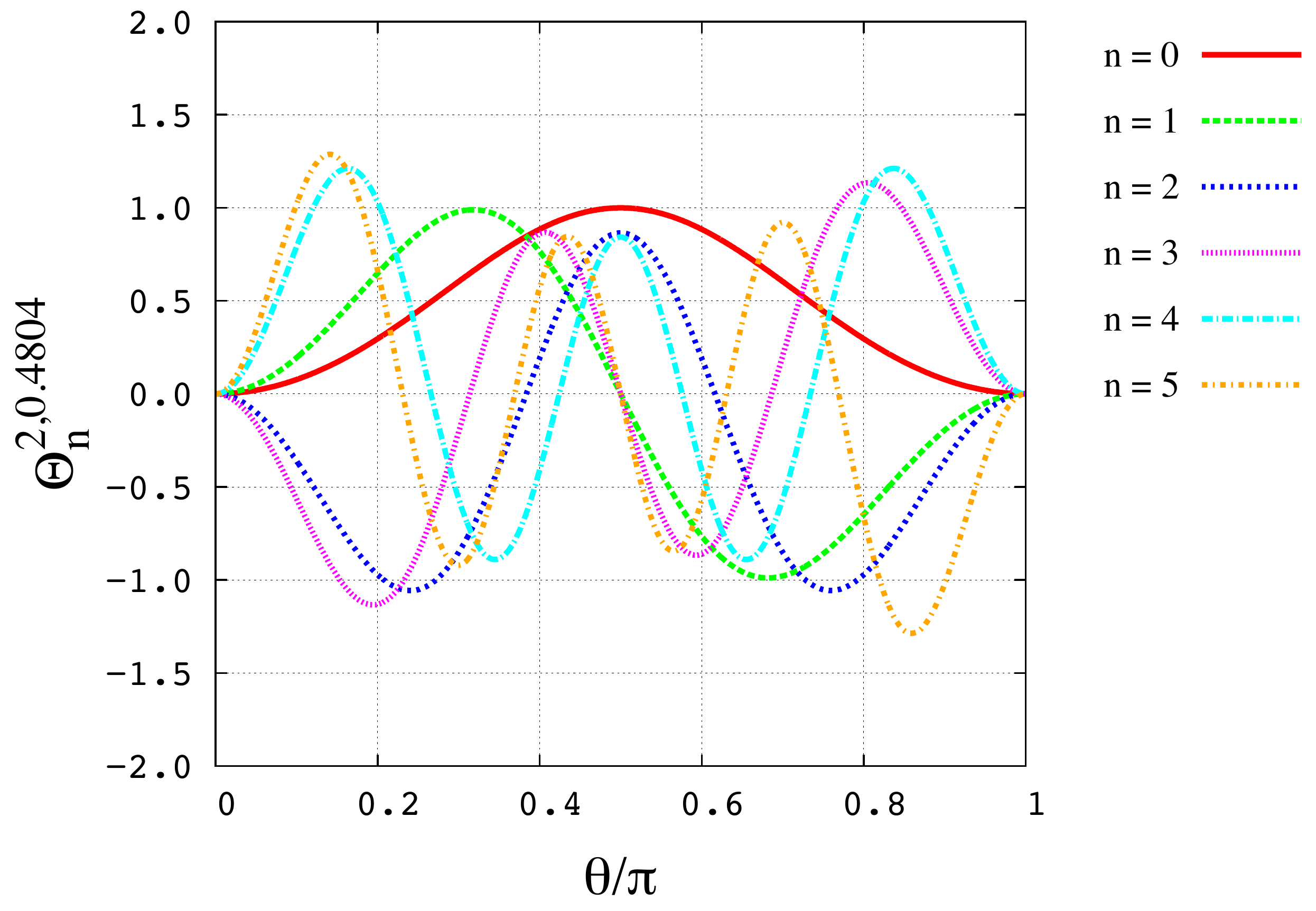}\\%
 \includegraphics[width=0.96\textwidth,clip]{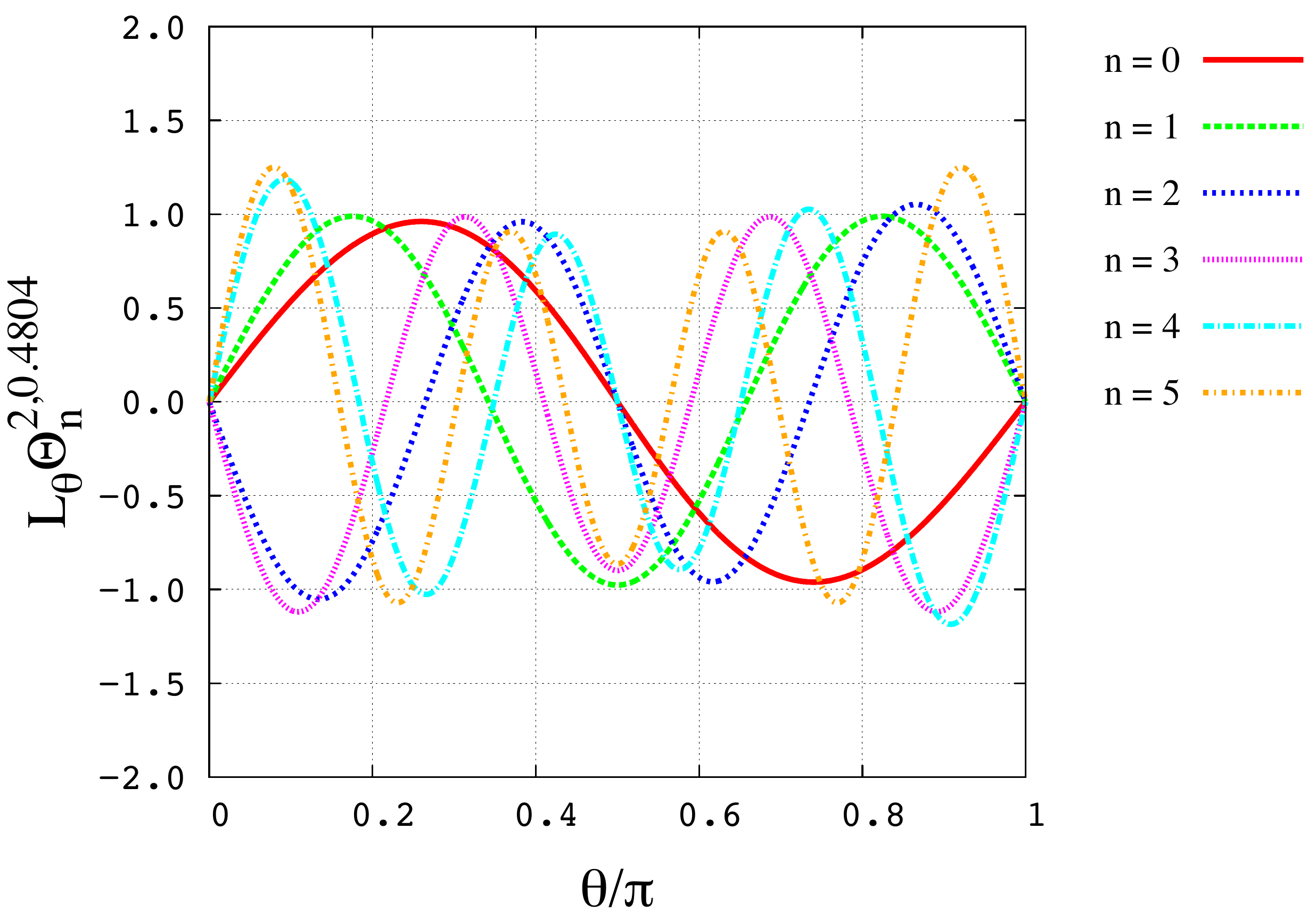}\\%
   \includegraphics[width=0.96\textwidth,clip]{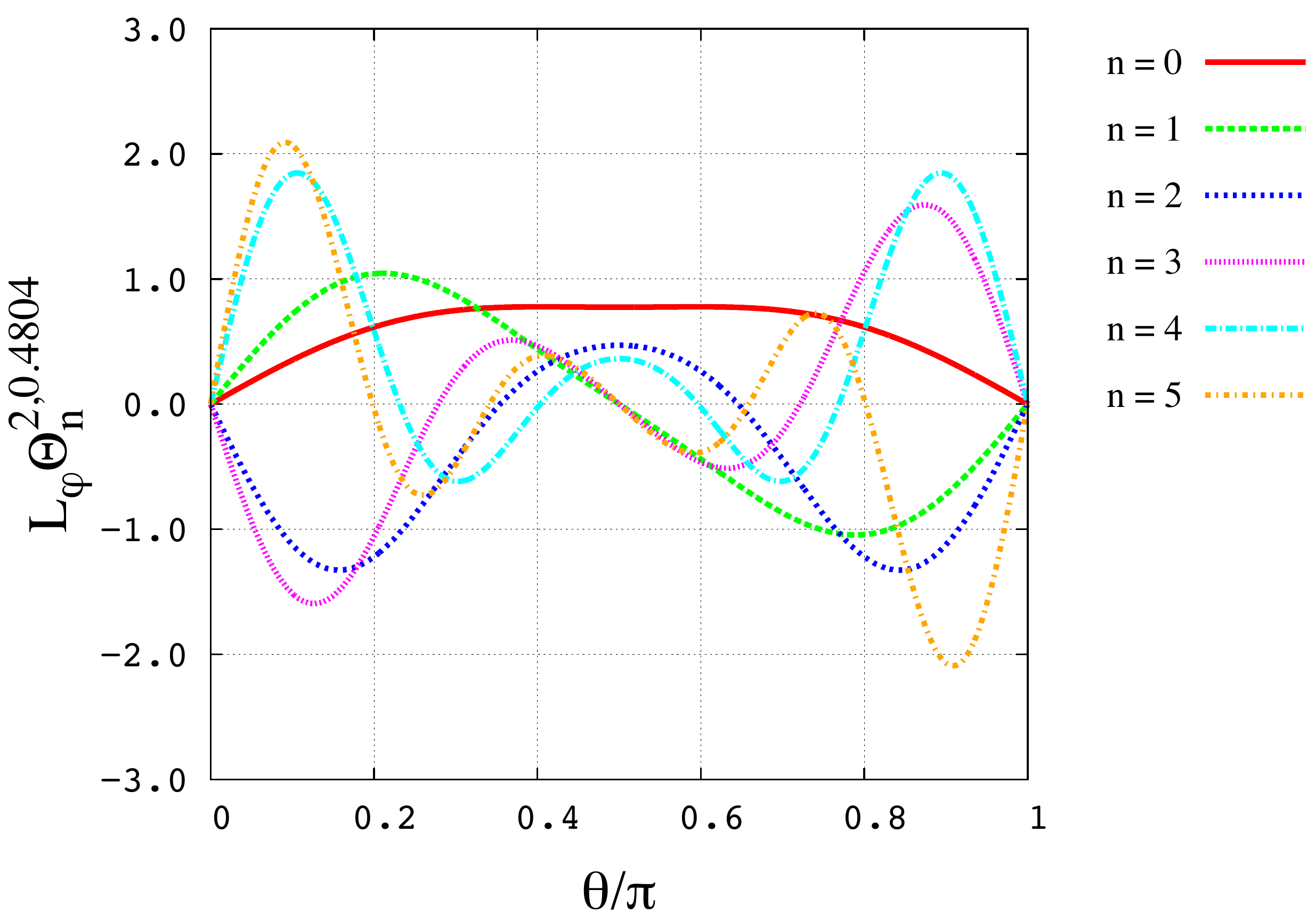}%
  \textsf{ \caption{\label{fig:fonctions_Hough_Venus}  First gravity modes ($ 0 \leq n \leq 5 $) of normalized Hough functions $ \Theta_{n}^{m,\nu} $, $ \mathcal{L}_\theta^{m,\nu}  \Theta_{n}^{m,\nu}  $ and $ \mathcal{L}_\varphi^{m,\nu} \Theta_{n}^{m,\nu} $ corresponding to the semidiurnal tide of Venus-like exoplanets ($ m = 2 $ and $ \nu = 0.4804 $). \jcor{We recall that $ \nu = 2 \Omega / \sigma $.} For more details about the behaviour of Hough functions, see Figs.~\ref{fig:fonctions_Hough} in Appendix~\ref{app:HF}. }}
\end{figure}

Under the traditional approximation, solutions of Eqs.~(\ref{eq_1}), (\ref{eq_2}) and (\ref{eq_3}) can be sought under the form of series of functions of separated coordinates,

\begin{equation}
\begin{array}{l}
     \displaystyle J^{m,\sigma} = \sum_n J_n^{m,\sigma} \left( r \right) \Theta_n^{m,\nu} \left( \theta \right), \\
     \displaystyle U^{m,\sigma} = \sum_n U_n^{m,\sigma} \left( r \right) \Theta_n^{m,\nu} \left( \theta \right), \\
     \displaystyle \xi_r^{m,\sigma} = \sum_n \xi_{r;n}^{m,\sigma} \left( r \right) \Theta_n^{m,\nu} \left( \theta \right), \\
     \displaystyle \xi_\theta^{m,\sigma} = \sum_n \xi_{\theta;n}^{m,\sigma} \left( r \right) \mathcal{L}_\theta^{m,\nu} \left[ \Theta_n^{m,\nu} \left( \theta \right) \right], \\
     \displaystyle \xi_\varphi^{m,\sigma} = \sum_n \xi_{\varphi;n}^{m,\sigma} \left( r \right)  \mathcal{L}_\varphi^{m,\nu} \left[ \Theta_n^{m,\nu} \left( \theta \right) \right], \\
      \displaystyle V_r^{m,\sigma} = \sum_n V_{r ; n}^{m,\sigma} \left( r \right) \Theta_n^{m,\nu} \left( \theta \right), \\
     \displaystyle V_\theta^{m,\sigma} = \sum_n V_{\theta;n}^{m,\sigma} \left( r \right)  \mathcal{L}_\theta^{m,\nu} \left[ \Theta_n^{m,\nu} \left( \theta \right) \right],  \\
     \displaystyle V_\varphi^{m,\sigma} = \sum_n V_{\varphi;n}^{m,\sigma} \left( r \right) \mathcal{L}_\varphi^{m,\nu} \left[ \Theta_n^{m,\nu} \left( \theta \right) \right],  \\
     \displaystyle \delta p^{m,\sigma} = \sum_n \delta p_n^{m,\sigma} \left( r \right) \Theta_n^{m,\nu} \left( \theta \right), \\
     \displaystyle \delta \rho^{m,\sigma} = \sum_n \delta \rho_n^{m,\sigma} \left( r \right) \Theta_n^{m,\nu} \left( \theta \right), \\
     \displaystyle \delta T^{m,\sigma} = \sum_n \delta T_n^{m,\sigma} \left( r \right) \Theta_n^{m,\nu} \left( \theta \right) \\
     \displaystyle y^{m,\sigma} = \sum_n y_n^{m,\sigma} \left( r \right) \Theta_n^{m,\nu} \left( \theta \right),
\end{array}
\label{series_Hough}
\end{equation}

where the functions \jcor{$ \Theta_n^{m,\nu} $} are the eigenvectors of the Laplace's operator, the corresponding eigenvalues being denoted \jcor{$ \Lambda_n^{m,\nu} $ (with $\nu = 2 \Omega / \sigma$)}. The Laplace's tidal equation, expressed as

\begin{equation}
\mathcal{L}^{m,\nu} \left( \Theta_n^{m,\nu}  \right) = - \Lambda_n^{m,\nu} \Theta_n^{m,\nu},
\label{eq_Laplace}
\end{equation}

describes the horizontal structure of the perturbation. The solutions of Eq.~(\ref{eq_Laplace}), \jcor{$ \left\{ \Theta_n^{m,\nu} \right\}_{n \in \mathbb{Z} } $} are called Hough functions \citep{Hough1898}\footnote{The properties of Hough functions and the method used to compute them in this work are detailed in Appendix~\ref{app:HF}.}. They form a complete set of continuous functions on the interval $ \theta \in \left[ 0 , \pi \right] $, parametrized by $ m $ and $ \nu $. They also verify the same boundary conditions as associated Legendre polynomials (denoted $ P_l^m $), which are solutions of Eq.~(\ref{eq_Laplace}) for $ \nu = 0 $ with $ \Lambda_n^{m,\nu} \left( \nu = 0 \right) = l \left( l + 1 \right) $. On Figs.~\ref{fig:fonctions_Hough_Terre} and \ref{fig:fonctions_Hough_Venus}, the functions \jcor{$ \Theta_n^{m,\nu} $, $ \mathcal{L}_\theta^{m,\nu} \Theta_n^{m,\nu} $ and $ \mathcal{L}_\varphi^{m,\nu} \Theta_n^{m,\nu} $} are plotted for the Earth and Venus in the case of the Solar semidiurnal tide defined by $ m = 2 $, $ \sigma = 2 \left( \Omega - n_{\rm orb} \right) $ \jcor{and $\nu = \Omega / \left( \Omega - n_{\rm orb} \right)$}, where $ n_{\rm orb} $ stands for the mean motion of the planet. In the case of the Earth, the set is mainly composed of gravity modes. Since the spin parameter is slightly greater than $ 1 $, Rossby modes exist but are trapped at the poles. In the case of Venus, Hough functions are composed of gravity modes only because $ \left| \nu \right| < 1 $. These functions are very similar to associated Legendre polynomials $ P_l^2 $, with $ l \geq 2 $. \\

The system composed of Eqs.~(\ref{eq_1}) to (\ref{eq_3}) can then be reduced to the couple of first-order differential equations

\begin{equation}
\left\{
\begin{array}{l}
    \displaystyle \dfrac{d}{d r} \left( r^2 \xi_{r;n}^{m,\sigma} \right) = A_1 r^2 \xi_{r;n}^{m,\sigma} + B_1 y_n^{m,\sigma} + C_1, \\[0.5cm] 
    \displaystyle \dfrac{d y_n^{m,\sigma}}{d r} = A_2 r^2 \xi_{r;n}^{m,\sigma} + B_2 y_n^{m,\sigma} + C_2,
\end{array}
\right.
\label{EDP12}
\end{equation}

with the coefficients

\begin{equation}
\begin{array}{l}
     \displaystyle A_1 = - \frac{1}{\rho_0} \dfrac{d \rho_0}{dr} -  \frac{i \sigma}{i \sigma + \sigma_0} \frac{N^2}{g}, \\[0.5cm]
      \displaystyle B_1 = \frac{\Lambda_n^{m,\nu}}{\sigma^2} - \frac{i \sigma + \Gamma_1 \sigma_0}{i \sigma + \sigma_0} \frac{r^2}{c_s^2}, \\[0.5cm]
       \displaystyle C_1 = \frac{\kappa r^2 \rho_0}{p_0 \left( i \sigma + \sigma_0 \right)} J_n^{m,\sigma} + \frac{\Lambda_n^{m,\nu}   }{\sigma^2} U_n^{m,\sigma}, 
\end{array}
\label{coefficients1}
\end{equation}

and

\begin{equation}
\begin{array}{l}
         \displaystyle A_2 = \frac{1}{r^2} \left( \sigma^2 - \frac{i \sigma}{i \sigma + \sigma_0} N^2 \right), \\[0.5cm]
     \displaystyle B_2 = - \frac{i \sigma + \Gamma_1 \sigma_0}{i \sigma + \sigma_0} \frac{g}{c_s^2} - \frac{1}{\rho_0} \dfrac{d \rho_0}{dr},  \\[0.5cm]
      \displaystyle C_2 = \frac{\kappa g \rho_0}{p_0 \left( i \sigma + \sigma_0 \right)} J_n^{m,\sigma} - \dfrac{d U_n^{m,\sigma}}{d r},
\end{array}
\label{coefficients2}
\end{equation}

%

where we have introduced the sound velocity

\begin{equation}
c_s = \sqrt{ \frac{\Gamma_1 p_0 }{\rho_0}  }.
\end{equation}

The system of Eq.~(\ref{EDP12}) can itself be reduced to a single equation in $ r^2 \xi_{r;n}^{m,\sigma} $ alone

\begin{equation}
\dfrac{d^2}{d r^2} \left( r^2 \xi_{r;n}^{m,\sigma} \right) + A \dfrac{d}{d r} \left( r^2 \xi_{r;n}^{m,\sigma} \right) + B r^2 \xi_{r;n}^{m,\sigma} = C,
\label{EDPglobal}
\end{equation}

with the set of coefficients

\begin{equation}
\left\{
\begin{array}{l}
   \displaystyle A = - A_1 - B_2 - \frac{1}{B_1} \dfrac{d B_1}{d r}, \\[0.5cm]
   \displaystyle B = A_1 \left( B_2 + \frac{1}{B_1} \dfrac{d B_1}{d r}  \right) -  \dfrac{d A_1}{dr} - B_1 A_2, \\[0.5cm]
   \displaystyle C = \dfrac{d C_1}{d r} + B_1 C_2 - C_1 \left(  \frac{1}{B_1} \dfrac{d B_1}{d r} + B_2   \right).
\end{array}
\right.
\label{coeffsABC}
\end{equation}

Eq.~(\ref{EDPglobal}) gives us the vertical structure of tidal waves generated in the fluid shell by both gravitational and thermal forcings. We have:

\begin{equation}
A =  \frac{2}{\rho_0} \dfrac{d \rho_0}{dr} + \frac{i \sigma}{i \sigma + \sigma_0} \frac{N^2}{g} + \frac{i \sigma + \Gamma_1 \sigma_0}{i \sigma + \sigma_0} \frac{g \rho_0}{\Gamma_1 P_0} - \frac{1}{B_1} \dfrac{d B_1 }{d r}.
\label{A_epais}
\end{equation}

The last term, $ \left( d B_1 / dr \right) / B_1 $, is associated with curvature and can be ignored in the context of the "shallow atmosphere" approximation, as discussed in the next section. It is expressed as (see Eq.~\ref{epsilon_sn})

\begin{equation}
\frac{1}{B_1} \dfrac{d B_1 }{d r} = -\frac{1}{1 - \varepsilon_{s;n}}\frac{d \varepsilon_{s;n}}{dr}.
\end{equation}

We now transform Eq.~(\ref{EDPglobal}) in a Schr\"odinger-like equation by introducing the radial function 
 
\begin{equation}
\mathcal{F} \left( r \right) =  \int_R^r A\left(u \right) du,
\end{equation}

and the variable change

\begin{equation}
r^2 \xi_{r;n}^{m,\sigma} = e^{- \frac{1}{2} \mathcal{F} \left( r \right)} \Psi_n^{m,\sigma},
\end{equation}

where the volume \jcor{$ \Psi_n^{m,\sigma} $} is the wave function of tidal waves associated with the mode of degree $ n $. Note that fixing $ \sigma_0 = 0 $ allows us to retrieve the usual variable change, $ r^2 \xi_{r;n}^{m,\sigma} = \rho_0^{-1/2} \Psi_n^{m,\sigma} $ \citep[][]{Press1981,Zahn1997,Mathis2008}. With the general variable change, Eq.~(\ref{EDPglobal}) becomes

\begin{equation}
\dfrac{d^2 \Psi_n^{m,\sigma}}{d r^2} + \left[ B - \frac{1}{2} \left( \dfrac{d A}{dr}  + \frac{A^2}{2} \right) \right] \Psi_n^{m,\sigma} = C e^{ \frac{1}{2} \mathcal{F} \left( r \right) }.
\label{Shrodinger_epais}
\end{equation}

which gives us the vertical profile of the perturbation\footnote{To solve the vertical structure equation numerically in any case, we use the method described in Appendix~\ref{app:num_scheme}.}. The form of this equation is very common in the literature because it describes the radial propagation of waves in a spherical shell. The typical wavenumber of these waves, denoted $ \hat{k}_n $, is given by 

\begin{equation}
  \hat{k}_n^2 = B - \frac{1}{2} \left( \dfrac{d A}{dr}  + \frac{A^2}{2} \right).
  \label{kv2_epais}
\end{equation}

We introduce the corresponding normalized length scale of variations of the mode of degree $ n $,

\begin{equation}
L_{\rm V;n} = \frac{2 \pi}{H_{\rm atm} \left| \hat{k}_n \right| }.
\label{Lvn}
\end{equation}  

This parameter is the typical heigh over which spatial variations of the mode can be observed.

\subsection{Polarization relations}

The perturbed quantities are readily deduced from $ \Psi $. Before going further, let us introduce the \mybf{Lamb frequency (the cutoff frequency of acoustic waves)} associated with the mode ($ \nu,m,n $),

\begin{equation}
\sigma_{s;n} \left( r \right)  =  \left( \Lambda_n^{m,\nu} \right)^{\frac{1}{2}}  \frac{c_s}{r},
\label{Lamb_frequency}
\end{equation}

and the ratio  

\begin{equation}
\varepsilon_{s,n} \left( r \right) = \frac{i \sigma + \Gamma_1 \sigma_0}{i \sigma + \sigma_0} \left( \frac{\sigma}{\sigma_{s;n} } \right)^2.
\label{epsilon_sn}
\end{equation}

The parameter $ \left| \varepsilon_{s,n} \right| $ measures the relevance of the anelastic approximation, which consists in neglecting all the terms that correspond to an acoustic perturbation in Eq.~(\ref{EDP12}). When $  \left| \varepsilon_{s,n} \right| \ll 1 $ (i.e. $ \left| \sigma \right| \ll \left| \sigma_{s;n} \right| $), this hypothesis can be assumed. Else, one should be particularly cautious when ignoring some terms. For instance, since the first gravity mode of the Earth's semidiurnal tide is characterized by $ \left| \sigma \right| \approx \left| \sigma_{s;0} \right| \approx  10^{-4} \ {\rm s^{-1}} $, the anelastic approximation cannot be done. In order to be able to study cases of the same kind, the polarization relations that follow will be given in their most general form. Hence, by substituting the series of Eq.~(\ref{series_Hough}) in Eqs.~(\ref{Vtheta}), (\ref{Vphi}), (\ref{xitheta}), (\ref{xiphi}), (\ref{eq_1}), (\ref{eq_2}) and (\ref{eq_3}), we obtain the radial profiles \jcor{$ \xi_{r;n}^{m,\sigma} $, $ \xi_{\theta;n}^{m,\sigma} $, $ \xi_{\varphi;n}^{m,\sigma} $, $ V_{r;n}^{m,\sigma} $, $ V_{\theta;n}^{m,\sigma} $, $ V_{\varphi;n}^{m,\sigma} $, $ \delta p_n^{m,\sigma} $, $ \delta \rho_n^{m,\sigma} $, $ \delta T_n^{m,\sigma} $ and $ y_n^{m,\sigma} $} as functions of \jcor{$ \Psi_n^{m,\sigma} $} and its first derivative. We obtain for the Lagrangian displacement

\begin{equation}
\xi_{r;n}^{m,\sigma} \left( r \right) = \frac{1}{r^2} e^{- \frac{1}{2} \mathcal{F} } \Psi_n^{m,\sigma},
\label{xirn}
\end{equation}

\begin{equation}
\begin{array}{ll}
\xi_{\theta;n}^{m,\sigma} \left( r \right) = & \displaystyle \frac{1}{\sigma^2 r \left( 1- \varepsilon_{s;n}  \right)}  \left\{ \frac{\sigma^2}{\Lambda_n^{m,\nu}} e^{- \frac{1}{2} \mathcal{F}} \left[   \dfrac{d \Psi_n^{m,\sigma}}{dr} + \mathcal{A}_n  \Psi_n^{m,\sigma}  \right] \right. \\ 
   & \left. \displaystyle - \frac{\Gamma_1 - 1}{i \sigma + \sigma_0} \frac{\sigma^2 }{\sigma_{s;n}^2} J_n^{m,\sigma} - \varepsilon_{s;n} U_n^{m,\sigma} \right\},
\end{array}
\end{equation}

\begin{equation}
\begin{array}{ll}
\xi_{\varphi;n}^{m,\sigma} \left( r \right) = & \displaystyle \frac{i}{\sigma^2 r \left( 1- \varepsilon_{s;n}  \right)}  \left\{ \frac{\sigma^2}{\Lambda_n^{m,\nu}} e^{- \frac{1}{2} \mathcal{F}} \left[   \dfrac{d \Psi_n^{m,\sigma}}{dr} + \mathcal{A}_n  \Psi_n^{m,\sigma}  \right] \right. \\ 
   & \left. \displaystyle - \frac{\Gamma_1 - 1}{i \sigma + \sigma_0} \frac{\sigma^2 }{\sigma_{s;n}^2} J_n^{m,\sigma} - \varepsilon_{s;n} U_n^{m,\sigma} \right\},
\end{array}
\end{equation}

for the velocity

\begin{equation}
V_{r;n}^{m,\sigma} \left( r \right) = \frac{i \sigma}{r^2} e^{- \frac{1}{2} \mathcal{F} } \Psi_n^{m,\sigma},
\end{equation}

\begin{equation}
\begin{array}{ll}
V_{\theta;n}^{m,\sigma} \left( r \right) = & \displaystyle \frac{i}{\sigma r \left( 1- \varepsilon_{s;n}  \right)}  \left\{ \frac{\sigma^2}{\Lambda_n^{m,\nu}} e^{- \frac{1}{2} \mathcal{F}} \left[   \dfrac{d \Psi_n^{m,\sigma}}{dr} + \mathcal{A}_n  \Psi_n^{m,\sigma}  \right] \right. \\ 
   & \left. \displaystyle - \frac{\Gamma_1 - 1}{i \sigma + \sigma_0} \frac{\sigma^2 }{\sigma_{s;n}^2} J_n^{m,\sigma} - \varepsilon_{s;n} U_n^{m,\sigma} \right\},
\end{array}
\end{equation}

\begin{equation}
\begin{array}{ll}
V_{\varphi;n}^{m,\sigma} \left( r \right) = & - \displaystyle \frac{1}{\sigma r \left( 1- \varepsilon_{s;n}  \right)}  \left\{ \frac{\sigma^2}{\Lambda_n} e^{- \frac{1}{2} \mathcal{F}} \left[   \dfrac{d \Psi_n^{m,\sigma}}{dr} + \mathcal{A}_n  \Psi_n^{m,\sigma}  \right] \right. \\ 
   & \left. \displaystyle - \frac{\Gamma_1 - 1}{i \sigma + \sigma_0} \frac{\sigma^2 }{\sigma_{s;n}^2} J_n^{m,\sigma} - \varepsilon_{s;n} U_n^{m,\sigma} \right\},
\end{array}
\end{equation}

and for scalar quantities

\begin{equation}
\begin{array}{ll}
\delta p_n^{m,\sigma} \left( r \right) = & \displaystyle  \frac{\rho_0}{1- \varepsilon_{s;n}}  \left\{ \frac{\sigma^2}{\Lambda_n^{m,\nu}} e^{- \frac{1}{2} \mathcal{F}} \left[   \dfrac{d \Psi_n^{m,\sigma}}{dr} + \mathcal{A}_n  \Psi_n^{m,\sigma}  \right]  \right.\\[0.3cm]
  & \displaystyle \left. - \frac{\Gamma_1 - 1}{i \sigma + \sigma_0} \frac{\sigma^2 }{\sigma_{s;n}^2} J_n^{m,\sigma} - U_n^{m,\sigma} \right\},
\end{array}
\end{equation}

\begin{equation}
\begin{array}{rl}
\delta \rho_n^{m,\sigma} \left( r \right) = &  \displaystyle \frac{i \sigma + \Gamma_1 \sigma_0}{ i \sigma + \sigma_0} \frac{\rho_0}{c_s^2 \left( 1 - \varepsilon_{s;n} \right) } \left\{ \frac{\sigma^2}{\Lambda_n^{m,\nu}} e^{- \frac{1}{2} \mathcal{F}} \left[ \dfrac{d \Psi_n^{m,\sigma}}{dr} \right. \right. \\
 & \left. \left. \displaystyle + \mathcal{B}_n \Psi_n^{m,\sigma} \right]  - \displaystyle \frac{\Gamma_1 - 1}{i \sigma + \Gamma_1 \sigma_0} J_n^{m,\sigma} - U_n^{m,\sigma} \right\},
 \end{array}
\end{equation}

\begin{equation}
\begin{array}{rl}
\delta T_n^{m,\sigma} \left( r \right) = & \displaystyle  \frac{i \sigma T_0 \left( \Gamma_1 - 1 \right) }{c_s^2 \left( i \sigma + \sigma_0 \right) \left( 1 - \varepsilon_{s;n}  \right)} \left\{  \frac{\sigma^2}{\Lambda_n^{m,\nu}} e^{- \frac{1}{2} \mathcal{F}  } \left[     \dfrac{d \Psi_n^{m,\sigma}}{dr}  \right.    \right. \\
&  \left. \left. \displaystyle + \mathcal{C}_n \Psi_n^{m,\sigma}  \right] + \displaystyle \frac{1}{i \sigma} \left( 1 -  \frac{\Gamma_1 \sigma^2 }{\sigma_{s;n}^2} \right) J_n^{m,\sigma} - U_n^{m,\sigma} \right\}.
\end{array}
\label{deltaTn}
\end{equation}

%

The coefficients in these expressions are given by

\begin{equation}
\begin{array}{lcl}
   \displaystyle \mathcal{A}_n \left( r \right) & = & \displaystyle \frac{1}{2} \frac{i \sigma}{i \sigma + \sigma_0} \left[  \frac{N^2}{g} - \frac{g}{c_s^2} \left( 1 - i \frac{\Gamma_1 \sigma_0}{\sigma} \right) \left( 1 + K_{\rm curv}  \right)  \right],  \\[0.5cm]
   \displaystyle  \mathcal{B}_n \left( r \right) & = & \displaystyle \frac{1}{2} \frac{i \sigma}{i \sigma + \sigma_0} \left[  \frac{N^2}{g} \left( \frac{2}{\varepsilon_{s;n}} - 1  \right) \right. \\[0.3cm]
    &  & \displaystyle \left. - \frac{g}{c_s^2} \left( 1 - i \frac{\Gamma_1 \sigma_0}{\sigma} \right) \left( 1 + K_{\rm curv} \right) \right], \\[0.5cm]
   \displaystyle \mathcal{C}_n \left( r \right) & = & \displaystyle \frac{N^2}{g \left( \Gamma_1 -  1 \right)} \left[ \frac{\left( \Gamma_1 + 1 \right) i \sigma + 2 \Gamma_1 \sigma_0 }{2 \left( i \sigma + \sigma_0  \right)}  - \frac{\sigma_{s;n}^2}{\sigma^2} \right]\\[0.3cm]
   &  & \displaystyle - \frac{g}{2 c_s^2} \frac{i \sigma }{i \sigma + \sigma_0} \left( 1 - i \frac{\Gamma_1 \sigma_0 }{\sigma} \right) \left( 1 + K_{\rm curv} \right),
\end{array}
\label{coeffsRP}
\end{equation}

where we have introduced the curvature term 

\begin{equation}
K_{\rm curv} = \frac{\sigma^2 c_s^2}{g \Lambda_n \left( 1 - \varepsilon_{s;n} \right)} \dfrac{d}{dr} \left( \frac{r^2}{c_s^2} \right),
\end{equation}

which will be ignored in the thin atmosphere approximation. These relations and Eq.~(\ref{Shrodinger_epais}) define a linear operator, denoted $ \Upsilon^\sigma $, which determines the vertical structure of the atmospheric tidal response. Let be $ Y_n^{m,\sigma} = \left( \Psi_n^{m,\sigma}, \boldsymbol{\xi}_n^{m,\sigma}, \textbf{V}_n^{m,\sigma}, \delta p_n^{m,\sigma}, \delta \rho_n^{m,\sigma} , \delta T_n^{m,\sigma}, y_n^{m,\sigma} \right) $ the output vector of the model and $ X_n^{m,\sigma} = \left(  U_n^{m,\sigma},J_n^{m,\sigma}  \right) $ the input vector. Eqs.~(\ref{xirn}) to (\ref{deltaTn}) can be reduced to 

\begin{equation}
Y_n^{m,\sigma} = \Upsilon_n^{m,\sigma}  X_n^{m,\sigma}.
\end{equation}

For any quantity $ f$, the component of $ \Upsilon_n^{m,\sigma} $ describing the vertical profile of $ f $ is denoted $ \Upsilon_{f ; n}^{m,\sigma} $ (for example $ \Upsilon_{p;n}^{m,\sigma} $ for the pressure, $ \Upsilon_{\rho ; n}^{m,\sigma} $ for the density, etc.).  \\

Polarization relations can be naturally divided into two distinct contributions. The first contribution, which is called ``\mybf{horizontal part}'' in this work, is a linear combination of the forcings \jcor{$ U_n^{m,\sigma} $ and $ J_n^{m,\sigma} $}. It corresponds to the \scor{component} of the response that does not depend on the \mybf{\scor{tidal} vertical displacement}. The length scale of \mybf{this horizontal component} is the typical thickness of the atmosphere ($ H_{\rm atm} $). The other part, expressed as a function of \jcor{$ \Psi_n^{m,\sigma} $} and called ``\mybf{vertical part}'', \scor{thus} results from \mybf{the \scor{tidal} fluid motion along the vertical direction}. In the following, we use the subscripts \mybf{$ _{\rm H} $ ant $ _{\rm V} $} to designate \mybf{the horizontal and vertical components}. \\

As may be noted, polarization relations point out the necessity of taking into account dissipative processes to study tidal regimes close to synchronization. Indeed, without dissipation \scor{($ \sigma_0 = 0 $)}, synchronization ($ \sigma = 0 $) is characterized by a singularity. The terms associated with the \mybf{horizontal component}, which are directly proportional to \jcor{$ U_n^{m,\sigma} $ and $ J_n^{m,\sigma} $}, tend to infinity at $ \sigma \rightarrow 0 $. \mybf{Terms in \jcor{$ \Psi_n^{m,\sigma} $ and  $ {\rm d} \Psi_n^{m,\sigma} / {\rm dr} $} behave similarly, as demonstrated in the next section.} These terms are highly oscillating along the vertical direction over the whole depth of the atmosphere in this frequency range because $ \left| \hat{k}_n \right| \rightarrow + \infty $ when $ \sigma \rightarrow 0 $. Hence, radiation regularizes the behaviour around $ \sigma = 0 $ and damps the vertical oscillations of waves associated with the \mybf{vertical component}. Diffusion acts in the same way by flattening the oscillations of smallest length scales ($ L_{\rm V;n} $) \citep[][]{Press1981,Zahn1997,Mathis2008}. We return to this point when we solve analytically the case of the isothermal atmosphere with constant profiles for the forcings (Sect. 3). \\

\begin{figure*}[htb]
\centering
{\includegraphics[width=0.9\textwidth]
{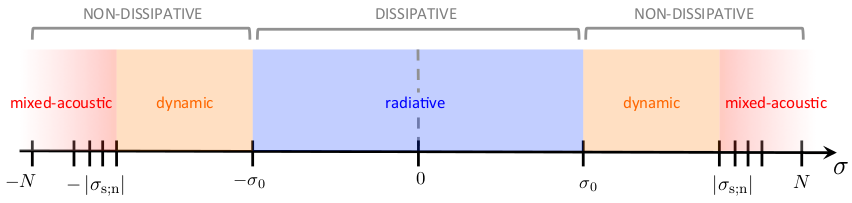}
\textsf{\caption{\label{fig:spectre} Frequency spectrum of the atmospheric tidal response. The identified regimes are given as functions of the tidal frequency ($ \sigma $). } }}
\end{figure*}

The whole spectrum of possible tidal regimes is represented on Fig.~\ref{fig:spectre}. \mybf{The radiative regime} (blue) characterizes a flow where thermal losses due to the linear Newtonian cooling (see the heat transport equation, Eq.~\ref{transport_chaleur_2}) predominate. In the dynamic regime (orange), dissipation can be neglected because the system is governed by \scor{the} Coriolis \scor{acceleration}. Finally, the mixed-acoustic regime (red) corresponds to high tidal frequencies comparable to the \mybf{Lamb} frequency, given by \mybf{Eq.~(\ref{Lamb_frequency})}. This regime marks the limit of the traditional approximation.


\subsection{Tidal potential, Love numbers and tidal torque}

The new mass distribution resulting from tidal waves generates a tidal gravitational potential $ \mathcal{U} $ which is a linear perturbation of the spherical gravitational potential of the planet. This atmospheric potential presents disymmetries with respect to the direction of the perturber. The tidal torque thus induced will affect the rotationnal dynamics of the planet over secular time scales and determine the possible equilibrium states of the spin, as demonstrated by \cite{CL01}. This torque is deduced from the Poisson's equation expressed in the co-rotating frame

\begin{equation}
\Delta  \mathcal{U} = - 4 \pi \mathscr{G} \delta \rho,
\end{equation} 

the notation $ \mathscr{G} $ designating the gravitational constant. Like $ \delta \rho $, the potential is expanded in series of functions of separated variables

\begin{equation}
\mathcal{U} = \sum_{m,\sigma,l} \mathcal{U}_l^{m,\sigma} \left( r \right) P_l^m \left( \cos \theta \right) e^{i \left( \sigma t + m \varphi  \right)}
\end{equation}

where the $ P_l^m $ (with $ l \in \mathbb{N} $ such as $ l \geq \left| m \right| $) are the normalized associated Legendre polynomials. Decomposing $ \delta \rho^{m,\sigma} $ on this basis, we get the equation describing the vertical structure of $ \mathcal{U}_l^{m,\sigma} $,

\begin{equation}
\dfrac{d}{dr} \left( r^2 \dfrac{d \mathcal{U}_l^{m,\sigma}}{dr}  \right) - l \left( l +1 \right) \mathcal{U}_l^{m,\sigma} = - 4 \pi \mathscr{G} r^2 \delta \rho_l^{m,\sigma} .
\label{Poisson2}
\end{equation}

For the upper boundary condition, one requires that the tidal potential shall remain bounded at $ r \rightarrow \infty $. At $ r = 0 $, we impose the same condition. Therefore, the solution of Eq.~(\ref{Poisson2}) is

\begin{equation}
\tilde{\mathcal{U}}_l^{m,\sigma} \left( r \right) =  \frac{4 \pi \mathscr{G}}{2 l + 1} \left[   r^l  \mathcal{F}_{l}^{m,\sigma} \left( r \right) + r^{- \left( l +1  \right)}  \mathcal{H}_{l}^{m,\sigma} \left( r \right)  \right],
\label{potentiel_atmo}
\end{equation}

where $ \mathcal{F}_{l}^{m,\sigma} $ and $ \mathcal{H}_{l}^{m,\sigma} $ are the functions

\begin{equation}
\left\{
\begin{array}{l}
     \displaystyle \mathcal{F}_{l}^{m,\sigma} \left( r \right) = \int_r^{+ \infty} u^{-l +1} \delta \rho_l^{m,\sigma} \left( u \right)  du, \\[0.5cm]
     \displaystyle \mathcal{H}_{l}^{m,\sigma} \left( r \right) = \int_R^{r} u^{l + 2} \delta \rho_l^{m,\sigma} \left( u \right) du.
\end{array}
\right.
\end{equation}

Love numbers are defined as the ratio between the gravitational potential due to the tidal response of the atmosphere and the forcing potential at the upper boundary of the layer. Let us denote $ R_{\rm atm} = R + H_{\rm atm} $ this upper boundary. Then, at the upper boundary, the atmospheric tidal potential given by Eq.~(\ref{potentiel_atmo}) is simply expressed:

\begin{equation}
\mathcal{U}_l^{\sigma , m}  \left(  R_{\rm atm} \right) = \frac{4 \pi \mathscr{G}}{\left(2 l + 1 \right) R_{\rm atm}^{l+1}}  \int_R^{R_{\rm atm}} r^{l + 2} \delta \rho_l^{m,\sigma} \left( r \right) {\rm dr}.
\end{equation}

Considering Eq.~(\ref{UJ_Legendre}) and using the linearity property of $ \Upsilon_{n}^{m,\sigma} $, the density component $ \delta \rho_l^{m,\sigma} $ can be expanded 

\begin{equation}
\delta \rho_l^{m,\sigma} = \sum_{n \in \mathbb{Z}} \sum_{k \geq \left| m \right|} c_{l,n,k}^{m,\sigma} \delta \rho_{n,k}^{m,\sigma} 
\end{equation}

with $ \delta \rho_{n,k}^{m,\sigma} = \Upsilon_{\rho ; n}^{m,\sigma} \left(  U_k^{m,\sigma} , J_k^{m,\sigma} \right) $ and the change-of-basis coefficients:

\begin{equation}
c_{l,n,k}^{m,\sigma} = \langle P_l^m , \Theta_n^{m,\sigma}  \rangle \langle P_k^m , \Theta_n^{m,\sigma}  \rangle,
\label{clnk}
\end{equation}

which allows us to write the atmospheric tidal potential

\begin{equation}
\mathcal{U}_l^{\sigma , m}  \left(  R_{\rm atm} \right) =  \frac{4 \pi \mathscr{G}}{\left(2 l + 1 \right) R_{\rm atm}^{l+1}} \sum_{n \in \mathbb{Z}} \sum_{k \geq \left| m \right|} c_{l,n,k}^{m,\sigma} \int_R^{R_{\rm atm}} r^{l + 2} \delta \rho_{n,k}^{m,\sigma} \left( r \right) dr.
\end{equation}

Let us treat the case of a simplified planet-star system, where the planet $ P $ circularly orbits around its host star, of mass $ M_S $. The semi-major axis, obliquity, and mean motion of the planet are denoted $ a $, $ \varepsilon $, and $ n_{\rm orb} $. In this case, the modes $ \left( \sigma , m \right) $ of the gravitational and thermal forcings applied on the atmosphere can be written (Appendix~\ref{app:thermal_forcing})

\begin{equation}
\left\{
\begin{array}{l}
     \displaystyle U^{m,\sigma}  =  \sum_{l \geq \left| m \right|} U_l^{m,\sigma} \left( r \right) P_l^m \left( \cos \theta \right) e^{i \left( \sigma t + m \varphi \right)} , \\[0.5cm]
     \displaystyle J^{m,\sigma}  =  \sum_{l \geq \left| m \right|} J_l^{m,\sigma} \left( r \right) P_l^m \left( \cos \theta \right) e^{i \left( \sigma t + m \varphi \right)}.
\end{array}
\right.
\label{UJ_Legendre}
\end{equation}

Hence the complex Love numbers can be written generically

\begin{equation}
k_{l}^{m,\sigma} = \left. \frac{\mathcal{U}_l^{\sigma , m} }{U_{l}^{m,\sigma} } \right|_{r = R_{\rm atm}},
\label{k}
\end{equation}

In celestial dynamics, the second-order Love number ($ l = 2 $) is commonly used to quantify the tidal response of a body. Therefore, we will illustrate the expression given by Eq.~(\ref{k}) by computing this coefficient for $ m = 2 $. For the sake of simplicity, we set \jcor{the obliquity} $ \varepsilon = 0 $ (the spin of the planet is supposed to be aligned with its orbital angular momentum). The tidal frequency is $ \sigma_{\rm SD} = 2 \left( \Omega - n_{\rm orb} \right) $ and the series of Eq.~(\ref{UJls}) are reduced to the terms characterized by $ \left( l , m , j , p , q \right) = \left( 2,2,2,0,0 \right) $\footnote{The parameters $ j $, $ p $ and $ q $ are the usual indexes of the Kaula's expansion of the tidal gravitational potential, recalled in Appendix~\ref{app:thermal_forcing}.}

\begin{equation}
\begin{array}{lcl}
    U_{2,2,2,0,0} = \displaystyle \frac{1}{2} \sqrt{\frac{3}{2}} \frac{\mathscr{G} M_S }{a^3} r^2 & \mbox{and} & J_{2,2,2,0,0} = \displaystyle \frac{1}{2} \sqrt{\frac{3}{2}} \frac{\mathcal{J}_2 \left( r \right)}{a^2}. 
\end{array}
\label{UJ22200}
\end{equation}

The tidal potential becomes

\begin{equation}
\left. \mathcal{U}_2^{2,\sigma_{\rm SD}} \right|_{R_{\rm atm} }  =  \frac{4 \pi \mathscr{G}}{5 R_{\rm atm}^{3}} \sum_{n \in \mathbb{Z}} \sum_{k \geq 2} c_{2,n,k}^{2,\sigma_{\rm SD}} \int_R^{R_{\rm atm}} r^{4} \delta \rho_{n,k}^{2,\sigma_{\rm SD}} \left( r \right) dr.
\label{tidalU2}
\end{equation}

If the star may be considered as a point-mass perturber ($ R \ll a $), then $ \mathcal{U}_2^{2,\sigma_{\rm SD}}  $ rapidly decays with $ k $. So, the terms of orders $ k > 2 $ are generated by the thermal forcing only. We also assume that the horizontal pattern of $ J $ is well represented by the function $ P_2^2 $, which allows us to ignore other terms. At the end, we consider the case $ \left| \nu \right| \sim 1 $ ( $ \left| n_{\rm orb} \right| \ll \left| \Omega \right| $\jcor{, see Eq.~\ref{nu}}), where $ c_{2,0,2}^{2,\sigma_{\rm SD}} \approx 1 $ and $  0 < c_{2,n,2}^{2,\sigma_{\rm SD}}  \ll 1 $ for $ n > 0 $ ( $ \Theta_0^{2,\sigma_{\rm SD}} \approx P_2^2 $). In this simplified framework, the second-order Love number can be approximated in magnitude order by

\begin{equation}
k_2^{2,\sigma_{\rm SD}} \sim \frac{4 \pi}{5}  \frac{a^3}{M_S R_{\rm atm}^5} \int_R^{R_{\rm atm}} r^4 \Upsilon_{\rho ; 2}^{2,\sigma_{\rm SD}} \left[ \frac{\mathscr{G} M_S}{a^3} r^2 , \frac{\mathcal{J}_2 \left( r \right)}{a^2}  \right] dr.
\label{k2_simple}
\end{equation}

To conclude this section, we compute the tidal torque exerted on the atmosphere with respect to the spin axis. The monochromatic torque associated with the frequency $ \sigma $ writes \citep{Zahn1966a}
 
 \begin{equation}
\mathcal{T}^{m,\sigma} = \Re \left\{ \frac{1}{2} \int_{ \mathscr{V} } \dfrac{\partial U^{m,\sigma}}{\partial \varphi} \left( \delta \rho^{m,\sigma} \right)^* {\rm d \mathscr{V} } \right\},
\label{torque_Zahn}
\end{equation}

where the notation $ \mathscr{V} $ designates the volume of the atmospheric shell. Denoting $ \Delta \varphi_l^{m,\sigma} = {\rm arg} \left( \delta \rho_l^{m,\sigma} \right) - {\rm arg} \left( U_l^{m,\sigma} \right) $ the phase lag between the forcing and the response, we obtain

\begin{equation}
\mathcal{T}^{m,\sigma} = \pi m \sum_{l \geq \left| m \right|}  \int_R^{R_{\rm atm}} r^2 \left| U_l^{m,\sigma} \right|  \left| \delta \rho_l^{m,\sigma} \right| \sin \left( \Delta \varphi_l^{m,\sigma} \right) dr,
\label{torque}
\end{equation} 

and the total tidal torque,

\begin{equation}
\mathcal{T} = \sum_{ \left( m , s \right) \in \mathbb{Z}^2 } \mathcal{T}^{\sigma_{m,s},m}.
\end{equation}

As previously done for the Love numbers, $ \mathcal{T}^{m,\sigma} $ is expanded using Hough modes

\begin{equation}
\mathcal{T}^{m,\sigma} = \pi m \sum_{l,n,k}  c_{l,n,k}^{m,\sigma} \int_R^{R_{\rm atm}} r^2 \left| U_l^{m,\sigma} \right|  \left| \delta \rho_{n,k}^{m,\sigma} \right| \sin \left( \Delta \varphi_{l,n,k}^{m,\sigma} \right) dr.
\label{torque_sum}
\end{equation}

with $ l \geq \left| m \right| $, $ n \in \mathbb{Z} $ and $ k \geq \left| m \right| $. Indeed, we recall here that $ \delta \rho_{n,k}^{\sigma , m} $ is the component of density variations represented by $ P_l^m $ caused by the component of the excitation represented by $ P_k^m $ and projected on $ \Theta_n $. Hence, the parameter $ \Delta \varphi_{l,n,k}^{m,\sigma} = {\rm arg} \left( \delta \rho_{n,k}^{m,\sigma} \right) - {\rm arg} \left( U_l^{m,\sigma} \right) $ is the phase difference between the component of degree $ \left( n , k \right) $ of the response and the component of degree $ l $ of the excitation. When the tidal gravitational potential is quadrupolar ($ R \ll a $), the terms of orders higher than $ l = 2 $ can be neglected. Denoting $ U_2^{m,\sigma} $  and $ J_2^{m,\sigma} $ the quantities $ U_l^{m,\sigma} $ and $ J_l^{m,\sigma} $ for $ l = 2 $ (which shall not be confused with the projections $ U_n^{m,\sigma} $ and $ J_n^{m,\sigma} $ of the forcings on the set of Hough functions), it follows

\begin{equation}
\mathcal{T}^{m,\sigma} = \pi m \sum_{n \in \mathbb{Z}} c_{2,n,2}^{m,\sigma}  \int_R^{R_{\rm atm}} r^2 \left| U_2^{m,\sigma} \right|  \left| \delta \rho_{n,2}^{m,\sigma} \right| \sin \left( \Delta \varphi_{2,n,2}^{m,\sigma} \right) {\rm dr}.
\label{torque_epais_n}
\end{equation}

If the forcing $ U_2^{\sigma , m} $ is a positive real function, then the previous expression becomes 

\begin{equation}
\mathcal{T}^{m,\sigma} = \pi m  \sum_{n \in \mathbb{Z}} c_{2,n,2}^{m,\sigma}   \int_R^{R_{\rm atm}} r^2 U_2^{m,\sigma}  \Im \left\{ \delta \rho_{n,2}^{m,\sigma} \right\}  {\rm dr}.
\label{couple_quadru}
\end{equation}

Substituting the polarization relation of density in Eq.~(\ref{couple_quadru}), we obtain the torque as a sum of two contributions

\begin{equation}
\mathcal{T}^{m,\sigma} = \mathcal{T}^{m,\sigma}_{\rm \mybf{H}} + \mathcal{T}^{m,\sigma}_{\rm \mybf{V}},
\end{equation}

where

\begin{equation}
\begin{array}{ll}
   \displaystyle \mathcal{T}^{m,\sigma}_{\rm \mybf{H}}  = &  \displaystyle - \pi m  \sum_{n \in \mathbb{Z}} c_{2,n,2}^{m,\sigma}  \Im \left\{   \int_R^{R_{\rm atm}}  \mathcal{H}_n \frac{\Lambda_n^{m,\nu}}{\sigma^2}  \left( U_2^{\sigma , m} \right)^2 dr  \right. \\[0.3cm]
    & \displaystyle \left. \displaystyle + \int_R^{R_{\rm atm}}  \mathcal{H}_n \frac{\Lambda_n^{m,\nu}}{\sigma^2} \frac{\Gamma_1 - 1}{i \sigma + \Gamma_1 \sigma_0} U_2^{m,\sigma} J_2^{m,\sigma} \right\},
\end{array}
\end{equation}


\begin{equation}
\begin{array}{ll}
 \mathcal{T}^{m,\sigma}_{\rm \mybf{V}}  = &  \displaystyle \pi m  \sum_{n \in \mathbb{Z}} c_{2,n,2}^{m,\sigma}  \Im \left\{   \int_R^{R_{\rm atm}}  \mathcal{H}_n  e^{- \frac{\mathcal{F}}{2} } \left[ \dfrac{d \Psi_n^{m,\sigma}}{dr} + \mathcal{B}_n \Psi_n^{m,\sigma} \right]  \right. \\[0.3cm]
 & \left. \displaystyle \times U_2^{m,\sigma} dr \right\}
 \end{array}
\end{equation}


with $ \Psi_n^{m,\sigma} = \Psi_n^{m,\sigma} \left( U_2^{m,\sigma} , J_2^{m,\sigma} \right) $ and 

\begin{equation}
\mathcal{H}_n = \frac{i \sigma + \Gamma_1 \sigma_0}{ i \sigma + \sigma_0} \frac{\rho_0 r^2 \sigma^2}{c_s^2 \Lambda_n^{m,\nu} \left( 1 - \varepsilon_{s;n} \right) }.
\end{equation}

Assuming $ \left| \varepsilon_{s;n} \right| \ll 1 $ and noticing that $ \sum_{n \in \mathbb{Z}} c_{2,n,2}^{m,\sigma} = 1 $, we establish that $ \mathcal{T}^{m,\sigma}_{\rm \mybf{H}}  $ does not depend on the eigenvalues of the Laplace's tidal equation and may be written

\begin{equation}
\begin{array}{lcl}
   \mathcal{T}^{m,\sigma}_{\rm \mybf{H}} & = & \displaystyle - \pi m  \Im \left\{ \int_R^{R_{\rm atm}}  \frac{\rho_0 r^2}{c_s^2} \frac{\Gamma_1 - 1}{i \sigma + \Gamma_1 \sigma_0}   U_2^{m,\sigma} J_2^{m,\sigma} dr \right\}\\[0.3cm]
    & & \displaystyle - \pi m  \Im \left\{ \int_R^{R_{\rm atm}} \frac{\rho_0 r^2}{c_s^2} \frac{i \sigma + \Gamma_1 \sigma_0}{ i \sigma + \sigma_0} \left( U_2^{m,\sigma} \right)^2 dr \right\}.
\end{array}
\end{equation}

 The torque induced by this component is expanded in series of Hough functions and characterized by the solutions $ \Psi_n^{m,\sigma} $ of the vertical structure equation. The typical scales of the \mybf{vertical component} are given by the vertical wavenumbers (see Eq.~\ref{kv2_epais}) and can be very small compared to the scale of the \mybf{horizontal ones}. For example, in the vicinity of synchronization, $ \hat{k}_n \, \propto \, 1/\sigma $ if the radiative damping is ignored (CL70), which means that the characteristic scales of variations rapidly decay when $ \sigma \rightarrow 0 $. Finally, we may notice that the series of modes in Eq.~(\ref{torque_epais_n}) can often be reduced to one dominating term. For instance, in the case of the Earth's semi-diurnal tide, $ \Theta_0^{\sigma,2} \sim P_2^2 $ and the expression above (Eq.~\ref{torque_epais_n}) can be simplified

\begin{equation}
\mathcal{T}^{m,\sigma} \sim 2 \pi  \int_R^{R_{\rm atm}} r^2 \left| U_2^{\sigma,2} \right|  \left| \delta \rho_{0,2}^{\sigma,2} \right| \sin \left( \Delta \varphi_{2,0,2}^{\sigma,2} \right) dr.
\label{torque2_simple}
\end{equation}

\section{Tides in a thin \mybf{stably stratified} isothermal atmosphere}
\label{sec:atmos_isotT}

We establish in this section the analytical equations that describe the tidal perturbation in the case where $ H_{\rm atm} \ll R $. This theoretical approach has been exhaustively formalized for the Earth in ``Atmospheric tides'' (CL70) and corresponds to the case of thin atmospheres. The formalism developed in the previous section allows us to go beyond this pioneer work, and more particularly to study regimes near synchronization thanks to the inclusion of radiative losses.

\subsection{Equilibrium state}

For a thin atmosphere ($ r \approx R $), the equilibrium structure can be determined analytically. We introduce the altitude $ z $, such as $ r = R + z $, which is here a more appropriate coordinate than $ r $. \\

\jcor{The} fluid is supposed to be at hydrostatic equilibrium, that leads to

\begin{equation}
\frac{1}{\rho} \dfrac{\partial p}{ \partial z} = - g,
\label{equilibre_hydro}
\end{equation} 

The local acceleration in $ \mathscr{R}_{\rm E;T} $, denoted $ \textbf{a}_{\mathscr{R}_{\rm E;T}} $, is not equal to $ \textbf{g} $ because of the rotating motion. The total acceleration can be written

\begin{equation}
 \textbf{a}_{\mathscr{R}_{\rm E;T}} = - g \left(z \right) \textbf{e}_r + \textbf{a}_c,
\label{acceleration}
\end{equation}

where $ \textbf{a}_c $ is the centrifugal acceleration,

\begin{equation}
\textbf{a}_c = r \sin \theta \ \Omega^2 \left(  \sin \theta \ \textbf{e}_r + \cos \theta \ \textbf{e}_\theta  \right).
\end{equation}


Here, Coriolis acceleration does not intervene because the global circulation of the atmosphere is ignored. \jcor{The centrifugal acceleration (Eq.~\ref{acceleration}) can be neglected with respect to the gravity if the rotation rate of the planet satisfies the condition $ \Omega \ll \Omega_{\rm c} $, where $ \Omega_{\rm c} = \sqrt{g / R} $ is the critical Keplerian angular velocity. In the case of an Earth-like planet, with $ g \sim 10 \ {\rm m.s^{-2}} $ and $ R \sim 6.0 \times 10^{3} \ {\rm km} $, $ \Omega_{\rm c} \sim 20 $ rotations per day.} From Eq.~(\ref{gaz_parfait}) and (\ref{equilibre_hydro}), a typical depth can be identified,

\begin{equation}
H = \frac{p_0}{g \rho_0} = \frac{\mathscr{R}_s T_0}{g},
\end{equation}

which is the local characteristic pressure height scale. The parameter $ H $ represents the vertical scale of variations of basic fields. It is of the same magnitude order as $ H_{\rm atm} $ and allows us to introduce the reduced altitude

\begin{equation}
  x = \int_0^z \dfrac{dz}{H \left( z \right) },
\end{equation}

which will be used in the following. Assuming that $ p_0 $, $ \rho_0 $ and $ T_0 $ only depends on $ x $, we deduce their expressions from Eqs.~(\ref{gaz_parfait}) and (\ref{equilibre_hydro})

\begin{equation}
\begin{array}{ccc}
 p_0 \left( x \right) = p_0 \left( 0 \right) e^{-x}, & \rho_0 \left( x \right) = \displaystyle \frac{p_0 \left( x \right)}{g H \left( x \right)}, & T_0 \left( x \right) = \displaystyle \frac{g  H \left( x \right)}{\mathscr{R}_s}.
\end{array}
\end{equation}

In the fluid shell, the gravitational acceleration does not vary very much

\begin{equation}
\left|  \frac{g \left( x \right)  -  g \left( 0 \right)  }{ g \left( 0 \right)  }  \right| \sim 2 \frac{H}{R} x.
\end{equation}

So, $ g $ can be considered as a constant. Assuming that $ H $ does not vary with the altitude either, we obtain a constant profile of the temperature and exponentially decaying profiles of the pressure and density:

\begin{equation}
\begin{array}{ccc}
 p_0 \left( x \right) = p_0 \left( 0 \right) e^{-x}, & \rho_0 \left( x \right) = \displaystyle \frac{p_0 \left( 0 \right)}{g H } e^{-x}, & T_0 = \displaystyle \frac{g H}{\mathscr{R}_s}.
\end{array}
\end{equation}

In this case, which corresponds to an isothermal atmosphere, $ 95 \ \% $ of the mass of the gas is contained within the interval $ x \in \left[ 0, 3 \right] $. Therefore, we can write: $ H_{\rm atm} \approx 3 H $. For the Earth, $ H \sim 8 \ {\rm km} $ (CL70) and $ H_{\rm atm} \sim 24 \ {\rm km} $.

\subsection{Wave equation and polarization relations}

The constant $ H $ hypothesis is very useful to simplify the expressions of Sect.~2. Indeed, with this approximation, the typical frequencies of the system do not vary with the radial coordinate anymore. The Brunt-V\"ais\"al\"a frequency simply writes

\begin{equation}
N = \sqrt{\frac{\kappa g}{H}},
\label{N}
\end{equation} 

and the acoustic frequency associated with the mode $ \left( \nu,m,n \right) $,  

\begin{equation}
\sigma_{s;n} = \left(\Lambda_n^{m,\nu} \right)^{\frac{1}{2}} \frac{c_s}{R},
\label{Lamb_mince}
\end{equation}

with $ c_s = \Gamma_1 g H $. Moreover, the horizontal structure of tidal waves does not change because the Laplace's tidal equation is not modified. So, the vertical structure equation and the radial polarization relations only are affected by the constant $ H $ hypothesis. The coefficients defined by Eq.~(\ref{coeffsABC}) become

\begin{equation}
\begin{array}{ll}
    \displaystyle A = & \displaystyle - \frac{1}{H}, \\[0.5cm]
     \displaystyle B = & \displaystyle \frac{\Lambda_n^{m,\nu}}{R^2} \left[ \frac{i \sigma}{i \sigma + \sigma_0} \frac{N^2}{\sigma^2} + \varepsilon_{s;n} - 1  \right],\\[0.5cm]
    \displaystyle C = & \displaystyle \frac{\Lambda_n^{m,\nu}}{H \sigma^2} \left\{   \frac{\kappa}{i \sigma + \sigma_0} \left[ \frac{h_n}{H} \dfrac{d J_n^{m,\sigma}}{dx} + \mathcal{K}_n J_n^{m,\sigma} - i \sigma U_n^{m,\sigma}  \right] \right. \\[0.3cm]
    & \left. \displaystyle + \varepsilon_{s;n} \dfrac{d U_n^{m,\sigma}}{dx}  \right\}. 
\end{array}
\label{coeffsABC_mince}
\end{equation}

where

\begin{equation}
\mathcal{K}_n = 1 - \frac{h_n}{H}.
\label{coeffK}
\end{equation}

The parameter $ h_n $, which writes

\begin{equation}
 h_n = \frac{R^2 \sigma^2}{g \Lambda_n^{m,\nu}},
 \end{equation}
 
is usually called the ``equivalent depth'' in literature \citep[e.g.][]{Taylor1936,ChapmanLindzen1970} and represents a typical length scale associated with the mode of degree $ n $. Note that the curvature term in Eq.~(\ref{A_epais}) vanishes in the shallow atmosphere approximation, because the coefficient $ B_1 $ (Eq.~\ref{coefficients1}) is a constant in this case. This allows us to simplify the vertical structure equation (Eq.~\ref{Shrodinger_epais}). Contrary to the vertical wave number in the thick shell (Eq.~\ref{kv2_epais}), the vertical wavenumber $ \hat{k}_n $ does not vary with $ x $ in this case. It is expressed

\begin{equation}
\hat{k}_n^2 = \frac{1}{4} \left[  \varsigma_n \left( \frac{i \sigma}{i \sigma + \sigma_0} \frac{N^2}{\sigma^2} + \varepsilon_{s;n} - 1  \right) - 1  \right],
\label{kv2_1}
\end{equation}

with the scale ratio $ \varsigma_n = 4 H^2 \Lambda_n^{m,\nu} / R^2 $. It can also be written

\begin{equation}
\hat{k}_n^2 = \frac{1}{4} \left[  \frac{i \sigma}{i \sigma + \sigma_0} \frac{4 \kappa H}{h_n}  - 1  + \varsigma_n \left( \varepsilon_{s;n} - 1  \right)  \right].
\label{kv2_2}
\end{equation}

 The term $ \varsigma_n \left( \varepsilon_{s;n} - 1  \right) $ comes from the radial acceleration in the Navier-Stokes equation (Eq.~\ref{NS3}) and is negligible for usual values of $ \Lambda_n^{m,\nu} $ ($ \left| \varepsilon_{s;n} \right| \ll1 $ and $ \varsigma_n \, \propto \, H^2 / R^2 \ll 1$). Therefore, by considering the case where $ \sigma_0 \ll \left| \sigma \right| $, we recover the wave number obtained by CL70 for the Earth semidiurnal tide. In the general case, the expression of Eq.~(\ref{kv2_2}) can be approximated by

\begin{equation}
\hat{k}_n^2 \approx \frac{1}{4} \left[ \frac{i \sigma}{i \sigma + \sigma_0} \frac{4 \kappa H}{h_n} - 1 \right],
\label{kv2_3}
\end{equation}

which gives for the real and imaginary parts, introducing the characteristic depth $ h_c = 4 \kappa H $ and fixing $ \Re \left\{ \hat{k}_n \right\} > 0 $, 

\begin{equation}
\left\{
\begin{array}{l}
    \displaystyle \Re \left\{ \hat{k}_n \right\}  =  \frac{1}{2 \sqrt{2}} \left[  \frac{\sigma^2}{\sigma^2 + \sigma_0^2}  \frac{h_c }{h_n} - 1 + \sqrt{\Delta} \right]^{1/2}, \\[0.5cm]
    \displaystyle \Im \left\{ \hat{k}_n \right\}  =  \frac{1}{2 \sqrt{2}} \frac{\sigma_0 \sigma}{\sigma^2 + \sigma_0^2 } \frac{h_c}{h_n} \left[  \frac{\sigma^2}{\sigma^2 + \sigma_0^2}  \frac{h_c }{h_n} - 1 + \sqrt{\Delta} \right]^{-1/2},
\end{array}
\right.
\label{kv_reduit}
\end{equation}

with 

\begin{equation}
\Delta = \left( \frac{\sigma^2}{\sigma^2 + \sigma_0^2} \frac{h_c}{h_n} - 1 \right)^2 + \frac{\sigma_0^2 \sigma^2}{ \left( \sigma^2 +  \sigma_0^2 \right)^2 } \frac{h_c^2}{h_n^2}.
\end{equation}

\begin{figure*}[htb]
 \includegraphics[width=0.475\textwidth,clip]{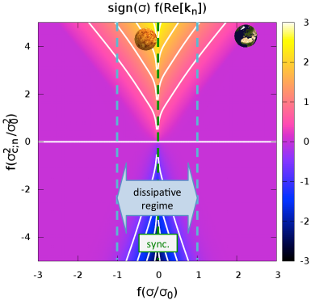}\hspace{5mm}
  \includegraphics[width=0.475\textwidth,clip]{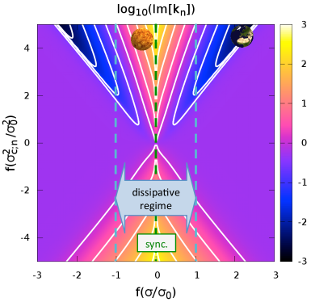}  
  \textsf{ \caption{\label{fig:kv_map}  \mybf{Real and imaginary part of the vertical wavenumber as functions of the tidal ($ \sigma $) and critical ($ \sigma_{\rm c;n} $) frequencies. {\it Left:} $ {\rm sign} \left( \sigma \right) f \left( \Re \left\{ \hat{k}_n \right\} \right) $}, with $ \hat{k}_n $ defined in Eq.~\ref{kv2_3}, as a function of \mybf{$ f \left( \sigma / \sigma_0 \right) $ (horizontal axis) and $ f \left( \sigma_{\rm c;n}^2 / \sigma_0^2  \right) $ (vertical axis) where  $ f $ is the function defined by $ f \left( X \right) = {\rm sign}\left( X \right) \log \left( 1 + \left| X \right| \right) $}. The color of the map, denoted $ c $, is defined by $ {\rm sign} \left( \sigma \right) f \left( \Re \left\{ \hat{k}_n \right\} \right) $. Hence, the purple region stands for \mybf{slowly oscillating waves}. The luminous (dark) regions correspond to \mybf{strongly oscillating waves} propagating \mybf{downward (upward)}. \mybf{ {\it Right:} $ \log \left( \Im \left\{ \hat{k}_n \right\}  \right) $ as a function of the same parameters. The luminous (dark) regions correspond to \mybf{strongly (weakly)} evanescent waves.} The positions of the planets (Earth $ _\Earth $, Venus $ _\Venus $) on the map are determined by the gravity mode of degree $ 0 $ in the case of the semidiurnal tide ($ \Lambda_{\Earth;0} = 11.1596 $, $ \Lambda_{\Venus ; 0} = 7.1485 $). The tidal frequencies of the planets are given by \mybf{$ \sigma_{\Earth} / \sigma_0 \approx 194 $ and $ \sigma_{\Venus} / \sigma_0 = 0.43 $, with $ \sigma_0 = 7.5 \times 10^{-7} \ {\rm s^{-1}} $}.  }}
\end{figure*}

The behaviour of the vertical wave number defines the possible regimes of the perturbation. These regimes are represented by the map of Figure~\ref{fig:kv_map}, which shows \mybf{the real and imaginary part of the vertical wavenumber as color functions} of the tidal frequency \mybf{and critical frequency, $ \sigma_{\rm c;n} $, given by $ \sigma_{\rm c;n}^2 = 4 \kappa H g \Lambda_n^{m,\nu} / r^2 $} \scor{(the frequency $\sigma_{\rm c ; n}$ is proportional to the Lamb frequency defined by Eq.~\ref{Lamb_mince})}. Propagative modes are characterized by $ \left| \Im \left\{ \hat{k}_n \right\}  \right| \ll 1  $ (\mybf{right panel, dark regions}), evanescent modes by $ \left| \Im \left\{ \hat{k}_n \right\}  \right| \gg 1 $ \mybf{(yellow regions)}. The transition between the radiative regime and the dynamic regime corresponds to $ \left| \sigma \right| \approx \sigma_0 $.  \\ 

In the case of the thin isothermal atmosphere, the equation giving the vertical profiles (Eq.~\ref{Shrodinger_epais}) writes

\begin{equation}
\dfrac{d^2 \Psi_n^{m,\sigma}}{dx^2} + \hat{k}_n^2 \Psi_n^{m,\sigma} =  H^2 e^{-x/2} C \left( x \right).
\label{Shrodinger_mince}
\end{equation}

where $ C $ is given by Eq.~(\ref{coeffsABC_mince}). Moreover, if the condition

\begin{equation}
\left|  \frac{U_n^{m,\sigma}}{J_n^{m,\sigma}} \right| \ll  \left|  \frac{H - h_n}{\sigma H}   \right|,
\label{condition_therm}
\end{equation}

is satisfied, the forcing in the right-hand member is dominated by thermal tides and the contribution of the gravitational tidal potential can be ignored in Eq.~(\ref{Shrodinger_mince}), which is rewritten:

\begin{equation}
\begin{array}{ll}
\displaystyle \dfrac{d^2 \Psi_n^{m,\sigma}}{dx^2} + \frac{1}{4} \left[  \frac{i \sigma}{i \sigma + \sigma_0} \frac{h_c}{h_n}  - 1  \right] \Psi_n^{m,\sigma} = & \! \! \! \!  \displaystyle \frac{\kappa H \Lambda_n^{m,\nu} e^{- \frac{x}{2}}}{i \sigma + \sigma_0} \left[  \frac{h_n}{H} \dfrac{d J_n^{m,\sigma}}{dx} \right. \\[0.3cm] 
 & \! \! \! \! \displaystyle \left. + \mathcal{K}_n J_n^{m,\sigma}  \right]  .
\end{array}
\end{equation}


The polarization relations of the thin atmospheric shell are deduced from Eq.~(\ref{xirn}) to (\ref{deltaTn})

\begin{equation}
\xi_{r;n}^{m,\sigma} = \frac{1}{R^2} e^{\frac{x}{2}  } \Psi_n^{m,\sigma},
\label{xir_mince}
\end{equation}

\begin{equation}
\begin{array}{ll}
\xi_{\theta;n}^{m,\sigma} = & \displaystyle \frac{1}{\sigma^2 R \left( 1- \varepsilon_{s;n}  \right)}  \left\{ \frac{\sigma^2}{H \Lambda_n^{m,\nu}} e^{ \frac{x}{2}} \left[   \dfrac{d \Psi_n^{m,\sigma}}{dx} + \mathcal{A}_n  \Psi_n^{m,\sigma}  \right] \right. \\[0.3cm]
 & \displaystyle \left. -  \mathcal{D}_n J_n^{m,\sigma} - \varepsilon_{s;n} U_n^{m,\sigma} \right\} ,
\end{array}
\end{equation}

\begin{equation}
\begin{array}{ll}
\xi_{\varphi;n} = & \displaystyle \frac{i}{\sigma^2 R \left( 1- \varepsilon_{s;n}  \right)}  \left\{ \frac{\sigma^2}{H \Lambda_n^{m,\nu}} e^{ \frac{x}{2} } \left[   \dfrac{d \Psi_n^{m,\sigma}}{dx} + \mathcal{A}_n  \Psi_n^{m,\sigma}  \right] \right. \\[0.3cm]
   & \left. \displaystyle - \mathcal{D}_n J_n^{m,\sigma} - \varepsilon_{s;n} U_n^{m,\sigma} \right\},
\end{array}
\end{equation}

\begin{equation}
V_{r;n}^{m,\sigma} = \frac{i \sigma}{R^2} e^{ \frac{x}{2} } \Psi_n^{m,\sigma},
\end{equation}

\begin{equation}
\begin{array}{ll}
V_{\theta;n}^{m,\sigma} = & \displaystyle \frac{i}{\sigma R \left( 1- \varepsilon_{s;n}  \right)}  \left\{ \frac{\sigma^2}{H \Lambda_n^{m,\nu}} e^{ \frac{x}{2} } \left[   \dfrac{d \Psi_n^{m,\sigma}}{dx} + \mathcal{A}_n  \Psi_n^{m,\sigma}  \right] \right. \\[0.3cm]
   & \left. \displaystyle -  \mathcal{D}_n J_n^{m,\sigma} - \varepsilon_{s;n} U_n^{m,\sigma} \right\},
\end{array}
\end{equation}

\begin{equation}
\begin{array}{ll}
V_{\varphi;n}^{m,\sigma} = & - \displaystyle \frac{1}{\sigma R \left( 1- \varepsilon_{s;n}  \right)}  \left\{ \frac{\sigma^2}{H \Lambda_n^{m,\nu}} e^{ \frac{x}{2} } \left[   \dfrac{d \Psi_n^{m,\sigma}}{dx} + \mathcal{A}_n  \Psi_n^{m,\sigma}  \right] \right. \\[0.3cm]
   & \left. \displaystyle -  \mathcal{D}_n J_n^{m,\sigma} - \varepsilon_{s;n} U_n^{m,\sigma} \right\},
\end{array}
\end{equation}

\begin{equation}
\begin{array}{ll}
   \delta p_n^{m,\sigma} = & \displaystyle  \frac{\rho_0}{1- \varepsilon_{s;n}}  \left\{ \frac{\sigma^2}{H \Lambda_n^{m,\nu}} e^{ \frac{x}{2}} \left[   \dfrac{d \Psi_n^{m,\sigma}}{dx} + \mathcal{A}_n  \Psi_n^{m,\sigma}  \right]  - \mathcal{D}_n J_n^{m,\sigma} \right. \\[0.3cm] 
   & \displaystyle \left.- U_n^{m,\sigma} \right\},
\end{array}
 \label{deltap_mince}
\end{equation}


\begin{equation}
\begin{array}{rl}
\delta \rho_n^{m,\sigma} = &  \displaystyle \frac{i \sigma + \Gamma_1 \sigma_0}{ i \sigma + \sigma_0} \frac{\rho_0}{c_s^2 \left( 1 - \varepsilon_{s;n} \right) } \left\{ \frac{\sigma^2}{H \Lambda_n^{m,\nu}} e^{ \frac{x}{2}} \left[ \dfrac{d \Psi_n^{m,\sigma}}{dx}  \right. \right. \\[0.3cm]
 & \left. \left. \displaystyle + \mathcal{B}_n \Psi_n^{m,\sigma} \right] - \displaystyle \frac{\Gamma_1 - 1}{i \sigma + \Gamma_1 \sigma_0} J_n^{m,\sigma} - U_n^{m,\sigma} \right\},
 \end{array}
 \label{deltaq_mince}
\end{equation}


\begin{equation}
\begin{array}{rl}
\delta T_n^{m,\sigma} = & \displaystyle  \frac{i \sigma T_0 \left( \Gamma_1 - 1 \right) }{c_s^2 \left( i \sigma + \sigma_0 \right) \left( 1 - \varepsilon_{s;n}  \right)} \left\{  \frac{\sigma^2}{H \Lambda_n^{m,\nu}} e^{\frac{x}{2} } \left[     \dfrac{d \Psi_n^{m,\sigma}}{dx}   \right. \right. \\[0.3cm]
&  \left. \left. \displaystyle + \mathcal{C}_n \Psi_n^{m,\sigma}  \right] \displaystyle  + \frac{1}{i \sigma} \left( 1 -  \frac{\Gamma_1 \sigma^2 }{\sigma_{s;n}^2} \right) J_n^{m,\sigma} - U_n^{m,\sigma} \right\},
\end{array}
\label{deltaT_mince}
\end{equation}


with 

\begin{equation}
 \mathcal{D}_n = \frac{\Gamma_1 - 1}{i \sigma + \sigma_0} \frac{\sigma^2 }{\sigma_{s;n}^2},
\end{equation}

and where $ \mathcal{A}_n $, $ \mathcal{B}_n $ and $ \mathcal{C}_n $ are the constant coefficients of Eq.~(\ref{coeffsRP})\footnote{These coefficients are the same as those given by Eq.~(\ref{coeffsRP}) to a $ H $ factor.},

\begin{equation}
\left\{
\begin{array}{l}
   \displaystyle \mathcal{A}_n = \frac{i \sigma}{i \sigma + \sigma_0} \left( \kappa - \frac{1}{2} + i \frac{\sigma_0}{2 \sigma}   \right), \\[0.5cm]
   \displaystyle \mathcal{B}_n = \frac{i \sigma}{i \sigma + \sigma_0} \left(  \frac{\kappa}{\varepsilon_{s;n}} - \frac{1}{2} + i \frac{\sigma_0}{2 \sigma}  \right), \\[0.5cm]
   \displaystyle \mathcal{C}_n  = \frac{1}{2} - \frac{\sigma_{s;n}^2}{\Gamma_1 \sigma^2 }.
\end{array}
\right.
\label{coeffs_ABC_mince}
\end{equation}

\subsection{Boundary conditions}

Solving Eq.~(\ref{Shrodinger_mince}) requires that we choose two boundary conditions. Following CL70, we fix $ \xi_r = 0 $ at the ground. This corresponds to a smooth rigid wall and is equivalent to $ \xi_{r;n}^{m,\sigma} = 0 $ at $ x = 0 $. In the case of Earth-like exoplanets, where the telluric surface is well defined, this condition is relevant. Thus, given the form of Eq.~(\ref{Shrodinger_mince}), $ \Psi_n^{m,\sigma} $ can be written

\begin{equation}
\Psi_n^{m,\sigma} = \mathscr{A} \left( x \right) e^{i \hat{k}_n x} + \mathscr{B} \left( x \right) e^{- i \hat{k}_n x},
\label{Psin}
\end{equation}

where $ \mathscr{A} $ and $ \mathscr{B} $ are the complex functions

\begin{equation}
\left\{
\begin{array}{l}
   \displaystyle \mathscr{A} \left( x \right) = K - i \frac{H^2}{2 \hat{k}_n} \int_0^x e^{- \left( i \hat{k}_n + 1/2 \right) u} C \left( u \right) du,\\[0.5cm]
   \displaystyle \mathscr{B} \left( x \right) = - K + i \frac{H^2}{2 \hat{k}_n} \int_0^x e^{ \left( i \hat{k}_n - 1/2 \right) u} C \left( u \right) du.
\end{array}
\right.
\label{fonctions_AB}
\end{equation}

The parameter $ K \in \mathbb{C} $ is an integration constant which is fixed by the upper boundary condition.\\

The upper border shall be treated with most carefulness, as this has been discussed by \cite{Green1965}. The fluid envelopes of stars and giant gaseous planets, which are considered as bounded fluid shells,  are usually treated with a stress-free condition applied at the upper limit $ x = x_{\rm atm} $ \mybf{\citep[e.g.][]{Unno1989}}, that is

\begin{equation}
\delta p_n^{m,\sigma} + \frac{1}{H}  \dfrac{d p_0}{dx} \xi_{r;n}^{m,\sigma} = 0.
\end{equation} 

Hence, 

\begin{equation}
\dfrac{d \Psi_n^{m,\sigma}}{dx} + \mathcal{P} \Psi_n^{m,\sigma} =  \mathcal{Q},
\end{equation}

with the complex coefficients

\begin{equation}
\left\{
\begin{array}{l}
     \displaystyle \mathcal{P} =  \mathcal{A}_n - \left( 1 - \varepsilon_{s;n} \right) \frac{\sigma_{s;n}^2}{\Gamma_1 \sigma^2}, \\[0.5cm]
     \displaystyle \mathcal{Q} = e^{-  \frac{x_{\rm atm}}{2}} \left[ \frac{\kappa R^2}{g \left( i \sigma + \sigma_0 \right)} J_n^{m,\sigma} + \frac{H \Lambda_n^{m,\nu}}{\sigma^2} U_n^{m,\sigma}  \right]_{x = x_{\rm atm}}.
\end{array}
\right.
\end{equation}

The atmosphere then behaves as a wave-guide of typical thickness $ x_{\rm atm} $ (cf. Appendix~\ref{app:boundary_conditions}). In this case, atmospheric tides are analogous to ocean tides \mybf{\citep[see][]{Webb1980}}, the amplitude of the perturbation being highly frequency-resonant. However, this condition appears to be inappropriate for the thin isothermal atmosphere because the fluid is not homogeneous and there is no discontinuous interface in this case. Setting the stress-free condition would give birth to unrealistic resonances depending on $ x_{\rm atm} $ (see Fig.~\ref{fig:couple_BC} in Appendix F). Therefore, we have to choose a condition consistent with the exponentially decaying density and pressure. \mybf{Following \cite{SZ1990}, we consider that there is no material escape when $ x \rightarrow + \infty $, i.e. $ \left. V_{r} \right|_{x \rightarrow \infty} = 0 $. This means that the amplitude of oscillations shall decrease with the altitude and, consequently, that we have to eliminate the diverging term in Eq.~(\ref{Psin}).}\footnote{\mybf{In the case of the semidiurnal tide treated by CL70 (no dissipative processes), this condition cannot be applied because $ \hat{k}_n \in \mathbb{R} $. To address this particular situation, CL70 proposes to apply a radiation condition at the upper boundary, that is to consider that the energy is propagating upward only and to eliminate the terms of the solution corresponding to an energy propagating downward. However, in the present work, whatever $ \sigma_0 > 0 $, $ \Im \left\{ \hat{k}_n  \right\} \neq 0 $. So we do not have to consider this case.}} \\

 Let us assume $ \Im \left\{ \hat{k}_n \right\} > 0 $. \mybf{As shown by CL70 (Eq. 88), the above condition requires that} 

\begin{equation}
\dfrac{d \Psi_n^{m,\sigma}}{dx} - i \hat{k}_n \Psi_n^{m,\sigma} = 0
\end{equation}

at the upper boundary ($ x = x_{\rm atm} $). To make this condition relevant, $ x_{\rm atm} $ must be chosen so that $ x_{\rm atm} /  x_{\rm crit} \gg 1  $, where $ x_{\rm crit} $ is the typical damping depth computed from the case where $ C $ is a constant, and given by

\begin{equation}
x_{\rm crit} = \frac{1}{  \Im \left\{ \hat{k}_n \right\} + 1/2  };
\label{xcrit}
\end{equation}

we usually fix $ x_{\rm atm} = 10 $ in computations. The integration constant of Eq.~(\ref{fonctions_AB}) is then obtained

\begin{equation}
K = i \frac{H^2}{2 \hat{k}_n} \int_0^{x_{\rm atm}} e^{ \left( i \hat{k}_n - 1/2 \right) u} C \left( u \right) du.
\label{Kconst}
\end{equation}

\subsection{Analytical expressions of the semidiurnal tidal torque, lag angle and amplitude of the pressure bulge}
\label{subsec:analytic_isoT}

If $ U_n^{m,\sigma} $ and $ J_n^{m,\sigma} $ are assumed to be constant (with $ U_n,J_n^{m,\sigma} \in \mathbb{C} $), $ C $ does not vary with $ x $ and the expression of Eq.~(\ref{Psin}) can be simplified significantly. Indeed, it may be be written in this case

\begin{equation}
\Psi_n^{m,\sigma} \left( x \right) = \frac{H^2 C}{\hat{k}_n^2 + \frac{1}{4}} \left[  e^{-\frac{x}{2}} - e^{i \hat{k}_n x} \right].
\label{Psi_ana}
\end{equation} 

We recall that we use the convention $ \Im \left\{ \hat{k}_n \right\} > 0 $, which implies $ \Psi_n^{m,\sigma} \rightarrow 0 $ at infinity. Substituting this simplified solution in polarization relations (Eqs.~\ref{xir_mince} to \ref{deltaT_mince}), we compute three-dimensional variations of the velocity field, displacement, pressure, density and temperature caused by the semidiurnal tide as explicit functions of the physical parameters of the atmosphere and tidal frequency. Let us introduce the function $ \mathscr{F}_{\mathcal{P}} $ of the altitude, parametrized by $ \mathcal{P} $ (which can be $ \mathcal{A}_n $, $ \mathcal{B}_n $ or $ \mathcal{C}_n $), and defined by

\begin{equation}
\mathscr{F}_{\mathcal{P}} \left( x \right) = \left( \mathcal{P} - \frac{1}{2} \right) e^{-x} - \left( \mathcal{P} + i \hat{k}_n \right) e^{ \left( i \hat{k}_n - \frac{1}{2} \right) x }.
\end{equation}

Then, polarization relations write

\begin{equation}
\xi_{r;n}^{m,\sigma} = \frac{\Lambda_n^{m,\nu}  H }{R^2 \sigma^2} \mathcal{E}_n \left[  1 - e^{\left( i \hat{k}_n + \frac{1}{2}  \right) x }  \right],
\end{equation}

\begin{equation}
    \xi_{\theta ; n}^{m,\sigma} =  \displaystyle \frac{1}{\sigma^2 R \left( 1 - \varepsilon_{s;n} \right)} \left\{ \mathcal{E}_n e^{x} \mathscr{F}_{\mathcal{A}_n} \left( x \right)  - \mathcal{D}_n J_n^{m,\sigma} - \varepsilon_{\rm s;n} U_n^{m,\sigma} \right\},
\end{equation}

\begin{equation}
    \xi_{\varphi ; n}^{m,\sigma} =  \displaystyle \frac{i}{\sigma^2 R \left( 1 - \varepsilon_{s;n} \right)} \left\{ \mathcal{E}_n e^{x} \mathscr{F}_{\mathcal{A}_n} \left( x \right) -\mathcal{D}_n J_n^{m,\sigma} - \varepsilon_{\rm s;n} U_n^{m,\sigma} \right\},
\end{equation}

\begin{equation}
V_{r;n}^{m,\sigma} = i \frac{ \Lambda_n^{m,\nu} H }{R^2 \sigma} \mathcal{E}_n \left[  1 - e^{\left( i \hat{k}_n + \frac{1}{2}  \right) x }  \right],
\end{equation}

\begin{equation}
    V_{\theta ; n}^{m,\sigma} =  \displaystyle \frac{ i }{\sigma R \left( 1 - \varepsilon_{s;n} \right)} \left\{ \mathcal{E}_n e^{x} \mathscr{F}_{\mathcal{A}_n} \left( x \right)  - \mathcal{D}_n J_n^{m,\sigma} - \varepsilon_{\rm s;n} U_n^{m,\sigma} \right\},
\end{equation}

\begin{equation}
    V_{\varphi ; n}^{m,\sigma} =  \displaystyle - \frac{1}{\sigma R \left( 1 - \varepsilon_{s;n} \right)} \left\{\mathcal{E}_n e^{x} \mathscr{F}_{\mathcal{A}_n} \left( x \right)  - \mathcal{D}_n J_n^{m,\sigma} - \varepsilon_{\rm s;n} U_n^{m,\sigma} \right\},
\end{equation}

\begin{equation}
   \delta p_n^{m,\sigma} =  \displaystyle  \frac{\rho_0 \left( 0 \right)}{  1 - \varepsilon_{\rm s;n}  } \left\{  \mathcal{E}_n \mathscr{F}_{\mathcal{A}_n} \left( x \right) - \left(\mathcal{D}_n J_n^{m,\sigma} + U_n^{m,\sigma}  \right) e^{-x} \right\},
\label{deltap_ana}
\end{equation}

\begin{equation}
\begin{array}{ll}
   \delta \rho_n^{m,\sigma} = & \displaystyle \frac{i \sigma + \Gamma_1 \sigma_0}{i \sigma + \sigma_0} \frac{\rho_0 \left( 0 \right)}{c_s^2 \left( 1 - \varepsilon_{\rm s;n} \right) } \left\{ \mathcal{E}_n  \mathscr{F}_{\mathcal{B}_n} \left( x \right)   \right.\\[0.3cm]
    & \left. - \displaystyle \left( \frac{\Gamma_1 - 1}{i \sigma + \Gamma_1 \sigma_0} J_n^{m,\sigma} + U_n^{m,\sigma}  \right) e^{-x} \right\},
\end{array}
\label{deltaq_ana}
\end{equation}

\begin{equation}
\begin{array}{ll}
   \delta T_n^{m,\sigma} = & \displaystyle  \frac{i \sigma \left( \Gamma_1 - 1 \right) T_0 }{c_s^2 \left( i \sigma + \sigma_0 \right) \left( 1 - \varepsilon_{\rm s;n} \right)} \left\{  \mathcal{E}_n e^{x} \mathscr{F}_{\mathcal{C}_n} \left( x \right)   \right.\\[0.3cm]
    & \left. + \displaystyle  \frac{1}{i \sigma} \left( 1 - \frac{\Gamma_1 \sigma^2}{\sigma_{\rm s;n}^2} \right) J_n^{m,\sigma} - U_n^{m,\sigma}  \right\}.
\end{array}
\end{equation}

with 

\begin{equation}
 \mathcal{E}_n =  \frac{\kappa \left( \mathcal{K}_n J_n^{m,\sigma} - i \sigma U_n^{m,\sigma} \right)}{ \left( i \sigma + \sigma_0 \right) \left( \hat{k}_n^2 + \frac{1}{4} \right)}.
\end{equation}

From the analytical solution and the polarization relation of density, given by Eq.~(\ref{deltaq_ana}), we then compute the second order Love number and the tidal torque. Integrating $ \delta \rho $ over the whole thickness of the atmosphere, we obtain

\begin{equation}
\begin{array}{ll}
\displaystyle \int_0^{x_{\rm atm}} \delta \rho_n^{m,\sigma} \left(x \right)dx = & \! \! \! \!  \displaystyle - \frac{i \sigma + \Gamma_1 \sigma_0}{i \sigma + \sigma_0} \frac{\rho_0 \left( 0 \right) }{c_s^2 \left( 1 - \varepsilon_{s;n} \right) } \left\{ \frac{ \left(\Gamma_1 - 1 \right)  J_n^{m,\sigma} }{i \sigma + \Gamma_1 \sigma_0}  \right. \\[0.3cm]
  & \! \! \! \! \displaystyle   +  \frac{ \kappa \left( \mathcal{B}_n + \frac{1}{2} \right) \left[ \mathcal{K}_n J_n^{m,\sigma} - i \sigma U_n^{m,\sigma}   \right] }{ \left( i \sigma + \sigma_0 \right) \left( i \hat{k}_n - \frac{1}{2} \right)^2 } \\[0.3cm] 
  & \! \! \! \!  \left. \displaystyle + U_n^{m,\sigma} \right\},
\end{array}
\label{int_deltaq}
\end{equation} 


where the expressions of $ \mathcal{B}_n $ and $ \mathcal{K}_n $ are given by Eqs.~(\ref{coeffs_ABC_mince}) and (\ref{coeffK}) respectively. The semidiurnal tide is supposed to be represented by spatial forcings of the form $ J \left( x, \theta , \varphi \right) = J_2 P_2^2 \left( \cos \theta \right) e^{ 2i \varphi} $ and $ U \left( x, \theta , \varphi \right) = U_2 P_2^2 \left( \cos \theta \right) e^{ 2i \varphi} $ with the tidal frequency $ \sigma = 2 \left( \Omega - n_{\rm orb} \right) $. The second order Love number, deduced Eq.~(\ref{tidalU2}), is thus expressed

\begin{equation}
k_2^{\sigma , 2} = \frac{4 \pi}{5}  \frac{\mathscr{G} H R_{\rm atm}}{U_2} \sum_{n \in \mathbb{Z}}  c_{2,n,2}^{\sigma , 2} \int_0^{x_{\rm atm}} \delta \rho_{n,2}^{\sigma , 2} \left( x \right) dx.
\end{equation}

Assuming $ \left| \varepsilon_{s;n} \right| \ll 1 $, we identify the different contributions. It follows

\begin{equation}
k_2^{\sigma , 2} = k_{\rm 2; \mybf{V} ; therm}^{\sigma , 2} + k_{\rm 2;\mybf{V} ; grav}^{\sigma , 2} + k_{\rm 2 ; \mybf{H} ; therm}^{\sigma , 2} + k_{\rm 2;\mybf{H};grav}^{\sigma , 2},
\end{equation}

where the subscripts $ _{\rm therm} $ and $ _{\rm grav} $ indicate the origin of a contribution (thermally or gravitationally forced) \scor{and $_{\rm H}$ and $ _{\rm V} $ its horizontal and vertical components}. The terms of this expansion are expressed as functions of the parameters of the atmosphere and tidal frequency

\begin{equation}
k_{\rm 2;\mybf{V} ; therm}^{\sigma , 2} = - \mathcal{W}  \sum_{n \in \mathbb{Z}}  c_{2,n,2}^{\sigma , 2} \frac{\kappa \left( i \sigma + \Gamma_1 \sigma_0 \right) \left( \mathcal{B}_n + \frac{1}{2} \right) \mathcal{K}_n}{ \left( i \sigma + \sigma_0 \right)^2 \left( i \hat{k}_n - \frac{1}{2} \right)^2 }  \frac{J_2}{U_2},
\end{equation}

\begin{equation}
k_{\rm 2 ; \mybf{V} ; grav}^{\sigma , 2} = \mathcal{W} \sum_{n \in \mathbb{Z}}  c_{2,n,2}^{\sigma , 2} i \sigma \frac{\kappa \left( i \sigma + \Gamma_1 \sigma_0 \right) \left( \mathcal{B}_n + \frac{1}{2} \right)}{ \left( i \sigma + \sigma_0 \right)^2 \left( i \hat{k}_n - \frac{1}{2} \right)^2 },
\end{equation}

\begin{equation}
k_{\rm 2 ; \mybf{H} ; therm}^{\sigma , 2} = - \mathcal{W}  \frac{\Gamma_1  - 1}{i \sigma + \sigma_0}  \frac{J_2}{U_2},
\end{equation}

\begin{equation}
k_{\rm 2 ; \mybf{H} ; grav}^{\sigma , 2} = - \mathcal{W}  \frac{ i \sigma + \Gamma_1 \sigma_0 }{i \sigma + \sigma_0},
\end{equation}

the parameter $ \mathcal{D}_n $ being a dimensionless constant given by

\begin{equation}
\mathcal{W} = \frac{4 \pi}{5}  \frac{\mathscr{G} H R_{\rm atm} \rho_0 \left( 0 \right) }{c_S^2}.
\end{equation}


Contrary to gravitational Love numbers ($ k_{\rm 2 ; \mybf{V} ; grav}^{\sigma , 2} $ and $ k_{\rm 2 ; \mybf{H} ; grav}^{\sigma , 2} $) which are intrinsic to the planet, thermal Love numbers ($ k_{\rm 2 ; \mybf{V} ; therm}^{\sigma , 2} $ and $ k_{\rm 2 ; \mybf{H} ; therm}^{\sigma , 2} $) are proportional to the forcings ratio $ J_2 / U_2 $ and thus depend of the properties of the whole star-planet system, particularly on the semi-major axis and the stellar luminosity \citep[e.g.][]{Correia2008}. 


In the same way as we obtained Love numbers, the torque exerted on the atmosphere may be computed by substituting Eq.~(\ref{int_deltaq}) in Eq.~(\ref{torque_sum}). This torque writes

\begin{equation}
\mathcal{T}^{\sigma,2} = 2 \pi R^2 H U_2 \sum_{n \in \mathbb{Z}} c_{2,n,2}^{\sigma,2} \Im \left\{  \int_0^{x_{\rm atm}} \delta \rho_{n,2}^{\sigma,2} \left( x \right)  dx  \right\}.
\label{torque_ana}
\end{equation}

Finally, assuming $ \left| \varepsilon_{s;n} \right| \ll 1 $ and introducing the factor 

\begin{equation}
\mathcal{H} = \frac{2 \pi R^2 H \rho_0 \left( 0 \right)}{c_s^2}
\end{equation}

allows us to write $ \mathcal{T}^{\sigma,2} $ under the form

\begin{equation}
     \mathcal{T}^{\sigma,2} =  \mathcal{T}^{\sigma,2}_{\rm \mybf{V}; therm} +  \mathcal{T}^{\sigma,2}_{\rm \mybf{V} ; grav} +  \mathcal{T}^{\sigma,2}_{\rm \mybf{H} ; therm} +  \mathcal{T}^{\sigma,2}_{\rm \mybf{H} ; grav},
\end{equation}

with 

\begin{equation}
    \mathcal{T}^{\sigma,2}_{\rm \mybf{V}; therm}  = - \mathcal{H} U_2 \sum_{n \in \mathbb{Z}} c_{2,n,2}^{\sigma,2} \mathcal{K}_n \Im \left\{ \frac{ \kappa \left( i \sigma + \Gamma_1 \sigma_0 \right) \left( \mathcal{B}_n + \frac{1}{2} \right)  }{ \left( i \sigma + \sigma_0 \right)^2 \left( i \hat{k}_n - \frac{1}{2} \right)^2 } J_2  \right\},
\label{couple_D_therm}
\end{equation}

\begin{equation}
 \mathcal{T}^{\sigma,2}_{\rm \mybf{V} ; grav} = \mathcal{H} U_2 \sum_{n \in \mathbb{Z}} c_{2,n,2}^{\sigma,2}  \Im \left\{ i \sigma \frac{ \kappa \left( i \sigma + \Gamma_1 \sigma_0 \right) \left( \mathcal{B}_n + \frac{1}{2} \right)  }{ \left( i \sigma + \sigma_0 \right)^2 \left( i \hat{k}_n - \frac{1}{2} \right)^2 } U_2  \right\},
\end{equation}

\begin{equation}
  \mathcal{T}^{\sigma,2}_{\rm \mybf{H} ; therm} =  - \mathcal{H} U_2 \Im \left\{ \frac{\Gamma_1 - 1}{i \sigma + \sigma_0} J_2 \right\} ,
\end{equation}

\begin{equation}
\mathcal{T}^{\sigma,2}_{\rm \mybf{H} ; grav} = - \mathcal{H} U_2 \Im \left\{ \frac{i \sigma + \Gamma_1 \sigma_0}{i \sigma + \sigma_0} U_2 \right\}.
\end{equation}

\mybf{Let us consider the case of the pure thermal tide ($U_2 \approx 0$) at the vicinity of synchronization. If $ \sigma \rightarrow 0 $, then $ \left| \hat{k}_n \right| \rightarrow + \infty $, $ \left| \mathcal{B}_n \right| \rightarrow + \infty $ and $ \mathcal{B}_n / \hat{k}_n^2 \, \propto \, \left( i \sigma + \sigma_0 \right) / \left( i \sigma + \Gamma_1 \sigma_0 \right) $. As a consequence $ k_{2 ; \rm V ; therm}^{\sigma,2} \approx - k_{2 ; \rm H ; therm}^{\sigma,2} $ and $\mathcal{T}_{\rm V ; therm}^{\sigma,2} \approx - \mathcal{T}_{\rm H ; therm}^{\sigma,2} $. The tidal torque exerted on the stably stratified isothermal atmosphere is thus very small compared to the horizontal thermal component, }

\begin{equation}
\mathcal{T}^{\sigma,2}_{\rm \mybf{H}} = 2 \pi \kappa \frac{ R^2 \rho_0 \left( 0 \right) }{g} U_2 J_2 \frac{\sigma}{\sigma^2 + \sigma_0^2},
\label{couple_equilibre}
\end{equation}

\scor{where we have assumed that the thermal forcing is in phase with the perturber ($ J_2 \in \mathbb{R} $).} \mybf{This behaviour had been identified before, for instance by \cite{AS2010} who studied thermal tides in hot Jupiters. According to \cite{AS2010}, at the vicinity of synchronization, the vertical displacement of fluid due to the restoring effect of the Archimedean force ($ N^2 V_r $ term in the heat transport equation, Eq.~\ref{transport_chaleur_1}) compensates exactly the local density variations generated by the thermal forcing, which annihilates the quadrupolar tidal torque. We will observe this effect in Section~\ref{sec:application_Earth_Venus} \scor{when} considering \jcor{the Earth's and a Venus-like planet atmospheres} isothermal and stably-stratified (see Fig.~\ref{fig:Terre_couple}). \scor{\jcor{For a stably-stratified structure,} the tidal torque exerted on the atmosphere is weak and cannot balance the solid torque, which leads the planet's spin to the synchronization configuration.} \pcor{We can note here that the importance of stable stratification has been pointed out \jcor{for} the case of Jupiter-like planets by \cite{IL1993b} \citep[see also][]{IL1993a,IL1994}, who showed that the atmospheric tidal response was very sensitive to variations of the Brunt-Väisälä frequency ($N$).}   }

\subsection{Comparison with CL70}

The vertical wavenumber is a key-parameter of the tidal perturbation. Hence, to highlight the interest of taking into account dissipative mechanisms such as radiation, we consider the relative difference between the $ \hat{k}_n $ given by Eq.~(\ref{kv2_2}) and the one established in the classical theory of tides without dissipation \citep[e.g.][CL70]{Wilkes1949}, denoted $ \hat{k}_{\rm CL;n} $ in reference to CL70 and given by

\begin{equation}
\hat{k}_{\rm CL ; n}^2  = \frac{1}{4} \left[ \frac{h_c}{h_n} - 1  \right].
\end{equation}

\begin{figure}[htb]
\centering
{\includegraphics[width=0.950\textwidth]
{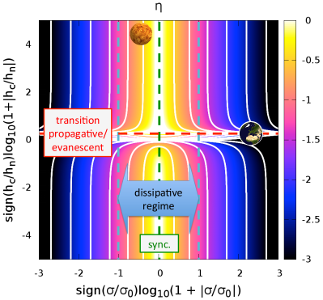}
\textsf{\caption{\label{fig:ecart_kv} Relative difference between the complex wavenumber obtained in this work (Eq.~\ref{kv2_2}) on the one hand, and the real one given by CL70 for Earth atmospheric tides ($ \hat{k}_{\rm CL;n} $) on the other hand. The difference $ \eta $, given by Eq.~(\ref{ecart_kv}), is plotted in logarithmic scale as a function of the normalized tidal frequency $ \sigma / \sigma_0 $ and of the ratio $ h_c / h_n $. The color of the map, denoted $ c $, is defined by the expression $ c = \log \left( \eta \right) $. The horizontal and vertical axis represent $ f \left( \sigma / \sigma_0 \right) $ and $ f \left( h_c / h_n \right) $ where $ f $ is the function defined by $ f \left( X \right) = {\rm sign}\left( X \right) \log \left( 1 + \left| X \right| \right) $ with $ X \in \mathbb{R} $. The positions of the planets are determined by the gravity mode of degree $ 0 $ in the case of the semidiurnal tide. Dark (luminous) regions correspond to the dynamic (dissipative) regime. Venus' vertical wavenumbers are strongly affected by dissipative processes contrary to those of the Earth.} }}
\end{figure}

This difference, written

\begin{equation}
\eta = 2 \left| \frac{\hat{k}_n - \hat{k}_{\rm CL;n}}{\hat{k}_n + \hat{k}_{\rm CL;n}} \right|,
\label{ecart_kv}
\end{equation}

is plotted on Fig~\ref{fig:ecart_kv} as a function of the reduced tidal frequency $ \sigma / \sigma_0 $, and of the height ratio $ h_c / h_n $. The supplementary term of Newtonian cooling in the heat transport equation induces an additional regime with respect to CL70. At low frequencies, around the radiation frequency, the vertical profiles are damped while they are highly oscillating in the case where $ \sigma_0 = 0 $. This thermal regime corresponds to the middle white-yellow region of the map. The two wavenumbers become similar for $ \left| \sigma \right|  \gg \sigma_0 $. This is the regime of fast rotating planets studied by CL70. Beyond the threshold materialized by the left and right edges of the map, which correspond to $ \left| \sigma \right| \sim N $, the traditional approximation assumed in Section 2 is not valid any more and the coupling between the three components of the Navier-Stokes equation shall be considered. At the end, the area located at $ h_c / h_n \approx 1 $ represents the discontinuous transition between propagative and evanescent waves in CL70, this transition being regular in our model.

\section{\mybf{Tides in a slowly rotating convective atmosphere}}
\label{sec:convective_atm}

\mybf{In this section, we treat the asymptotic case of \scor{a slowly rotating convective atmosphere}, for which the equations of tidal \scor{hydrodynamics} \scor{can be simplified} drastically. This situation typically corresponds to Venus-like planets. Indeed, the rotation period of Venus, 243 days, is of the same order of magnitude as its orbital period. As a consequence $ \left| \nu \right| < 1 $ for the Venus' semi-diurnal tide and the restoring force of inertial waves, the Coriolis \scor{acceleration, can be neglected as a first step}. The atmospheric layer located below 60 km is characterized by a strongly negative temperature gradient \citep[][]{Seiff1980}. In this layer, the temperature decreases from 750~K to 250~K, which make it subject to convective instability \citep[][]{Baker2000}. Therefore, the Brunt-Väisälä frequency of Venus is far lower than the one given by the isothermal approximation. As this frequency represents the ``stiffness'' of the Archimedean force, which restores gravity waves, these laters cannot propagate in the unstable region above the surface.} 


\subsection{\scor{Equilibrium distributions of} \mybf{pressure, density and temperature}}

\mybf{We assume $ \Omega = 0 $ and $ N = 0 $. According to Eq.~(\ref{sigmaBV}), the Brunt-Väisälä frequency can be expressed \scor{as}}

\begin{equation}
N^2 = \frac{g}{H} \left( \kappa + \frac{d H}{dz} \right). 
\end{equation}

\mybf{Therefore, the case $ N^2 = 0 $ corresponds to the adiabatic temperature gradient, given by }

\begin{equation}
\begin{array}{rcl}
 \displaystyle H \left( z \right) = H \left( 0 \right) - \kappa z & \mbox{or} &  \displaystyle T_0 \left( z \right) = T_0 \left( 0 \right) - \frac{\kappa g}{\mathscr{R}_{\rm s}} z. 
\end{array}
\end{equation} 

\mybf{For this structure, the reduced altitude is expressed \scor{by}}

\begin{equation}
x = \frac{1}{\kappa} \ln \left( \frac{H \left( 0 \right)}{H \left( 0 \right) - \kappa z} \right),
\end{equation}

\mybf{and we thus obtain the distributions of the basic pressure height, temperature, pressure and density, respectively}

\begin{equation}
\begin{array}{ll}
 \displaystyle  H \left( x \right) = H \left( 0 \right) e^{- \kappa x}, &  \displaystyle T_0 \left( x \right) = \frac{g H \left( 0 \right)}{\mathscr{R}_{\rm s}} e^{- \kappa x}, \\[3mm]
 \displaystyle  p_0 \left( x \right) = p_0 \left( 0 \right) e^{-x} , &  \displaystyle \rho_0 \left( x \right) = \rho_0 \left( 0 \right) e^{\left( \kappa - 1 \right) x}, 
\end{array}
\end{equation}

\mybf{with $ H \left( 0 \right) = p_0 \left( 0 \right) / \left[ g \rho_0 \left( 0 \right) \right] $.  }

\subsection{Tidal response}

\mybf{For the sake of simplicity, we neglect the contribution of the gravitational forcing in the Navier-Stokes equation and examine the case of pure thermal tides. When $ \sigma \rightarrow 0 $, the tidal response tends to its hydrostatic component, the equilibrium tide \citep[][]{Zahn1966a}. We thus assume the hydrostatic approximation. This allows \scor{us} to to reduce the radial projection of the Navier-Stokes equation (Eq.~\ref{NSbrut_3}) to }

\begin{equation}
\delta \rho = - \frac{1}{gH} \frac{\partial p}{\partial x}.
\label{NS_hydro}
\end{equation} 

\mybf{Now, by expanding the perturbed quantities in Fourier series and substituting Eq.~(\ref{NS_hydro}) in the heat transport equation \jcor{(Eq.~\ref{transport_chaleur_3})}, we get the simplified vertical structure equation of pressure}

\begin{equation}
\begin{array}{lcl}
\displaystyle \frac{d \delta p}{dx} + \tau_{\rm p} \delta p = \frac{\kappa \rho_0 J}{i \sigma + \Gamma_1 \sigma_0}, & \mbox{where} & \displaystyle \tau_{\rm p} \left( \sigma \right) = \frac{1}{\Gamma_1} \frac{i \sigma + \Gamma_1 \sigma_0}{i \sigma + \sigma_0},
\end{array}
\end{equation}  

\mybf{represents the complex damping rate of the perturbation. Following \cite{DI1980} and \cite{SZ1990}, we consider that thermal tides are generated by a heating at the ground and apply the profile of thermal forcing}

\begin{equation}
J \left( x \right) = J_0 \tau_{\rm J} e^{- \tau_{\rm J} x},
\end{equation}

\mybf{where \scor{$ 1 / \tau_{\rm J} \ll 1 $} is the thickness of the heated region \scor{(see Eq.~\ref{delta_skin} for a physical expression of this parameter)} and $ J_0 $ the mean absorbed power per unit mass. At the upper boundary $ x_{\rm atm} \gg 1 $, we apply the stress-free condition $ \delta p = 0 $ and get the pressure profile  }

\begin{equation}
\delta p \left( x \right) =  - \frac{\kappa \rho_0 \left( 0 \right) J_0}{i \sigma + \sigma_0} e^{- \tau_{\rm J} x},
\end{equation}

\mybf{from which we deduce the vertical profiles of the Lagrangian horizontal displacements}

\begin{equation}
\xi_\theta \left( x \right) = - \frac{\kappa J_0}{R \sigma^2 \left( i \sigma + \sigma_0 \right)} e^{- \tau_{\rm J} x}, 
\end{equation}

\begin{equation}
\xi_\varphi \left( x \right) = - i m \frac{\kappa J_0}{R \sigma^2 \left( i \sigma + \sigma_0 \right) } e^{- \tau_{\rm J}  x},
\end{equation}

\mybf{of the horizontal velocities }

\begin{equation}
V_\theta \left( x \right) = - i \frac{\kappa J_0 }{R \sigma \left( i \sigma + \sigma_0 \right) } e^{- \tau_{\rm J} x}, 
\end{equation}

\begin{equation}
V_\varphi \left( x \right) = m \frac{\kappa J_0}{R \sigma \left( i \sigma + \sigma_0 \right)} e^{- \tau_{\rm J} x},
\end{equation}

\mybf{and of the density and temperature}

\begin{equation}
\delta \rho \left( x \right) = - \frac{\kappa \rho_0 \left( 0 \right) \tau_{\rm J} J_0}{g H \left( 0 \right) \left( i \sigma + \sigma_0 \right)} e^{- \tau_{\rm J} x},
\end{equation}

\begin{equation}
\delta T \left( x \right) = - \frac{\kappa \tau_{\rm J} J_0}{\mathscr{R}_{\rm s} \left( i \sigma + \sigma_0 \right)} e^{- \tau_{\rm J} x}.
\end{equation}

\begin{figure}[htb]
\centering
{\includegraphics[width=0.95\textwidth]
{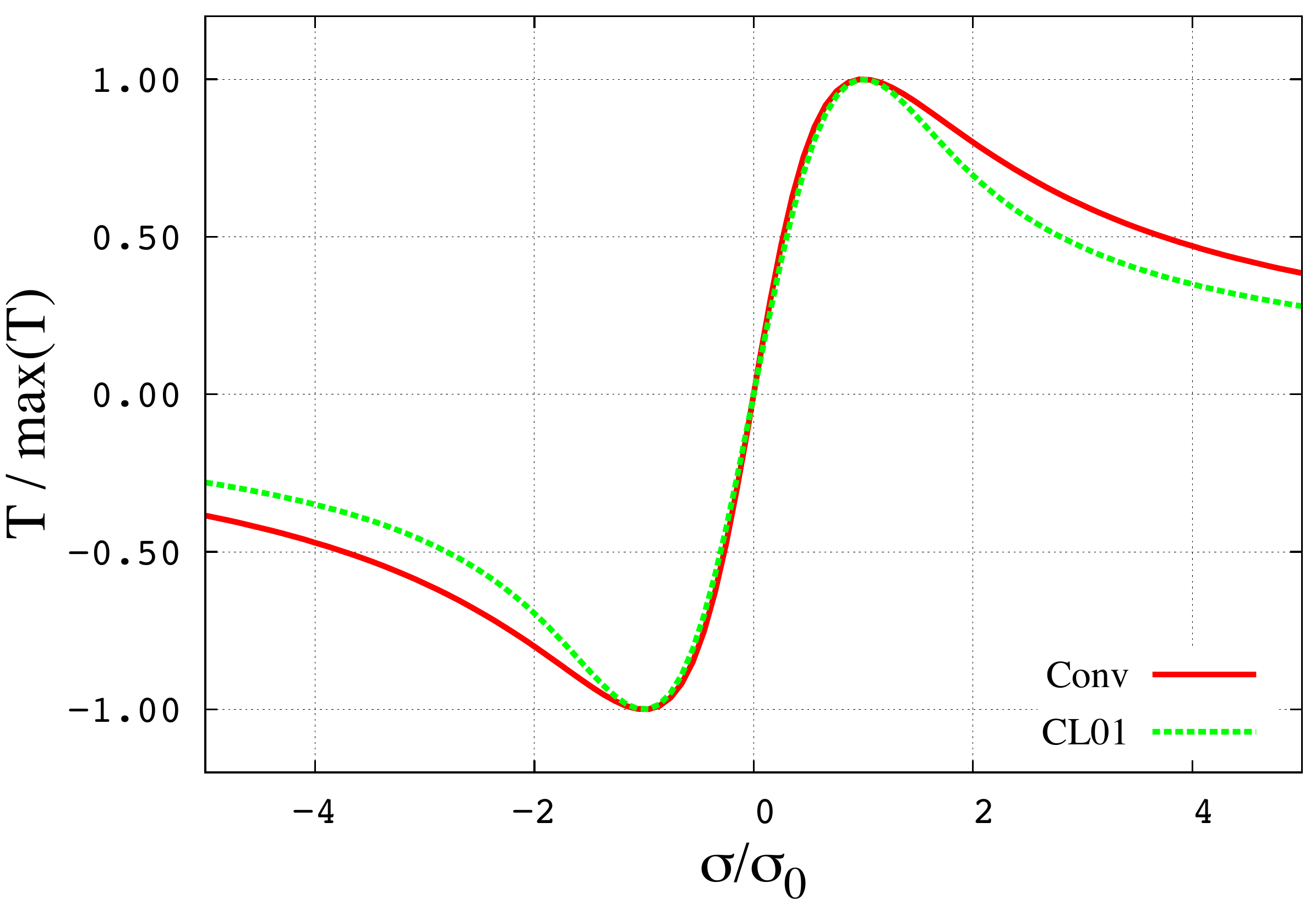}
\textsf{\caption{\label{fig:torque_conv} \mybf{Tidal torques exerted on the atmosphere in the case of the slowly rotating convective atmosphere (red line with the label ``\scor{Conv}'', Eq.~\ref{torque_nowave}) and in the model by \cite{CL01} (green dashed line with the label ``CL01'', Eq.~\ref{torque_CLNS}) as functions of the tidal frequency. Both functions are normalized by their respective maximum value, reached at $ \sigma = \sigma_0 $ ; the tidal frequency is normalized by the Newtonian cooling frequency ($\sigma_0$). The tidal forcing is quadrupolar, with $ U $ and $ J $ given by Section~\ref{subsec:k2_couple_nowave}. The parameters $ a $ and $ b $ of \scor{Eq.~(\ref{torque_CLNS})} are adjusted so that the peaks of the two functions are superposed. } } }}
\end{figure}

\subsection{\mybf{Second order Love number and tidal torque}}
\label{subsec:k2_couple_nowave}

\mybf{The previous results enable us to compute the tidal Love numbers and torque associated with the atmospheric tidal response. Due to the hydrostatic approximation, these quantities are directly proportional to $ \delta p \left( x = 0 \right) $. We consider that the semi-diurnal thermal tide can be reduced to its quadrupolar component, given by $ U \left( x, \theta , \varphi \right) = U_2 P_2^2 \left( \cos \theta \right) e^{i 2 \varphi} $ and $ J \left( x , \theta , \varphi \right) = J_2 \tau_{\rm J} e^{- \tau_{\rm J} x} P_2^2 \left( \cos \theta \right) e^{i 2 \varphi} $. Hence, using the expressions Eq.~(\ref{k2_simple}) for the second order Love number and Eq.~(\ref{couple_quadru}) for the tidal torque, we obtain}

\begin{equation}
k_{2 ; \rm conv}^{\sigma , 2} = - \frac{4 \pi}{5} \frac{ \mathscr{G} R H \kappa \rho_0 \left( 0 \right)}{ i \sigma + \sigma_0} \frac{J_2}{U_2},
\end{equation}

\mybf{and }

\begin{equation}
\mathcal{T}_{\rm conv}^{\sigma,2} = 2 \pi R^2 \frac{\kappa \rho_0 \left( 0 \right)}{g} U_2 J_2 \frac{\sigma}{\sigma^2 + \sigma_0^2}. 
\label{torque_nowave}
\end{equation}

\subsection{\mybf{Comparison with \cite{CL01} and \cite{Leconte2015}}}

\mybf{The tidal torque, plotted on Fig.~\ref{fig:torque_conv}, is identical to the the horizontal component of the torque exerted on the stably-stratified isothermal atmosphere, given by Eq.~(\ref{couple_equilibre}), \scor{owing to the fact that convection eliminates the component resulting from the fluid vertical displacement}. \pcor{In agreement with the early result of \cite{ID1978} (Eq.~4),} it is of the same form as the one given by the Maxwell model \citep[][]{Correia2014}; \scor{where} we identify the Maxwell time $ \mathscr{T}_0 = \sigma_0^{-1} $. It also corresponds to the torque computed by \cite{Leconte2015} for Venus-like planets with numerical simulations \scor{using GCM}, which suggests that the slow rotation and convective instability approximations are appropriate for this kind of planets.} \scor{This torque is stronger than the one computed in the case of the stably-stratified atmosphere. Consequently, it can lead to non-synchronized rotation states of equilibrium by counterbalancing the solid tidal torque.} \mybf{Finally, it must be compared to the} the model introduced by \cite{CL01} in early theoretical studies of the tidal torque exerted on Venus atmosphere. In this model, the tidal torque is given by

\begin{equation}
\mathcal{T} \left( \sigma \right) = \frac{a}{\sigma} \left[  1 - e^{- b \sigma^2}  \right],
\label{torque_CLNS}
\end{equation}

where $ a $ and $ b $ are two empirical positive real parameters. \mybf{As it can be observed on Fig~\ref{fig:torque_conv}} where (\ref{torque_nowave}) and (\ref{torque_CLNS}) are plotted as functions of the tidal frequency, $ \mathcal{T} \left( \sigma \right) $ looks like $ \mathcal{T}^{\sigma,2} \left( \sigma \right) $ of Eq.~(\ref{torque_nowave}) with amplitude maxima located at 

\begin{equation}
\sigma_0 \sim \pm \sqrt{\frac{2}{3 b}}.
\end{equation}

This strengthens the statement of \cite{CL01}, who expected that ``further studies of atmospheres of extra solar synchronous planets $ [...] $ may provide an accurate solution for the case $ \sigma \approx 0 $''. On Fig.~\ref{fig:Terre_couple}, the expression of Eq.~(\ref{torque_CLNS}) is plotted for the Earth and Venus in addition with the analytical results of the model presented here with parameters $ a $ and $ b $ adjusted numerically. \\


\section{Physical description of heat source terms}

In the left-hand side of Eq.~(\ref{Shrodinger_epais}) and in polarization relations, the perturbation is driven by $ U_n^{m,\sigma} $ and $ J_n^{m,\sigma} $. The tidal gravitational potential has been expanded in spherical harmonics since long and there is no need to come back on it\footnote{For instance, see the Kaula's expansion of the gravitational tidal potential \citep[][]{Kaula1964} detailed in Appendix~\ref{app:thermal_forcing} and \cite{MLP09}.}. Nevertheless, it is necessary to establish physical expressions of the heat power per unit mass to compute a solution. The different contributions must be clearly identified. \mybf{We detail them in the case of the optically thin atmosphere}. The most obvious \mybf{heat source}, but not necessary the main one in amplitude, is \mybf{the absorption of} the light flux coming from the star. \mybf{The non-absorbed part of this flux} reaches the ground and causes temperature oscillations at $ x = 0 $. This implies two contributions from the ground: a radiative emission in infrared and diffusive heating due to turbulences within the surface boundary layer.


%

\subsection{Insolation}

\begin{figure}[ht!]
 \centering
 \includegraphics[width=0.46\textwidth,clip]{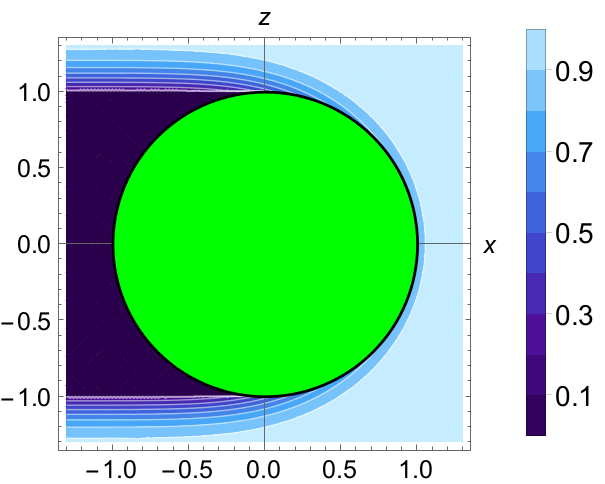}%
 \includegraphics[width=0.46\textwidth,clip]{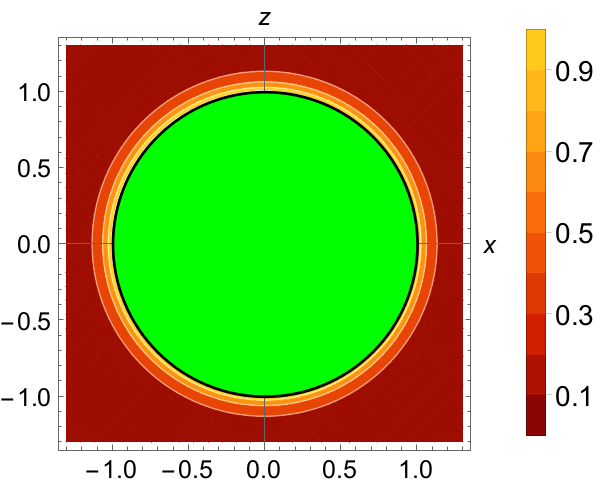} \\
  \includegraphics[width=0.46\textwidth,clip]{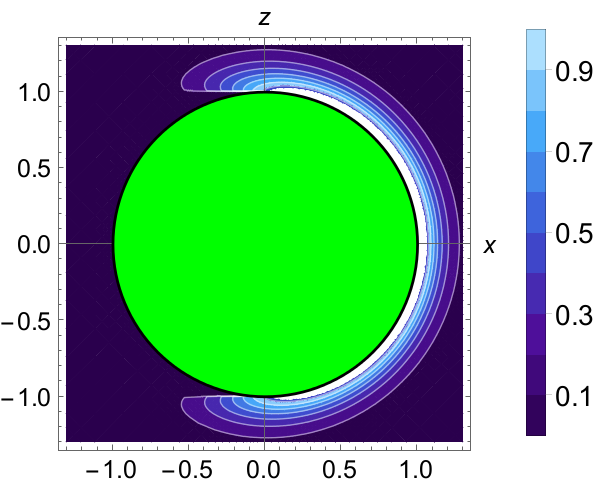}%
 \includegraphics[width=0.46\textwidth,clip]{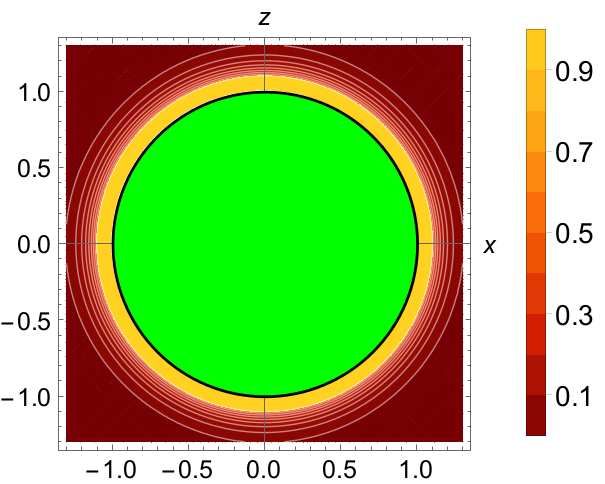} \\
  \textsf{ \caption{\label{fig:colormap_heating} \textit{Left, from top to bottom:} power per unit mass ($ J^{\rm inc} $, Eq.~\ref{Jinc}) and power per unit volume ($ \mathscr{J}^{\rm inc} $, Eq.~\ref{Jvolinc}) of a monochromatic incident flux normalized by their maxima and plotted as functions of the normalized spatial coordinates $ X $ (the direction of the star is along the horizontal axis) and $ Z $ (the spin axis is along the vertical axis). The planet is represented by a green disk of unit radius. The flux, of wavelength $ \lambda $, propagates in an atmosphere supposed homogeneous in composition ($ C_G = C_0 $) and characterized by the ratio $ H / R = 0.1 $. The \pcor{optical depth} is given by $ \tau_\lambda = 0.2 $. \textit{Right, from top to bottom:} power per mass unit ($ J^{\rm rad} $, Eq.~\ref{IJrad}) and power per volume unit ($ \mathscr{J}^{\rm rad} $) of a monochromatic flux radiated by the ground, normalized by their maxima and plotted as functions of $ X $ and $ Z $, with $ \tau_\lambda = 0.5 $.  }}
\end{figure}

The heating fluxes coming from the ground are all oriented radially and proportional to the local insolation flux at $ x = 0 $. Therefore, they can be easily written as a product of separable functions: $ \mathcal{F} \left( \theta \right) \mathcal{G} \left( x \right) $. This is not the case of the insolation flux, which goes through the spherical shell along the star-planet direction (see \jcor{Fig.}~\ref{fig:colormap_heating}). Particularly, the atmosphere is partly enlightened by stellar rays in the dark side. We note $ I_{\rm S} $ the bolometric flux transmitted to the atmosphere. If we assume that the star radiates like a black body, this flux is given by the integral of the spectral radiance $ I_{\rm S;\lambda} $ over a large range of wavelengths  

\begin{equation}
I_{\rm S} = \int_\lambda I_{\rm S;\lambda} \left( \lambda \right) { d \lambda},
\end{equation}

where $ I_{\rm S;\lambda} $ is given by the Plancks's law \citep{Planck1901}:

\begin{equation}
I_{\rm S;\lambda} \left( \lambda \right) =  \frac{R_{\rm S}^2 }{d^2}  \frac{ 2 h c^2 \pi \lambda^{-5}}{ e^{  \frac{hc}{ k_{\rm B} T_{\rm S} \lambda }  } - 1 },
\end{equation}

the parameter $ c $ being the speed of light in vacuum, $ h $ the Planck constant, $ k_{\rm B} $ the Boltzmann constant, $ d $ the star-planet distance, $ R_{\rm S} $ the radius of the star and $ T_{\rm S} $ its surface temperature.  The flux is partly reflected by the atmosphere with the albedo $ A_{\rm atm ; \lambda} $ for the wavelenght $ \lambda $, which gives the effective heating flux

\begin{equation}
I_{0;\lambda} \left( \lambda \right) =   \left( 1 - A_{\rm atm ; \lambda} \right) I_{\rm S;\lambda}.
\end{equation}

\begin{figure}[htb]
 \centering
 \includegraphics[width=0.8\textwidth,clip]{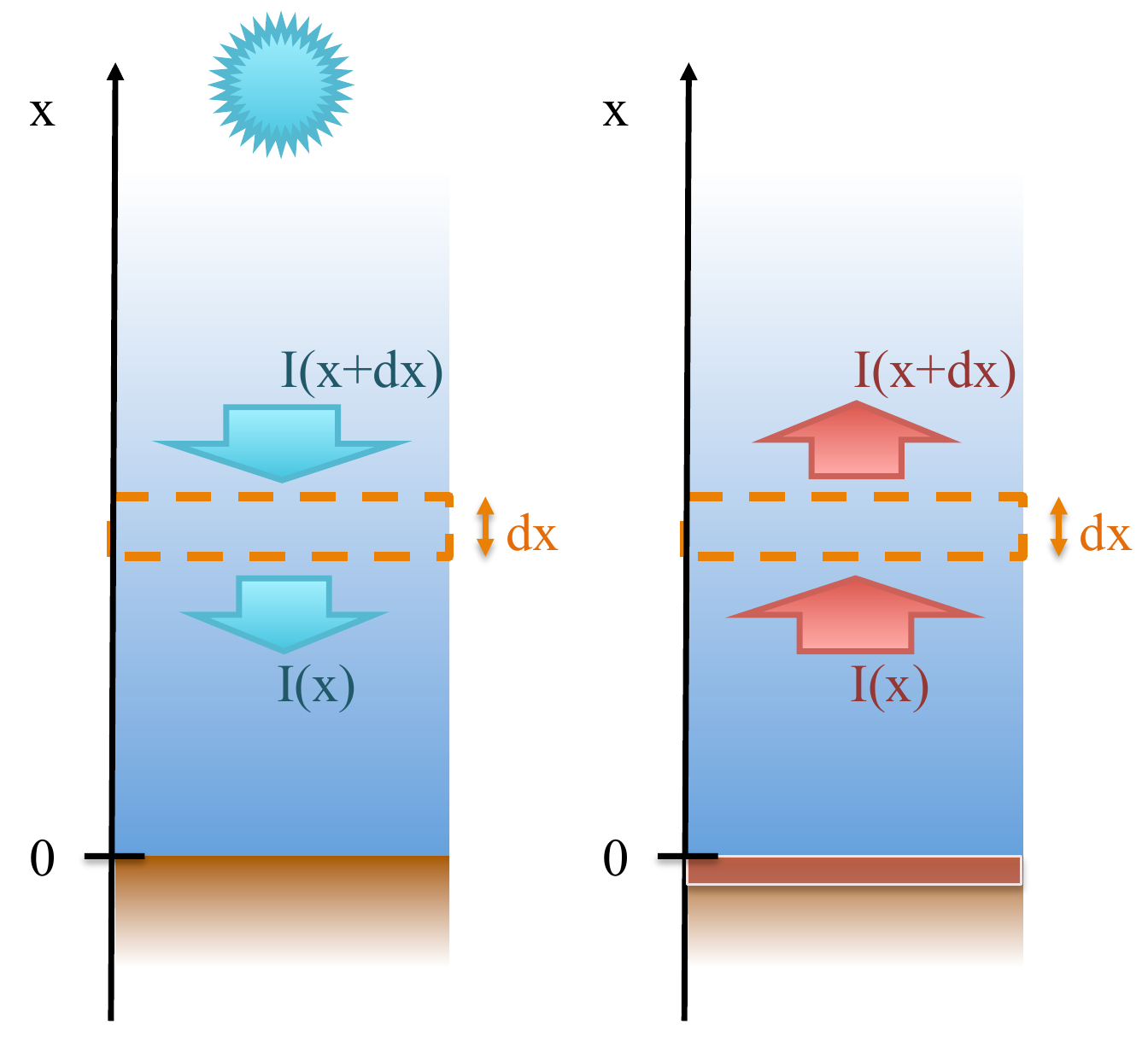}  
  \textsf{ \caption{\label{fig:heat_transfert} {\bf Left:} Absorption of the flux emitted by the star along the star-planet direction. {\bf Right:} Absorption of the flux emitted by the ground.}}
\end{figure}

To describe the absorption of $ I_0 $ by the atmosphere, we use the Beer-Lambert law \citep{Bouguer1729,K1760,Beer1852}. Light rays propagate along a straight line defined by the linear spatial coordinate $ l $. So, the absorption of the incident spectral density $ I^{\rm inc}_\lambda $ by a gas $ G $ of molar concentration $ C_{\rm G} $, illustrated by Fig.~\ref{fig:heat_transfert}, can be written

 \begin{equation}
\dfrac{d I^{\rm inc}_\lambda}{I^{\rm inc}_\lambda} = -  \varepsilon_{{\rm G};\lambda}  C_{\rm G} \left( l \right) dl,
\label{dIinc}
\end{equation}

the parameter $ \varepsilon_{{\rm G};\lambda} $ being the molar extinction density coefficient of the gas, supposed to depend on the light wavelength ($ \lambda $) only. One shall take this dependence into account because $ \varepsilon_{{\rm G};\lambda} $ can vary over several orders of magnitude; typically, $ \varepsilon_{{\rm G};\lambda} \, \propto \, \lambda^{-4} $ in the regime of Rayleigh diffusion (molecules are small compared to the wavelength). From Eq.~(\ref{dIinc}), we get

\begin{equation}
I^{\rm inc}_{\lambda} = I_{0;\lambda} e^{- \varepsilon_{{\rm G};\lambda} \int_l C_{\rm G} dl},
\label{Iinc_lambda}
\end{equation}

and the incident flux

\begin{equation}
I^{\rm inc} = \int_\lambda I^{\rm inc}_\lambda d \lambda.
\end{equation}

Then, derivating Eq.~(\ref{Iinc_lambda}), the heat power per unit volume is obtained,

\begin{equation}
\mathscr{J}^{\rm inc} = C_{\rm G} \int_\lambda \varepsilon_{{\rm G};\lambda}  I_{0;\lambda} e^{- \varepsilon_{{\rm G};\lambda} \int_l C_{\rm G} {\rm dl}} d \lambda,
\label{Jvolinc}
\end{equation}

and the heat power per unit mass comes straightforwardly

\begin{equation}
J^{\rm inc} = \frac{C_{\rm G}}{\rho_0}  \int_\lambda \varepsilon_{{\rm G};\lambda}   I_{0;\lambda} e^{- \varepsilon_{{\rm G};\lambda} \int_l C_{\rm G} dl} d \lambda.
\label{Jinc}
\end{equation}

We retrieve the formulation of the thermal forcing proposed by CL70. Considering that the gaz is homogeneous in composition leads us to $ C_{\rm G} = C_0 = \rho_0 / M $. Therefore, the incident power flux and heat power per unit mass become

\begin{equation}
\left\{
\begin{array}{l}
    \displaystyle I^{\rm inc} = \int_\lambda I_{0;\lambda} e^{- \varepsilon_{\lambda} \int_l C_0 dl} d \lambda, \\[0.5cm]
    \displaystyle J^{\rm inc} = \frac{1}{M}  \int_\lambda \varepsilon_{\lambda}   I_{0;\lambda} e^{- \varepsilon_{\lambda} \int_l C_0 dl} d \lambda.
\end{array}
\right.
\label{IJinc}
\end{equation}

Near the planet-star axis, they are simply expressed

\begin{equation}
\left\{
\begin{array}{l}
    \displaystyle I^{\rm inc} = \int_\lambda I_{0;\lambda} e^{- \tau_\lambda e^{-x}} d \lambda, \\[0.5cm]
    \displaystyle J^{\rm inc} = \frac{1}{M}  \int_\lambda \varepsilon_{\lambda}   I_{0;\lambda} e^{- \tau_\lambda e^{-x}} d \lambda,
\end{array}
\right.
\label{IJ_inc}
\end{equation}

where $ \tau_\lambda $ is the \pcor{dimensionless optical depth for} the radiance of wavelength $ \lambda $,

\begin{equation}
\tau_\lambda = \frac{\rho_0 \left( 0 \right) H }{M} \varepsilon_\lambda.
\label{coeff_absorb}
\end{equation}

\begin{figure*}[ht!]
 \centering
 \includegraphics[width=0.42\textwidth,clip]{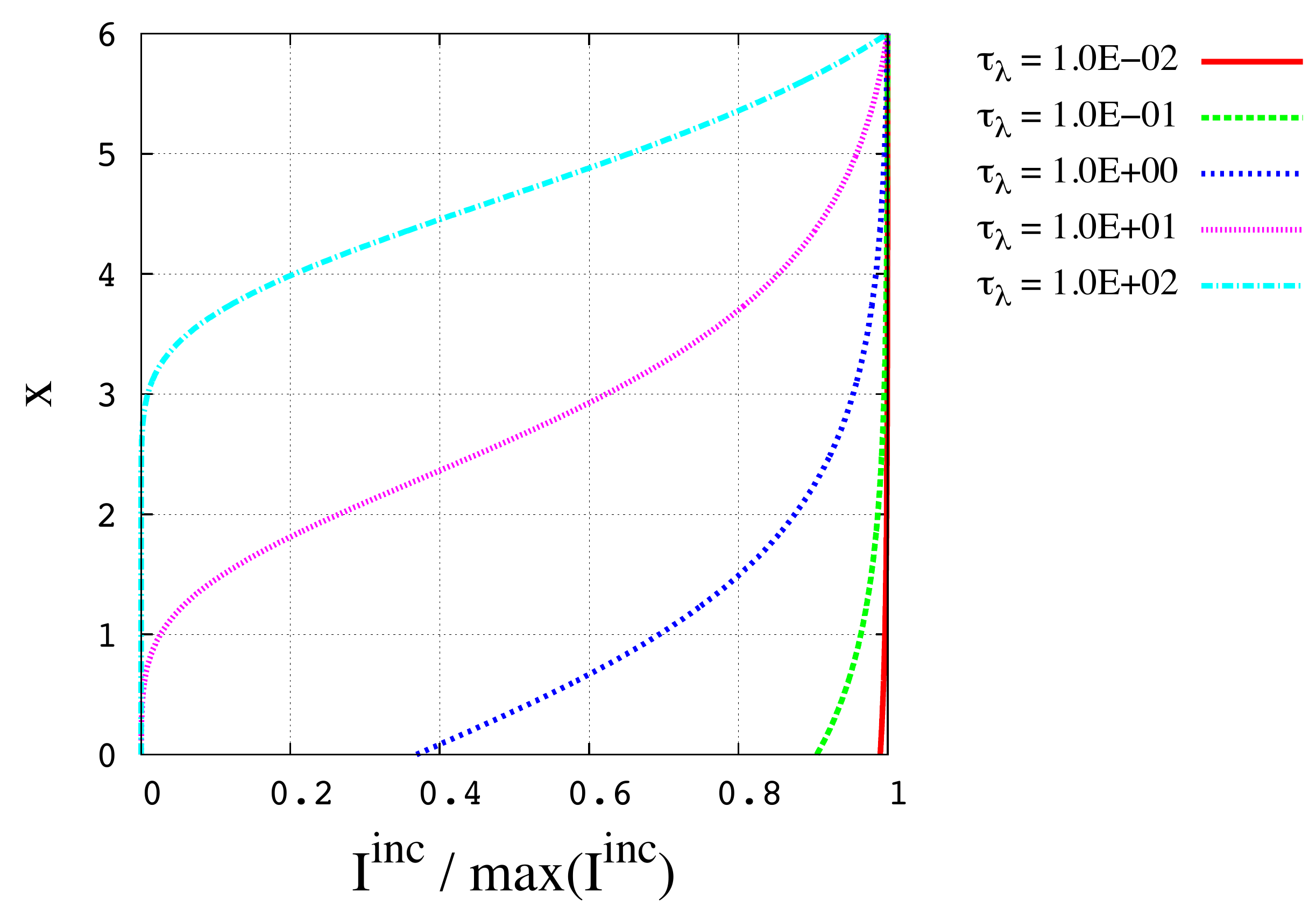} \hspace{1cm}      
 \includegraphics[width=0.42\textwidth,clip]{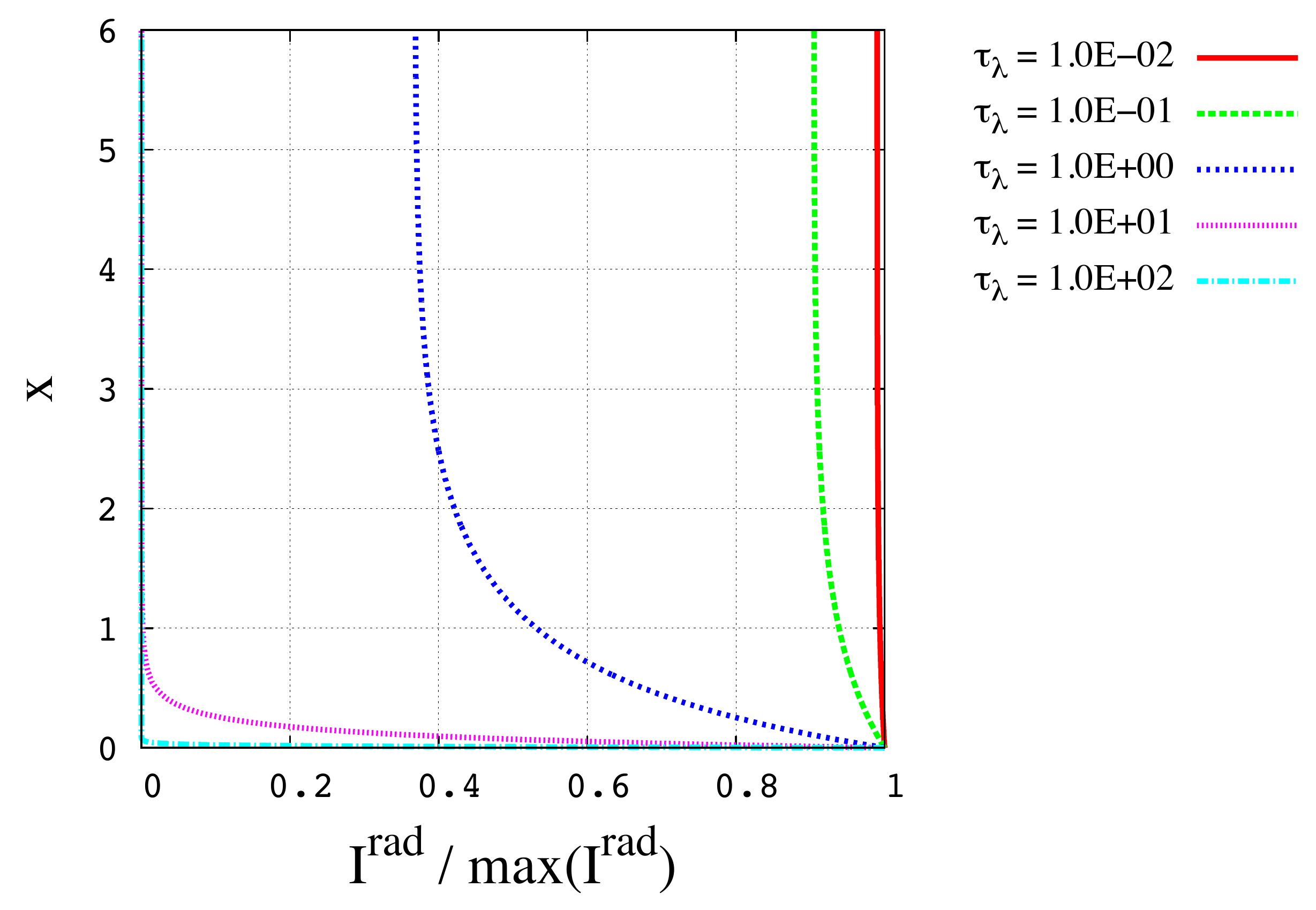} \\
  \includegraphics[width=0.42\textwidth,clip]{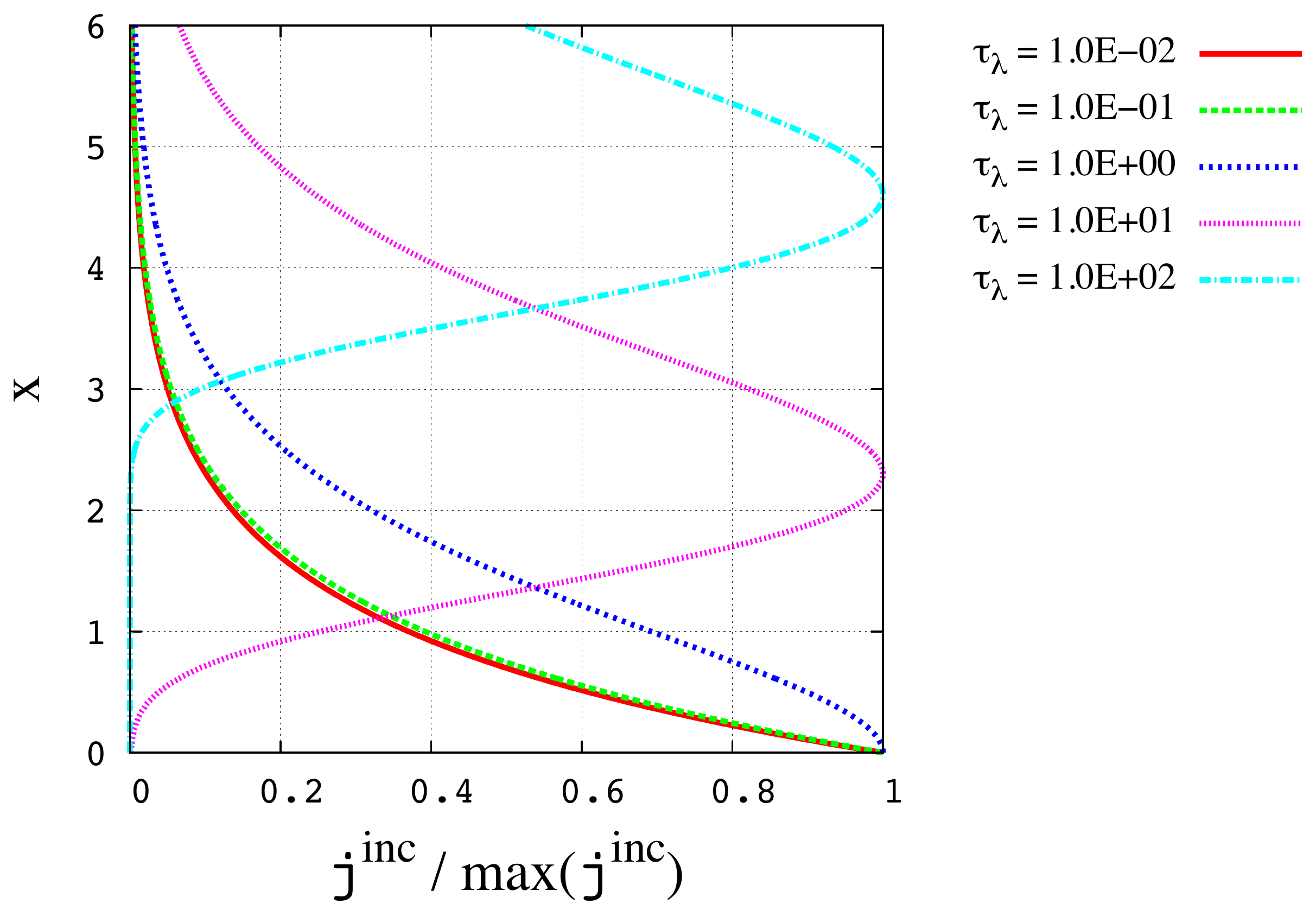} \hspace{1cm}      
 \includegraphics[width=0.42\textwidth,clip]{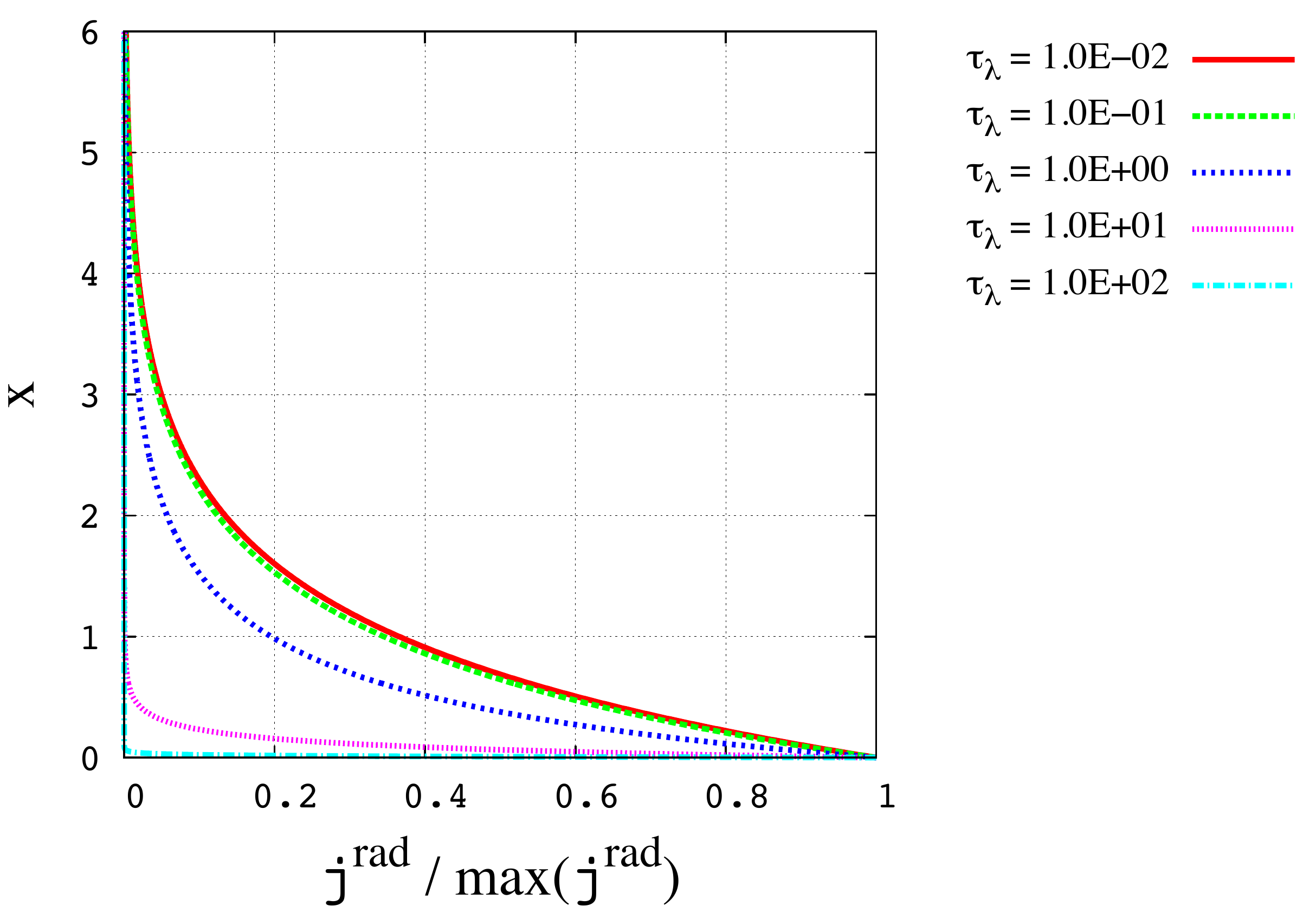} \\
  \includegraphics[width=0.42\textwidth,clip]{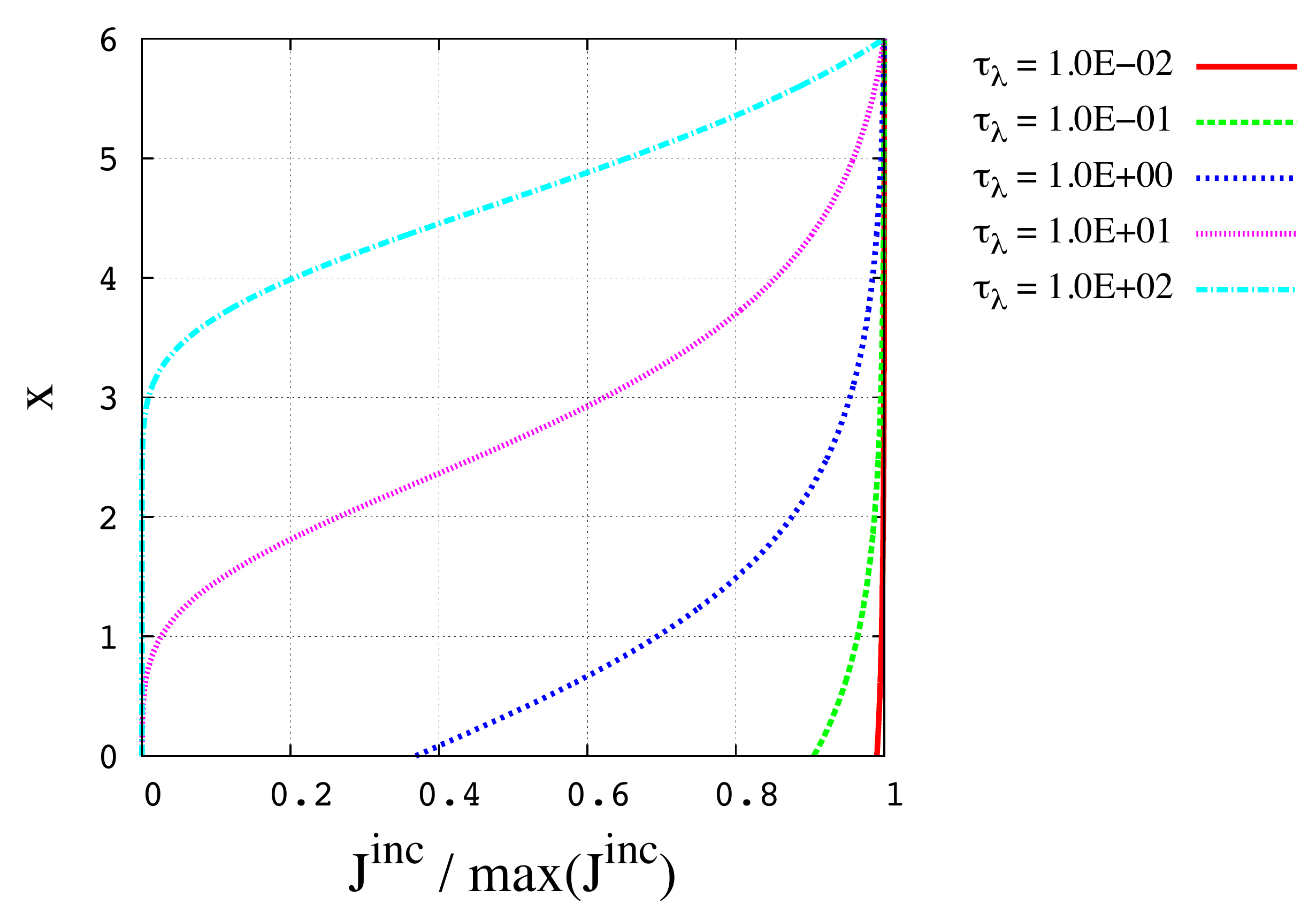} \hspace{1cm}     
 \includegraphics[width=0.42\textwidth,clip]{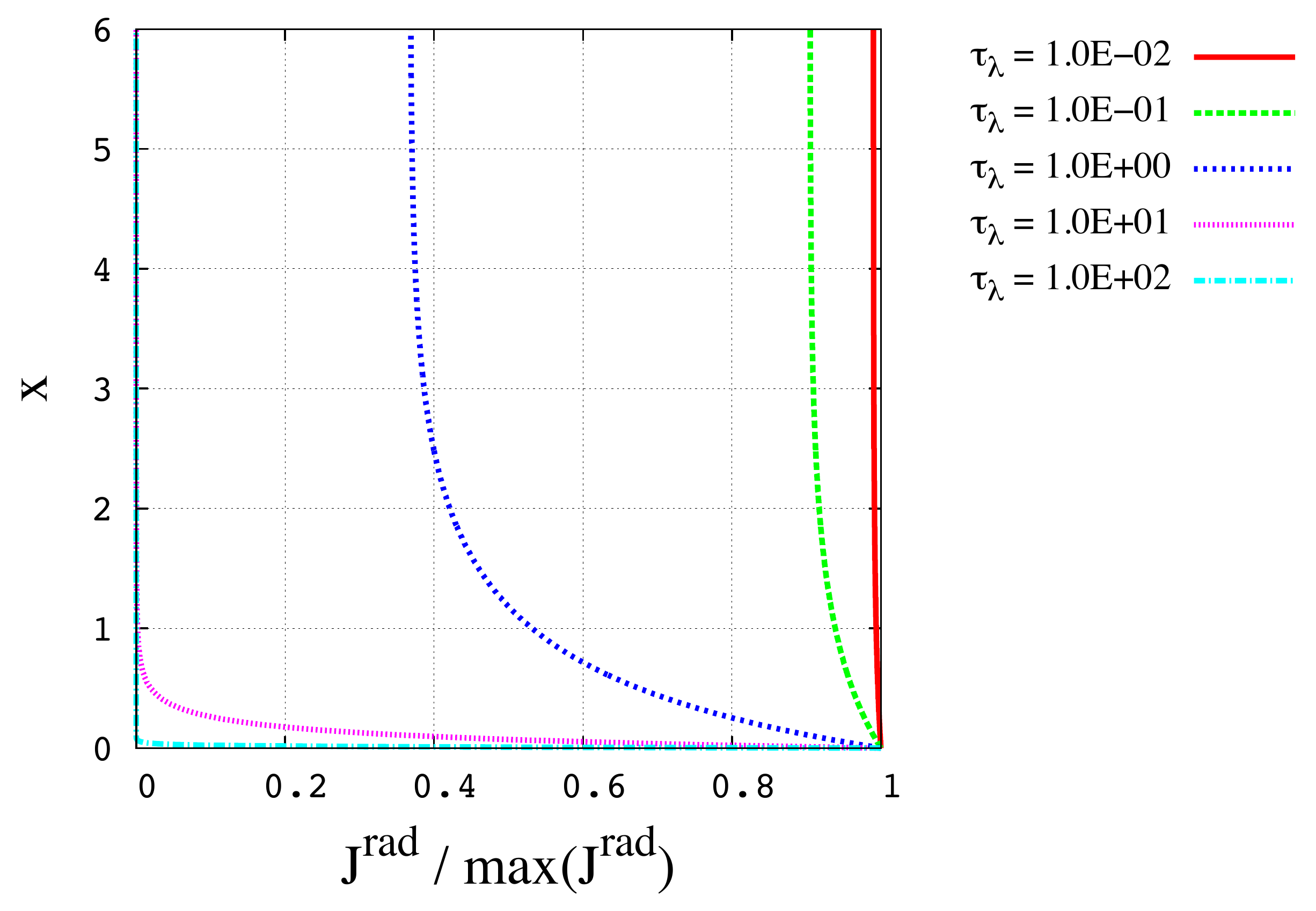} 
  \textsf{ \caption{\label{fig:profils} Vertical profiles of radiative fluxes ($ I $), power per unit volume ($ \mathscr{J} $) and power per unit mass ($ J $) for various values of the \pcor{dimensionless optical depth} $ \tau_{\lambda} $ (Eq.~\ref{coeff_absorb}). As usual, the reduced altitude, $ x = z / H $, is on the vertical axis. {\bf Left:} Incident flux, defined by Eq.~(\ref{IJ_inc}), at $ \psi = 0 $. If $ \tau_\lambda \ll1 $ the ground receives almost $ 100 \ \% $ of the total power. On the contrary, if $ \tau_\lambda \gg 1 $, all the power is absorbed by the atmosphere. {\bf Right:} Radiative flux emitted by the ground, defined by Eq.~(\ref{IJrad}). For small $ \tau_\lambda $, the flux goes trough the atmosphere without being absorbed. For high values of the parameter, it is absorbed by the lowest layers. }}
\end{figure*}

The parameter $ 1 / \tau_\lambda $ corresponds to an absorption depth of the flux in a homogeneous fluid normalized by the pressure height scale ($ H $). If $ \tau_\lambda \ll 1 $, the heating power of a wavelength $ \lambda $ is almost totally transmitted to the ground. On the contrary, if $ \tau_\lambda \gg 1 $, the flux does not reach the surface because it is entirely absorbed by the atmosphere (see Fig.~\ref{fig:profils}). In this case, the atmosphere is not optically thin, one cannot assume that its radiative losses are proportional to $ \delta T $ as done in Sect.~2, and thermal diffusion has to be considered. Therefore, the case treated by the present work corresponds to $ \tau_\lambda \ll 1 $. \\

The proportion of flux density transmitted to the ground $ \alpha_\lambda $ is deduced from Eq.~(\ref{IJ_inc}) taken at $ x = 0 $,

\begin{equation}
\alpha_\lambda = e^{- \tau_\lambda}.
\end{equation}

Like the atmosphere, the ground has an albedo denoted $ A_{\rm gr;\lambda} $, which implies that the upcoming flux is partly reflected. Therefore, we can write the absorption equation

\begin{equation}
\dfrac{d I^{\rm ref}_\lambda}{dx}  +  \tau_\lambda e^{-x} I^{\rm ref}_\lambda = 0,
\end{equation}

and compute the reflected surface power

\begin{equation}
I^{\rm ref}_\lambda  = A_{\rm gr;\lambda} \alpha_\lambda I^{\rm inc}_\lambda \left( R , \psi \right) \cos \psi e^{\tau_\lambda e^{-x} },
\end{equation}

the coordinate $ \psi $ being the angle between the star-planet direction and the position vector of the current point. The heat power per unit mass and wavelength provided by the reflected flux is derived from the previous equation

\begin{equation}
J_{\lambda}^{\rm ref}  = A_{\rm gr ; \lambda} \alpha_\lambda \frac{\varepsilon_\lambda}{M} I^{\rm inc}_\lambda \left( R , \psi \right) \cos \psi  e^{\tau_\lambda e^{-x}}.
\end{equation}

So, the contribution of the reflected flux is finally obtained

\begin{equation}
\left\{
\begin{array}{l}
   I^{\rm ref} = \displaystyle \int_\lambda A_{\rm gr;\lambda} \alpha_\lambda I^{\rm inc}_\lambda \left( R , \psi \right) \cos \psi e^{\tau_\lambda e^{-x} }  {\rm d \lambda} \\[0.5cm]
    J^{\rm ref}  = \displaystyle \int_\lambda A_{\rm gr ; \lambda} \alpha_\lambda \frac{\varepsilon_\lambda}{M} I^{\rm inc}_\lambda \left( R , \psi \right) \cos \psi  e^{\tau_\lambda e^{-x}} {\rm d \lambda}.
\end{array}
\right.
\end{equation}

\subsection{Radiative heating from the ground}

The part of the incident flux which is not reflected causes surface temperatures oscillations. This effect has been studied in meteorological works along the twentieth century. We will follow here the theoretical approach proposed by \cite{Bernard1962} for its simplicity and the physical landmarks that it brings. The ground is considered as a black body of temperature $ T_{\rm gr} $, emitting the power flux $ I_{\rm gr ; BB} $ given by the Stefan-Boltzmann law,

\begin{equation}
I_{\rm gr ; BB} = \epsilon_{\rm gr} \mathscr{S} T_{\rm gr}^4,
\end{equation}

the parameter $ \epsilon_{\rm gr} $ being the emissivity of the ground and $ \mathscr{S} = 5.670373 \times 10^{-8} \ {\rm W.m^{-2}.K^{-4} } $ the Stefan-Boltzmann constant \citep[][]{codata2010}. Moreover, the atmosphere emits a counter radiation to the surface, which implies that the effective flux is less than $ I_{\rm gr ; BB} $. According to \cite{Bernard1962}, this counter-radiation can be assumed to be proportional to $ I_{\rm gr;BB} $, and expressed by the semi-empirical formula

\begin{equation}
I_{\rm atm} = \epsilon_{\rm atm} \mathscr{S} T_{\rm gr}^4,
\end{equation}

which allows us to take it into account without studying the whole coupled system ground-atmosphere. The factor $ \epsilon_{\rm atm} $ corresponds to an effective emissivity. So, introducing the spectral radiance of the atmosphere $ I_{\rm atm ; \lambda} $, the albedo of the ground in the infrared $ A_{\rm gr ; IR} $ can be defined by the relationship

\begin{equation}
A_{\rm gr ; IR} I_{\rm atm} = \int_\lambda A_{\rm gr;\lambda} I_{\rm atm;\lambda} d \lambda,
\end{equation}

and the effective flux emitted by the telluric surface can be written

\begin{equation}
I_{\rm gr} = I_{\rm gr;BB} - A_{\rm gr}^{IR} I_{\rm atm}.
\label{Ieff_1}
\end{equation}

Given that $ A_{\rm gr}^{IR} \approx 1 - \epsilon_{\rm gr} $, Eq.~(\ref{Ieff_1}) becomes

\begin{equation}
I_{\rm gr}^{\rm rad} = \epsilon \mathscr{S} T_{\rm gr}^4,
\label{Ieff}
\end{equation}

where $ \epsilon = \epsilon_{\rm gr} \left( 1 - \epsilon_{\rm atm} \right) \leq 1 $ is the effective emissivity of the ground. The terrestrial atmosphere being rather thin and transparent, $ \epsilon \approx 1 $ for the Earth. This expression will be used further to establish the heating by turbulent diffusion. We also introduce here the corresponding spectral radiance of the ground,

\begin{equation}
I_{\rm gr ; \lambda}^{\rm rad} = \epsilon \frac{2 h c^2 \pi \lambda^{-5}}{ e^{hc/ \left( k_{\rm B} \lambda T_{\rm gr} \right)  } - 1 }. 
\end{equation}

The radiative flux emitted by the ground, denoted $ I^{\rm rad} $, is solution of the absorption equation (similar to the one used to compute $ I^{\rm inc} $),

\begin{equation}
\dfrac{d I_\lambda^{\rm rad}}{ dx} + \tau_\lambda e^{-x} I_\lambda^{\rm rad} = 0,
\end{equation}

with the boundary condition $ I_\lambda^{\rm rad} = I_{\rm gr ; \lambda}^{\rm rad} $ at $ x = 0 $. Therefore, 

\begin{equation}
I_\lambda^{\rm rad} = \alpha_\lambda e^{\tau_\lambda e^{-x}} I_{\rm gr ; \lambda}^{\rm rad},
\end{equation}

and finally,

\begin{equation}
I^{\rm rad} = \int_\lambda I_{\rm gr ; \lambda}^{\rm rad} \alpha_\lambda e^{\tau_\lambda e^{-x}} d \lambda.
\end{equation}

The temperature of the ground $ T_{\rm gr} $ depends on the power reaching the surface,

\begin{equation}
I_{\rm gr}^{\rm inc} = I^{\rm inc} \left( R , \psi \right) \cos \psi.
\end{equation}

Both quantities are linearized near the equilibrium, $ I_{\rm gr}^{\rm inc} = I_{\rm gr ; 0}^{\rm inc} + \delta I_{\rm gr }^{\rm inc} $ and $ T_{\rm gr} = T_{\rm gr ; 0} + \delta T_{\rm gr} $, and the perturbation is expanded in series

\begin{equation}
\begin{array}{l}
   \displaystyle \delta I_{\rm gr}^{\rm inc} = \sum_{\sigma , m} \delta I_{\rm gr}^{{\rm inc ;} m,\sigma} e^{i \left( \sigma t + m \varphi \right)}, \\[0.5cm]
   \displaystyle \delta T_{\rm gr} = \sum_{\sigma ,m} \delta T_{\rm gr}^{m,\sigma} e^{i \left( \sigma t + m \varphi \right)}.
\end{array}
\end{equation}

The spatial functions $ \delta I_{\rm gr}^{{\rm inc ;} m,\sigma} $ and $ \delta T_{\rm gr}^{m,\sigma} $ are related by a transfer function of the tidal frequency, denoted $ \mathscr{B}_{\rm gr} $, that will be given explicitly in the next subsection

\begin{equation}
\delta T_{\rm gr}^{m,\sigma} = \mathscr{B}_{\rm gr}^{\sigma}  \delta I_{\rm gr}^{{\rm inc ;} m,\sigma}.
\label{deltaTgr}
\end{equation}

So, the perturbation of the spectral radiance emitted by the ground can be written

\begin{equation}
\delta I_\lambda^{{\rm rad ;} m,\sigma} = \alpha_\lambda e^{\rm \tau_\lambda e^{-x}} \left. \dfrac{\partial I_{\rm gr ; \lambda}^{\rm rad}}{\partial T_{\rm gr}} \right|_{T_{\rm gr} = T_{\rm gr;0}}  \mathscr{B}_{\rm gr}^{\sigma} \delta I_{\rm gr}^{{\rm inc ;} m,\sigma},
\end{equation}

and the flux and heat power per mass unit,

\begin{equation}
\left\{
\begin{array}{l}
   \displaystyle \delta I^{{\rm rad ;} m,\sigma} = \mathscr{B}_{\rm gr}^{\sigma} \delta I_{\rm gr}^{{\rm inc ;} m,\sigma} \int_\lambda  \alpha_\lambda e^{\rm \tau_\lambda e^{-x}} \left. \dfrac{\partial I_{\rm gr ; \lambda}^{\rm rad}}{\partial T_{\rm gr}} \right|_{T_{\rm gr} = T_{\rm gr;0}} d \lambda, \\[0.5cm]
   \displaystyle \delta J^{{\rm rad ;} m,\sigma} =  \frac{\varepsilon_\lambda}{M} \mathscr{B}_{\rm gr}^{\sigma} \delta I_{\rm gr}^{{\rm inc ;} m,\sigma} \int_\lambda  \alpha_\lambda e^{\rm \tau_\lambda e^{-x}} \left. \dfrac{\partial I_{\rm gr ; \lambda}^{\rm rad}}{\partial T_{\rm gr}} \right|_{T_{\rm gr} = T_{\rm gr;0}} d \lambda.
\end{array}
\right.
\label{IJrad}
\end{equation}

The partial derivative of $ I_{\rm gr ; \lambda}^{\rm rad} $ is explicitly given by

\begin{equation}
\left. \dfrac{\partial I_{\rm gr ; \lambda}^{\rm rad}}{\partial T_{\rm gr}} \right|_{T_{\rm gr} = T_{\rm gr;0}} = \epsilon \frac{2 \pi h^2 c^3 \lambda^{-6}}{k_{\rm B} T_{\rm gr ; 0}^2 } \frac{ e^{ hc / \left( k_{\rm B} \lambda T_{\rm gr ; 0}  \right) }  }{ \left[  e^{ hc / \left( k_{\rm B} \lambda T_{\rm gr ; 0}  \right) } - 1 \right]^2 }.
\end{equation}

Here, we note that the response of the ground to the tidal forcing induces a dependence on the tidal frequency through the oscillations of $ T_{\rm gr} $ and the transfer function $ \mathscr{B}_{\rm gr}^{\sigma} $. Thus, the spatial functions describing the contribution of the ground are parametrized by $ \sigma $ contrary to those associated with the incident flux.


\subsection{Boundary layer turbulent heat diffusion}

\begin{figure}[htb]
 \centering
 \includegraphics[width=0.96\textwidth,clip]{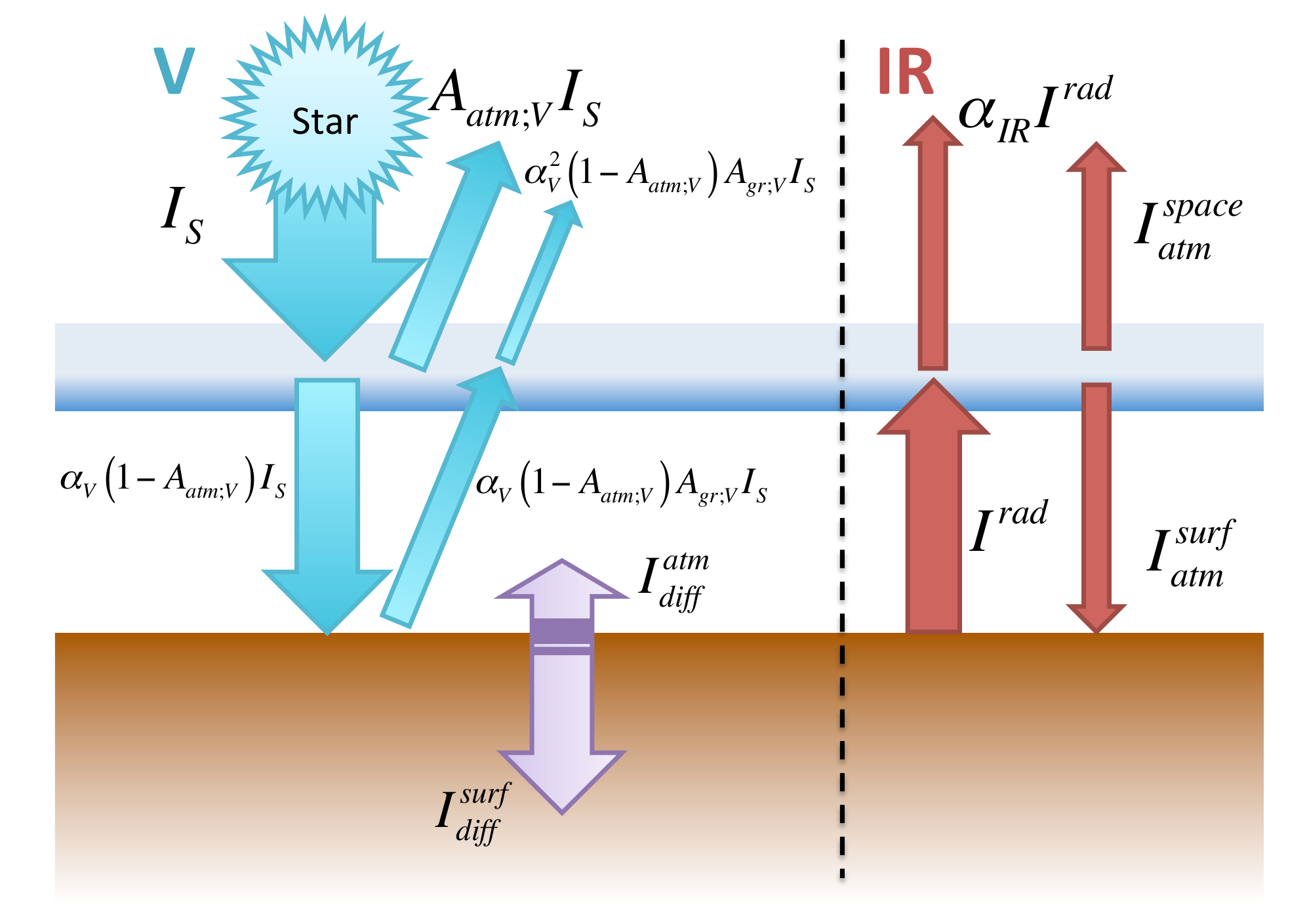} \\     
 \includegraphics[width=0.96\textwidth,clip]{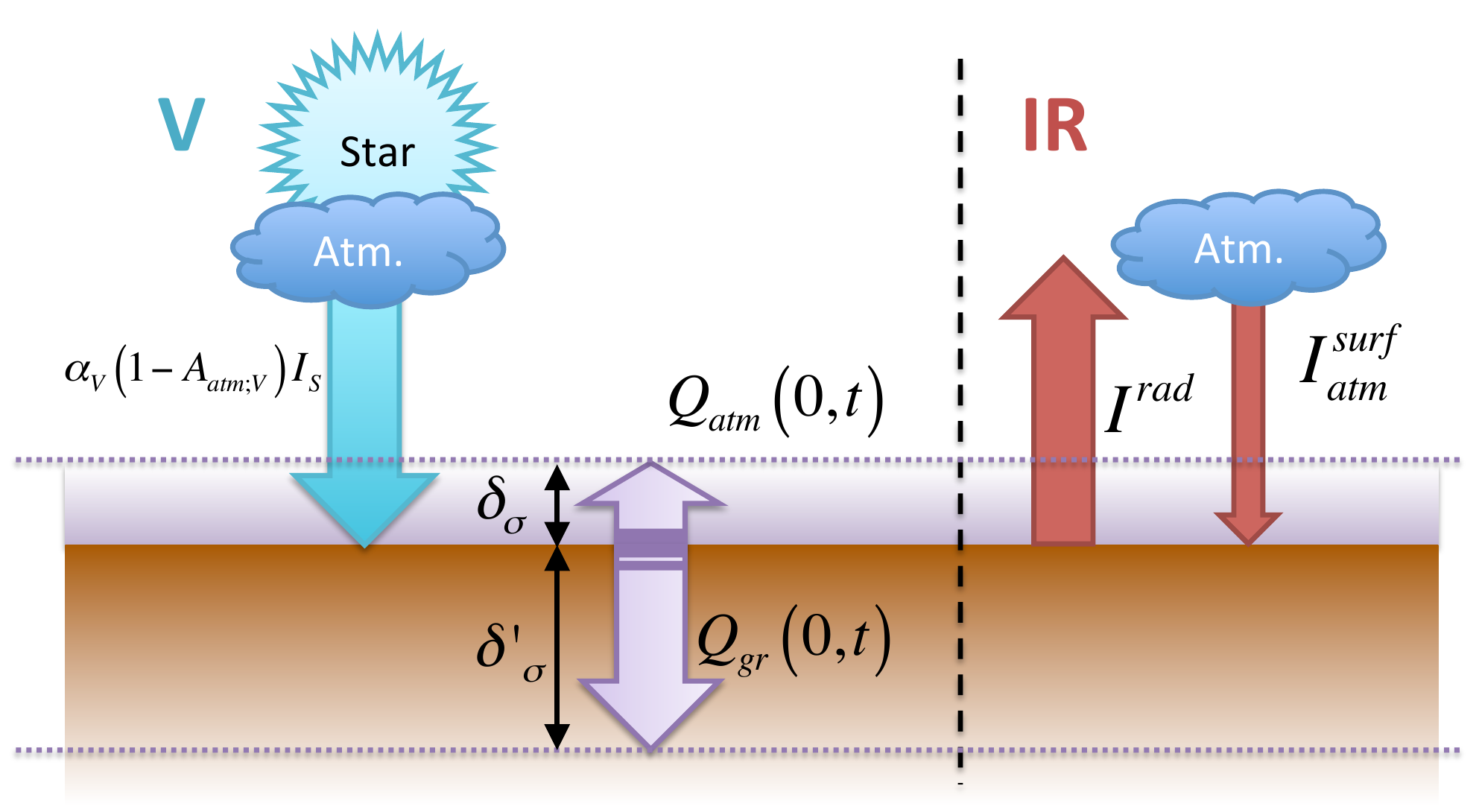} 
  \textsf{ \caption{\label{fig:power_balance} Bichromatic case. {\it Top:} Heat transfers between the atmosphere, space and the ground. {\it Bottom:} Power balance of the interface ground-atmosphere.}}
\end{figure}

Near the ground, heat transfers are dominated by turbulent mechanisms and diffusion. This is the so-called planetary boundary layer, illustrated by Fig.~\ref{fig:power_balance}. Thus, for $ x \ll 1 $, thermal diffusion cannot be ignored like in Sect. 2 and drives the oscillations of temperature. \mybf{However, note that it is only valid at the vicinity of the surface, where friction with the ground is important. After CL70 and \cite{Siebert1961}, we neglect \scor{as a first step} the heat transport through the troposphere due to advection \scor{to provide a first analytical treatment of the problem}. At the interface, the diffusive fluxes in the ground and in the atmosphere write}

\begin{equation}
\begin{array}{lll}
\displaystyle Q_{\rm gr} \left(t \right) = \frac{k_{\rm gr}}{H}  \left(  \dfrac{\partial T}{\partial x} \right)_{x = 0^-} & \! \! \! \! \mbox{and} & \! \!  \displaystyle Q_{\rm atm} \left( t \right) = - \frac{k_{\rm atm}}{H} \left(  \dfrac{\partial T}{\partial x} \right)_{x = 0^+},
\end{array}
\label{powerQ}
\end{equation}

the parameters $ k_{\rm gr} $ and $ k_{\rm atm} $ representing the thermal conductivities of the ground and of the lowest layer of the atmosphere. Besides, denoting the associated thermal capacities $ c_{\rm gr} $ and $ c_{\rm atm} $, we introduce the thermal diffusivities,

\begin{equation}
\begin{array}{ccc}
   K_{\rm gr} = \displaystyle  \frac{k_{\rm gr}}{\rho_{0} \left( 0^- \right) c_{\rm gr} } & \mbox{and} & K_{\rm atm} = \displaystyle \frac{k_{\rm atm}}{\rho_0 \left( 0^+ \right) c_{\rm atm} },
\end{array}
\end{equation}

the parameter $ K_{\rm atm} $ being the vertical turbulent thermal diffusivity. Therefore, temperature variations near the surface are described by the  equations of heat transport \mybf{(e.g. CL70)},

\begin{equation}
\left\{
\begin{array}{lcl}
 \displaystyle  \dfrac{\partial T}{ \partial t } = \frac{K_{\rm gr}}{H^2} \dfrac{\partial^2 T}{\partial x^2} & \mbox{for} & x < 0, \\
   \vspace{0.01mm}\\
 \displaystyle   \dfrac{\partial T}{ \partial t } = \frac{K_{\rm atm} }{H^2} \dfrac{\partial^2 T}{\partial x^2} & \mbox{for} & x > 0, \\ 
\end{array}
\right.
\label{heat_transport_bl}
\end{equation}

where the diffusive terms only are taken into account. The power balance of the surface is deduced from Eqs. (\ref{powerQ}) and (\ref{Ieff}),

\begin{equation}
I_{\rm gr}^{\rm inc} - \epsilon \mathscr{S} T_{\rm gr}^4 - Q_{\rm gr} - Q_{\rm atm} = 0,
\label{bilan}
\end{equation}

and provides the temperature $ T_{\rm gr;0} $ of the ground at the equilibrium. Linearized with variables of the form given by Eq.~(\ref{perturbation}), Eq.~(\ref{heat_transport_bl}) can be written

\begin{equation}
\left\{
\begin{array}{lcl}
  \delta T^{m,\sigma} = \mathscr{B}_{\rm gr}^{\sigma}  e^{ \left[ 1 + {\rm sign} \left( \sigma \right) i \right]  x / \delta_{\rm gr}^{\sigma}  } \delta I_{\rm gr}^{\rm inc ; {m,\sigma}}  & \mbox{for} & x < 0, \\
   \vspace{0.01mm}\\
  \delta T^{m,\sigma} = \mathscr{B}_{\rm gr}^{\sigma} e^{- \left[ 1 + {\rm sign} \left( \sigma \right) i \right]  x / \delta_{\rm atm}^\sigma} \delta I_{\rm gr}^{\rm inc ; {m,\sigma}} & \mbox{for} & x > 0, \\ 
\end{array}
\right.
\label{deltaTbl}
\end{equation}

where $ \delta_{\rm gr}^\sigma $ and $ \delta_{\rm atm}^\sigma $ are the tidal frequency-dependent skin thicknesses of diffusive heat transport in the solid and fluid parts respectively. Their expressions are given by

\begin{equation}
\begin{array}{ccc}
\displaystyle  \delta_{\rm gr}^{\sigma} = \frac{1}{H} \sqrt{ \displaystyle \frac{2 K_{\rm gr}}{ \left| \sigma \right|} }  & \mbox{and} & \displaystyle \delta_{\rm atm}^{\sigma} = \frac{1}{H} \sqrt{ \displaystyle \frac{2 K_{\rm atm}}{\left| \sigma \right|} }.
\end{array}
\label{delta_skin}
\end{equation}

\begin{figure}[htb]
\centering
{\includegraphics[width=0.95\textwidth]
{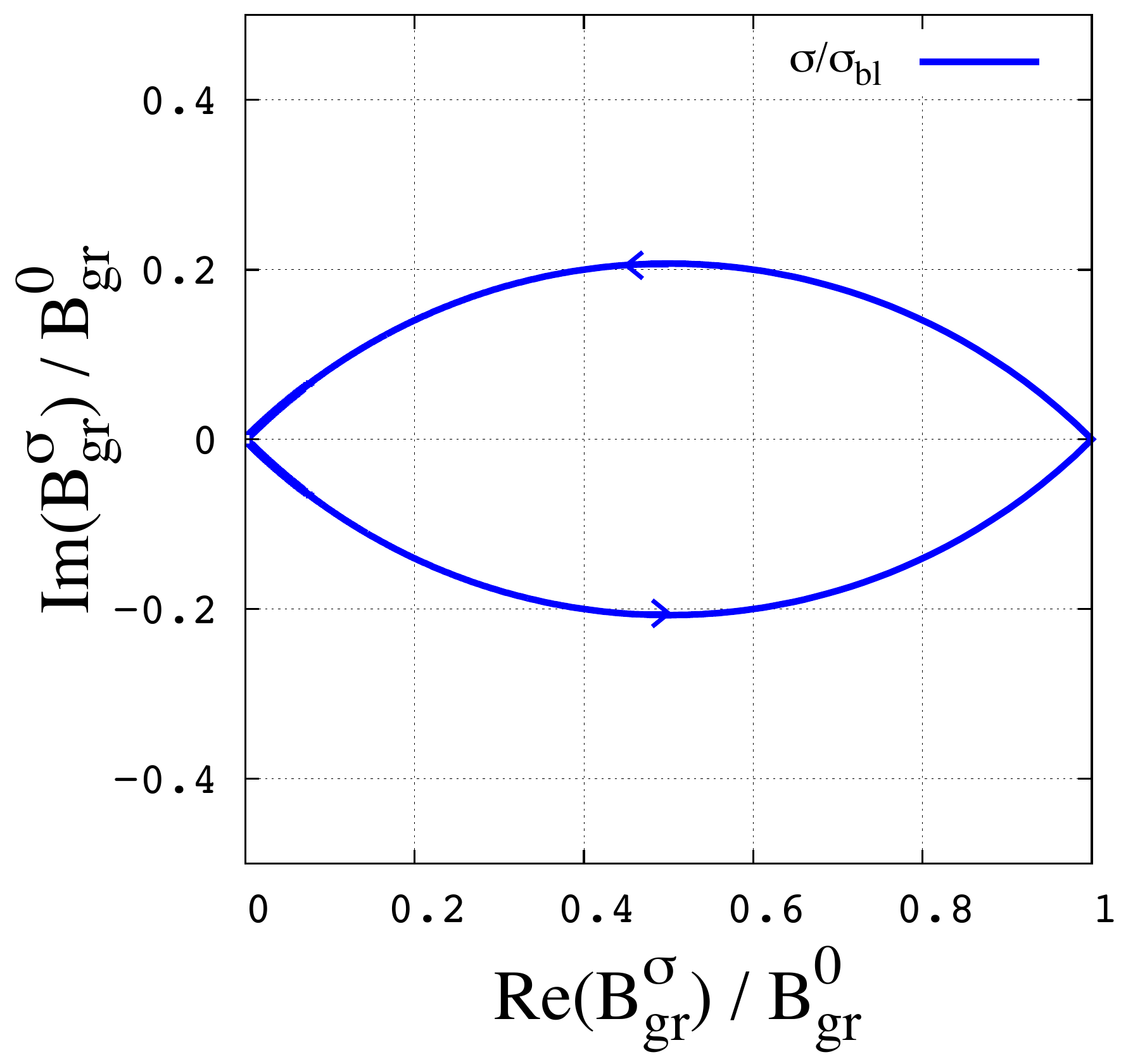}
\textsf{\caption{\label{fig:Bgr} Nyquist diagram of the boundary layer transfer function $ \mathscr{B}_{\rm gr}^{\sigma} $ in the frequency range $ -10^3 \leq \sigma/\sigma_{\rm bl} \leq 10^3 $ ($ \sigma $ being the tidal frequency and $ \sigma_{\rm bl} $ the frequency characterizing the boundary layer introduced in Eq.~\ref{sigmabl}). The parametrized complex transfer function $  \mathscr{B}_{\rm gr}^{\sigma} $ (Eq.~\ref{Bgr}) reduced by its value at $ \sigma = 0 $ is plotted as a function of the tidal frequency ($ \sigma $) in the complex plane, where the horizontal axis corresponds to the real part of the function, $ \Re \left\{ \mathscr{B}_{\rm gr}^{\sigma}  \right\} / \mathscr{B}_{\rm gr}^{0} $, and the vertical axis to the imaginary part $ \Im \left\{ \mathscr{B}_{\rm gr}^{\sigma}  \right\} / \mathscr{B}_{\rm gr}^{0} $. The arrow indicates the direction of increasing $ \sigma $. The gain of the transfert function is given by the distance of a point $ P $ of the curve to the origin $ O $ of coordinates $ \left( 0 , 0 \right) $, and its phase by the angle of the  vector $ \textbf{OP} $ to the horizontal axis. This plot shows that the response of the ground is always delayed with respect to the thermal forcing except at synchronization, where $ \arg \left( \mathscr{B}_{\rm gr} \right) = 0 $ and the gain is maximal. It also highlights the two identified regimes of the ground response: the lag tends to $ 0 $ (response in phase with the perturbation) and the gain increases at $ \sigma \rightarrow 0 $, while the lag tends to $ \pm \pi/4 $ and the gain to $ 0 $ at $ \sigma \rightarrow \pm \infty $, the critical transition frequency being $ \sigma_{\rm bl} $.  } }}
\end{figure}

Note that $  \delta_{\rm gr}^{\sigma}  $ and $ \delta_{\rm atm}^{\sigma} $ both decay when the tidal frequency increases. Since $ \delta_{\rm atm}^\sigma $ corresponds to the typical depth of the boundary layer, it has to satisfy $ \delta_{\rm atm}^\sigma \ll 1 $. Otherwise, the diffusive effects could not be ignored in the dynamical core of the model (Sect. 2). Let us denote $ \varsigma_{\rm gr} = 4 \mathscr{S} T_{\rm gr ; 0}^3 $ the radiative impedance of the ground considered as a perfect black body and submitted to a perturbation in temperature. The transfer function $ \mathscr{B_{\rm gr}^{\sigma}} $ introduced in Eq.~(\ref{deltaTgr}) can be deduced from the linearized power balance,

\begin{equation}
\delta I_{\rm gr}^{{\rm inc} ; {m,\sigma}}  - \epsilon \varsigma_{\rm gr} \delta T_{\rm gr}^{m,\sigma} - \frac{k_{\rm gr}}{H} \left. \dfrac{\partial \delta T^{m,\sigma}}{\partial x} \right|_{0^-}  + \frac{k_{\rm atm}}{H} \left.  \dfrac{\partial \delta T^{m,\sigma}}{\partial x} \right|_{ 0^+} = 0,
\end{equation}

in which the expressions of Eq.~(\ref{deltaTbl}) are substituted. We finally get (see Fig.~\ref{fig:Bgr})

\begin{equation}
\mathscr{B}_{\rm gr}^{\sigma} = \frac{1}{\epsilon \varsigma_{\rm gr}} \frac{1}{  1 + \left[ 1 + {\rm sign} \left( \sigma \right) i  \right] \sqrt{\frac{ \left| \sigma \right| }{\sigma_{\rm bl}}} },
\label{Bgr}
\end{equation}

expression that can be decomposed in its modulus and argument

\begin{equation}
\left\{
\begin{array}{l}
   \displaystyle \left| \mathscr{B}_{\rm gr}^{\sigma} \right| = \frac{1}{ \epsilon \varsigma_{\rm gr} \sqrt{ 1 + 2 \sqrt{\frac{ \left| \sigma \right| }{\sigma_{\rm bl}}} + 2 \frac{ \left| \sigma \right| }{\sigma_{\rm bl}} }}, \\[0.5cm]
    \displaystyle \arg \left( \mathscr{B}_{\rm gr}^{\sigma} \right) =  - {\rm sign} \left( \sigma \right) \arctan \left[ \frac{1}{1 + \sqrt{\frac{ {\sigma_{\rm bl}}}{\left| \sigma \right|}}  } \right]. 
\end{array}
\right.
\label{mod_arg_B}
\end{equation}

 The frequency $ \sigma_{\rm bl} $, which parametrizes $ \mathscr{B}_{\rm gr}^{\sigma}  $, is a characteristic frequency reflecting the thermal properties of the diffusive boundary layer, and is expressed

\begin{equation}
\sigma_{\rm bl} = 2 \left(  \frac{\epsilon \varsigma_{\rm gr}}{ \beta_{\rm atm} + \beta_{\rm gr} }  \right)^2,
\label{sigmabl}
\end{equation}

where the parameters $ \beta_{\rm gr} $ and $ \beta_{\rm atm} $ are the thermal conductive capacities of the ground and of the atmosphere near the telluric surface

\begin{equation}
\begin{array}{ccc}
  \beta_{\rm gr} = \rho_0 \left( 0^- \right) c_{\rm gr} \sqrt{K_{\rm gr}}  & \mbox{and} & \beta_{\rm atm} = \rho_0 \left( 0^+ \right) c_{\rm atm} \sqrt{K_{\rm atm}}.
\end{array}
\end{equation}

According to \citep{Bernard1962}, turbulent diffusion plays a more important role than thermal conduction in the ground, which means $ \beta_{\rm atm} \gg \beta_{\rm gr} $. However, the turbulent diffusivity $ K_{\rm atm} $ may vary over a large range of sizes of magnitude, taking extremal values above oceans (typically $ \beta_{\rm atm} \sim 10^4 \ {\rm J.m^{-2}.K^{-1}.s^{-1/2}} $). For these reasons, \cite{Bernard1962} prescribes for the Earth the effective mean value $ K_{\rm atm} \sim 10 \ {\rm m^2.s^{-1}} $, which was prescribed before by \cite{Wilkes1949} and leads to $ \sigma_{\rm bl} \sim 10^{-6} \ {\rm s^{-1}} $. \\

As shown by Eq.~(\ref{mod_arg_B}), the frequency ratio $ \left| \sigma \right| / \sigma_{\rm bl} $ determines the angular delay of the ground temperature variations. If $ \left| \sigma \right| \ll \sigma_{\rm bl} $, the response is in phase with the excitation. It corresponds to low conductive capacities. On the contrary, if $ \left|  \sigma \right| \gg \sigma_{\rm bl} $, then the delay tends to its asymptotical limit, $ \pi / 4 $. By using the frequency $ \sigma_{\rm bl} \sim 10^{-6} \ {\rm s^{-1}} $ derived from the prescriptions of \cite{Wilkes1949} and \cite{Bernard1962}, we remark that the Earth Solar diurnal tide belongs to the second category. This explains the position of the diurnal peak observed in surface temperature oscillations, three hours late with respect to the subsolar (midday) point. The gain of $ \mathscr{B}_{\rm gr}^{\sigma} $ is damped by the conductive capacities of the material at the interface. Therefore, a very diffusive interface tends to attenuate the temperature oscillations of the ground, as expected. \\

Now, we consider the profile of temperature variations in the boundary layer obtained in Eq.~(\ref{deltaTbl})

\begin{equation}
\delta T_{\rm bl} \left( x \right) = \mathscr{B}_{\rm gr}^{\sigma} e^{- \left[ 1 + {\rm sign} \left( \sigma \right) i \right] \tau_{\rm atm}^{\sigma} x } \delta I_{\rm gr}^{{\rm inc}; m,\sigma},
\label{dTsn}
\end{equation}

where $ \tau_{\rm atm}^{\sigma} = 1 / \delta_{\rm atm}^\sigma $ is the vertical damping ratio of the diffusive perturbation. CL70 shows that a heat power due to diffusion $ J^{\rm diff} $ can be deduced from the previous expression. Indeed, by writing the heat transport equation:

\begin{equation}
c_{\rm atm} \dfrac{\partial \delta T}{ \partial t} = \delta J^{\rm diff}, 
\end{equation} 

and linearizing it, we compute the expression of the tidal diffusive forcing term,

\begin{equation}
\delta J^{{\rm diff ;} m,\sigma} \left( x \right) = i \sigma c_{\rm atm} \mathscr{B}_{\rm gr}^{\sigma}  e^{- \left[ 1 + {\rm sign} \left( \sigma \right) i \right] \tau_{\rm atm}^{\sigma} x }  \delta I_{\rm gr}^{{\rm inc}; m,\sigma}.
\label{Jdiff}
\end{equation}

\subsection{The simplified bichromatic case}

The expressions of the thermal forcings obtained in the previous section may appear of some complexity because of the dependence of albedos and \mybf{optical depths} on the wavelength. They can be simplified notably if each source is assumed to be monochromatic. In this case, the star emits in the frequency range of visible light, and the planet in the range of infrared. Consequently, the wavelength-dependent $ A_{\rm gr ; \lambda} $, $ A_{\rm atm ; \lambda} $, $ \varepsilon_{\lambda} $, $ \epsilon_{\lambda} $, $ \alpha_\lambda $ and $ \tau_\lambda $ are reduced to the constant coefficients $ A_{\rm gr ; V} $, $ A_{\rm atm ; V} $, $ \varepsilon_{\rm V} $, $ \epsilon_{\rm V} $, $ \alpha_{\rm V} $ and $ \tau_{\rm V} $ in the range of visible light and $ A_{\rm gr ; IR} $, $ A_{\rm atm ; IR} $, $ \varepsilon_{\rm IR} $, $ \epsilon_{\rm IR} $, $ \alpha_{\rm IR} $ and $ \tau_{\rm IR} $ in the infrared. Hence, the bolometric flux is given by

\begin{equation}
I_{\rm S} = \frac{R_{\rm S}^2}{d^2} \mathscr{S} T_{\rm S}^4
\end{equation}

where $ T_{\rm S} $ is the temperature of the star. This flux is partly reflected back to space by the atmosphere, so that the effective flux $ I_0 $ getting through the layer can be written

\begin{equation}
I_0 = \left( 1 - A_{\rm atm ; V}  \right) I_{\rm S}.
\end{equation}

The power of the incident flux $ I^{\rm inc} $ decays along the path of a light beam, absorbed by the gaz. The power absorbed per unit mass is denoted $ J^{\rm inc} $ and the expressions of Eq.~(\ref{IJinc}) become

\begin{equation}
\left\{
\begin{array}{l}
    \displaystyle I^{\rm inc} = I_0 e^{- \varepsilon_{\rm V} \int_l C_0 dl} , \\[0.5cm]
    \displaystyle J^{\rm inc} = \frac{ \varepsilon_{\rm V}}{M}    I_0 e^{- \varepsilon_{\rm V} \int_l C_0 dl}.
\end{array}
\right.
\end{equation}

From $ I^{\rm inc} $, we deduce the flux reflected by the ground $ I^{\rm ref}  $ and the power absorbed per mass unit $ J^{\rm ref}  $, given by

\begin{equation}
\left\{
\begin{array}{l}
    \displaystyle I^{\rm ref} = A_{\rm gr ; V} \alpha_{\rm V} I^{\rm inc} \left( 0 , \psi \right) \cos \psi e^{\tau_{\rm V} e^{-x}} , \\[0.5cm]
    \displaystyle J^{\rm ref} = A_{\rm gr ; V} \alpha_{\rm V} \frac{\varepsilon_{\rm V}}{M} I^{\rm inc} \left( 0 , \psi \right) \cos \psi e^{\tau_{\rm V} e^{-x}} .
\end{array}
\right.
\end{equation}

At the end, the expressions of Eq.~(\ref{IJrad}) are simply replaced by 

\begin{equation}
\left\{
\begin{array}{l}
   \displaystyle \delta I^{{\rm rad ;} m,\sigma} = \epsilon \varsigma_{\rm gr} \mathscr{B}_{\rm gr}^{\sigma} \delta I_{\rm gr}^{{\rm inc ;} m,\sigma}  \alpha_{\rm IR} e^{\rm \tau_{\rm IR} e^{-x}}, \\[0.5cm]
   \displaystyle \delta J^{{\rm rad ;} m,\sigma} =  \frac{\varepsilon_{\rm IR}}{M} \epsilon \varsigma_{\rm gr} \mathscr{B}_{\rm gr}^{\sigma} \delta I_{\rm gr}^{{\rm inc ;} m,\sigma}  \alpha_{\rm IR} e^{ \tau_{\rm IR} e^{-x}}.
\end{array}
\right.
\end{equation}

The expressions of the powers per unit mass of the incident and radiated flux ($ J^{\rm inc} $ and $ J^{\rm rad} $), and the corresponding powers per volume unit ($ \mathscr{J}^{\rm inc} $ and $ \mathscr{J}^{\rm rad} $), are plotted on Fig.~\ref{fig:colormap_heating}. The heating due to the turbulent boundary layer remains unchanged (see Eq.~\ref{Jdiff}). 

\begin{figure}[htb]
\centering
{\includegraphics[width=0.95\textwidth]
{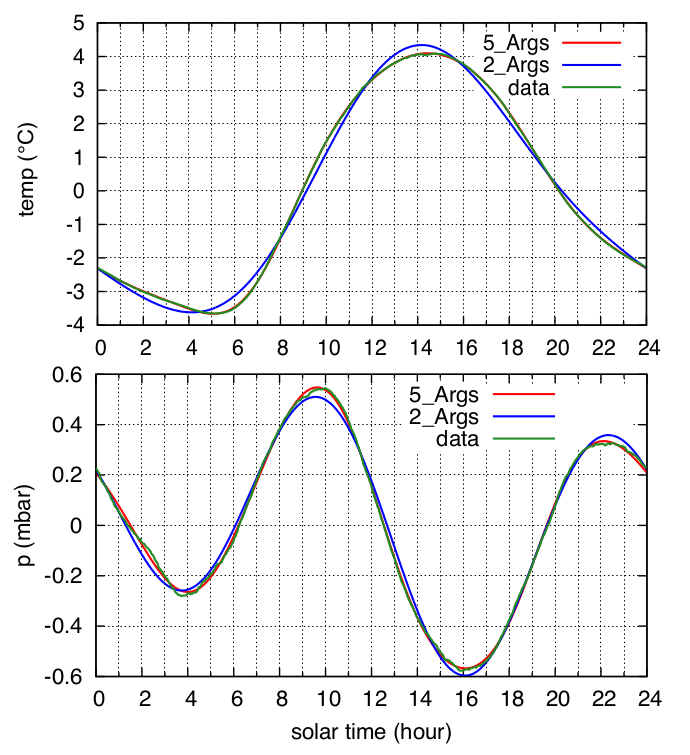}
\textsf{\caption{\label{fig:mesures_pression} Average daily variation of temperature in degree (top) and pressure in mbar (bottom) over one year, plotted over solar time (in hour),  the sub-solar point corresponding to 12h. The data has been recorded every minute over the full year 2013 with a Vantage Pro 2 weather station at latitude $48.363^\circ$N. The time is converted in solar time and the data averaged over an exact count of 365 days. The average value of temperature ($10.8973^\circ$ C) and pressure (1015.83 mbar) has been removed to display only the variable part. In both figures, the raw data (in green) has been decomposed in a Fourier series over the 24 h period, limited to  2 harmonics (blue), or 5 harmonics (red). Temperature is  dominated by the diurnal component, with a maximum value around 14h26mn. The largest harmonics of the pressure variations is the semi-diurnal term, followed by the diurnal term. Two pressure maxima can be observed, at 9h38mn and 22h07mn.} }}
\end{figure}

\section{Application to the Earth and Venus}
\label{sec:application_Earth_Venus}

To \mybf{illustrate the regimes identified and the dependence of the atmospheric response on the forcing frequency}, we apply \mybf{the model} to the cases of the Earth's and Venus semidiurnal thermal tides assuming \scor{first} for both planets isothermal \mybf{stably-stratified} atmospheres, as described by the equations of Sect.~3. \mybf{Let us remind here that the atmosphere of Venus is not stably stratified but convective in regions close to the surface \citep[][]{Seiff1980,Baker2000}, where the tidal mass redistribution is the most important \citep[][]{DI1980,SZ1990}. \scor{\jcor{Nevertheless}, examining the case of a stably-stratified isothermal atmosphere for a Venus-like planet will allow us to better understand in an academic framework the role of stratification and radiative losses in the tidal response.}} \jcor{To make a clear difference between Venus and the studied stably-stratified Venus-like planet, we call the later ``VenusX''.} For the sake of simplicity, we \mybf{assume} that \mybf{planets orbit} in \mybf{their} equatorial plane circularly. Hence, the tidal frequency\footnote{This frequency corresponds to the term defined by $ \left( l , m , j , p , q \right) = \left( 2 , 2 , 2 , 0 ,0  \right) $ in the multipole expansion of the forcings detailed in Appendix~\ref{app:thermal_forcing}.} is given by $ \sigma = 2 \left( \Omega - n_{\rm orb} \right) $. \mybf{One of the most crucial parameters of the model is the Newtonian cooling frequency ($ \sigma_0 $). This parameter varies over several orders of magnitude with the altitude \citep[][]{PY1975}. However, our purpose here is not to compute a \scor{quantitative} tidal perturbation but to illustrate \scor{qualitatively the non-dissipative (Earth) and dissipative (Venus) regimes}. Therefore, following \cite{LM1967}, we consider the interesting case where the radial profile $ \sigma_0 $ is \scor{assumed to be} constant. For both planets, we arbitrarily set $ \sigma_0 = 7.5 \times 10^{-7} \ {\rm s^{-1}} $, which is the effective value of $ \sigma_0 $ \scor{computed} by \cite{Leconte2015} with a GCM for Venus. This value is such that $ \sigma \sim \sigma_0 $ for Venus and $ \sigma \gg \sigma_0 $ for the Earth. In the case of an optically thin atmosphere, the Newtonian cooling frequency can be estimated using Eq.~\ref{sigma0}.} Moreover, we \mybf{impose} academic quadrupolar perturbations of the form 

\begin{equation}
\begin{array}{l} 
   \displaystyle U \left( x, \theta , \varphi , t \right) = U_2 P_2^2 \left( \cos \theta \right) e^{ i \left( \sigma t + 2 \varphi \right)},\\[0.5cm]
   \displaystyle J \left( x, \theta , \varphi , t \right) = J_2 P_2^2 \left( \cos \theta \right) e^{ i \left( \sigma t + 2 \varphi \right)},
\end{array}
\end{equation}

where $ U_2 $ and $ J_2 $ are fixed constant radial profiles. The tidal potential $ U_2 $ is computed from the Kaula's multipole expansion detailed in Appendix D, and explicitly given by the expression of Eq.~(\ref{UJ22200}). A zero-order approximation of the thermal power $ J_2 $ can be quantified by writing the total surface power absorbed by the atmosphere as a function of $ \theta $ and $ \varphi $ which is then expanded in Fourier series of longitude and associated Legendre polynomials. Given that atmospheric tides are considered in this model as a linear perturbation around the equilibrium state, the amplitudes of the perturbed quantities are proportional to the forcing.  \\

\begin{figure*}[htb]
\centering
{\includegraphics[width=0.45\textwidth]{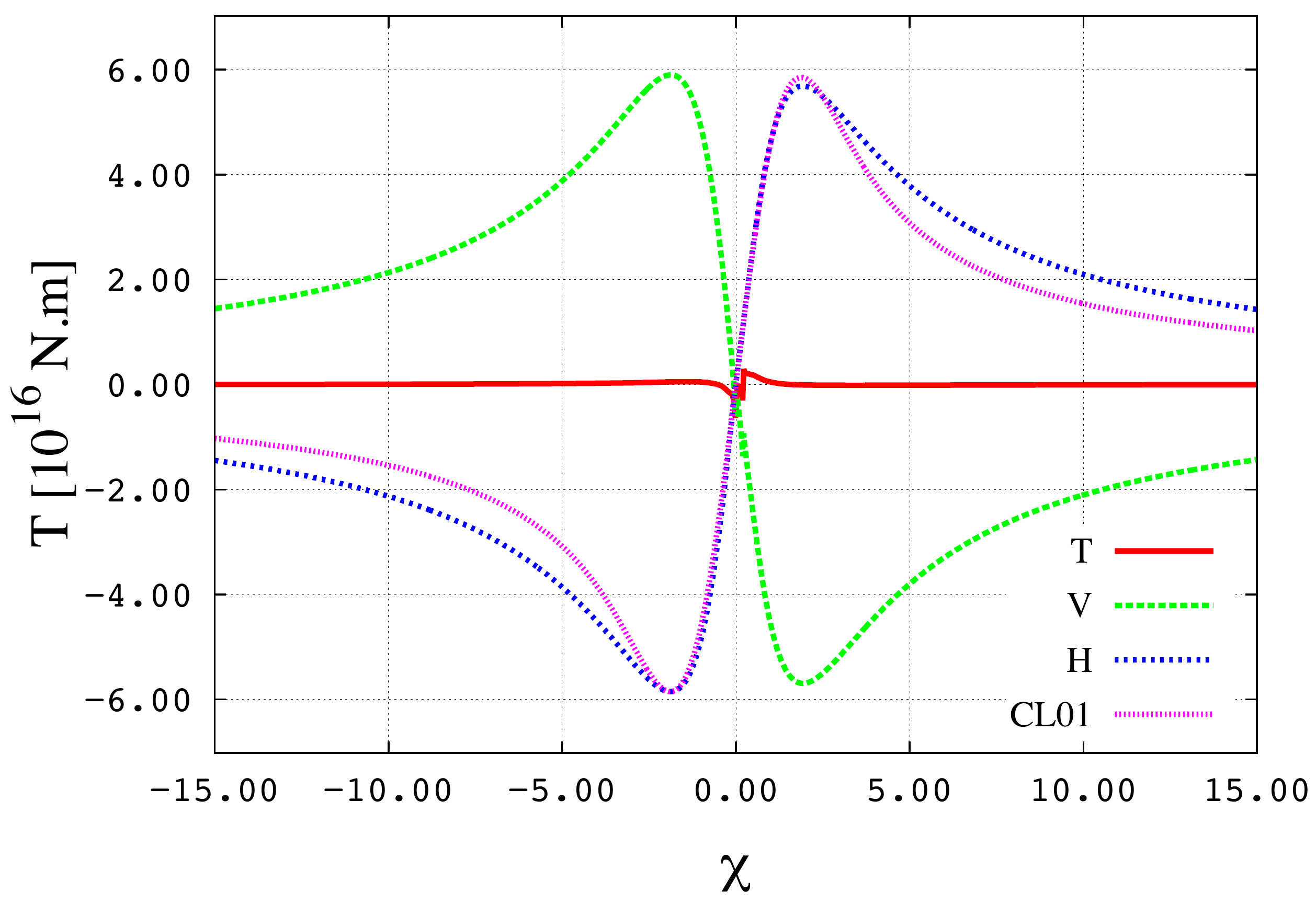} \hspace{1cm} 
\includegraphics[width=0.45\textwidth]{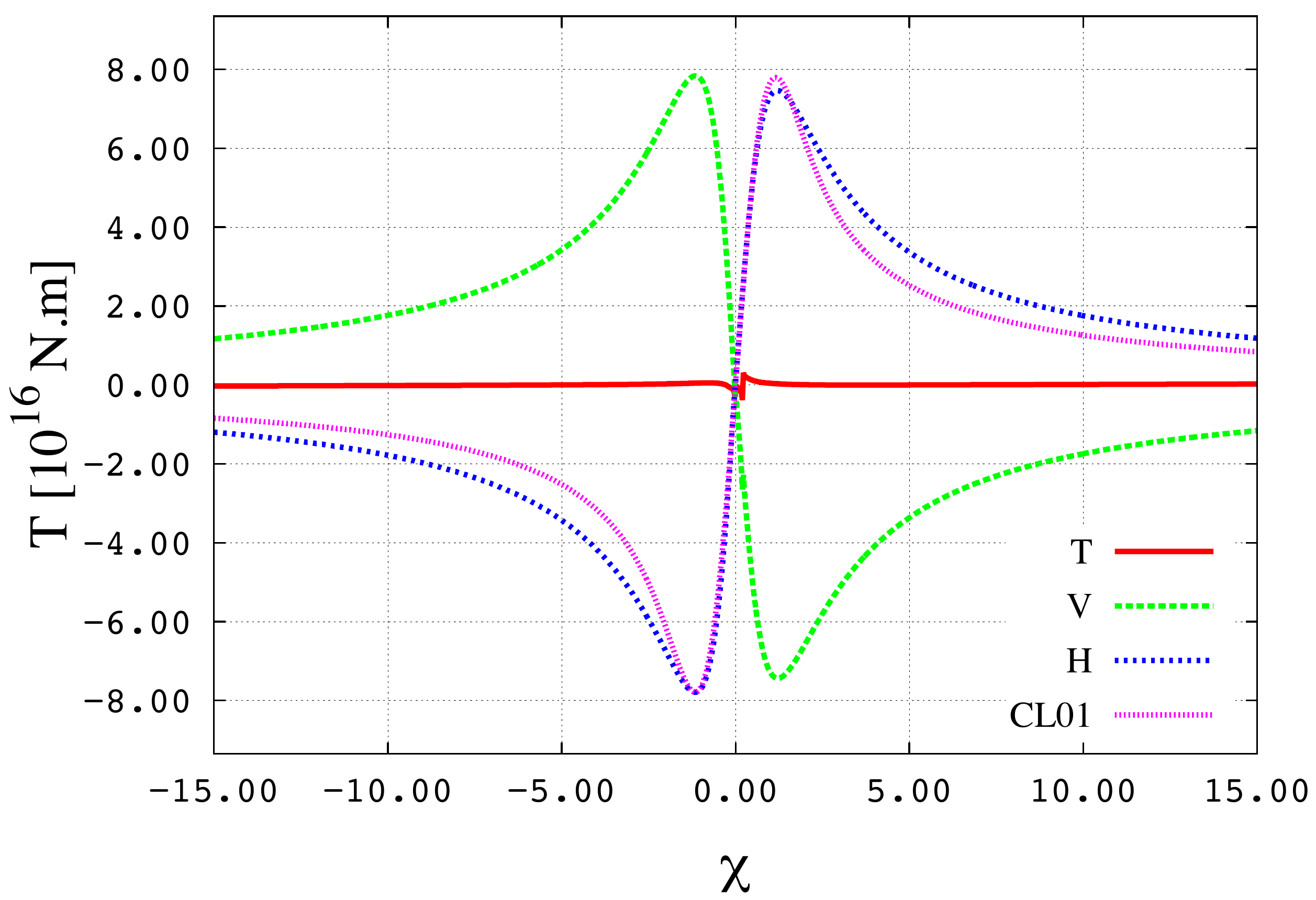}
\textsf{\caption{\label{fig:Terre_couple} Tidal torque applied to the Earth's (left panel) and Venus' (right panel) atmospheres by the thermal semidiurnal tide and its components as functions of the reduced frequency $ \chi = \left( \Omega - n_{\rm orb} \right) / n_{\rm orb} $ (horizontal axis). The forcings consist in academic quadrupolar perturbations expressed as $ U \left( x , \theta , \varphi , t \right) = U_2 P_2^2 \left( \cos \theta \right) e^{i \left( \sigma t + 2 \varphi \right) } $ and $ J \left( x , \theta , \varphi , t \right) = J_2 P_2^2 \left( \cos \theta \right) e^{i \left( \sigma t + 2 \varphi \right) } $ with $ U_2 = 0.985 \ {\rm m^2.s^{-2}} $ \mybf{and $ J_2 = 1.0 \times 10^{-2} \ {\rm m^2.s^{-3}} $ for the Earth and $ U_2 = 2.349 \ {\rm m^2.s^{-2}} $ and \jcor{$ J_2 = 1.0 \times 10^{-4} \ {\rm m^2.s^{-3}} $} for Venus}. \mybf{Labels T, H and V} refer to the total response and \mybf{its horizontal and vertical components} respectively \scor{(we recall that $\mathcal{T}_{\rm H} \sim \mathcal{T}_{\rm conv} $)}. \mybf{The label CL01 refers to the model by \cite{CL01}, given by Eq.~(\ref{torque_CLNS}).}
 } }}
\end{figure*}

\begin{table}[htb]
\centering
    \begin{tabular}{ l  l  l  l }
      \hline
      \hline
      \textsc{Parameters} & \textsc{Earth} & \textsc{Venus} \\
      \hline
       $ R $ [km] & $ 6371.0 $ & $ 6051.8 $\\
       $ g $ $ {\rm  [m.s^{-2}]} $ & $ 9.81 $ & $ 8.87 $ \\
       $ H $ [km] & $ 8.5 $ & $ 15.9 $ \\
       $ M $ $ {\rm [g.mol^{-1}]} $ & $ 28.965 $ & $ 43.45 $ \\
       $ \Gamma_1 $ & $ 1.4 $ & $ 1.4 $ \\
       $ \sigma_0 $ $ {\rm [ 10^{-7} \ s^{-1}]} $ & \mybf{$ 7.5 $} & \mybf{$ 7.5 $}\\
       $ P_0 \left( 0 \right) $ $ {\rm [10^5  \ Pa]} $ & $ 1.0 $ & $ 93 $ \\
       $ \Omega $ $ {\rm [ 10^{-7} \ s^{-1}]} $ & $ 729.21 $ & $ - 2.9927 $ \\
       $ n_{\rm orb} $ $ {\rm [10^{-7} \ s^{-1}]} $ & $ 1.9910 $ & $ 3.2364 $\\
       $ U_2 $ $ {\rm [m^2.s^{-2}]} $ & $ 0.985 $ & $ 2.349 $ \\
       $ J_2 $ $ {\rm [W.kg^{-1}]} $ & \mybf{$ 10^{-2} $} & \jcor{$ 10^{-4} $} 
      \vspace{0.1mm}\\
       \hline
    \end{tabular}
    \textsf{\caption{\label{Valeurs_para} Values of physical parameters used in simulations. Most of them are given by NASA fact sheets\protect\footnotemark ~and IMCCE databases. \mybf{For both planets,} the radiation frequency is arbitrary fixed at \mybf{$ \sigma_0 = 7.5 \times 10^{-7} \ {\rm s^{-1}} $}, \mybf{which is the effective Newtonian cooling frequency obtained by \cite{Leconte2015} for Venus}. This order of magnitude can be computed using Eq.~(\ref{sigma0}). The quadrupolar tidal potentials are computed from the expression of the gravitational potential given by \cite{MLP09}, recalled in Appendix~\ref{app:thermal_forcing}.}}
 \end{table}

\footnotetext{Link: \textcolor{blue}{ \url{http://nssdc.gsfc.nasa.gov/planetary/factsheet/venusfact.html}}}

The horizontal structure equation is solved using the spectral method described in Appendix~\ref{app:num_scheme}. The vertical structure equation is integrated numerically on the domain $ x \in \left[ 0 , x_{\rm atm} \right] $, using a regular mesh with element of size $ \delta x $. As pointed out by CL70, the number of points of the mesh, denoted $ N $, must be sufficiently large to obtain vertical profiles of the perturbation with a good accuracy \mybf{(see Appendix~\ref{app:spatial_resolution})}. To fix $ N $ correctly in the case of the Earth, CL70 suggests to use a criterion that we adapt for other configurations, i.e.

\begin{equation}
N > 100 \frac{x_{\rm atm} \left| \hat{k}_n \right| }{2 \pi},
\label{Ncrit}
\end{equation}

considering that the scale of the vertical variations is defined by the module of the complex wavenumber given in Eq.~(\ref{Shrodinger_mince}). \\

 For each planet, we give here the spatial distribution of perturbed quantities and the evolution of the tidal torque with the tidal frequency. The numerical values used in these simulations are summarized in Table~\ref{Valeurs_para}.

\begin{figure*}[htb!]
 \centering
 \includegraphics[width=0.40\textwidth,clip]{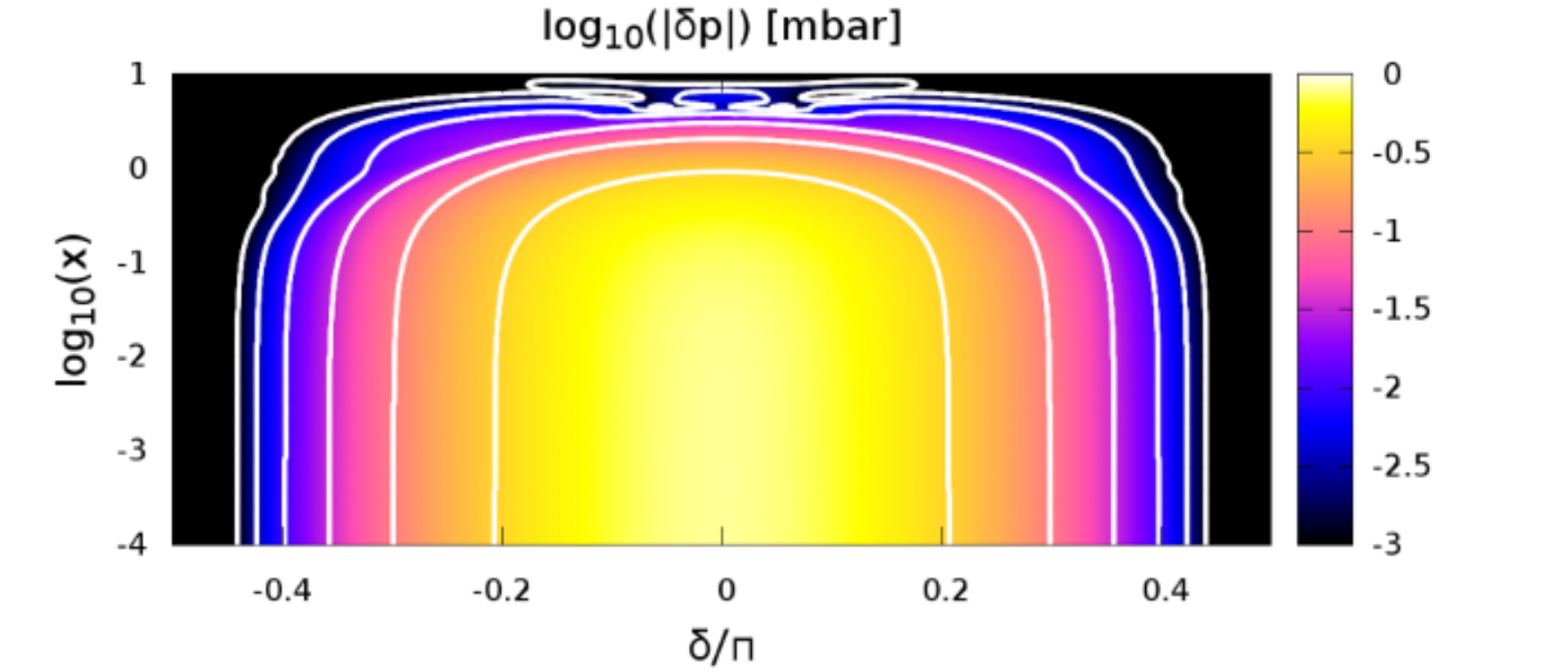} \hspace{1cm}       
 \includegraphics[width=0.40\textwidth,clip]{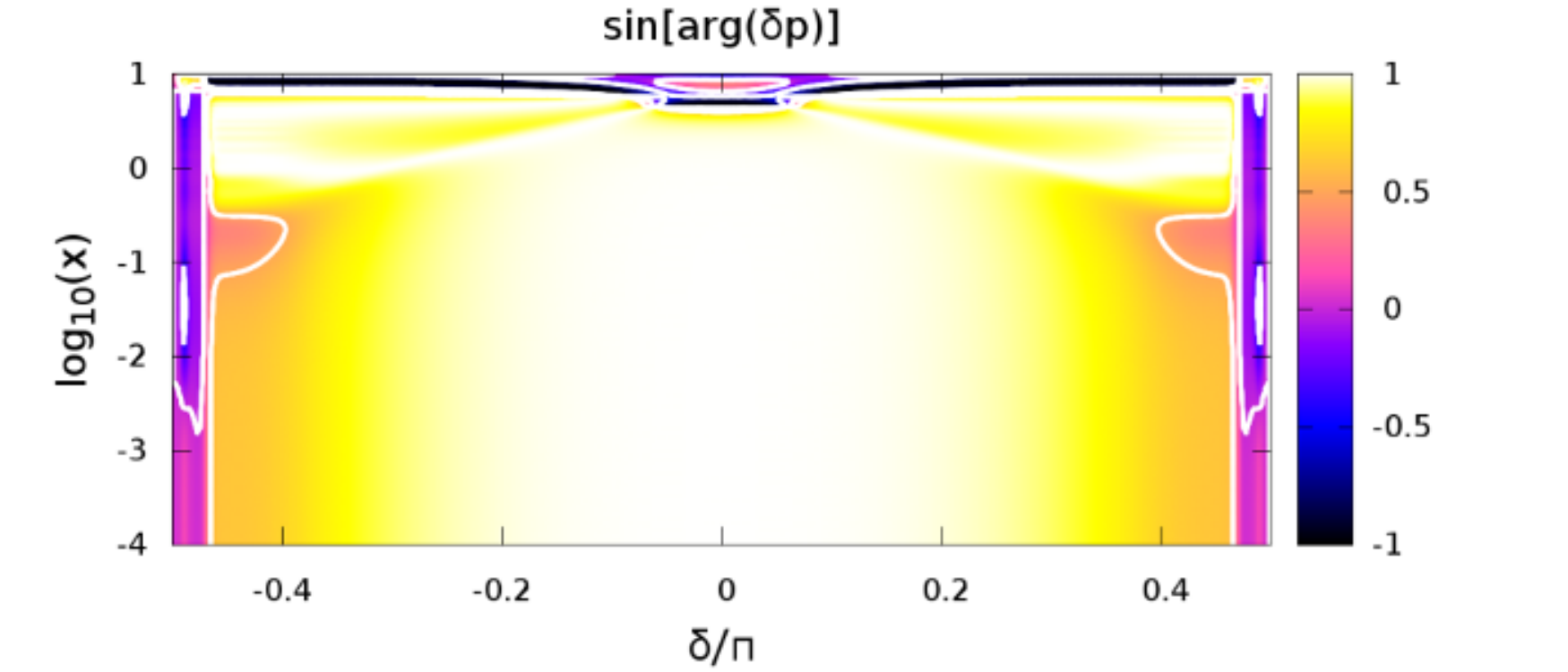} \\
  \includegraphics[width=0.40\textwidth,clip]{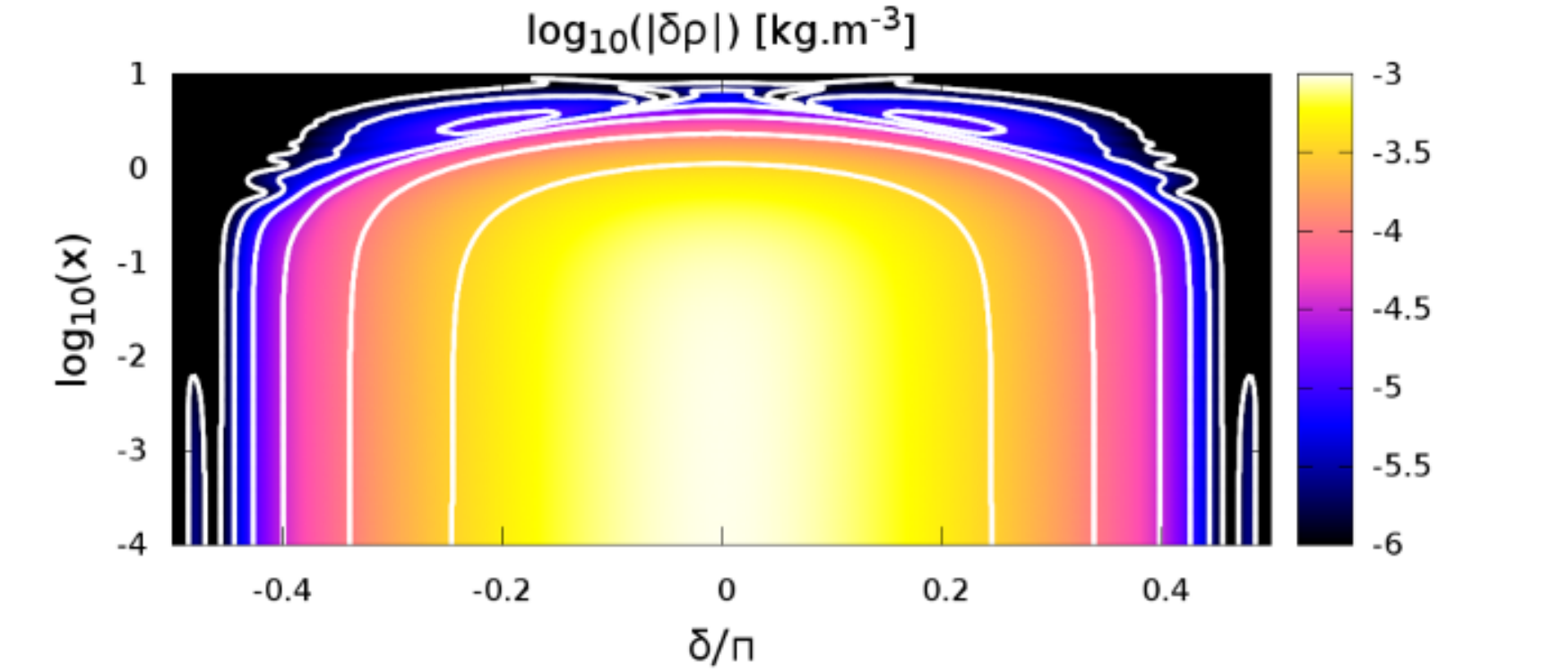} \hspace{1cm}       
 \includegraphics[width=0.40\textwidth,clip]{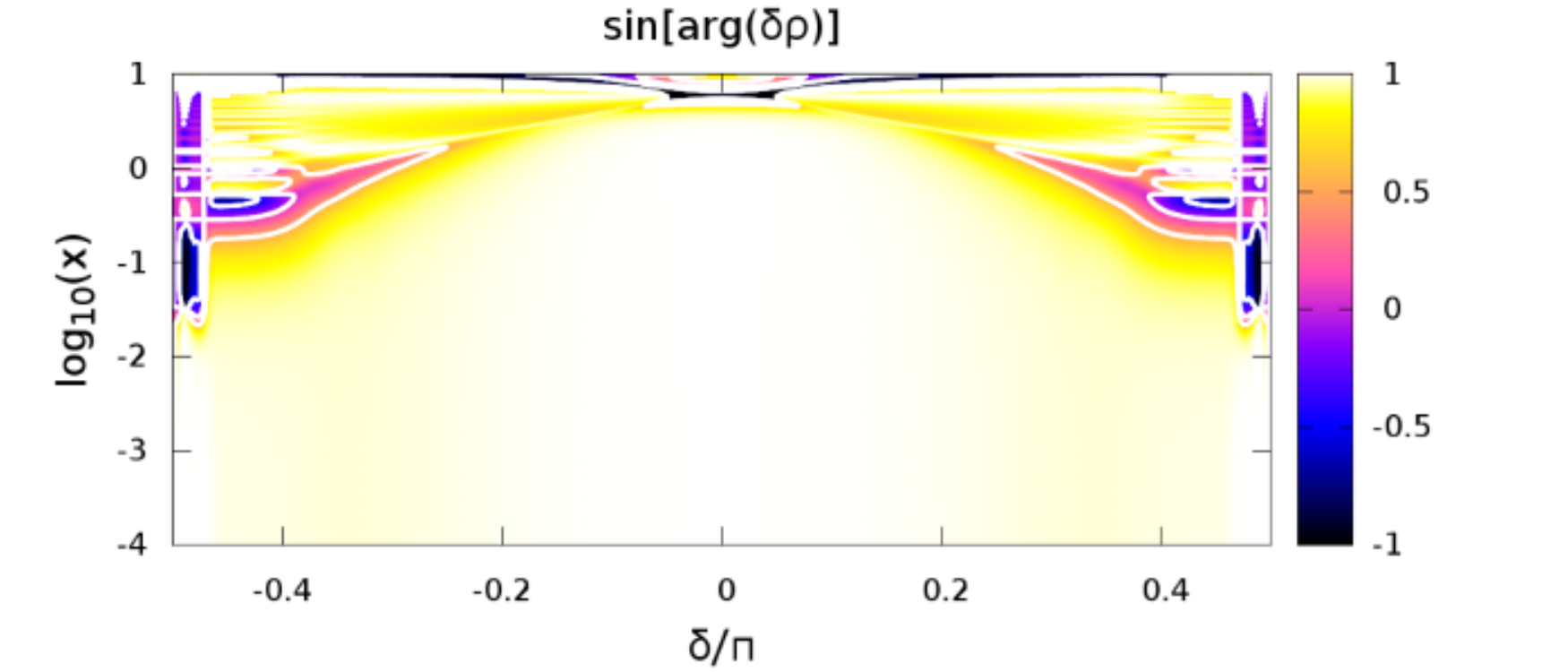} \\
   \includegraphics[width=0.40\textwidth,clip]{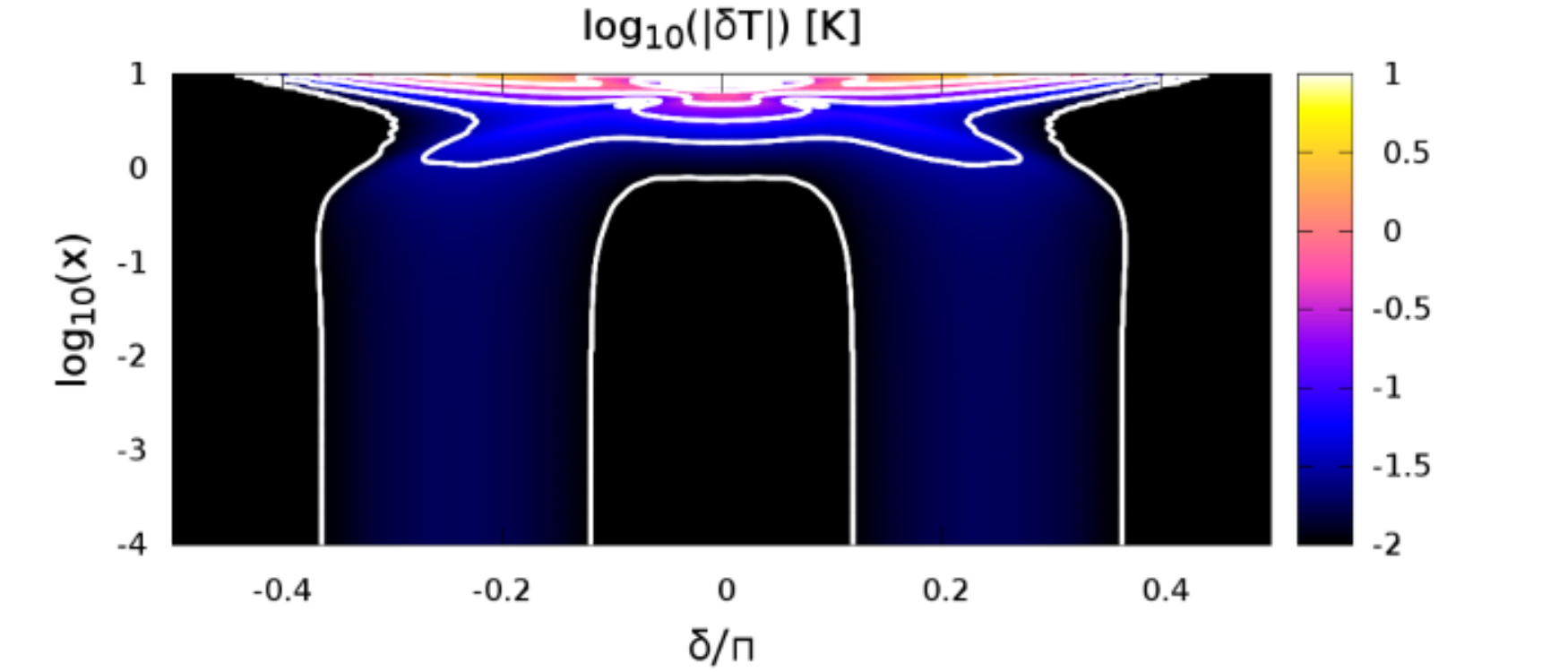} \hspace{1cm}       
 \includegraphics[width=0.40\textwidth,clip]{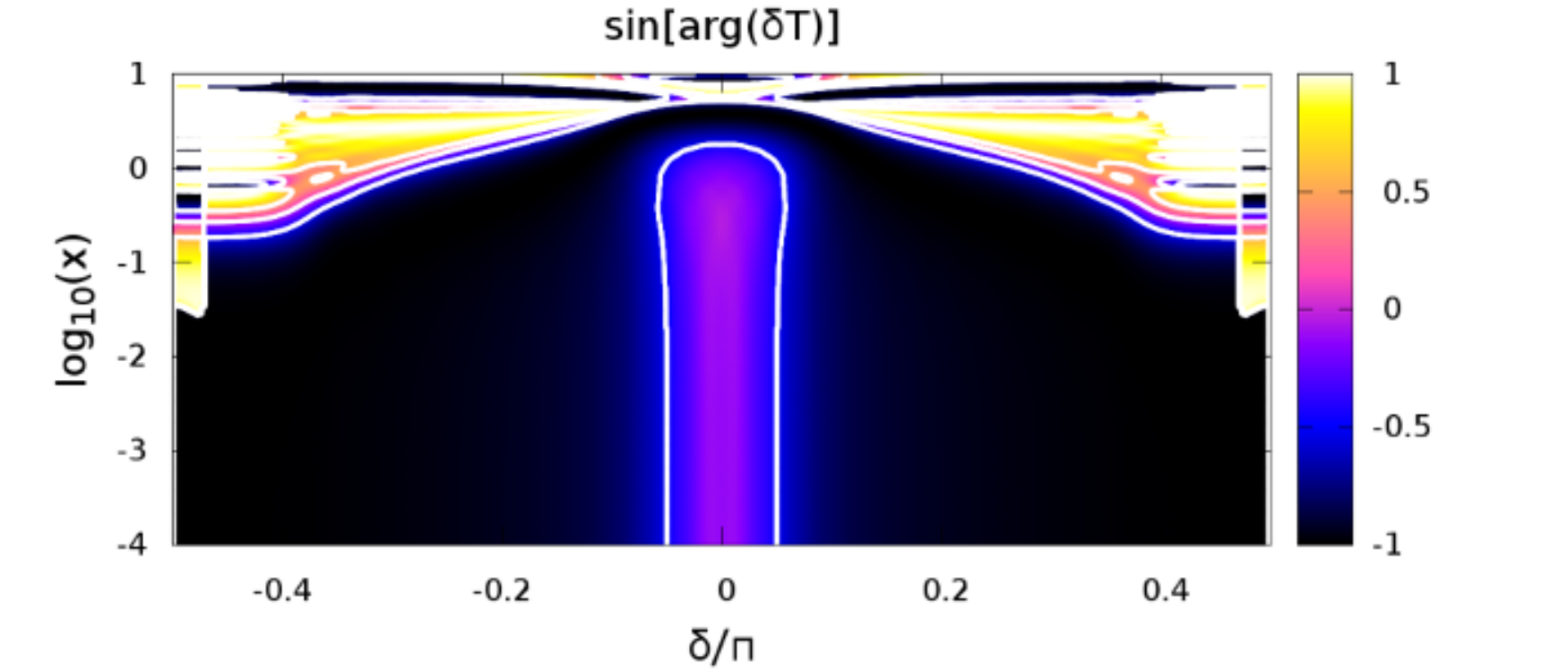}\\
    \includegraphics[width=0.40\textwidth,clip]{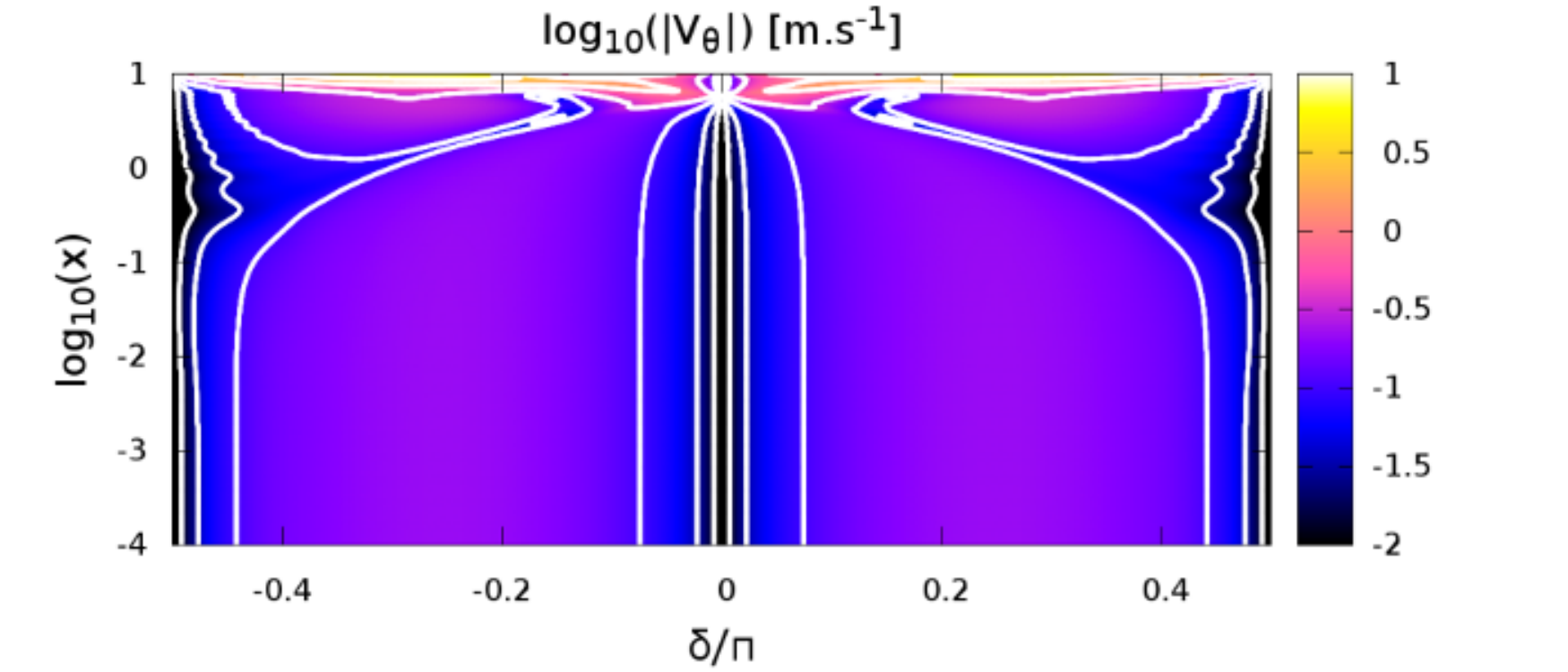} \hspace{1cm}      
 \includegraphics[width=0.40\textwidth,clip]{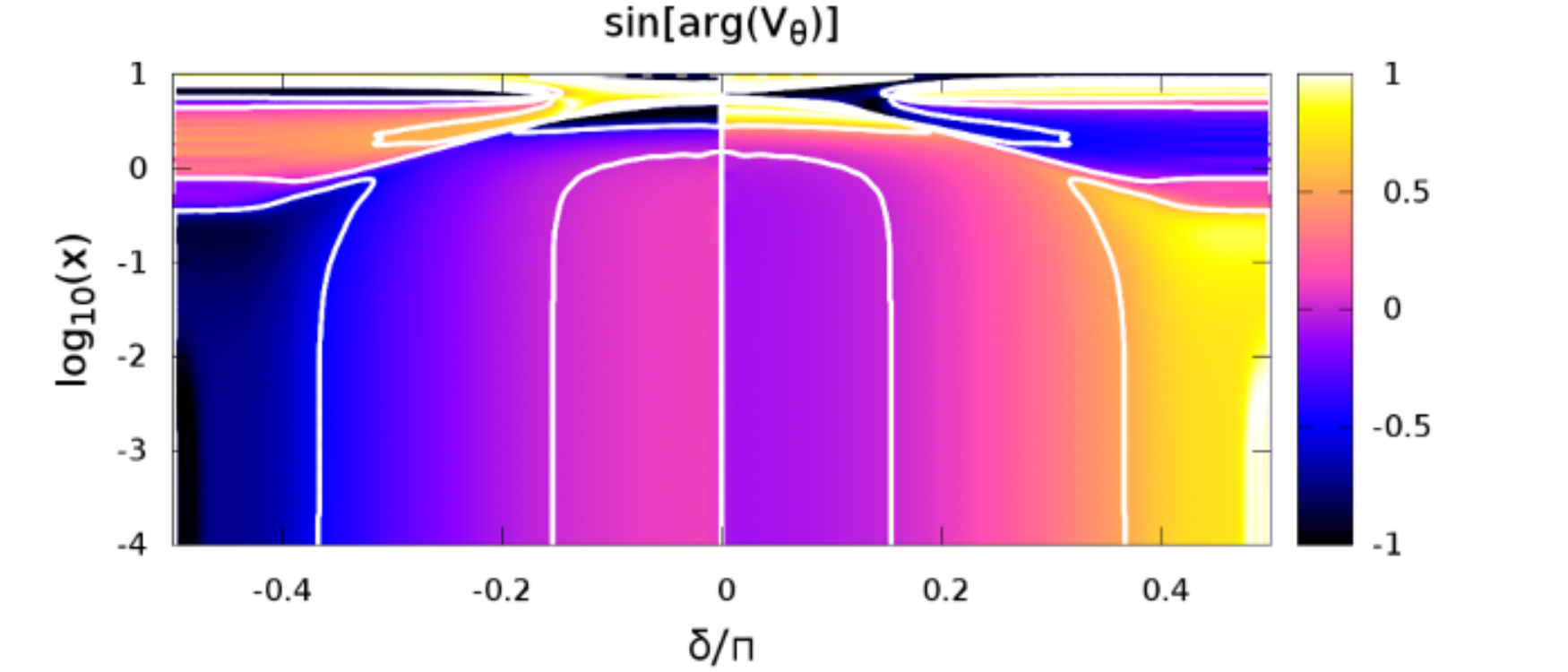}\\
    \includegraphics[width=0.40\textwidth,clip]{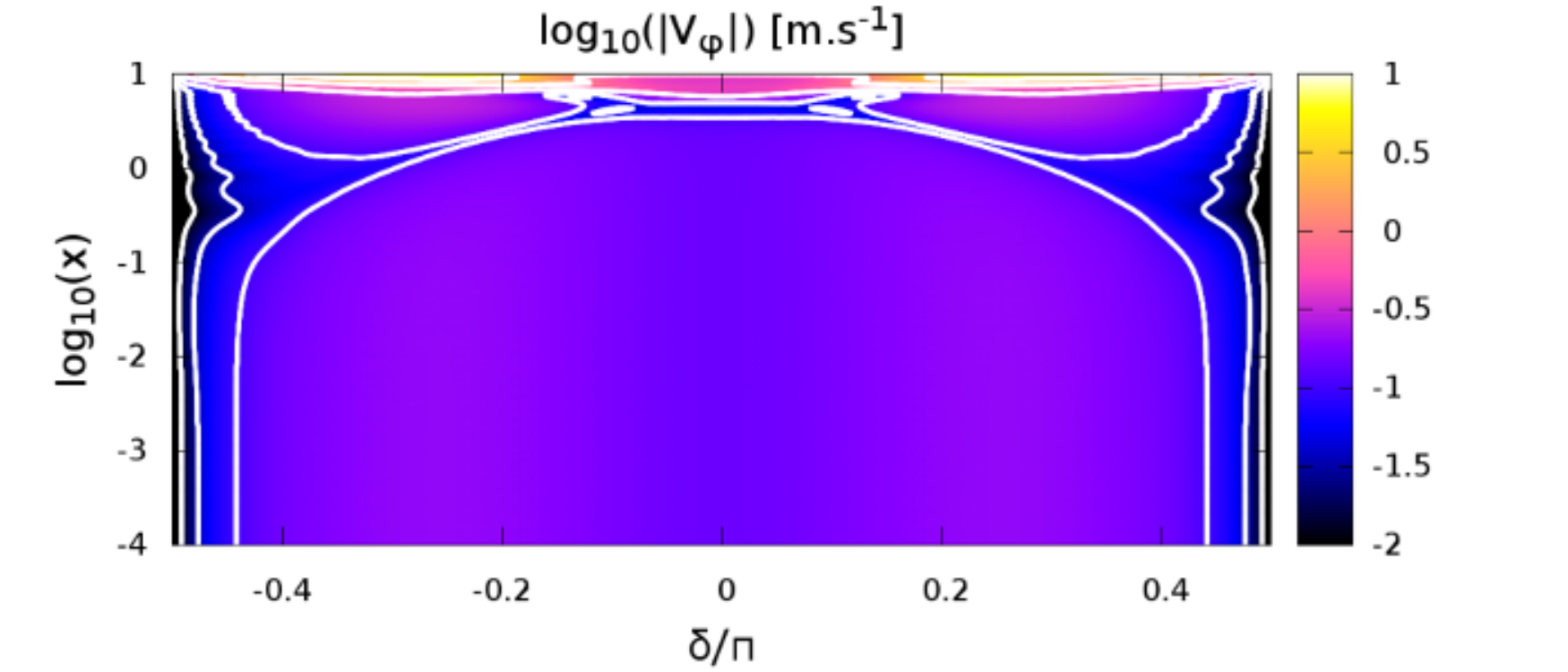} \hspace{1cm}       
 \includegraphics[width=0.40\textwidth,clip]{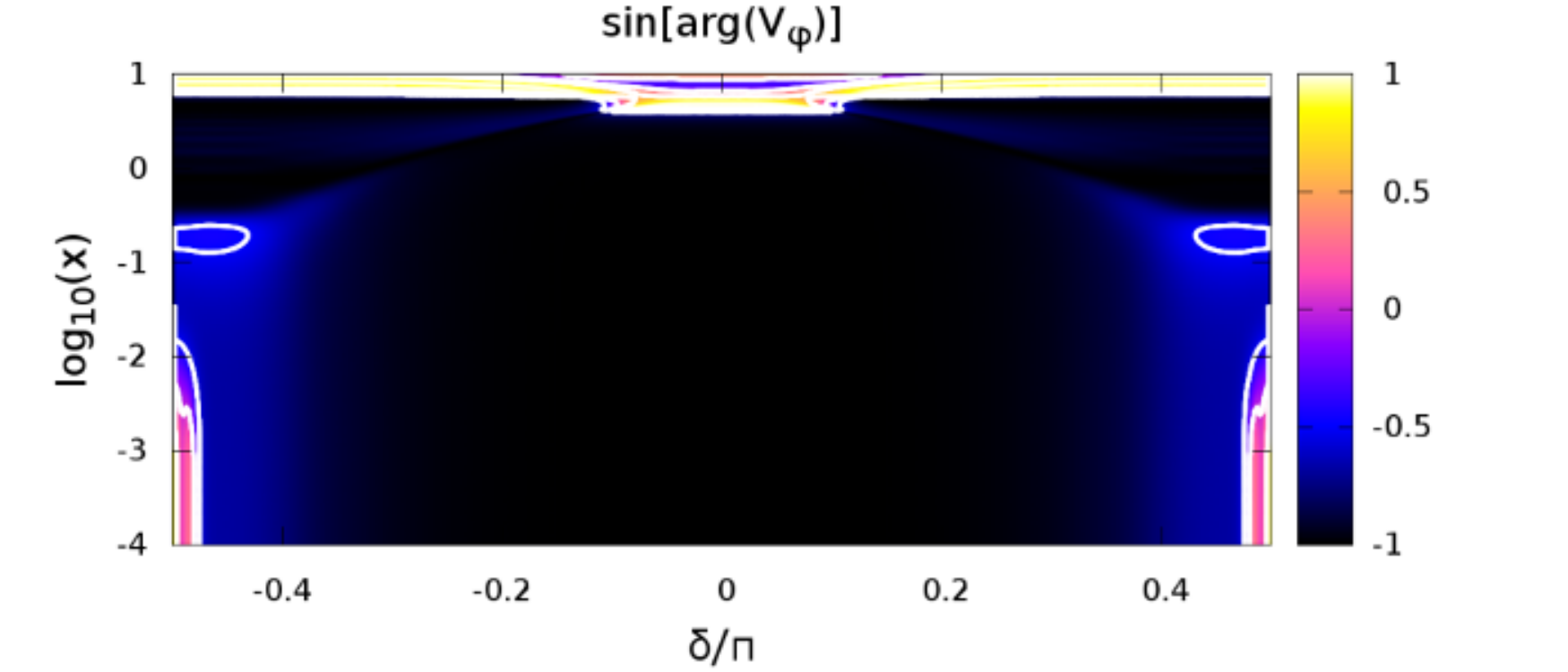}\\
     \includegraphics[width=0.40\textwidth,clip]{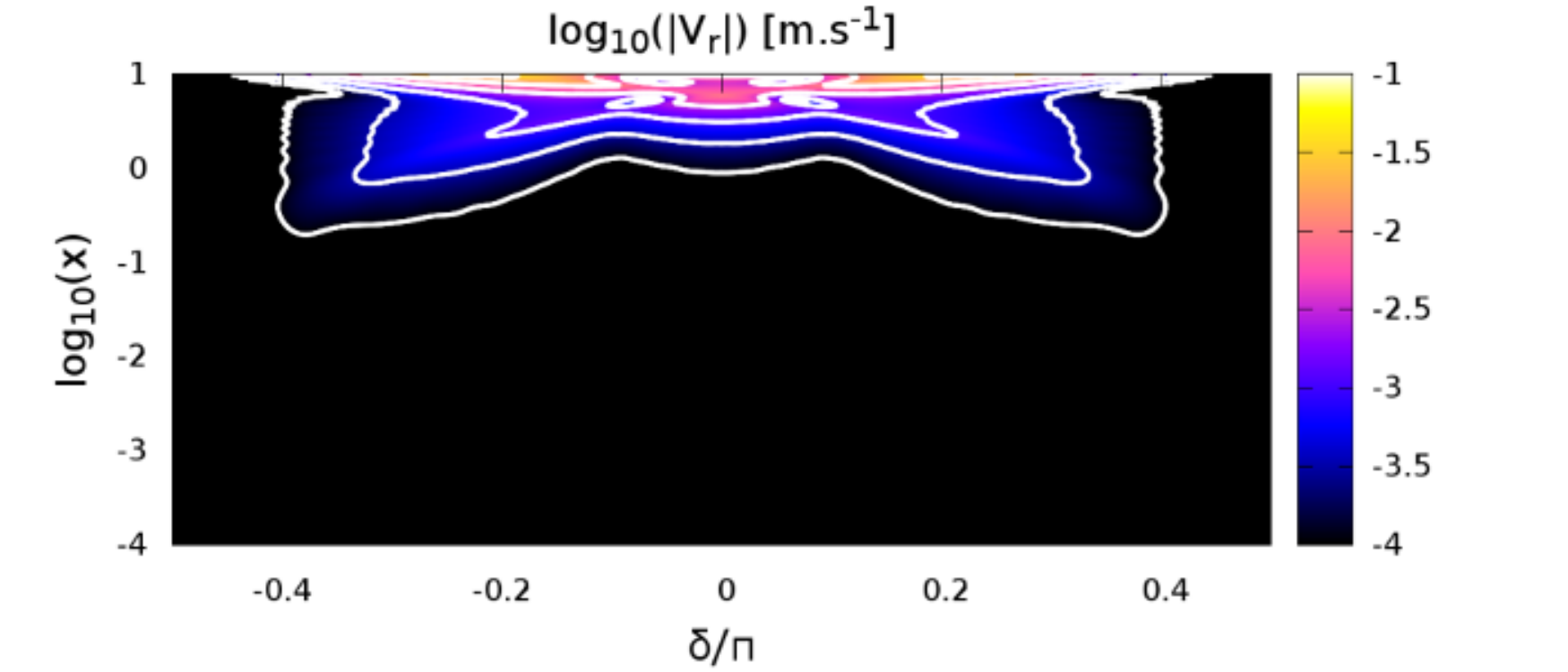} \hspace{1cm}      
 \includegraphics[width=0.40\textwidth,clip]{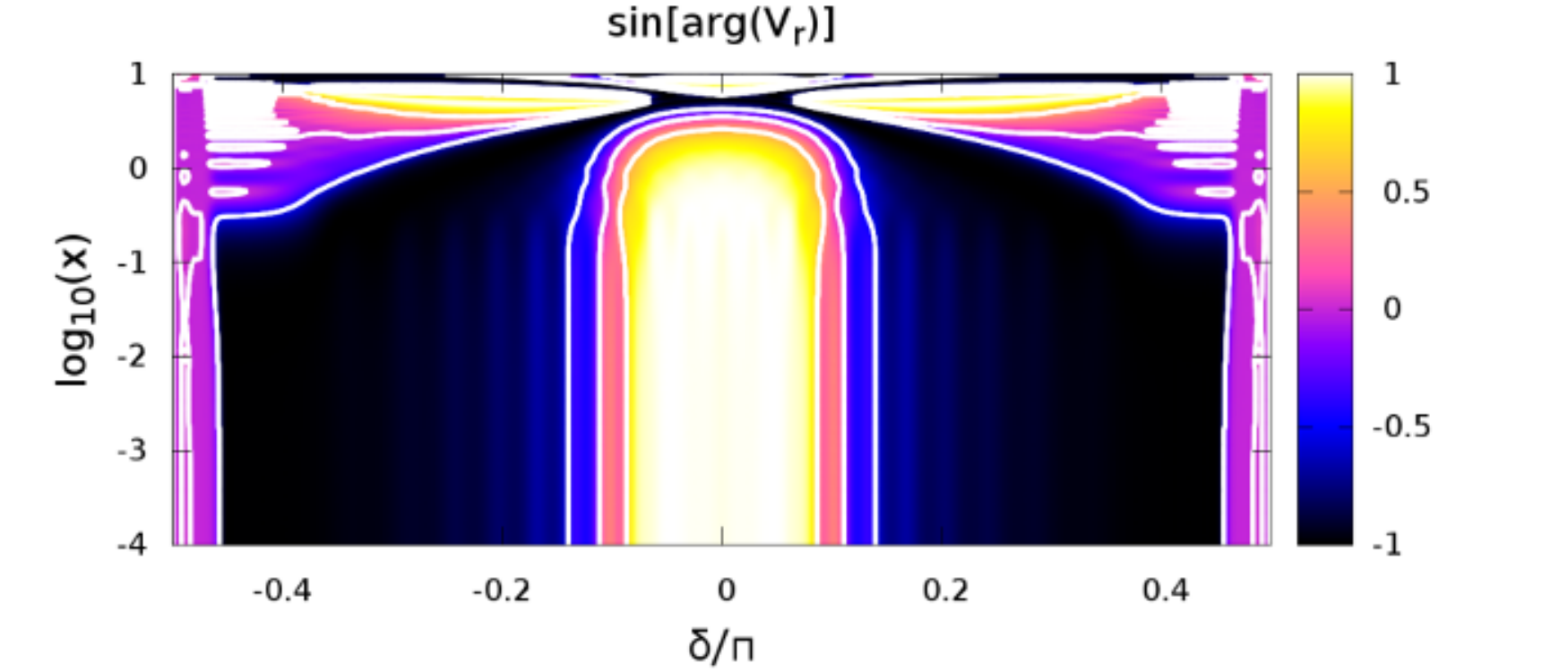}
  \textsf{ \caption{\label{fig:Earth_section}  Total tidal response of the Earth's atmosphere to the solar semidiurnal perturbation caused by the quadrupolar academic forcings $ U \left( x , \theta , \varphi , t \right) = U_2 P_2^2 \left( \cos \theta \right) e^{i \left( \sigma t + 2 \varphi \right) } $ and $ J \left( x , \theta , \varphi , t \right) = J_2 P_2^2 \left( \cos \theta \right) e^{i \left( \sigma t + 2 \varphi \right) } $ with $ U_2 = 0.985 \ {\rm m^2.s^{-2}} $ and $ J_2 = 1.0 \times 10^{-2} \ {\rm m^2.s^{-3}} $. {\it Left, from top to bottom:} amplitudes of $ \delta p $, $ \delta \rho $, $ \delta T $, $ V_\theta $, $ V_\varphi $ and $ V_r $ in logarithmic scales as functions of the reduced latitude $ \delta / \pi $ (horizontal axis) and altitude $ x = z / H $ in logarithmic scale (vertical axis). The color function $ c $ is given by $ c = \log \left( \left| \delta f \right| \right) $ for any quantity $ \delta f $. {\it Right, from top to bottom:} sinus of the argument of the same quantities as functions of the reduced latitude and altitude. In these plots, $ c = \sin \left[  \arg \left( \delta f \right) \right] $. }}
\end{figure*}

\begin{figure*}[htb!]
 \centering
 \includegraphics[width=0.40\textwidth,clip]{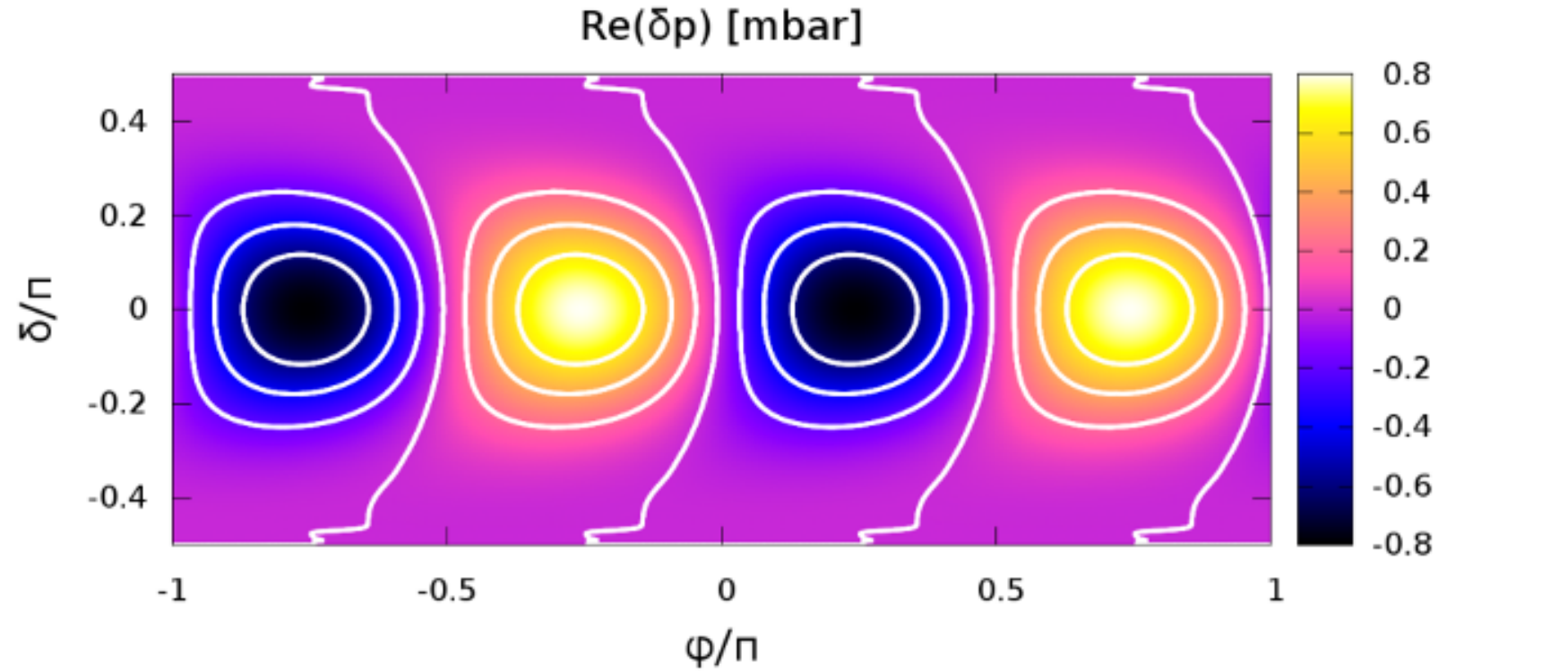} \hspace{1cm}       
 \includegraphics[width=0.40\textwidth,clip]{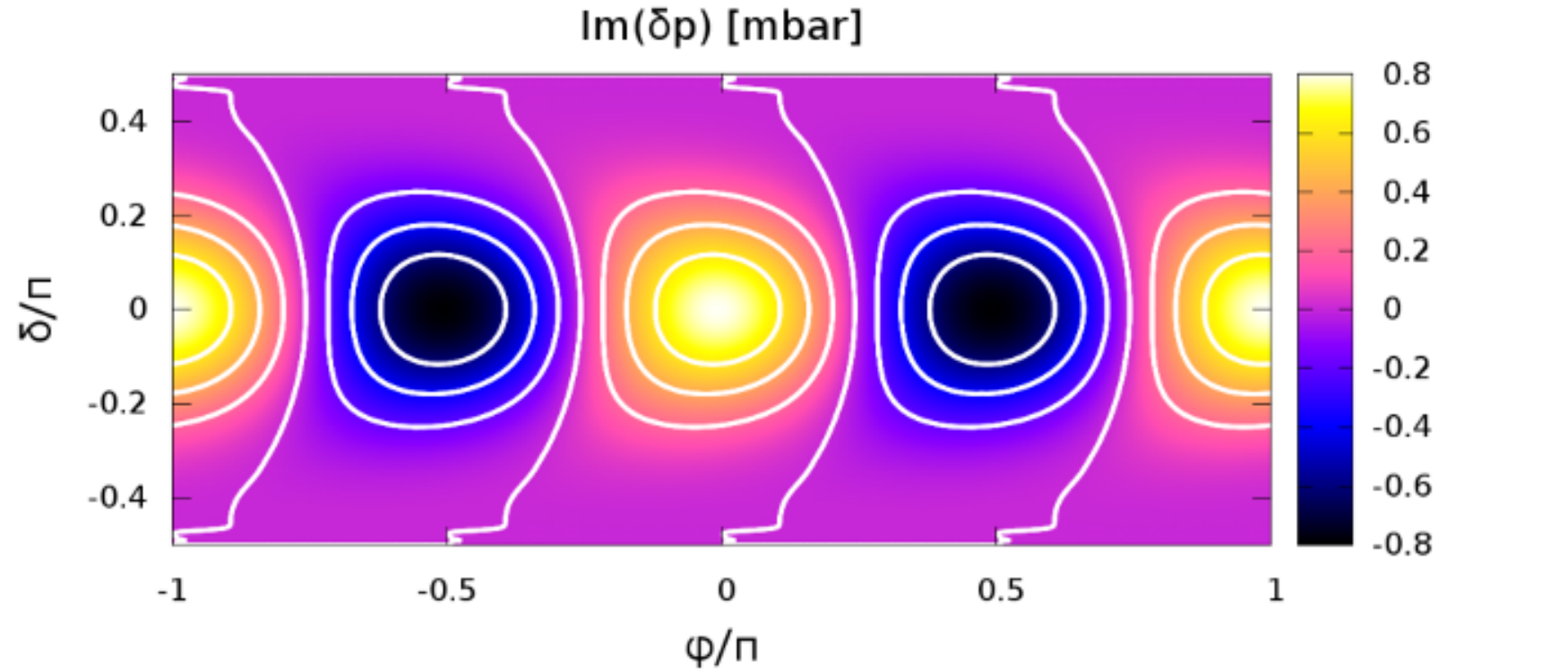} \\
  \includegraphics[width=0.40\textwidth,clip]{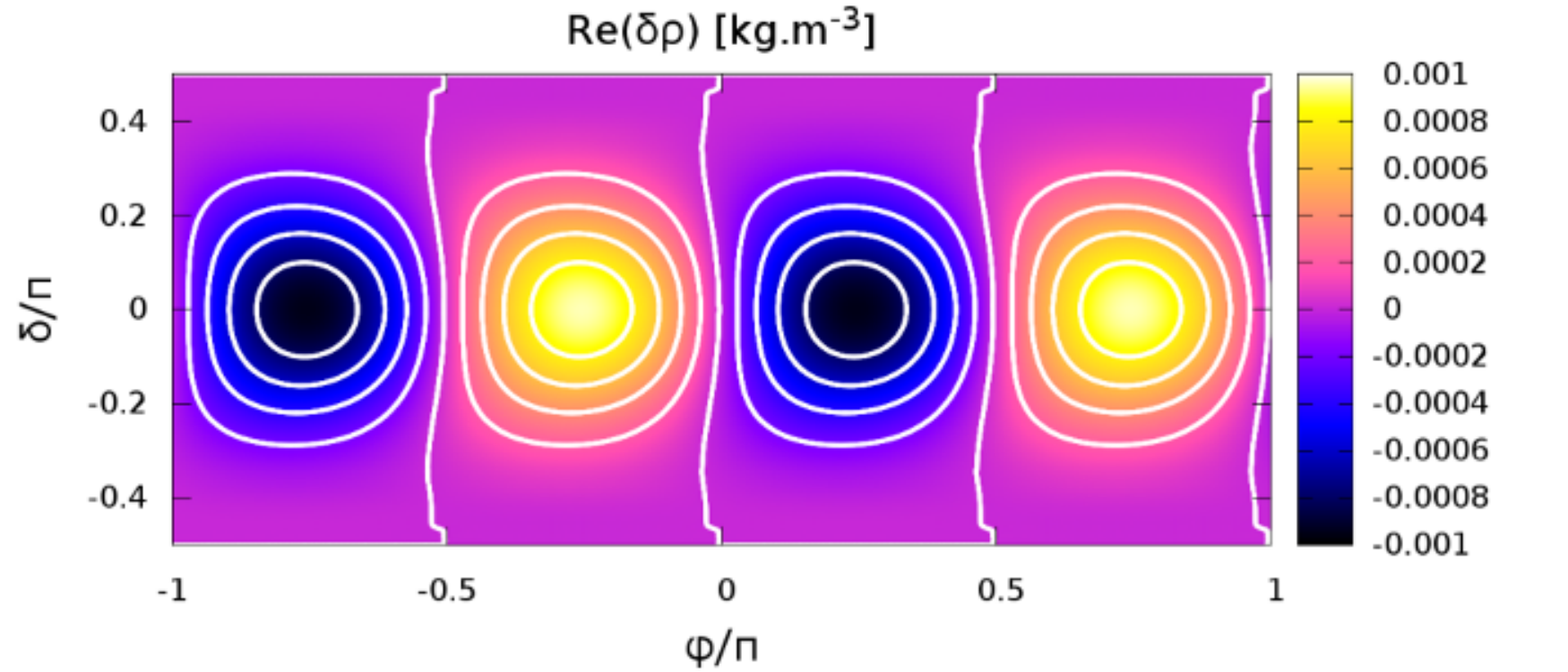} \hspace{1cm}       
 \includegraphics[width=0.40\textwidth,clip]{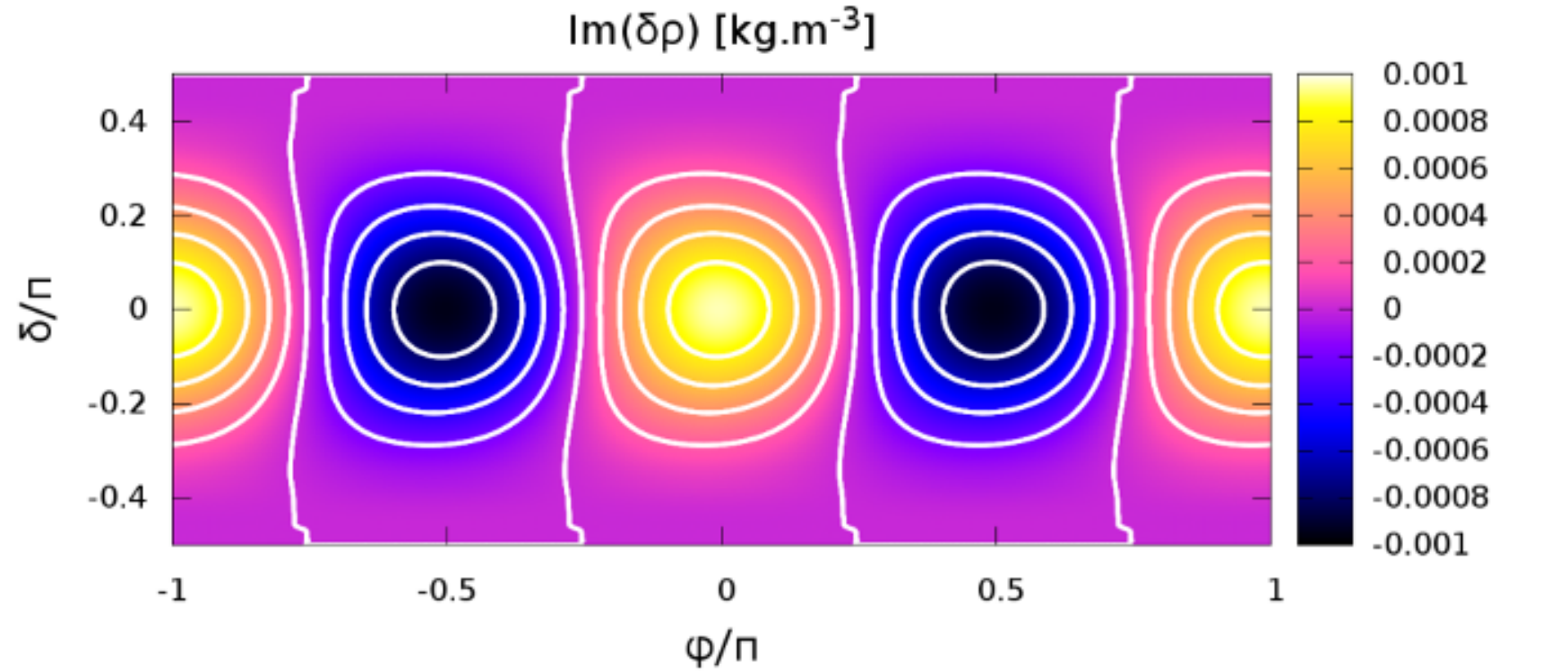} \\
   \includegraphics[width=0.40\textwidth,clip]{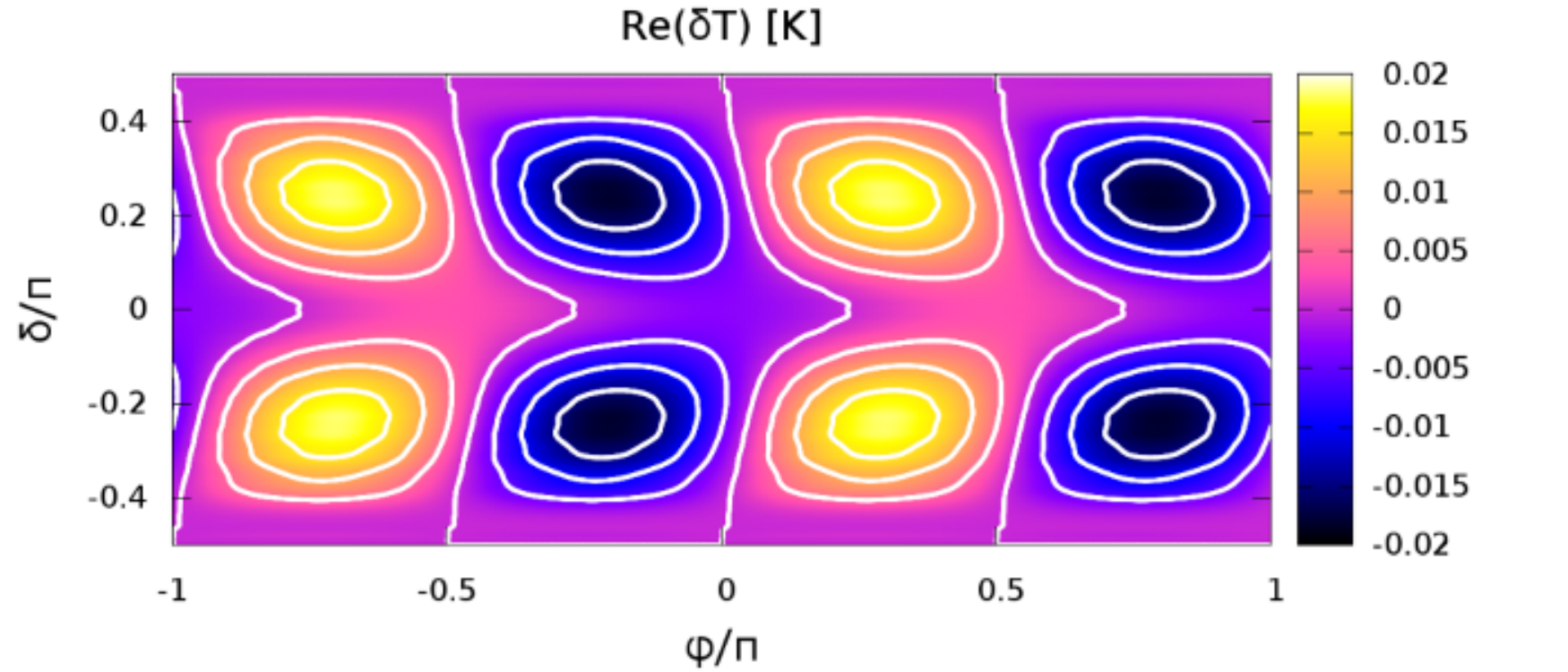} \hspace{1cm}       
 \includegraphics[width=0.40\textwidth,clip]{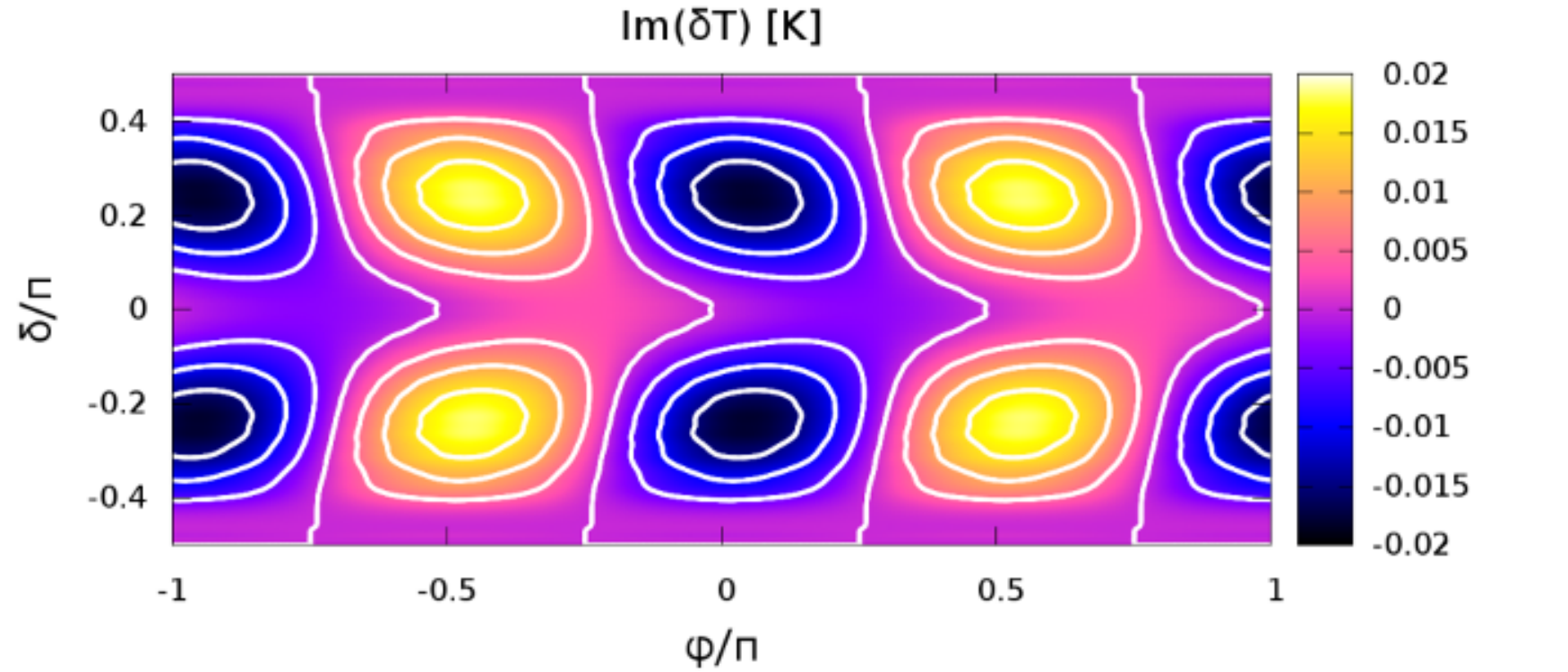}\\
    \includegraphics[width=0.40\textwidth,clip]{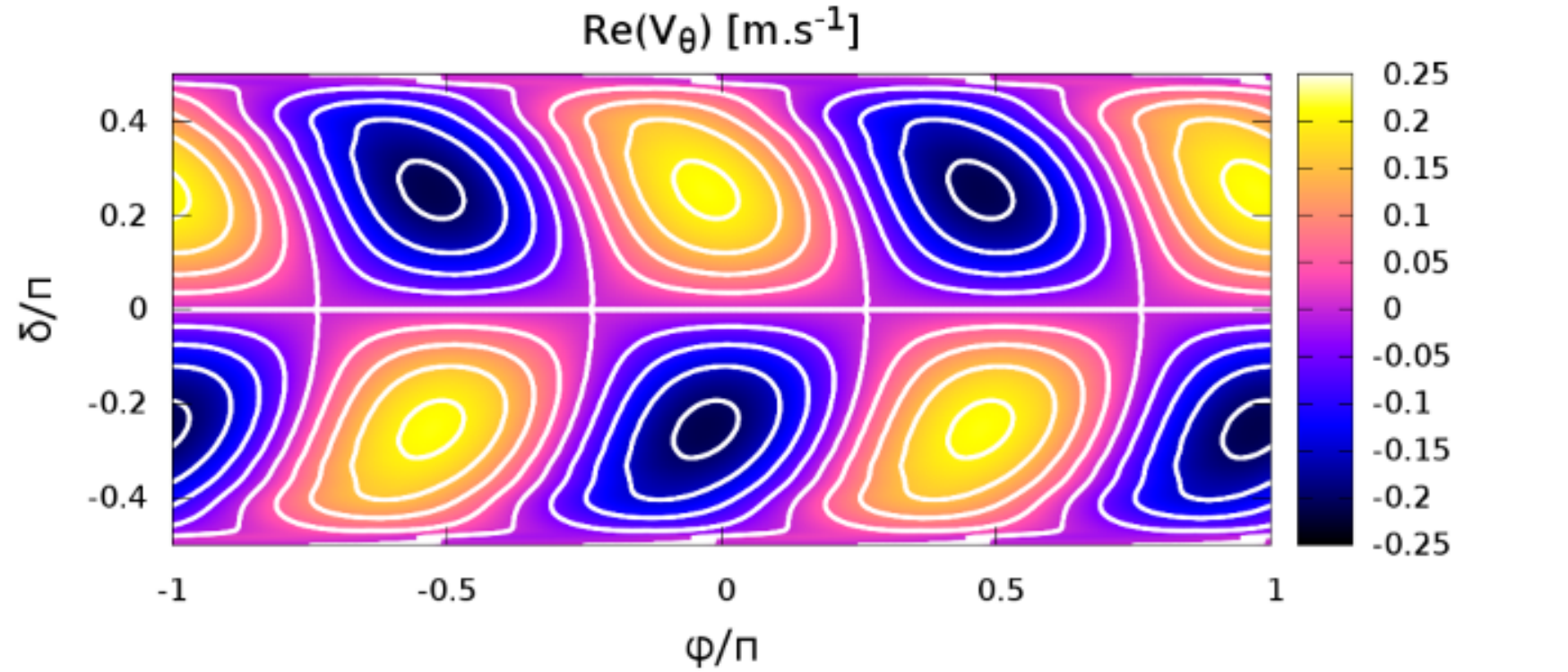} \hspace{1cm}       
 \includegraphics[width=0.40\textwidth,clip]{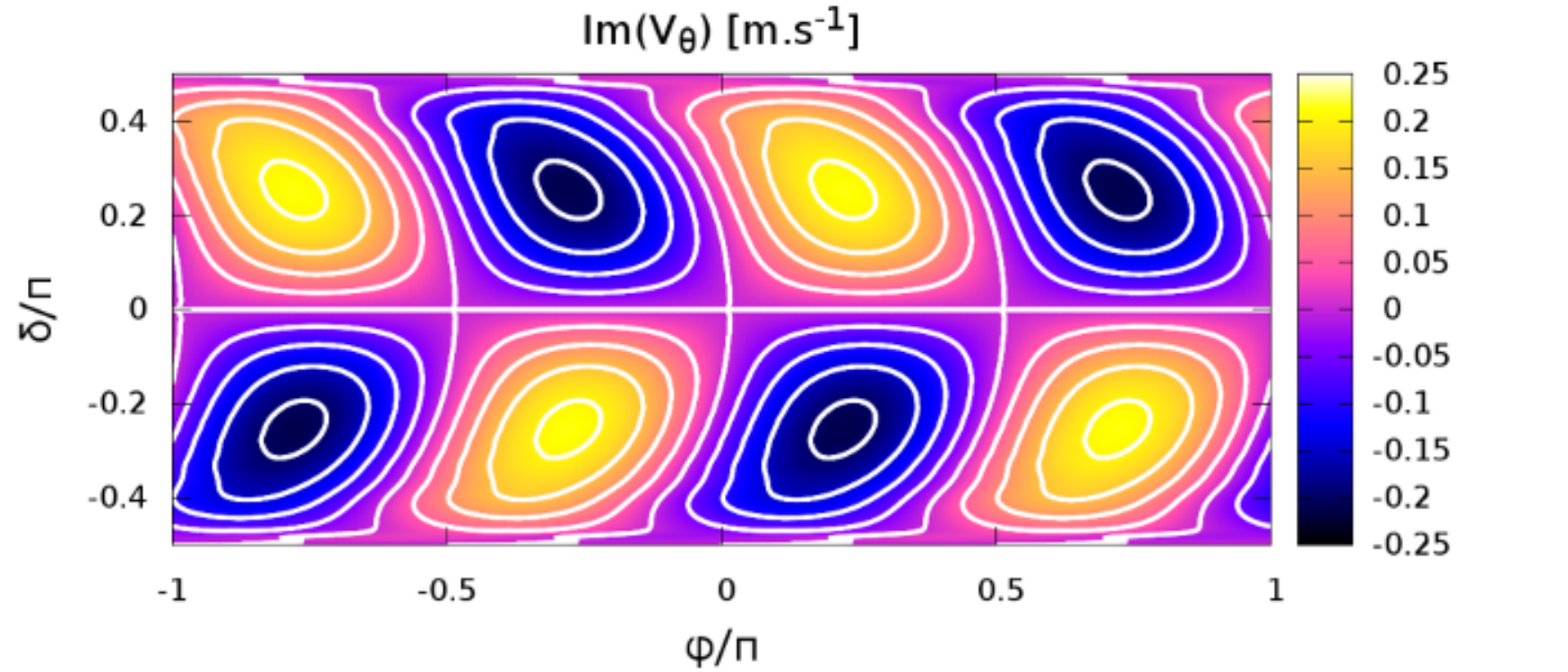}\\
    \includegraphics[width=0.40\textwidth,clip]{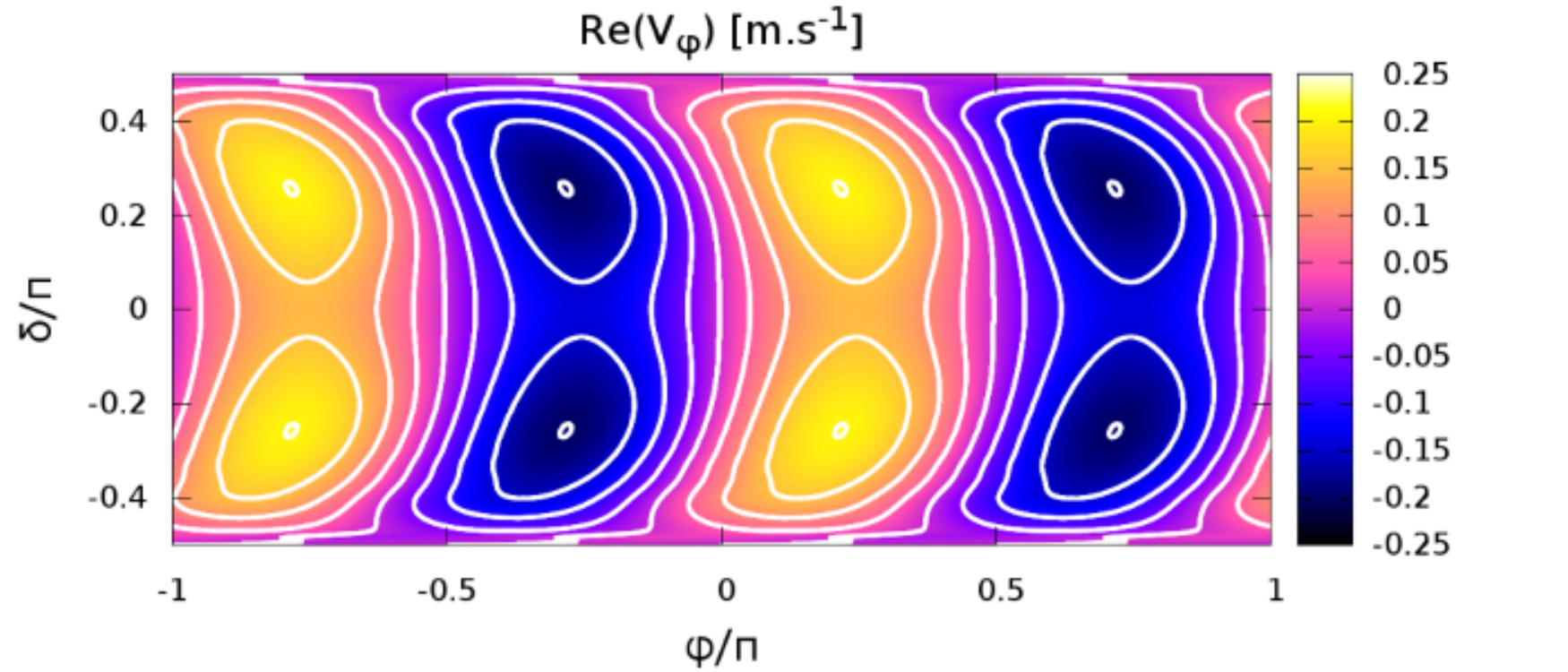} \hspace{1cm}       
 \includegraphics[width=0.40\textwidth,clip]{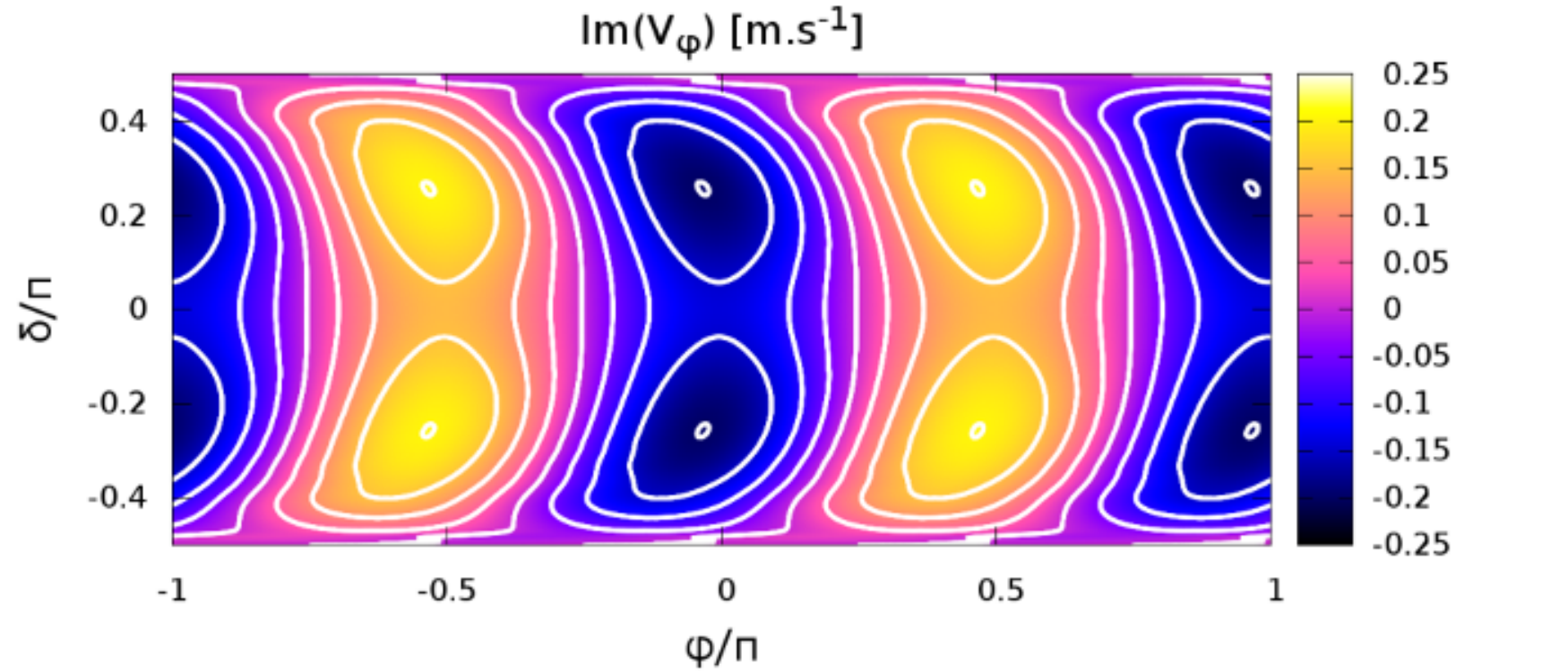}
  \textsf{ \caption{\label{fig:Earth_planis} Total tidal response of the Earth's atmosphere to the solar semidiurnal perturbation caused by the quadrupolar academic forcings $ U \left( x , \theta , \varphi , t \right) = U_2 P_2^2 \left( \cos \theta \right) e^{i \left( \sigma t + 2 \varphi \right) } $ and $ J \left( x , \theta , \varphi , t \right) = J_2 P_2^2 \left( \cos \theta \right) e^{i \left( \sigma t + 2 \varphi \right) } $ with $ U_2 = 0.985 \ {\rm m^2.s^{-2}} $ and $ J_2 = 1.0 \times 10^{-2} \ {\rm m^2.s^{-3}} $. {\it Left, from top to bottom:} real parts of $ \delta p $, $ \delta \rho $, $ \delta T $, $ V_\theta $ and $ V_\varphi $ taken at the ground as functions of the reduced longitude $ \varphi / \pi $ (horizontal axis) and latitude $ \delta / \pi $ (vertical axis). The color function $ c $ is given by $ c = \Re \left\{ \delta f \right\} $ for any quantity $ \delta f $. {\it Right, from top to bottom:} imaginary parts of the same quantities as functions of the reduced longitude and latitude. In these plots, $ c = \Im \left\{ \delta f \right\} $. }}
\end{figure*}

\begin{figure*}[htb!]
 \centering
 \includegraphics[width=0.40\textwidth,clip]{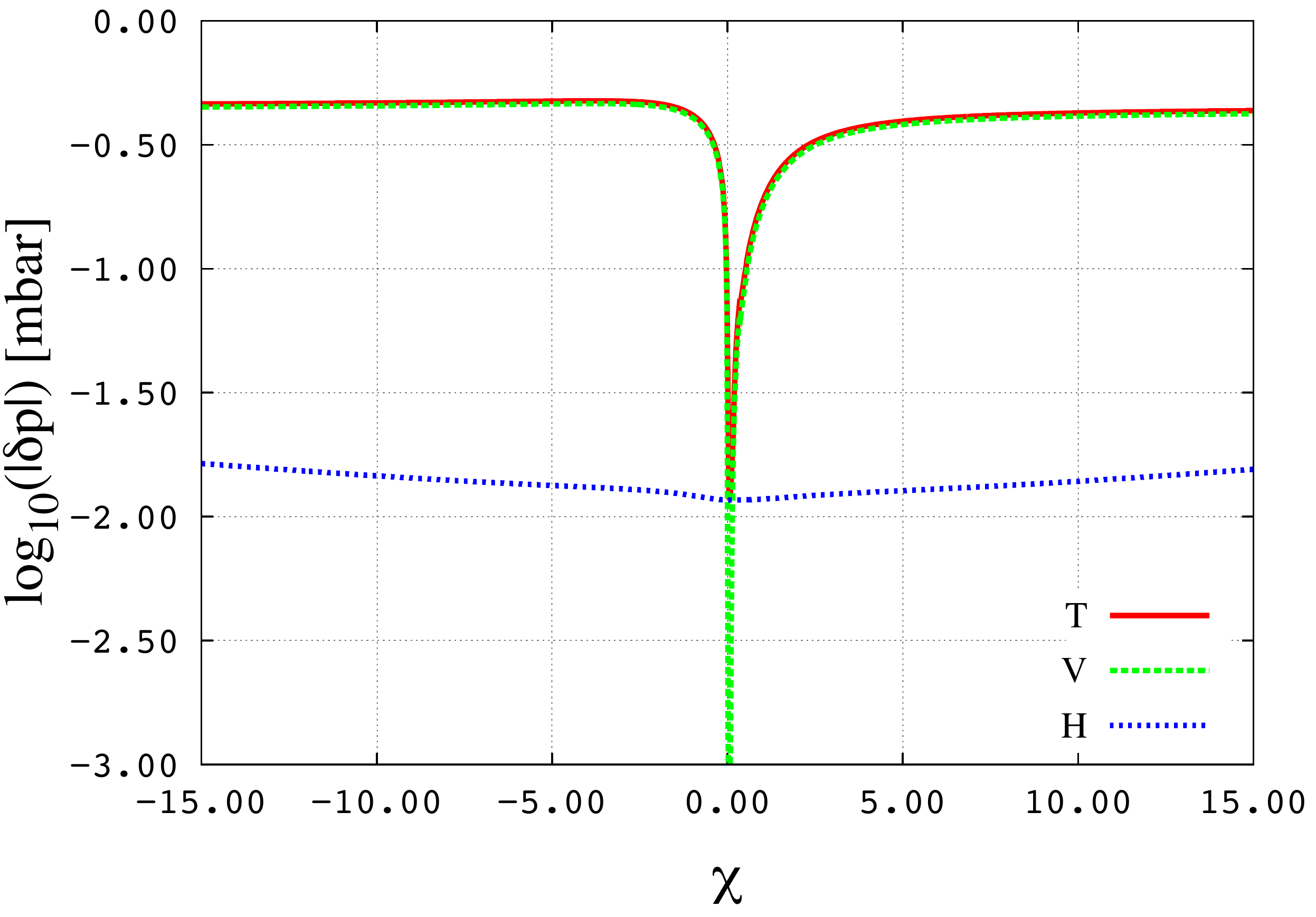} \hspace{1cm}       
 \includegraphics[width=0.40\textwidth,clip]{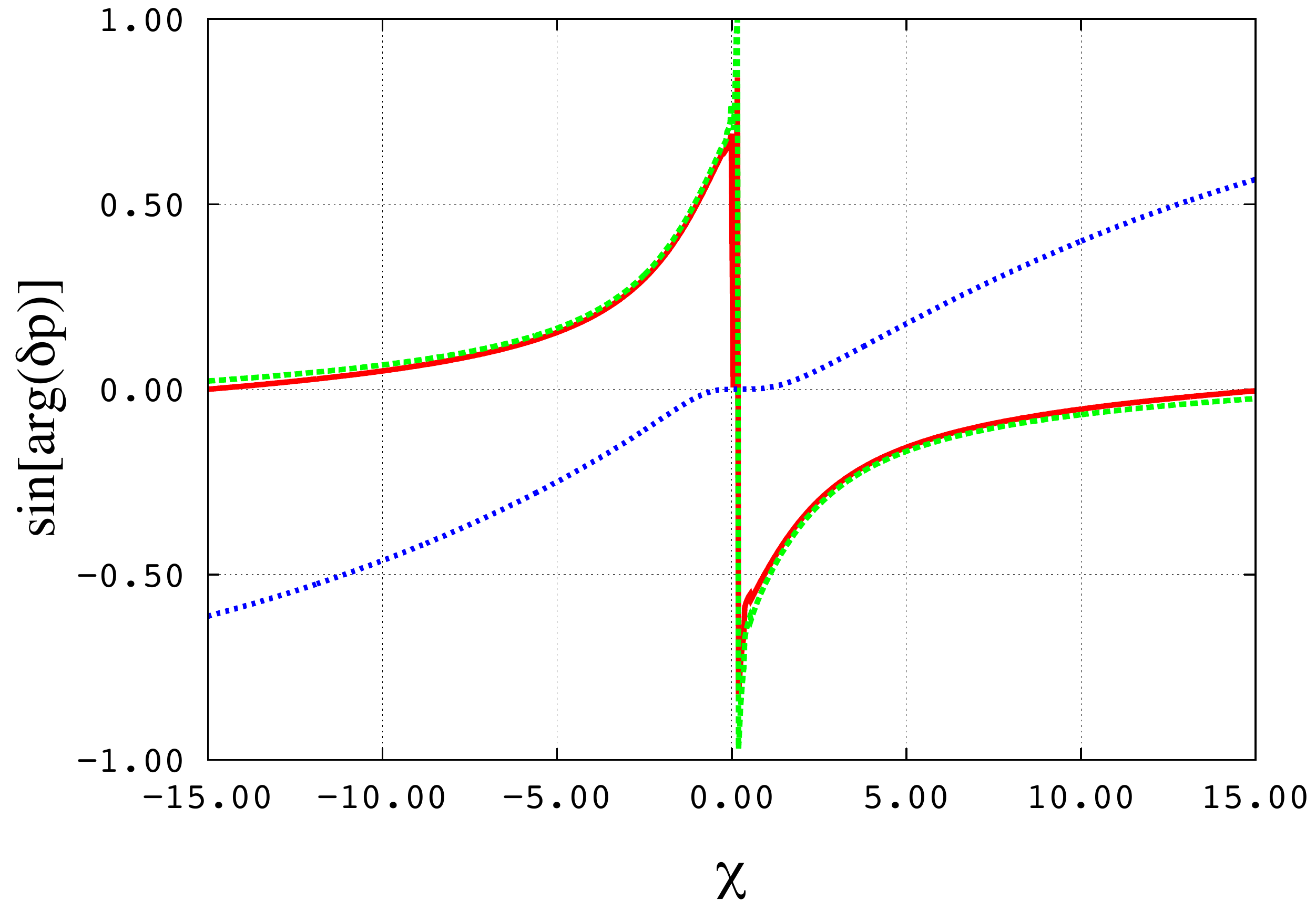} \\
  \includegraphics[width=0.40\textwidth,clip]{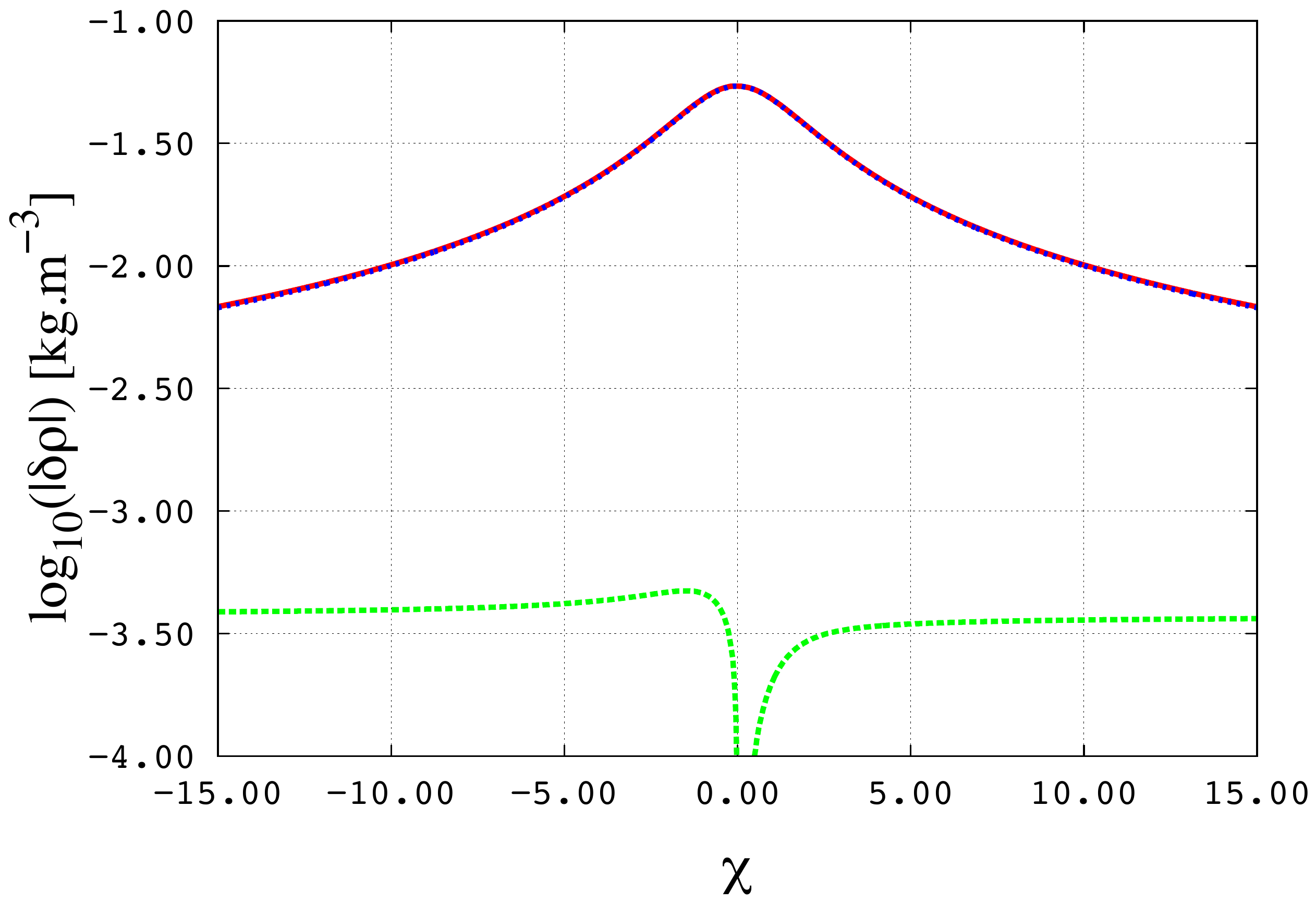} \hspace{1cm}       
 \includegraphics[width=0.40\textwidth,clip]{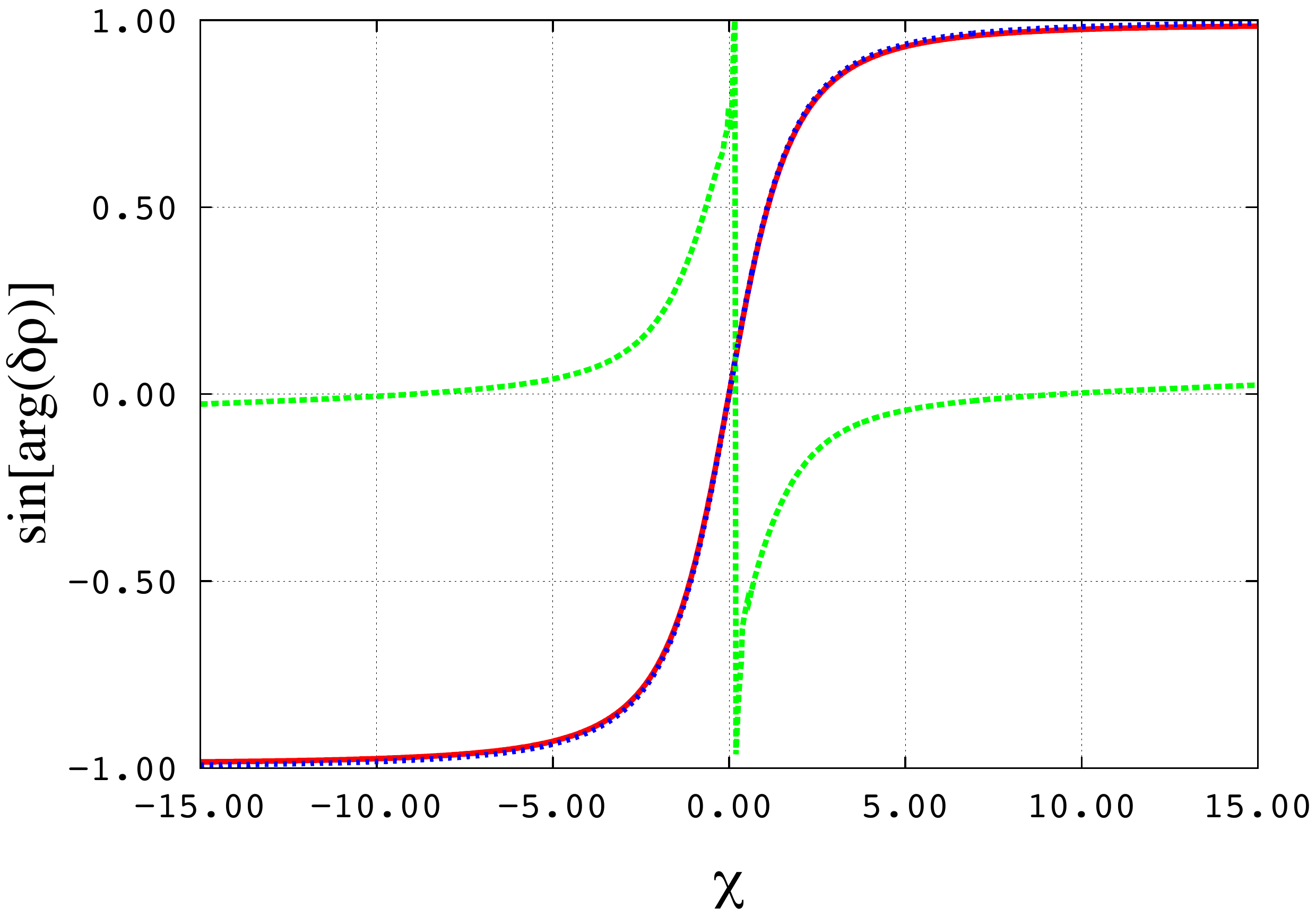} \\
   \includegraphics[width=0.40\textwidth,clip]{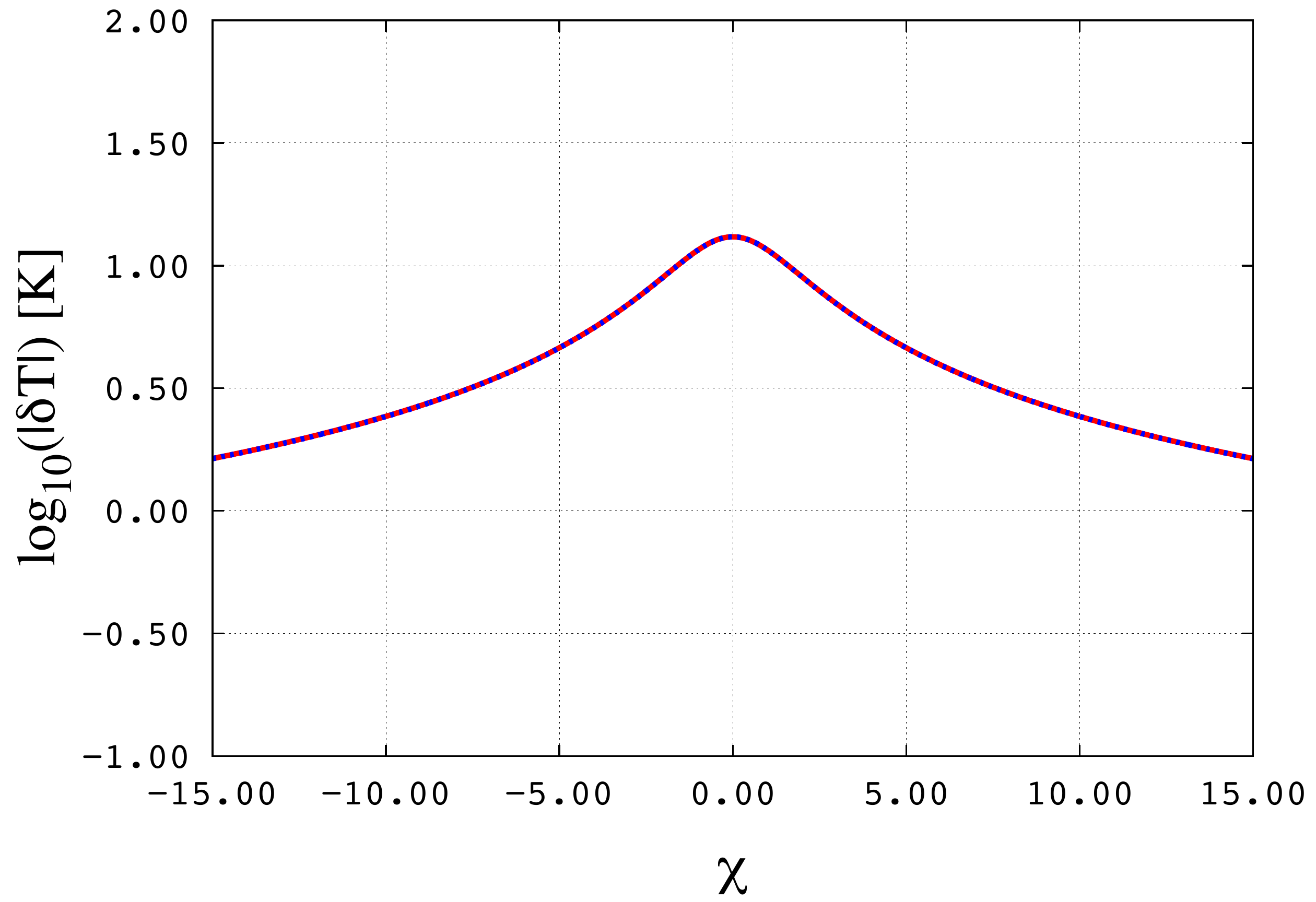} \hspace{1cm}       
 \includegraphics[width=0.40\textwidth,clip]{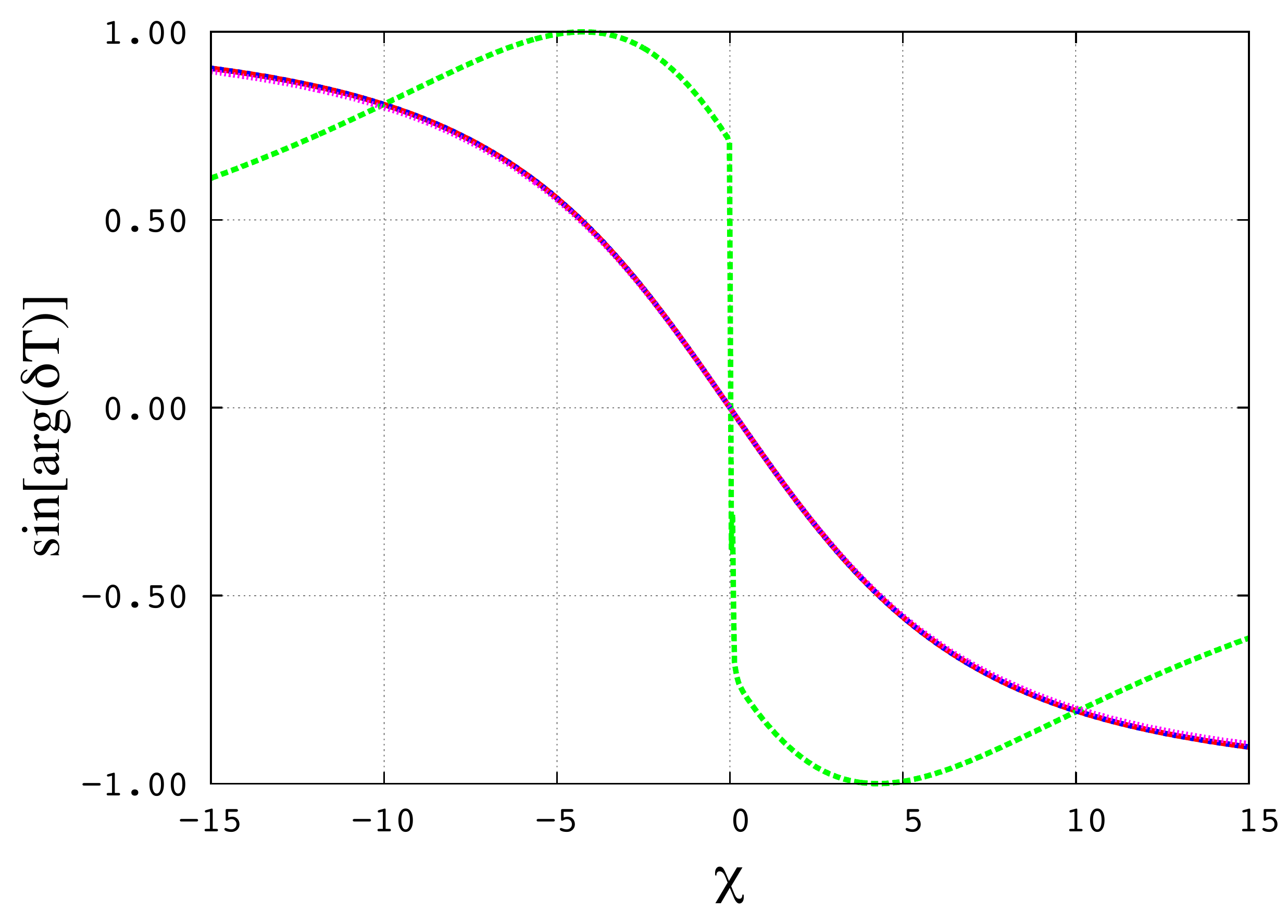}    
  \textsf{ \caption{\label{fig:Terre_variations_sol} Tidal surface oscillations of pressure, density, temperature and their components at the equator of the Earth as functions of the reduced tidal frequency $ \chi = \left( \Omega - n_{\rm orb} \right) / n_{\rm orb} $ (horizontal axis). The perturbation is defined by the academic quadrupolar forcings $ U \left( x , \theta , \varphi , t \right) = U_2 P_2^2 \left( \cos \theta \right) e^{i \left( \sigma t + 2 \varphi \right) } $ and $ J \left( x , \theta , \varphi , t \right) = J_2 P_2^2 \left( \cos \theta \right) e^{i \left( \sigma t + 2 \varphi \right) } $ with $ U_2 = 0.985 \ {\rm m^2.s^{-2}} $ and $ J_2 = 1.0 \times 10^{-2} \ {\rm m^2.s^{-3}} $. Labels \scor{T}, $ \mybf{H} $ and $ \mybf{V} $ refer to the total response and \mybf{its horizontal and vertical components} respectively.  }}
\end{figure*}


\subsection{The Earth}

The Earth semidiurnal thermal tide corresponds to one of the cases detailed in CL70. Here, the radiation frequency is clearly negligible compared to the tidal frequency $ \sigma \approx 2 \Omega $. Therefore, the tidal response of the Earth's atmosphere is driven by dynamical effects only and the radiative losses added in the heat transport equation could be ignored. This case is typical of a pure dynamic behaviour (see Figs.~\ref{fig:kv_map} and \ref{fig:ecart_kv}). Given that $  \nu $ is slightly larger than $ 1 $, the horizontal structure of the perturbations is essentially described by gravity modes (Fig.~\ref{fig:fonctions_Hough_Terre}). Rossby modes are trapped at the poles. Figures~\ref{fig:Earth_section} and \ref{fig:Earth_planis} show maps of the perturbed quantities in latitude (denoted $ \delta $), longitude ($ \varphi $) and altitude ($ x $), the subsolar point being indicated by the coordinates $ \left( \delta , \varphi  \right) = \left( 0 , 0 \right) $. \mybf{On these figure, we can observe that the pressure and density variations present very similar spatial distributions. Particularly, they are in phase with each other and the phase lag is approximatively the same in the whole layer, which materializes the tidal bulge. The lag angle $ \gamma = \pi/4 $ of the bulge with respect to the subsolar point is given by Eq.~(\ref{deltap_ana}) taken at $ x = 0 $ for the gravity mode of lowest degree ($n=0$) and is not very sensitive to temperature structure \citep[][]{Lindzen1968}. For the Earth, these semidiurnal pressure peaks are measured around 9h38mn and 22h07mn (see Fig.~\ref{fig:mesures_pression}).    }

The patterns of the velocities \jcor{$ V_{\theta}^{\sigma,2} $ and $ V_{\varphi}^{\sigma,2} $} are also well explained by the first horizontal Hough functions, \jcor{$ \mathcal{L}_\theta^{2,\nu} \Theta_0^{2,\nu} $ and $ \mathcal{L}_\varphi^{2,\nu} \Theta_0^{2,\nu} $} (see Fig.~\ref{fig:fonctions_Hough_Terre}, middle and bottom left). In constrast, the behaviour of the temperature involves other modes. 


We then compute the evolution of the tidal torque with the tidal frequency by changing the value of the spin frequency ($ \Omega $) in simulations. Hence, the reduced tidal frequency $ \chi = \left( \Omega - n_{\rm orb} \right) / n_{\rm orb} $ varies within the interval $ \left[ - 15 , 15  \right] $. The torque is computed using the analytical formulae of Eqs.~(\ref{int_deltaq}) and (\ref{torque_ana}). We also compute the \mybf{horizontal and vertical components} isolated in Sect.~\ref{sec:atmos_isotT}.The results are plotted on Fig.~\ref{fig:Terre_couple}. \\


\begin{figure*}[htb!]
 \centering
 \includegraphics[width=0.40\textwidth,clip]{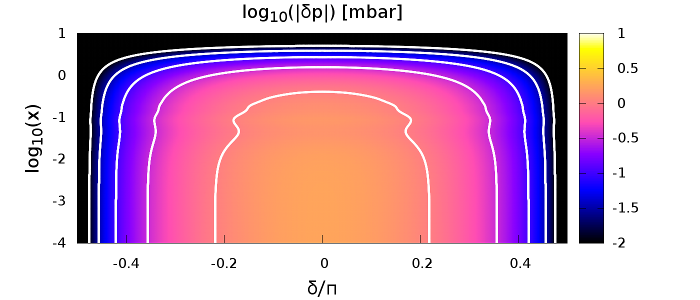} \hspace{1cm}      
 \includegraphics[width=0.40\textwidth,clip]{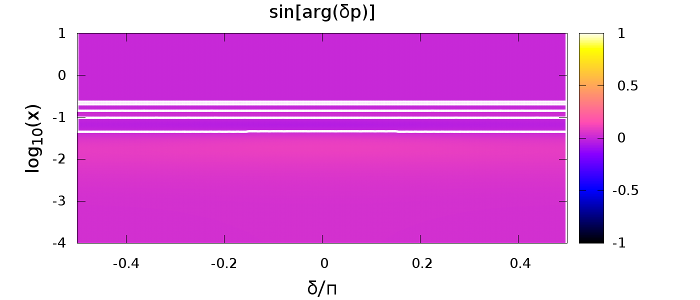} \\
  \includegraphics[width=0.40\textwidth,clip]{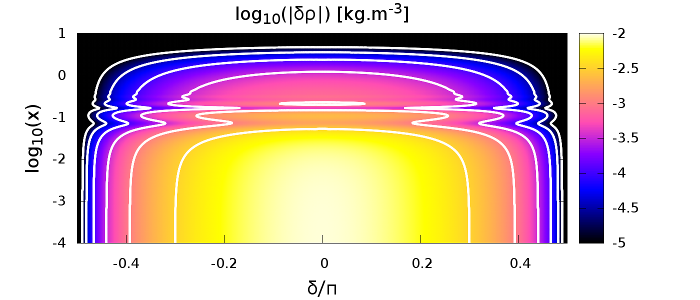} \hspace{1cm}      
 \includegraphics[width=0.40\textwidth,clip]{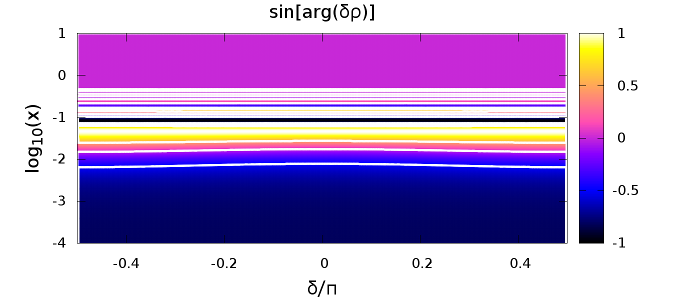} \\
   \includegraphics[width=0.40\textwidth,clip]{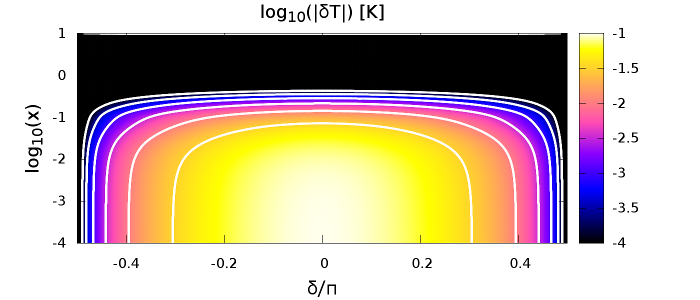} \hspace{1cm}       
 \includegraphics[width=0.40\textwidth,clip]{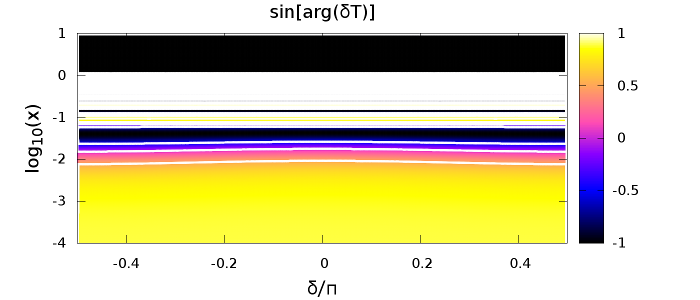}\\
    \includegraphics[width=0.40\textwidth,clip]{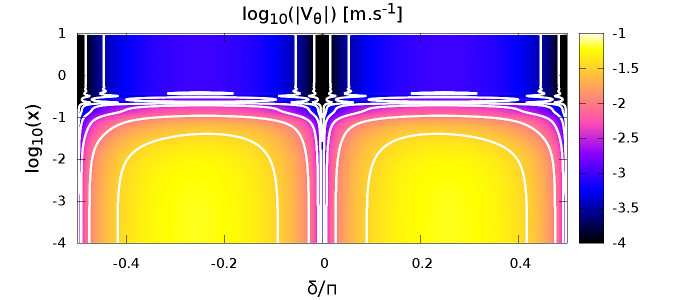} \hspace{1cm}       
 \includegraphics[width=0.40\textwidth,clip]{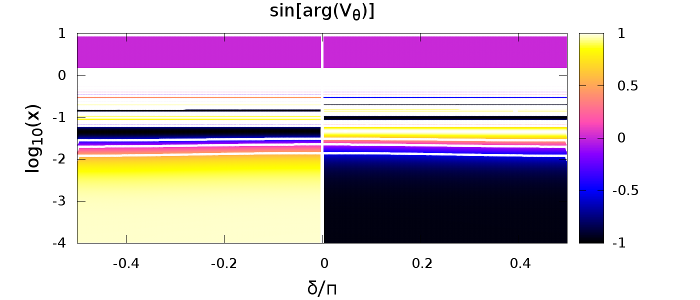}\\
    \includegraphics[width=0.40\textwidth,clip]{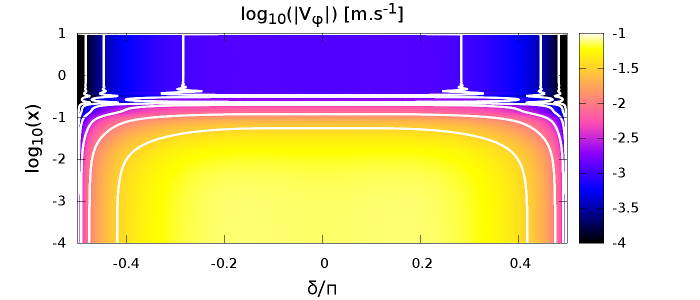} \hspace{1cm}      
 \includegraphics[width=0.40\textwidth,clip]{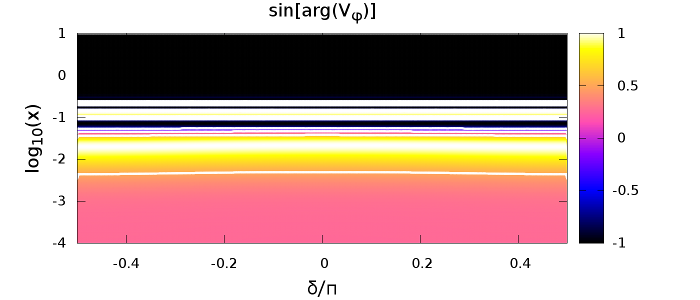}\\
     \includegraphics[width=0.40\textwidth,clip]{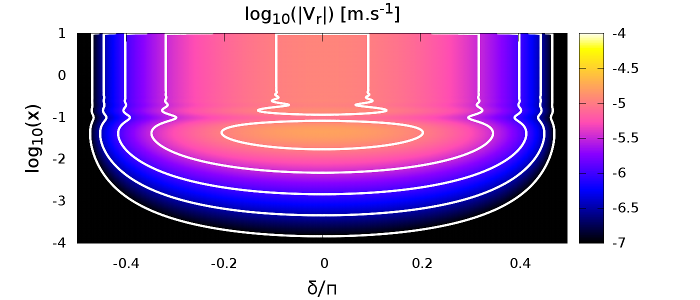} \hspace{1cm}       
 \includegraphics[width=0.40\textwidth,clip]{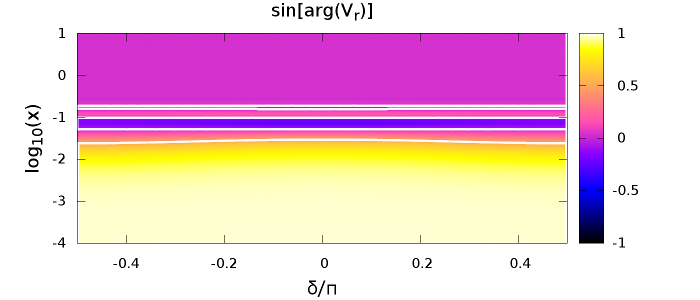}
  \textsf{ \caption{\label{fig:Venus_section} Total tidal response of Venus' atmosphere to the solar semidiurnal perturbation caused by the quadrupolar academic forcings $ U \left( x , \theta , \varphi , t \right) = U_2 P_2^2 \left( \cos \theta \right) e^{i \left( \sigma t + 2 \varphi \right) } $ and $ J \left( x , \theta , \varphi , t \right) = J_2 P_2^2 \left( \cos \theta \right) e^{i \left( \sigma t + 2 \varphi \right) } $ with $ U_2 = 2.349 \ {\rm m^2.s^{-2}} $ and \jcor{$ J_2 = 1.0 \times 10^{-4} \ {\rm m^2.s^{-3}} $}. {\it Left, from top to bottom:} amplitudes of $ \delta p $, $ \delta \rho $, $ \delta T $, $ V_\theta $, $ V_\varphi $ and $ V_r $ in logarithmic scales as functions of the reduced latitude $ \delta / \pi $ (horizontal axis) and altitude $ x = z / H $ in logarithmic scale (vertical axis). The color function $ c $ is given by $ c = \log \left( \left| \delta f \right| \right) $ for any quantity $ \delta f $. {\it Right, from top to bottom:} sinus of the argument of the same quantities as functions of the reduced latitude and altitude. In these plots, $ c = \sin \left[  \arg \left( \delta f \right) \right] $. }}
\end{figure*}

\begin{figure*}[htb!]
 \centering
 \includegraphics[width=0.40\textwidth,clip]{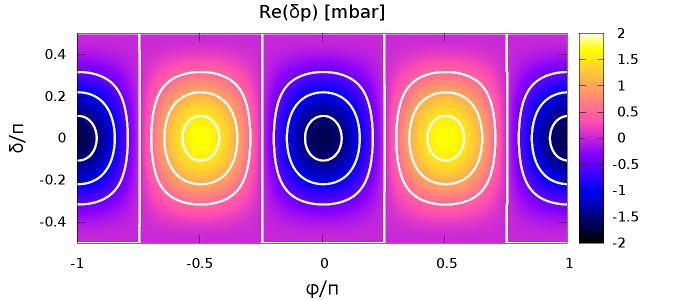} \hspace{1cm}     
 \includegraphics[width=0.40\textwidth,clip]{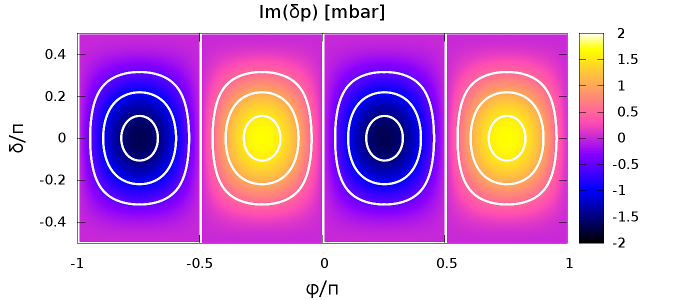} \\
  \includegraphics[width=0.40\textwidth,clip]{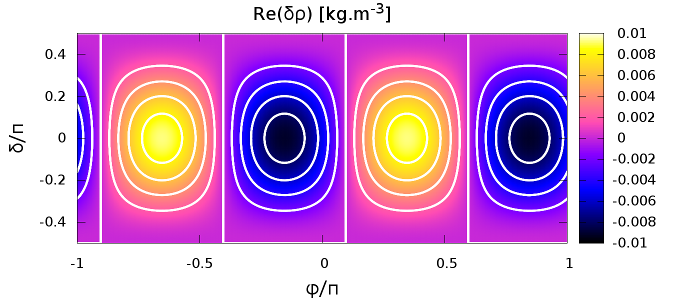} \hspace{1cm}      
 \includegraphics[width=0.40\textwidth,clip]{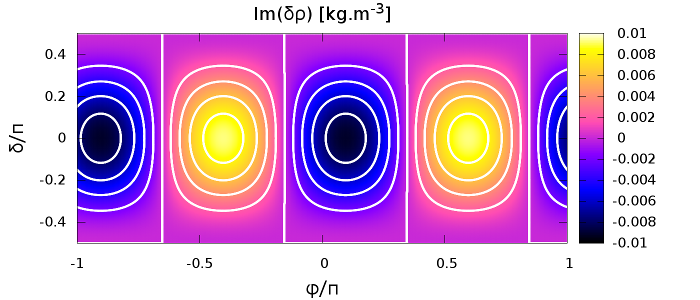} \\
   \includegraphics[width=0.40\textwidth,clip]{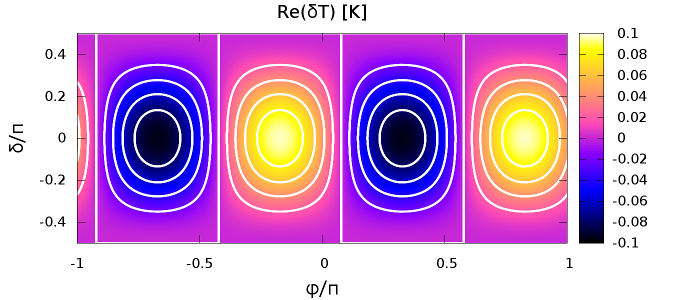} \hspace{1cm}      
 \includegraphics[width=0.40\textwidth,clip]{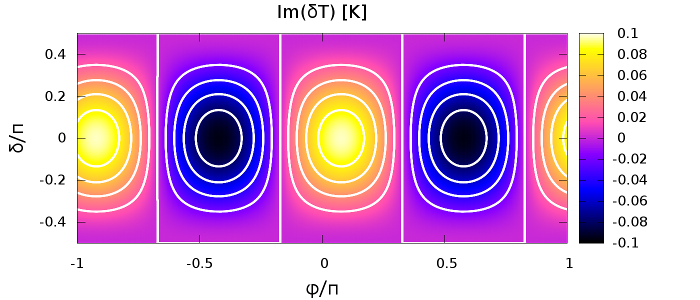}\\
    \includegraphics[width=0.40\textwidth,clip]{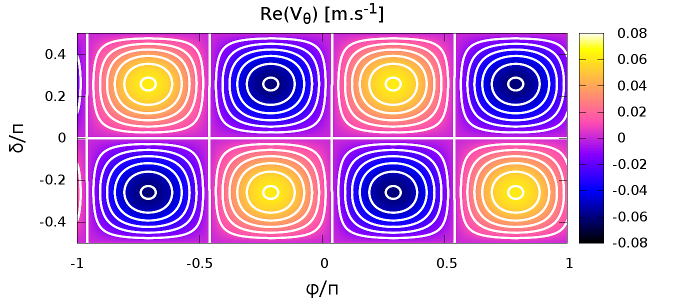} \hspace{1cm}       
 \includegraphics[width=0.40\textwidth,clip]{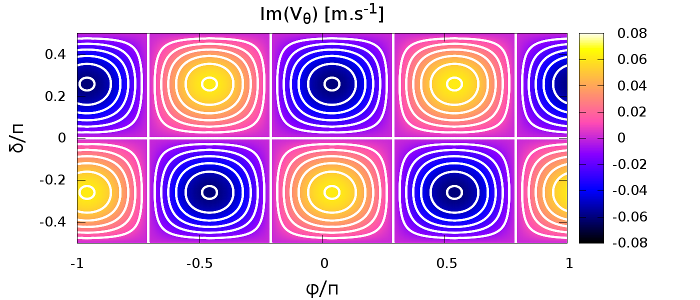}\\
    \includegraphics[width=0.40\textwidth,clip]{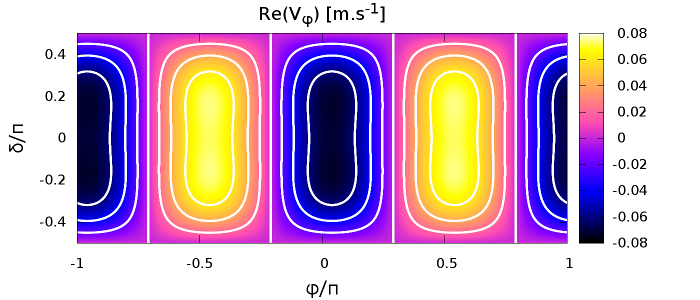} \hspace{1cm}       
 \includegraphics[width=0.40\textwidth,clip]{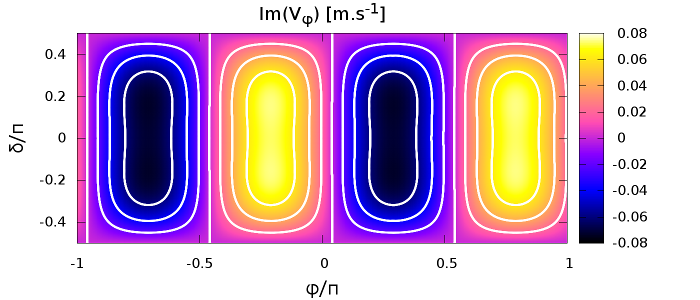} 
  \textsf{ \caption{\label{fig:Venus_planis} Total tidal response of Venus' atmosphere to the solar semidiurnal perturbation caused by the quadrupolar academic forcings $ U \left( x , \theta , \varphi , t \right) = U_2 P_2^2 \left( \cos \theta \right) e^{i \left( \sigma t + 2 \varphi \right) } $ and $ J \left( x , \theta , \varphi , t \right) = J_2 P_2^2 \left( \cos \theta \right) e^{i \left( \sigma t + 2 \varphi \right) } $ with $ U_2 = 2.349 \ {\rm m^2.s^{-2}} $ and \jcor{$ J_2 = 1.0 \times 10^{-4} \ {\rm m^2.s^{-3}} $}. {\it Left, from top to bottom:} real parts of $ \delta p $, $ \delta \rho $, $ \delta T $, $ V_\theta $ and $ V_\varphi $ taken at the ground as functions of the reduced longitude $ \varphi / \pi $ (horizontal axis) and latitude $ \delta / \pi $ (vertical axis). The color function $ c $ is given by $ c = \Re \left\{ \delta f \right\} $ for any quantity $ \delta f $. {\it Right, from top to bottom:} imaginary parts of the same quantities as functions of the reduced longitude and latitude. In these plots, $ c = \Im \left\{ \delta f \right\} $.}}
\end{figure*}

\begin{figure*}[htb!]
 \centering
 \includegraphics[width=0.40\textwidth,clip]{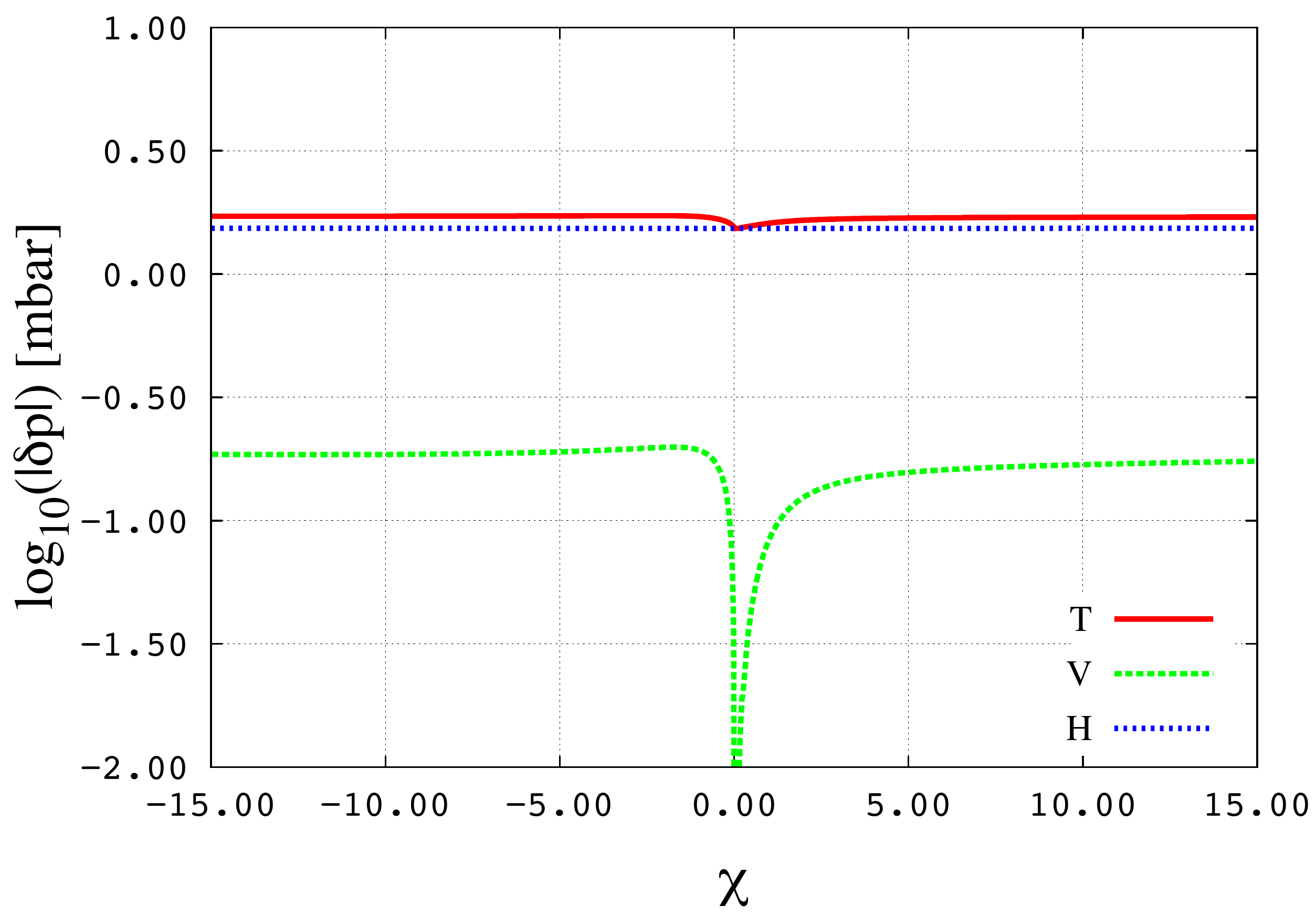} \hspace{1cm}      
 \includegraphics[width=0.40\textwidth,clip]{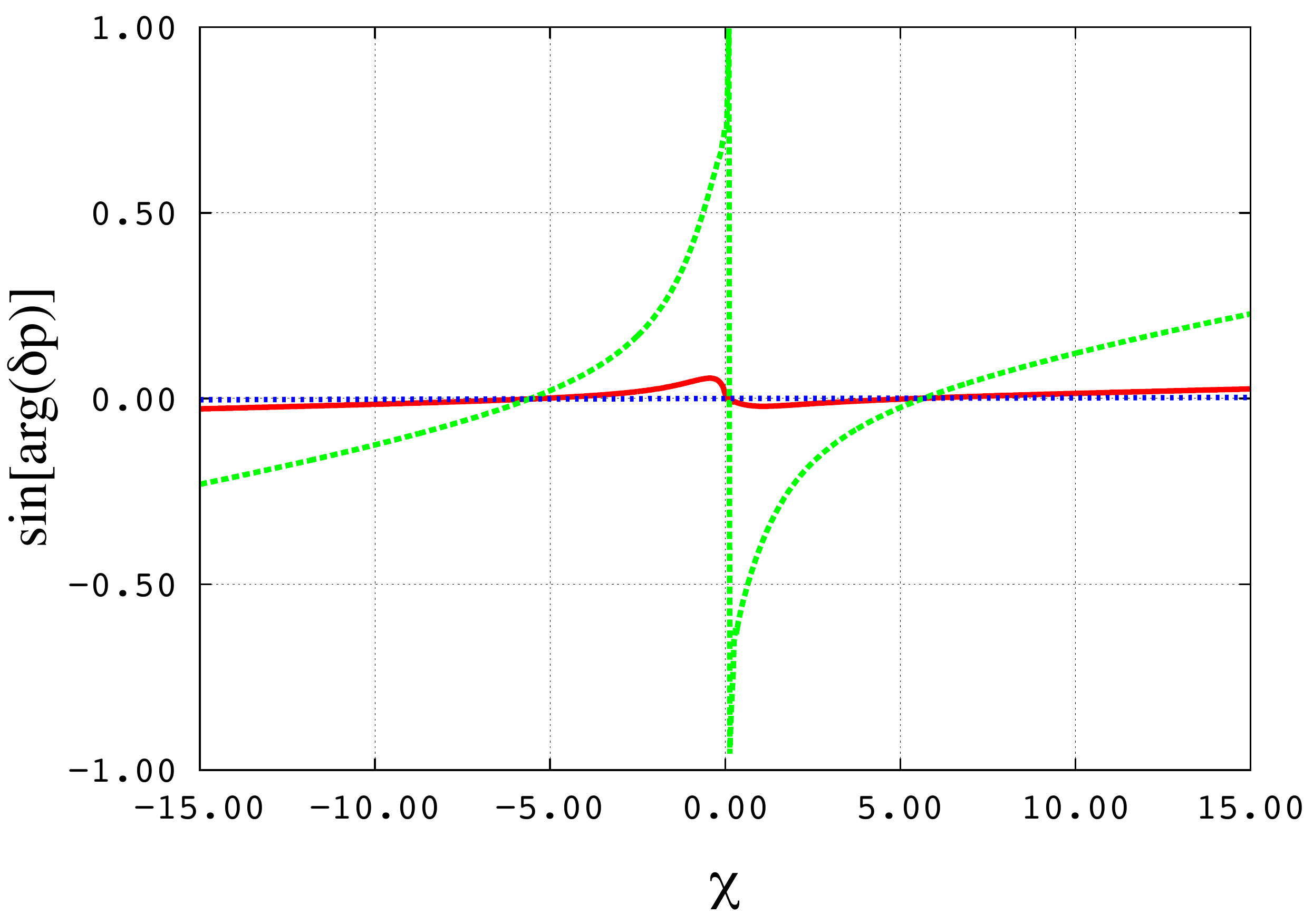} \\
  \includegraphics[width=0.40\textwidth,clip]{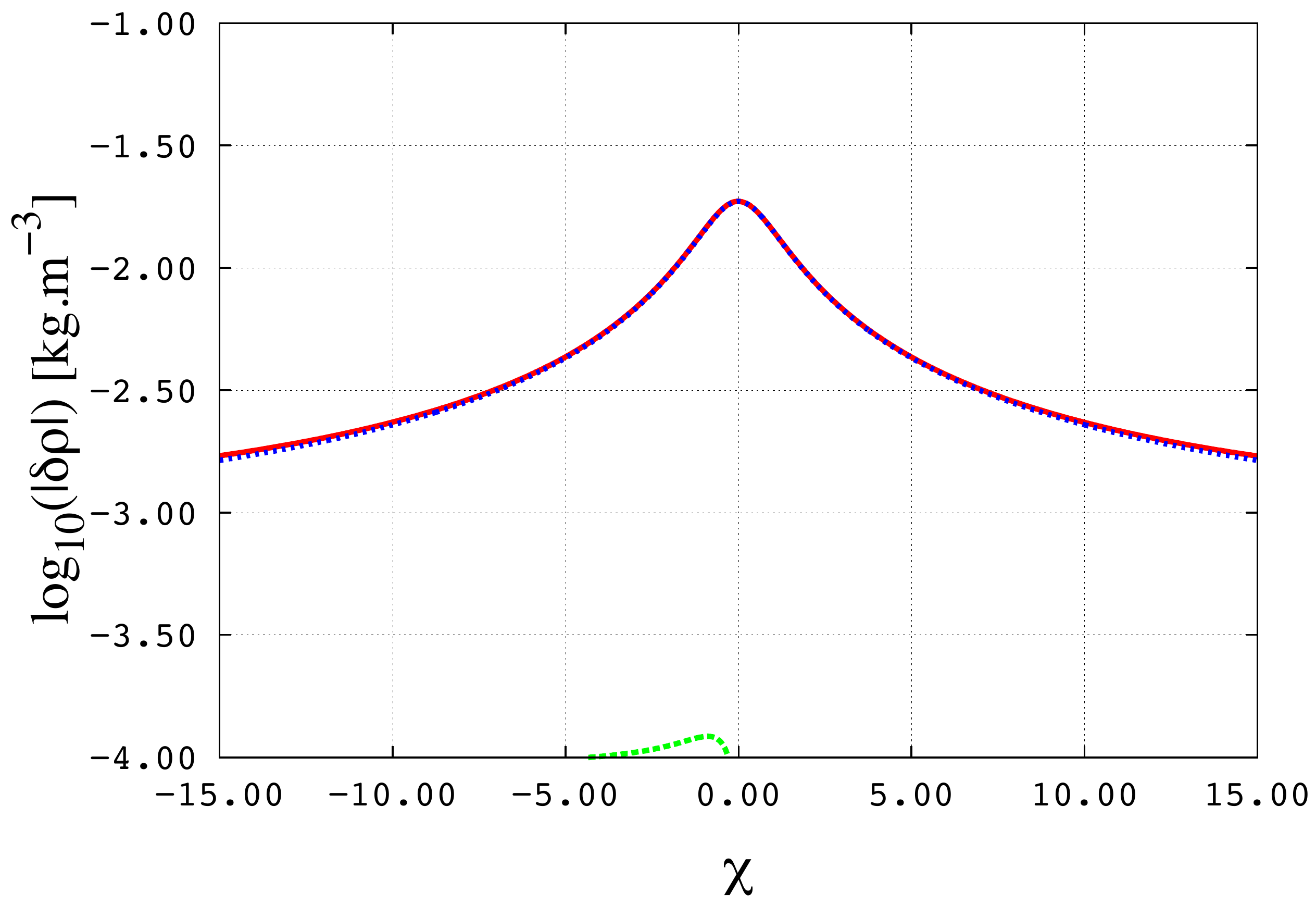} \hspace{1cm}      
 \includegraphics[width=0.40\textwidth,clip]{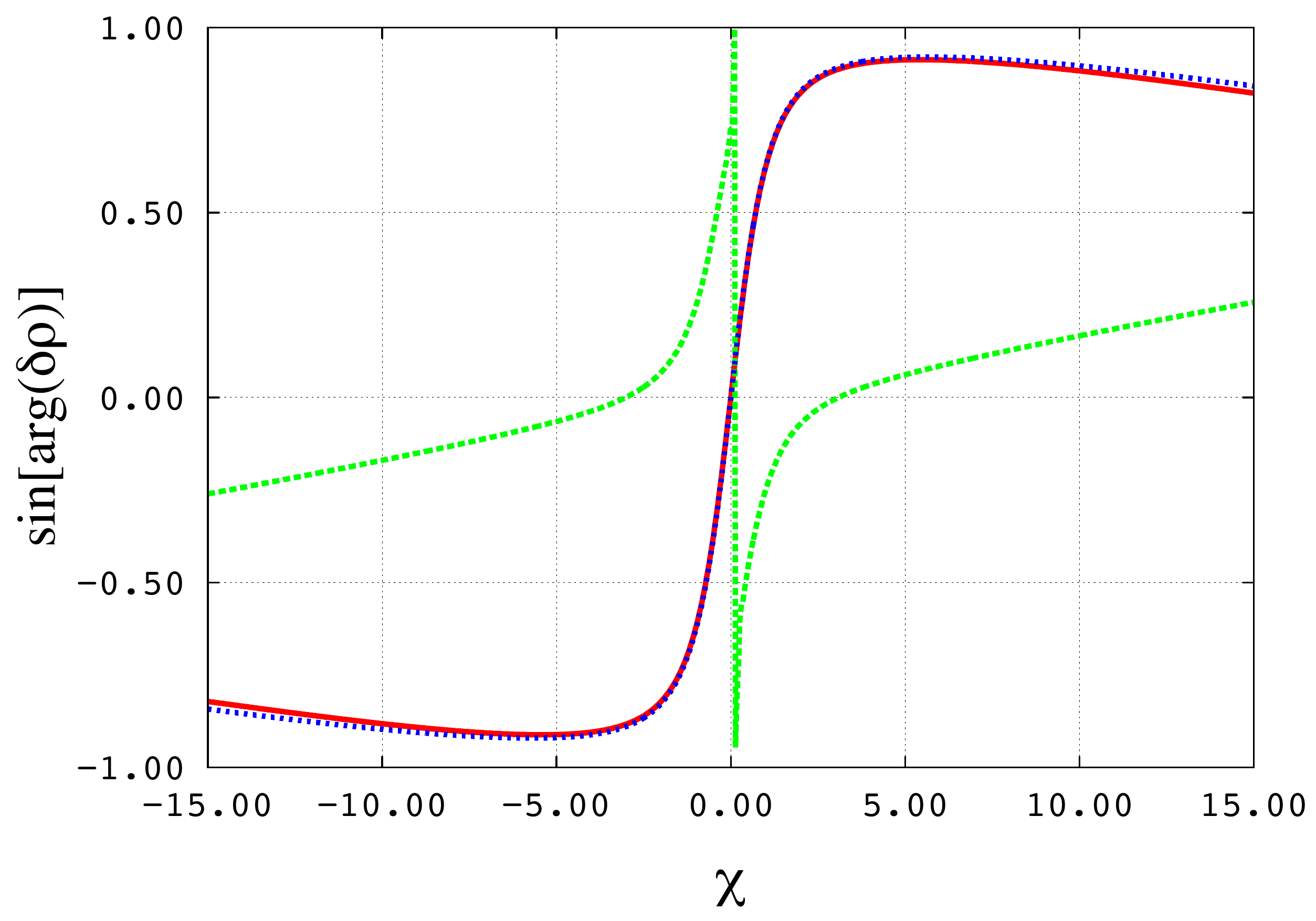} \\
   \includegraphics[width=0.40\textwidth,clip]{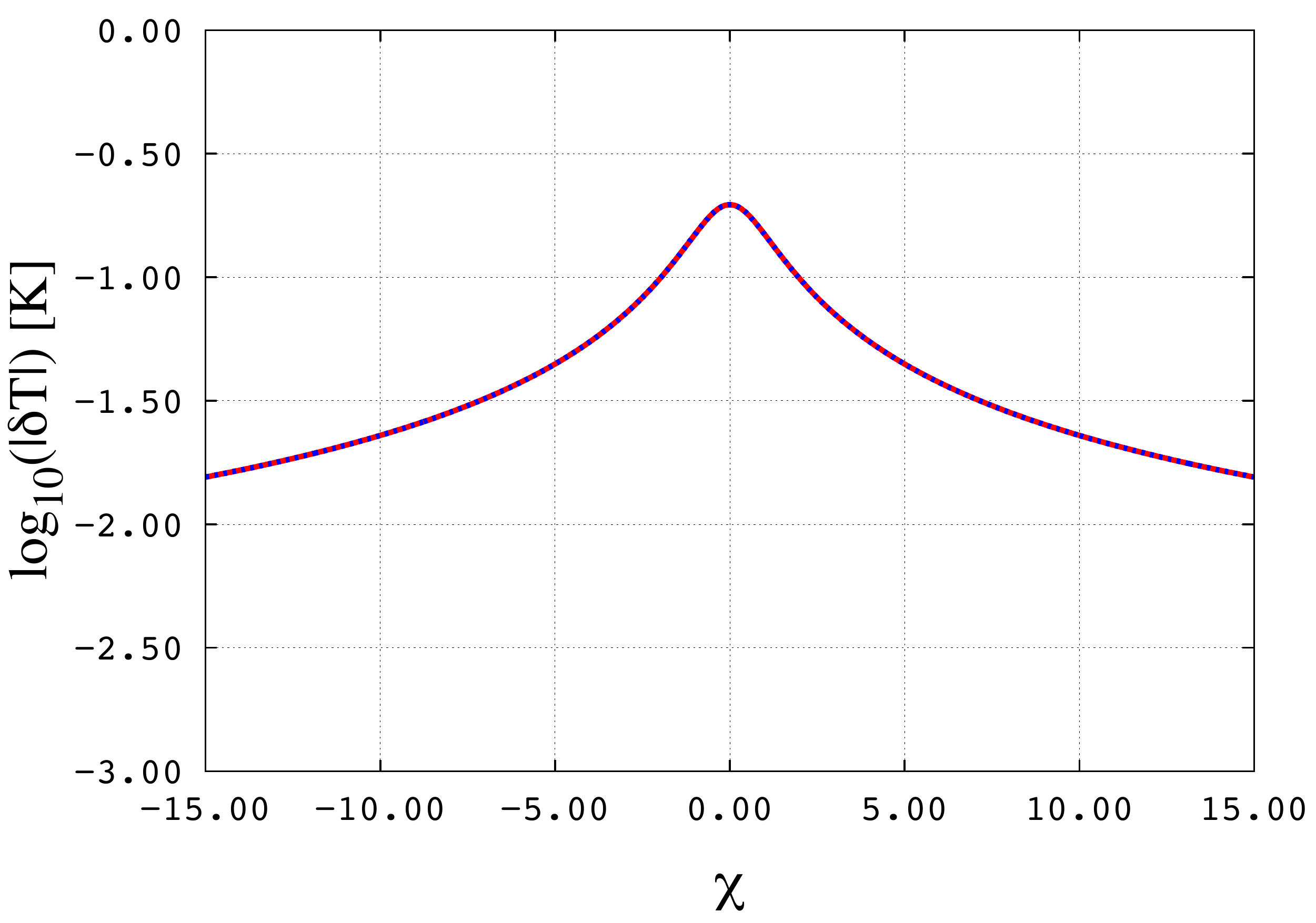} \hspace{1cm}       
 \includegraphics[width=0.40\textwidth,clip]{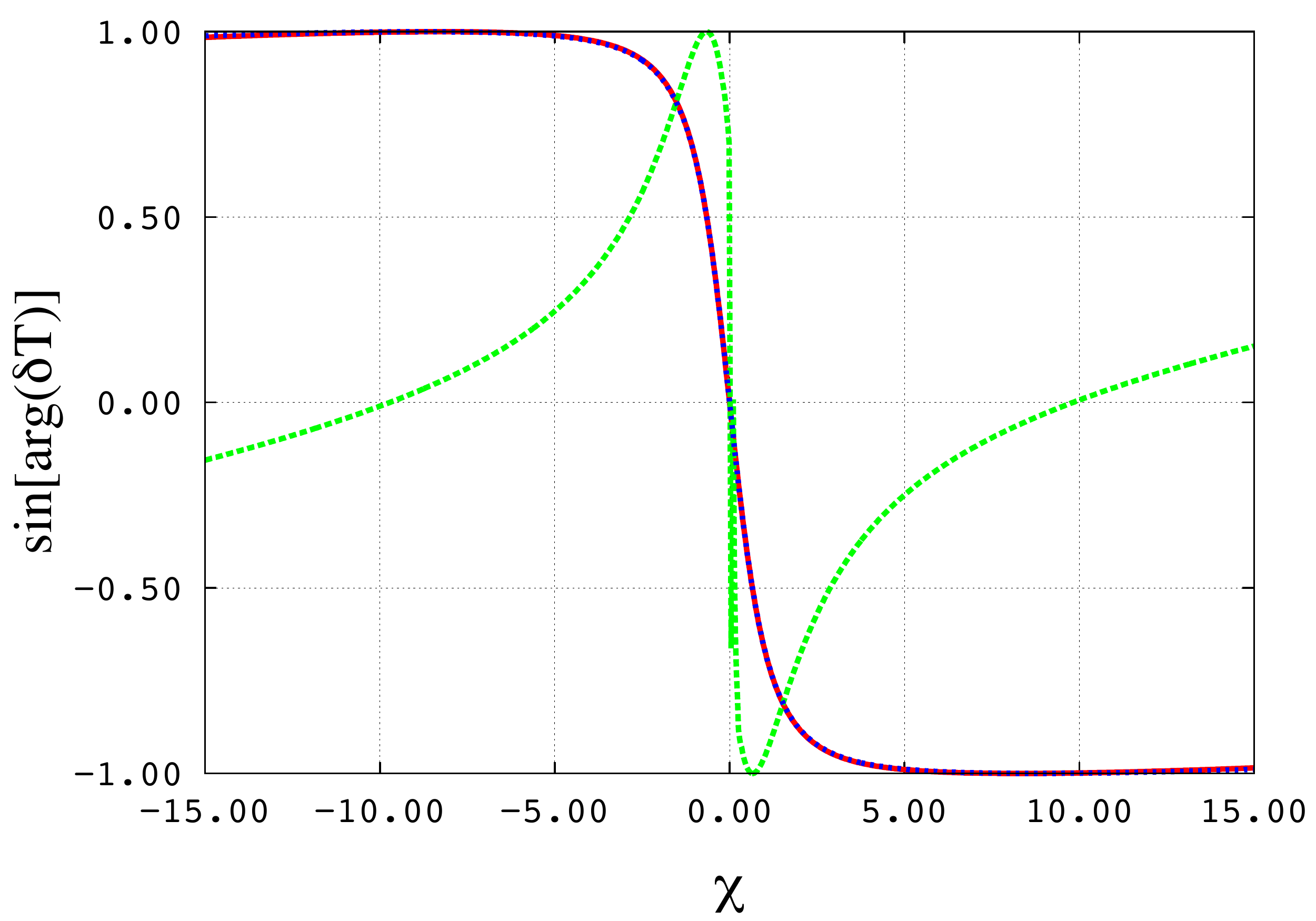}    
  \textsf{ \caption{\label{fig:Venus_variations_sol} Tidal surface oscillations of pressure, density, temperature and their components at the equator of Venus as functions of the reduced tidal frequency $ \chi = \left( \Omega - n_{\rm orb} \right) / n_{\rm orb} $ (horizontal axis). The perturbation is defined by the academic quadrupolar forcing $ U \left( x , \theta , \varphi , t \right) = U_2 P_2^2 \left( \cos \theta \right) e^{i \left( \sigma t + 2 \varphi \right) } $ and $ J \left( x , \theta , \varphi , t \right) = J_2 P_2^2 \left( \cos \theta \right) e^{i \left( \sigma t + 2 \varphi \right) } $ with $ U_2 = 2.349 \ {\rm m^2.s^{-2}} $ and \jcor{$ J_2 = 1.0 \times 10^{-4} \ {\rm m^2.s^{-3}} $}. Labels T, $ \mybf{H} $ and $ \mybf{V} $ refer to the total response and \mybf{its horizontal and vertical components} respectively.  }}
\end{figure*}


At first, we note that the \mybf{horizontal and vertical components of the torque} are of the same order of magnitude and in opposition to one another, as shown by Eq.~(\ref{int_deltaq}) taken at the limit $ \sigma \rightarrow 0 $. This behaviour \mybf{was identified previously in Sect.~\ref{subsec:analytic_isoT}. It results from the fact that vertical displacement of fluid \scor{induced by the stable stratification} compensates the local density variations in this frequency regime} and strongly flattens the total tidal torque. \mybf{As demonstrated in Sect.~\ref{sec:convective_atm}, this is not \scor{the case in general and it depends on the strength of the stratification}. In the case of a slowly rotating \scor{convective} atmosphere ($N^2 \approx 0$) forced by a heating \scor{located} at the ground, only the horizontal component remains. This \scor{leads to} a much stronger tidal torque, similar to \scor{those} obtained by \cite{Leconte2015} with a GCM (given by \scor{Eq.~\ref{torque_nowave}) and \cite{CL01} (see Eq.~\ref{torque_CLNS})}.   } \\

On Fig.~\ref{fig:Terre_variations_sol}, we plot the surface variations of pressure, density, temperature and their components as functions of the tidal frequency. For each quantity, \mybf{horizontal component} varies smoothly while the \mybf{vertical one} is discontinuous at the synchronization. \mybf{This discontinuity} is \mybf{a consequence of} the dissymmetry between prograde and retrograde Hough modes due to the Coriolis acceleration (Fig.~\ref{fig:lambda_nu}), particularly as regards the gravity mode of lowest degree which is the most important one. \\


Assuming that the tidal torque is entirely transmitted to the telluric core of the planet, it is possible to estimate the timescale of the spin evolution induced by the atmospheric semidiurnal tide. Introducing the moment of inertia of the planet, $ \mathcal{I} $, we write:

\begin{equation}
\frac{1}{\Omega} \dfrac{d \Omega}{dt} = \frac{\mathcal{T}}{\mathcal{I} \Omega}.
\end{equation}

Thus, the timescale of the spin evolution is given by

\begin{equation}
\mathscr{T} = \left|  \frac{\mathcal{I} \Omega}{\mathcal{T}} \right|.
\end{equation}

For the Earth's semidiurnal tide, $ \chi \approx 365  $ and the \mybf{horizontal component of the corresponding torque} is $ \mathcal{T} \approx 10^{15} \ {\rm N.m} $. With $ \mathcal{I} = 8.02 \times 10^{37} \ {\rm kg.m^{2}} $ (NASA fact sheets\footnote{Link: \textcolor{blue}{\url{http://nssdc.gsfc.nasa.gov/planetary/factsheet/earthfact.html}}}), $ \mathscr{T}_{\Earth} \sim 180 $ Gyr. Therefore, the spin frequency of the Earth is not affected by atmospheric tides.


\subsection{Venus}

Geometrically speaking, Venus can be seen as an Earth-like planet, with scales of the same magnitude order for the telluric core and the thickness of the atmosphere. However, there are some important differences compared to the case of the Earth. The Venusian atmosphere does not have the same properties as the Earth's atmosphere. \scor{First}, the fluid layer is a hundred times more massive (denoting $ \mathscr{M} $ the mass of the atmosphere, one has $ \mathscr{M}_\Venus \sim 4.8 \times 10^{20} $ kg while $ \mathscr{M}_\Earth \sim 5.1 \times 10^{18} $ kg). \mybf{Thus, the surface pressure is around 93 bar}. \mybf{Second, it is} optically thicker. \mybf{\scor{The} major part of the heating flux is \scor{thus} deposited at high altitudes. The absorbed flux near the planet surface at the subsolar point $ F_{\rm abs} $ is estimated at $ F_{\rm abs} \sim 100 - 200 \ {\rm W.m^{-2}} $ \citep[][]{Avduevsky1970,Lacis1975,DI1980}, which implies $ J_2 \sim 10^{-4}  \ {\rm W.kg^{-1}} $.} \mybf{Finally, layers below 60 km are characterized by a strongly negative temperature gradient \citep[][]{Seiff1980} and are, therefore, weakly stratified, or convective. As discussed in Sect.~\ref{sec:atmos_isotT} and \ref{sec:convective_atm},} these properties have a strong impact on the nature of the tidal response. \mybf{Particularly, the tidal torque may be much stronger in the convective case than in the \scor{stably} stratified one. \jcor{In the present section, we study a Venus-like planet with a stably stratified isothermal atmosphere (VenusX) in order to understand the effects of stratification and radiative losses in a general context.}}


\pcor{The other important difference between the Earth and Venus is} their rotational dynamics. While the spin frequency of the Earth is far much higher than its mean  motion and the radiation frequency of its atmosphere, these three frequencies are of the same order of magnitude for Venus. Therefore, the Venusian semidiurnal thermal tide is characterized by a tidal frequency $ \left| \sigma \right| \sim \sigma_0  $, which is typical of the thermal regime. The term describing radiative losses in the heat transport equation (Eq.~\ref{transport_chaleur_2}) plays an important role in this regime by damping the profiles in altitude of the tidal response; figure~\ref{fig:kv_map} illustrates that point. This figure shows that the imaginary part of $ \hat{k}_n $, which is responsible for the damping, is always comparable to the real part in this frequency range. \\

Contrary to the case of the Earth, the semidiurnal thermal tide \jcor{VenusX} belongs to the family of super-inertial waves, with $ \nu \approx 0.48 $. So the horizontal component of the tidal response is composed of gravity modes only. We can note that the surface variations of perturbed quantities, plotted on Figure~\ref{fig:Venus_planis}, are very well represented by the first mode. The patterns of the Hough functions \jcor{$ \Theta_0^{2,\nu} $, $ \mathcal{L}_\theta^{2,\nu} \Theta_0^{2,\nu} $ and $ \mathcal{L}_\varphi^{2,\nu} \Theta_0^{2,\nu} $} (see Fig.~\ref{fig:fonctions_Hough_Venus}, red curves) clearly appear on the maps. Moreover, the lags of $ \delta p $ and $ \delta \rho $ are different from the one observed in the Earth's semidiurnal tide. In particular, it is interesting to note that the pressure and density peaks are not superposed in this case. The effect of the damping caused by radiative losses can be observed in vertical cross-sections of Fig.~\ref{fig:Venus_section}, where the variations of pressure and density are located close to the ground. Given that the real and imaginary part of the vertical wavenumber are both very high, \mybf{vertical component is both} strongly oscillating and strongly damped. \\

As previously done for the Earth, we draw the variations of the tidal torque (Fig.~\ref{fig:Terre_couple}) and perturbed quantities at the ground (Fig.~\ref{fig:Venus_variations_sol}) around the synchronization, for $ \chi \in \left[ -15 , 15 \right] $. The behaviour of the torque and its components is the same as the one observed for the Earth previously. The maxima of \mybf{horizontal and vertical components} are located at $ \chi = \pm \chi_\Venus $ with $ \chi_\Venus < \chi_\Earth $ owing to the difference of mean motions between the two planets. Since the tidal torque is proportional to the surface density of the atmosphere, it is a hundred time stronger in the case of Venus than in the case of the Earth for a given thermal forcing. This is also verified by the surface variations of the perturbed quantities which, except their amplitudes, are comparable to those of the Earth.  \\

 To conclude this section, let us compute the timescale of the spin evolution. The Venusian semidiurnal tide is characterized by the frequency $ \chi \approx -1.92 $. With the moment of inertia $ \mathcal{I} = 5.88 \times 10^{37} \ {\rm kg.m^2} $ (\textcolor{blue}{\href{http://nssdc.gsfc.nasa.gov/planetary/factsheet/venusfact.html}{NASA fact sheets}}) and the \mybf{horizontal component of the torque} given by Fig.~\ref{fig:Terre_couple}, i.e. \jcor{$ \mathcal{T} \sim - 5 \times 10^{16} \ {\rm N.m} $}, we obtain \jcor{$\mathscr{T}_{\Venus} \sim 10 $ Myr. As the thermal forcing corresponds approximately to the mean level of the heating profile used by \cite{DI1980} ($J_2 \sim 10^{-4} \ {\rm W.kg^{-1}}$), the obtained torque is of the same order of magnitude as the one given by this early work, $ \mathcal{T} \sim - 1.8 \times 10^{16} \ {\rm N.m} $. Note here that the amplitude of the response is directly proportional to the amplitude of the excitation if one of its two components, $ J $ or $ U $, may be neglected (Sect.~\ref{sec:thick_atmos}).   }

\subsection{Discussion of physical assumptions}

We have developed a new analytic model for atmospheric tides by generalizing the classical tidal theory detailed in CL70 to dissipative atmospheres. \\

This work leans on several assumptions which contribute to make computations simpler. The most important of them are listed thereafter.

\begin{figure}[htb]
\centering
{\includegraphics[width=0.9\textwidth]
{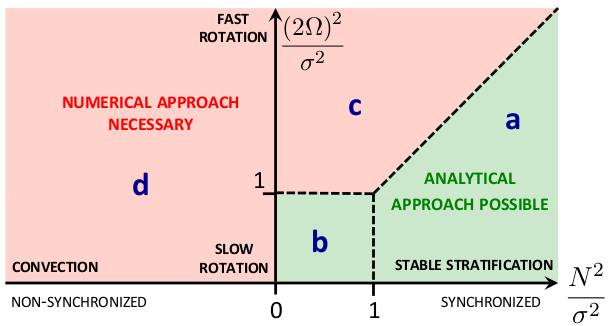}
\textsf{\caption{\label{fig:domaines_ana} \mybf{Asymptotic domains of parameters where the analytical approach presented in this work can be applied (green) and where the numerical approach is necessary (red) as functions of the frequency ratios $ N^2 / \sigma^2 $ (horizontal axis) and $ \left( 2 \Omega \right)^2 / \sigma^2 $ (vertical axis). Green regions: stably stratified atmospheres satifiying the hierarchy of frequencies $ \left( 2 \Omega \right)^2 \ll N^2 $ necessary to apply the traditional approximation (\textbf{a});  weakly stratified or convective atmospheres of slowly rotating planets (\textbf{b}). Red regions: stably stratified atmospheres where the hierarchy of frequencies $ \left( 2 \Omega \right)^2 \ll N^2 $ is not satisfied (\textbf{c}); \scor{super-adiabatic rotating convective atmospheres} (\textbf{d}). } } }}
\end{figure}

\begin{itemize}
\item[$\bullet$] \textit{The traditional approximation}. This hypothesis, commonly used in the literature of geophysics, is very useful because it allows \scor{us} to separate the horizontal and vertical dependences in the Navier-Stokes equation. Hence, the horizontal and vertical structure can be computed separately. This approximation \scor{is} relevant in strongly \scor{stably-}stratified fluids where $ 2 \Omega < \left| \sigma \right| \ll N $, but less appropriate if $ \left| \sigma \right| \sim 2 \Omega $, as discussed in \cite{M09}. \mybf{Its validity domain, not well defined, remains essentially limited to the regime of super-inertial waves \citep[][]{Prat2016}. \scor{It is represented on Fig.~\ref{fig:domaines_ana}, which shows the cases that can be treated analytically (regions a and b).} Concerning the importance of the neglected Coriolis terms, we refer the reader to} \cite{GS2005}, who explored the physics of gravito-inertial waves without the traditional approximation \citep[see also][]{Tort2014}. \mybf{The traditional approximation appears to be a relatively strong hypothesis because it prevents to treat the case of \scor{rotating} convective atmospheres \scor{\citep[$ N^2 \approx 0 $, e.g.][]{OL2004}}. Convection is an important feature of the Earth's nearly adiabatic layer (altitudes below 2 km) as well as the Venus' atmosphere \citep[][]{Linkin1986}. It has strong implications on the propagations of waves: gravity waves cannot exist in convective layers, which may significantly modify the tidal response, particularly if the convective turnover time is close to the period of the tidal forcing. \scor{In the general case,} the effect of convection on the structure of waves can be studied either numerically with GCM (Fig.~\ref{fig:domaines_ana}, d) or analytically with simplified models (e.g. based on a Cartesian geometry). \scor{The only exception for a global analytical study of a convective atmosphere is the case treated in Section~\ref{sec:convective_atm}, where stable stratification and rotation are rather weak to be neglected (Fig.~\ref{fig:domaines_ana}, b).}}   \\
\item[$\bullet$] \textit{The linearization around the equilibrium}. In Sect. 2 and 3, the equations have been linearized around the equilibrium. This method is only relevant in cases where the tidal response of the atmosphere can be considered as a small perturbation with respect to the equilibrium state. This condition is verified for the Earth and Venus. \\
\item[$\bullet$] \textit{The solid rotation approximation}. We supposed that the atmosphere rotates with the solid core uniformly. Thus, the atmospheric circulation (both meridional and zonal) is ignored. This approximation is justified by the fact that the most important pressure and density variations are concentrated in the densest layers, near the ground. Given that these layers are bound to the solid core by viscous friction, they rotate at the same spin frequency. However, tidal waves might \scor{interact with} to the mean flow in cases where $ \sigma \approx 0 $ \scor{\citep[][]{GN1989}} and it will be interesting to take into account the global atmospheric circulation in further works, as done in GCM simulations.  \\
\item[$\bullet$] \scor{\textit{The use of simplified atmospheric vertical structures}. In order to lighten calculations, we considered the cases of uniform background temperature (isothermal stably-stratified atmosphere) and temperature gradient (convective atmosphere), and supposed that some parameters, such as $ \sigma_0 $, do not vary with altitude. These simplified structures are a first step towards more sophisticated ones. Owing to the importance of the stratification in the tidal response, it will be necessary to treat the cases of observed structures in future works; for instance, those of the Earth, Venus, and Mars atmospheres. } \\
\item[$\bullet$] \textit{The optically thin layer hypothesis}. Radiative processes are introduced in the dynamics of atmospheric tides through a Newtonian cooling. The linear dependence of the corresponding radiative term on the temperature perturbation can be interpreted as an optically thin layer approximation. Indeed, this term corresponds to the thermal behaviour of an atmosphere that would emit a grey body radiation towards space and the planet surface. Actually, the flux emitted is partly absorbed by the atmosphere itself, which induces a complex thermal coupling between the different layers of the atmosphere. \mybf{This is true particularly for optically thick atmosphere. In these cases, the Newtonian cooling shall be seen as a convenient approximation of thermal radiation \scor{for an} analytical approach, where the dissipative processes cannot be treated \scor{without approximations}. \scor{This approach} seems to be well validated by numerical simulations using GCM \citep[][]{Leconte2015}. }
\end{itemize}

\begin{figure*}[htb]
\centering
{\includegraphics[width=0.8\textwidth]
{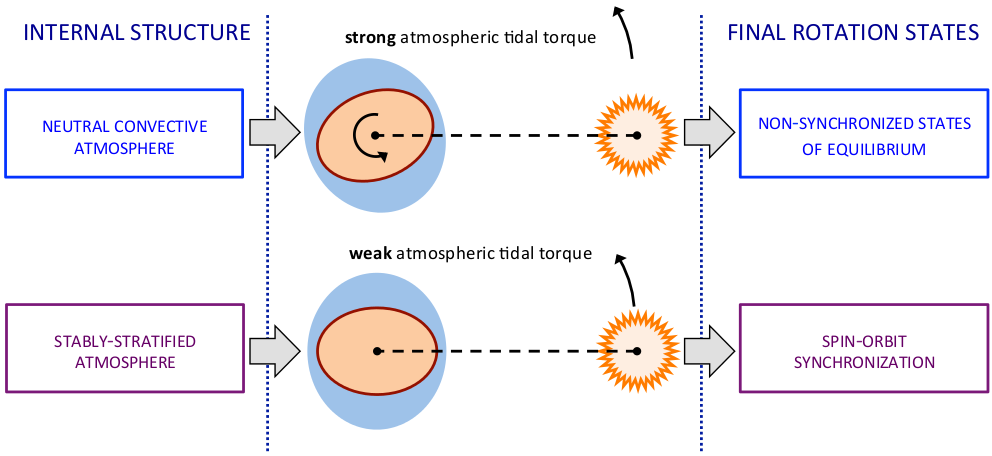}
\textsf{\caption{\label{fig:prediction_strat} \jcor{Predictive aspects related to the dependence of the atmospheric tidal torque on stratification. The neutral convective one implies a strong torque, which can lead to non-synchronized states of equilibrium. The stably-stratified case implies a weak torque letting the planet evolve towards spin-orbit synchronization.  } } }}
\end{figure*}

\section{Conclusions}

We have set the bases of a new theoretical framework to study the atmospheric tides of Earth-like exoplanets analytically. This work is complementary with numerical simulations. The main objective is to explain the qualitative behavior of an atmosphere submitted to both thermal and gravitational tidal perturbations and to explore the domain of key physical parameters. The laters, highlighted by the model, are
\begin{itemize}
\item the characteristic frequencies of the system (e.g. inertia, Brunt-V\"ais\"al\"a, radiative, boundary layer, tidal frequencies),\\ 
\item the involved characteristic length scales (e.g. planetary radius, pressure height scale of the atmosphere, length scales of the vertical variations of tidal waves), \\
\item the physical parameters of the planet (e.g. the molar mass of the atmosphere, the adiabatic exponent, the surface gravity, the surface pressure).
\end{itemize}
Using the equations of hydrodynamics, we have adapted the theoretical approach detailed in CL70 to thick dissipative atmospheric shells. We computed the analytic expressions of the horizontal and vertical structure equations, of the perturbed quantities (pressure, density, temperature, velocity, displacement), of Love numbers and tidal torques exerted on the atmosphere. These expressions have been used to identify the possible regimes of atmospheric tides: a non-dissipative Earth-like dynamical regime, characterizing fast rotating bodies and a dissipative Venus-like thermal regime, for planets near synchronization. The behaviours involved in both cases are very different. The dynamical regime corresponds to the case studied by CL70. In this regime, tidal waves are either oscillating or evanescent, depending of the equivalent depth ($ h_n $) associated with the mode. In the radiative regime, they are both oscillating and evanescent and the perturbation can be strongly damped.\\ 

The tidal response of the atmosphere will also be affected by the nature of the tidal excitation. A predominating thermal forcing generates thermal tides. On the contrary, if the thermal energy brought to the fluid layer is negligible compared to the effect of the tidal gravitational potential, the  perturbation belongs to the family of gravitational tides. This case has not been studied in this work but the vertical structure equation and polarization relations of the model clearly attest of a behaviour qualitatively different from the one observed for thermal tides. Finally, the tidal response of the atmospheric shell is obviously impacted by lower and upper boundary conditions. We showed for instance that a stress-free condition and a \mybf{no-material-escape} condition would not give the same vertical profiles of the perturbation. Similarly, the ground may affect temperature oscillations by introducing a delayed thermal forcing locally. \\

Finally, the results obtained for the tidal torque are in agreement with the GCM simulations by \cite{Leconte2015}. This torque appears to be strongly dependent \mybf{both} on the tidal frequency \mybf{and on the atmospheric structure \jcor{(Fig.~\ref{fig:prediction_strat})}. In the case of stably-stratified atmosphere, the tidal bulge vanishes at the vicinity of synchronization. As a consequence, the tidal torque exerted on the atmosphere is very weak and cannot balance the solid tidal torque, which \jcor{should} lead the planet to synchronization. On the contrary, if the stable stratification is sufficiently weak (weakly stratified or convective atmosphere), the \scor{atmospheric} tidal torque can \scor{be strong}. We recover in this case the Maxwell behaviour obtained by \cite{Leconte2015} numerically and described by the early model of \cite{CL01}. \scor{This would lead planets to non-synchronized states.} These results highlight the predictive interest of the modeling of atmospheric tides by showing that it is possible to constrain the atmospheric structure of terrestrial planets from their rotation states of equilibrium.  }


\begin{acknowledgements}
\mybf{We thank the anonymous referee for comments that helped to improve the paper.} We thank Richard Lindzen for the very interesting discussions we had about the analytic modeling of the Earth's atmospheric tides. We are also very grateful to Umin Lee for his help in understanding how to compute Hough functions efficiently. \mybf{Finally}, we thank Jeremy Leconte, who shared with us his helpful approach of the physical modeling of super-Earths atmospheres from the point of view of numerical simulations. P. Auclair-Desrotour and S. Mathis acknowledge funding by the European Research Council through ERC grant SPIRE 647383. This work was also supported by the Programme National de Plan\'etologie (CNRS/INSU) and CoRoT/Kepler and PLATO CNES grant at CEA-Saclay. 
\end{acknowledgements}

\bibliographystyle{aa}  
\bibliography{ADLM2016} 

\appendix

\section{Hough functions}
\label{app:HF}

Hough functions, named after Hough's initial work \citep{Hough1898}, are the solutions of Eq.~(\ref{eq_Laplace}) integrated as an eigenvalue problem of $ \Lambda_n^{m,\nu} $ with appropriate boundary conditions \citep{Longuet1968,ChapmanLindzen1970,LS1997}. Several ways are possible to compute these functions. Here, we present two possible approaches: the finite differences method proposed in the \textit{Numerical recipes} \citep{Press1986} and detailed in \cite{LS1997} and a spectral method developed from the works of \cite{Hough1898}, \cite{Longuet1968} and \cite{ChapmanLindzen1970}.

\subsection{Finite differences method}

The operator $ \mathcal{L}^{m,\nu} $ of Eq.~(\ref{Laplace}) writes

\begin{equation}
\begin{array}{ll}
\mathcal{L}^{m,\nu}   \equiv  & \displaystyle  \dfrac{d }{d \mu} \left[ \frac{1 - \mu^2}{1 - \mu^2 \nu^2} \dfrac{d}{d \mu}  \right]   \\[0.3cm]
& \displaystyle - \frac{1}{1 - \mu^2 \nu^2} \left( \frac{m^2}{1 - \mu^2} + m \nu \frac{1 + \mu^2 \nu^2}{1 - \mu^2 \nu^2}  \right),
\end{array}
\label{Laplace_mu}
\end{equation}


where $ \mu = \cos \theta $. The new Laplace's tidal equation is not modified when $ \mu $ is replaced by $ - \mu $. Therefore, the domain over which this equation is integrated can be reduced to $ 0~\leq~\mu~\leq~1 $. Like the associated Legendre polynomials, the solutions are either even or odd with respect to the equatorial plane $ \mu = 0 $. Even solutions satisfy $ \Theta_n^{m,\nu}  \left( - \mu \right) = \Theta_n^{m,\nu}  \left( \mu \right)  $, odd solutions $ \Theta_n^{m,\nu}  \left( - \mu \right) = - \Theta_n^{m,\nu}  \left( \mu \right) $. For the sake of simplicity, the parity of the subscript used in the notations corresponds to the parity of the associated function. The solutions belonging to one of these two families are all defined by the boundary condition applied at $ \mu = 0 $: $ d \Theta_n^{m,\nu}  / d \mu = 0 $ for even solutions and $ \Theta_n^{m,\nu}  = 0 $ for odd solutions. The operator $ \mathcal{L}^{m,\nu}  $ has two regular singular points, a permanent one at $ \mu = 1 $ and an other one at $ \mu = 1/\left| \nu \right| $ in the sub-inertial regime, characterized by $ \left| \nu \right| > 1 $ and, therefore, depending on the spin parameter. Both of them are sources of numerical difficulties. To tackle the problem of the singular point at $ \mu = 1 $, we introduce the variable $ y \left( \mu \right) $ defined by \cite{Press1986} and \cite{LS1997},

\begin{equation}
\Theta_n^{m,\nu}  \left( \mu \right) = \left( 1 - \mu^2 \right)^{ \left| m \right| / 2} y \left( \mu \right).
\end{equation}

The Laplace's tidal equation becomes

\begin{equation}
\dfrac{d^2 y}{d \mu} + \mathcal{P} \dfrac{d y}{d \mu} + \mathcal{Q} y = 0,
\label{laplace_y}
\end{equation}

where 

\begin{equation}
\left\{
\begin{array}{ll}
   \mathcal{P} = & \displaystyle 2 \mu \left(  \frac{\nu^2}{1 - \mu^2 \nu^2} - \frac{1 + \left| m \right|}{1 - \mu^2} \right), \\[0.5cm]
   \mathcal{Q} = & \displaystyle \frac{1}{1 - \mu^2} \left[ \Lambda_n^{m,\nu} \left( 1 - \mu^2 \nu^2 \right) - \left| m \right| - m^2 \right. \\
                          & \displaystyle \left. - \frac{2 \left| m \right| \mu^2 \nu^2 + m \nu \left( 1 + \mu^2 \nu^2 \right)}{1 - \mu^2 \nu^2} \right].
\end{array}
\right.
\label{PQ_Hough}
\end{equation}

The boundary condition applied at $ \mu = 0 $ depends on the parity of the sought solution, i.e. $ d y / d \mu = 0 $ at the equator for even modes and $ y = 0 $ for odd modes. We also apply at $ \mu = 0 $ the normalization condition $ y = 1 $ for even modes and $ dy / d \mu = 1 $ for odd modes. Since $ \mathcal{P} $ and $ \mathcal{Q} $ are not defined at $ \mu = 1 $, the boundary condition at the poles is applied at $ \mu = 1 - \epsilon $, where $ 0 < \epsilon \ll 1 $. This condition is derived from the Sturm-Liouville theory, which requires that $ \Theta_n^{m,\nu}  $ shall be regular at $ \mu = 1 $, and is given explicitly by $ \left( 1 - \mu^2 \right) \mathcal{P} dy / d \mu + \left( 1 - \mu^2 \right) \mathcal{Q} y = 0 $ \citep[see][]{LS1997}. \\

In order to solve Eq.~(\ref{laplace_y}) numerically, the domain is discretized. Hence, the equation is reduced to a linear system of dimension $ N_S = 2 N $, where $ N $ is the number of elements of the mesh. In the regime of sub-inertial waves, the singular point at $ \mu = 1 / \left| \nu \right| $ corresponds to a turning point, in the vicinity of which the resolution of the mesh must be sufficiently high. \cite{LS1997} showed that a regular solution could be locally expanded in series of the form $ y = \sum_{j=0}^{+ \infty} a_j x^j $, where $ x = \mu - 1 / \left| \nu \right| $, and satisfying the condition

\begin{equation}
a_1 + \frac{ \left| \nu \right|  \left(  \left| m \right| + m \nu \right) }{1 - \nu^2  } a_0  = 0.
\end{equation}

This allows us to compute the solution properly in the sub-inertial regime. To construct the set of Hough functions, we use as predictors the asymptotic results given by \cite{Longuet1968} and \cite{Townsend2003}. Two types of modes can be identified. \\

\begin{figure}[htb]
\centering
\includegraphics[width=0.89\textwidth,clip]{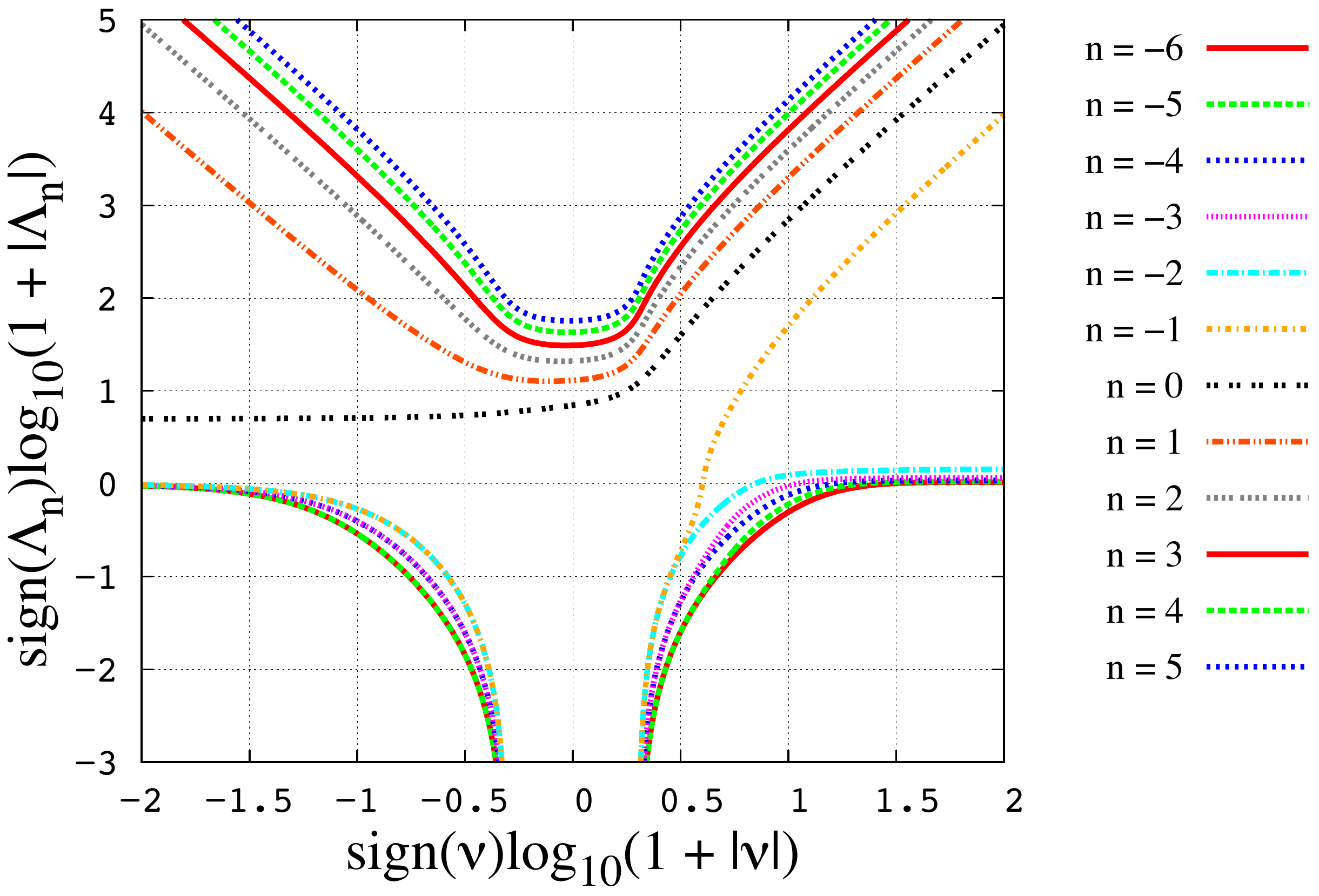}\\
\includegraphics[width=0.950\textwidth,clip]{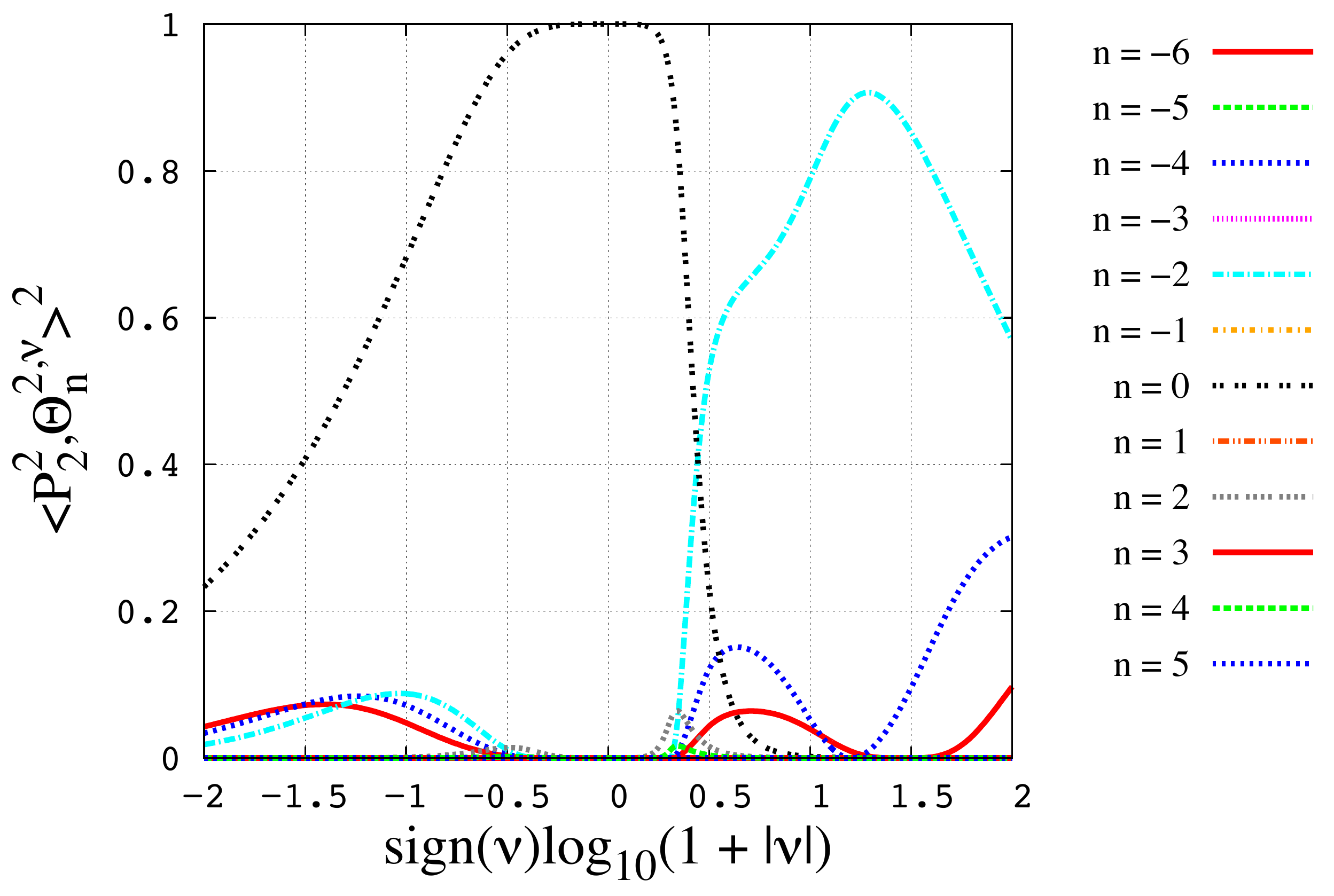}
\textsf{\caption{\label{fig:lambda_nu} \textit{Top:} Eigenvalues ($ \Lambda_n^{m,\nu} $) of the gravity and Rossby modes of lowest degrees ($ -6 \leq n \leq 5 $) as functions of the spin parameter ($ \nu $). The horizontal axis represents the function of $ \nu $ given by $ {\rm sign} \left( \nu \right) {\rm log} \left( 1 + \left| \nu \right| \right)  $, and the vertical axis the function of $ \Lambda_n^{m,\nu} $ given by $ {\rm sign} \left( \Lambda_n^{m,\nu} \right) {\rm log} \left( 1 + \left| \Lambda_n^{m,\nu} \right| \right)  $. \textit{Bottom:} Projection coefficients $ c_{2,n,2} =  \langle P_2^2 , \Theta_n^{2,\nu} \rangle^2 $ of the normalized Legendre polynomial $ P_2^2 $ on normalized Hough functions associated to these modes as functions of the spin parameter (see Eq.~\ref{clnk}).} }
\end{figure}

The modes of the first type, called gravity modes, are defined for $ \nu \in \mathbb{R} $ and characterized by positive eigenvalues (Fig.~\ref{fig:lambda_nu}, top panel). In the non-rotating case, where $ \nu = 0 $, they correspond to the associated Legendre polynomials (Fig.~\ref{fig:lambda_nu}, bottom panel). The gravity modes of Hough functions are ordered with positive $ n $, by analogy with this particular case, for which the degrees $ l $ of Legendre polynomials are given by  $ l = \left| m \right| + n $. In the asymptotic cases where $ \left| \nu \right| \rightarrow + \infty $, analytic solutions using Hermite polynomials can be computed \citep{Townsend2003}. In the sub-inertial regime, the oscillations of gravity modes are trapped in an equatorial belt, in the interval given by $ 0 \leq \mu \leq 1 / \left| \nu \right| $ (see Fig.~\ref{fig:fonctions_Hough}). This explains why these modes are not well represented by associated Legendre polynomials for high values of $ \left| \nu \right| $. \\

The modes of the second type have various denominations in literature \citep[see][]{Longuet1968}. We call them Rossby waves in reference to the early studies by \cite{Rossby1939}. These modes are defined for $ \left| \nu \right| > 1 $ only. Most of them are characterized by negative $ \Lambda_n^{m,\nu} $ and identified in this work by subscripts $ n < 0 $ such as $ \Lambda_{-1} \geq \Lambda_{-2} \geq \Lambda_{-3} \geq \ldots $, the parity of $ n $ corresponding to the parity of the associated function. \cite{Longuet1968} established the analytic expressions of the functions in the vicinity of the boundaries $ \nu = \pm 1 $ using Laguerre polynomials. In contrast with gravity modes, the oscillations of Rossby modes are concentrated around the poles, in a region defined by $ 0 \leq \mu \leq 1 / \left| \nu \right| $. As pointed out by \cite{Lindzen1966}, Rossby modes have to be taken into account in the regime of sub-inertial waves. Otherwise, the set of eingenfunctions is not complete. \\

On Fig.~(\ref{fig:lambda_nu}), we can observe that the eigenvalues ($ \Lambda_n^{m,\nu} $) strongly depend on the spin parameter. This dependence will affect the propagation of tidal waves along the vertical direction through their vertical wavenumbers. Besides, the Coriolis acceleration induces a dissymmetry between prograde ($ \nu < 0 $) and retrograde ($ \nu > 0 $) modes. Particularly, the gravity mode of lowest degree present two different asymptotic behaviours in the sub-inertial regime. The projection of the associated Legendre polynomial $ P_2^2 $, which is representative of the semidiurnal tidal forcing, is also tightly related to $ \nu $. We retrieve for $ \langle P_2^2 , \Theta_n^{2,\nu} \rangle $ the dissymmetry observed for the eigenvalues and note that the variations of the decomposition with $ \nu $ are not simple to describe. In the present study, this decomposition is computed numerically with the spectral method presented thereafter.

\subsection{Spectral method}

The early studies by \cite{Longuet1968} and CL70 provide a detailed theoretical approach of Hough functions and present the very powerful general method introduced by \cite{Hough1898} to compute the eigenvalues and eigenvectors of $ \mathcal{L}^{m,\nu}  $. In this method, which will not be detailed here here, Hough functions are expanded in series of normalized associated Legendre polynomials \mybf{\citep[][]{AS1972}}. In the previous subsection, Eq.~(\ref{Laplace_mu}) highlights a symmetry between the two parameters defining these functions, $ m $ and $ \nu $. Hence, for $ m>0 $ and $ \nu \in \mathbb{R} $,

\begin{equation}
\begin{array}{lcl}
\Theta_n^{-m , \nu} = \Theta_n^{m,-\nu} & \mbox{and} & \Lambda_n^{-m,\nu} = \Lambda_n^{m , - \nu}.
\end{array}
\end{equation}

Therefore, Hough functions and eigenvalues for $ m < 0 $ are readily deduced from those obtained for $ m > 0 $. This allows us to write

\begin{equation}
\Theta_n^{s, \nu} \left( \theta \right) = \sum_{k = s}^{\infty} C_{n,k}^{s,\nu} P_k^s \left( \cos \theta \right),
\label{coeff_HPlm}
\end{equation}

where $ s = \left| m \right| $. A first possible way to compute the coefficients $ C_{n,k}^{s , \nu} $ and eigenvalues $ \Lambda_n^{s, \nu} $, described by CL70, consists in finding the roots of a high-degree polynomial (typically, the degree $ N $ of this polynomial corresponds to the number of coefficients of the series given by Eq.~\ref{coeff_HPlm}). With the current computing resources, $ N = 500 $ can be reached easily. However, this method rapidly becomes tricky to apply beyond $ N \sim 500 $ because of very large polynomial coefficients. Since the size of the $ \Lambda_n^{s,\nu} $ range depends on $ N $, this particularity can limit the number of gravity modes or Rossby modes found. For instance, in the case of the Earth's semidiurnal tide, where $ \nu $ is very close to $ 1 $, the first Rossby modes are characterized by strongly negative eigenvalues. Therefore, only few functions of this type can be computed. To address this issue, the previous problem may be written as an explicit eigenvalue problem. Hence, following \cite{Hough1898} and CL70, we expand the non-normalized Hough functions $ \tilde{\Theta}_n^{s,\nu} $ in non-normalized associated Legendre Polynomials, denoted $ \tilde{P}_k^s $:

\begin{equation}
\tilde{\Theta}_n^{s,\nu} = \sum_{k = s}^{\infty} A_{n,k}^{s, \nu} \tilde{P}_{k}^s \left( \mu \right).
\end{equation}

To lighten expressions, the $ A_{n,k}^{s, \nu} $ are simply denoted $ A_k^s $ in the following analytic development. Let us introduce $ X = 1 /  \left( \nu^2 \Lambda \right) $ and the coefficients

\begin{equation}
\begin{array}{ll}
    K_k = \displaystyle \frac{k-s}{ \left( 2 k - 1 \right) \left[ k \left( k - 1 \right) - s \nu \right] }, & 
    L_k = \displaystyle \frac{k-s -1}{2 k - 3},\\[0.5cm]
    M_k = \displaystyle \frac{\left( k - 1 \right)^2 \left( k + s \right) }{k^2 \left( 2 k +1 \right)}, &
    N_k = \displaystyle \frac{k \left( k + 1 \right) - s \nu }{\nu^2 k^2 \left( k + 1 \right)^2},\\[0.5cm]
    S_k = \displaystyle \frac{k + s +1}{\left( 2k+3 \right) \left[ \left( k +1 \right) \left( k + 2 \right) - s \nu \right]}, &
    U_k = \displaystyle \frac{k + s + 2}{2k + 5},\\[0.5cm]
    T_k = \displaystyle \frac{\left( k + 2 \right)^2 \left( k - s + 1 \right) }{\left( k +1 \right)^2 \left( 2k + 1 \right)}.
\end{array}
\end{equation}

\cite{Hough1898} demonstrated the following equation, which corresponds to Eq. (60) in CL70

\be
X A_k^s  = R_k  A_k^s  + Q_k  A_{k -2}^s + P_k A_{k +2}^s 
\label{pb_vp}
\ee 

with 
\be
\EQM{
P_k  &=  - S_k U_k, \crm 
Q_k  &= - K_k L_k, \crm 
R_k  &=  N_k - K_k M_k - S_k T_k.
} 
\ee

The  system of equations reduces then to the two eigenvalue problems 
\be
X
\begin{bmatrix}
\A_s \\\A_{s+2} \\ \A_{s+4}\\ \vdots\\ \A_{s+2p} \\ \vdots
\end{bmatrix} 
= \Pi_{\rm even}
\begin{bmatrix}
\A_s\\\A_{s+2} \\ \A_{s+4}\\ \vdots\\ \A_{s+2p} \\ \vdots
\end{bmatrix} 
,
\ee
and 

\be
X
\begin{bmatrix}
\A_{s+1}\\\A_{s+3} \\ \A_{s+5}\\ \vdots\\ \A_{s+2p+1} \\ \vdots
\end{bmatrix} 
 = \Pi_{\rm odd}
\begin{bmatrix}
\A_{s+1}\\\A_{s+3} \\ \A_{s+5}\\ \vdots\\ \A_{s+2p+1} \\ \vdots
\end{bmatrix} 
,
\ee

with the matrices 

\begin{equation}
\Pi_{\rm even} = 
\begin{bmatrix}
R_s     & P_s     & 0       &  \cdots & 0  & \cdots \\
Q_{s+2} & R_{s+2} & P_{s+2}   & \cdots & 0  & \cdots \\ 
0       & Q_{s+4} & R_{s+4}  & \cdots & 0  & \cdots \\ 
\vdots  &  \vdots   &   \vdots    &        &  P_{s+2p-2}  & \cdots \\ 
0       &         &         & Q_{s+2p} &R_{s+2p}& \cdots  \\ 
\vdots  &         &          &     \vdots&       &      \\
\end{bmatrix} 
,
\end{equation}

\begin{equation}
\Pi_{\rm odd} = 
\begin{bmatrix}
R_{s+1}     & P_{s+1}     & 0        & \cdots & 0  & \cdots \\
Q_{s+3} & R_{s+3} & P_{s+3}        & \cdots & 0  & \cdots \\ 
0       & Q_{s+5} & R_{s+5} &  \cdots & 0  & \cdots \\ 
\vdots  &  \vdots   &   \vdots     &        &  P_{s+2p-1}  & \cdots \\ 
0       &         &         &   Q_{s+2p+1} &R_{s+2p+1}& \cdots  \\ 
\vdots  &         &         &     \vdots&       &      \\
\end{bmatrix} 
.
\end{equation}

They are solved numerically by linear algebra algorithms with a great efficiency. For example, the case $ N = 1000 $ is computed on a laptop instantaneously. To make the comparison with \cite{LS1997} easier (see the previous subsection), we set $ \Theta_n^{s , \nu}  > 0 $ at the equatorial point ($ \theta = \pi /2 $) for even solutions and $ d \Theta_n^{s, \nu} / d \theta < 0 $ for odd solutions. Hence, the coefficients of Eq.~(\ref{coeff_HPlm}) are explicitly given by

\begin{equation}
C_{n,k}^{s, \nu} = \mathcal{S}_n^{s , \nu}  \frac{ \sqrt{ \frac{2 \left( k + s \right)! }{ \left( 2 k + 1 \right) \left( k - s \right)! }} A_{n,k}^{s , \nu} }{ \sqrt{ \sum_{j = s}^{+ \infty}   \frac{2 \left( j + s \right)! }{ \left( 2 j + 1 \right) \left( j - s \right)! } \left( A_{n,j}^{s , \nu} \right)^2 } },
\end{equation}

where 

\begin{equation}
\begin{array}{ll}
   \mathcal{S}_n^{s , \nu} = {\rm sign} \left\{ \tilde{\Theta}_n^{s,\nu} \left( \theta = \frac{\pi}{2} \right) \right\} & \mbox{for $ n $ even,} \\[0.5cm]
   \mathcal{S}_n^{s,\nu} = - {\rm sign} \left\{ \left. \dfrac{ d \tilde{\Theta}_n^{s,\nu}}{d \theta} \right|_{\theta = \pi/2}  \right\} & \mbox{for $ n $ odd.}
\end{array}
\end{equation}

\begin{figure*}[htb]
 \centering
 \includegraphics[width=0.42\textwidth,clip]{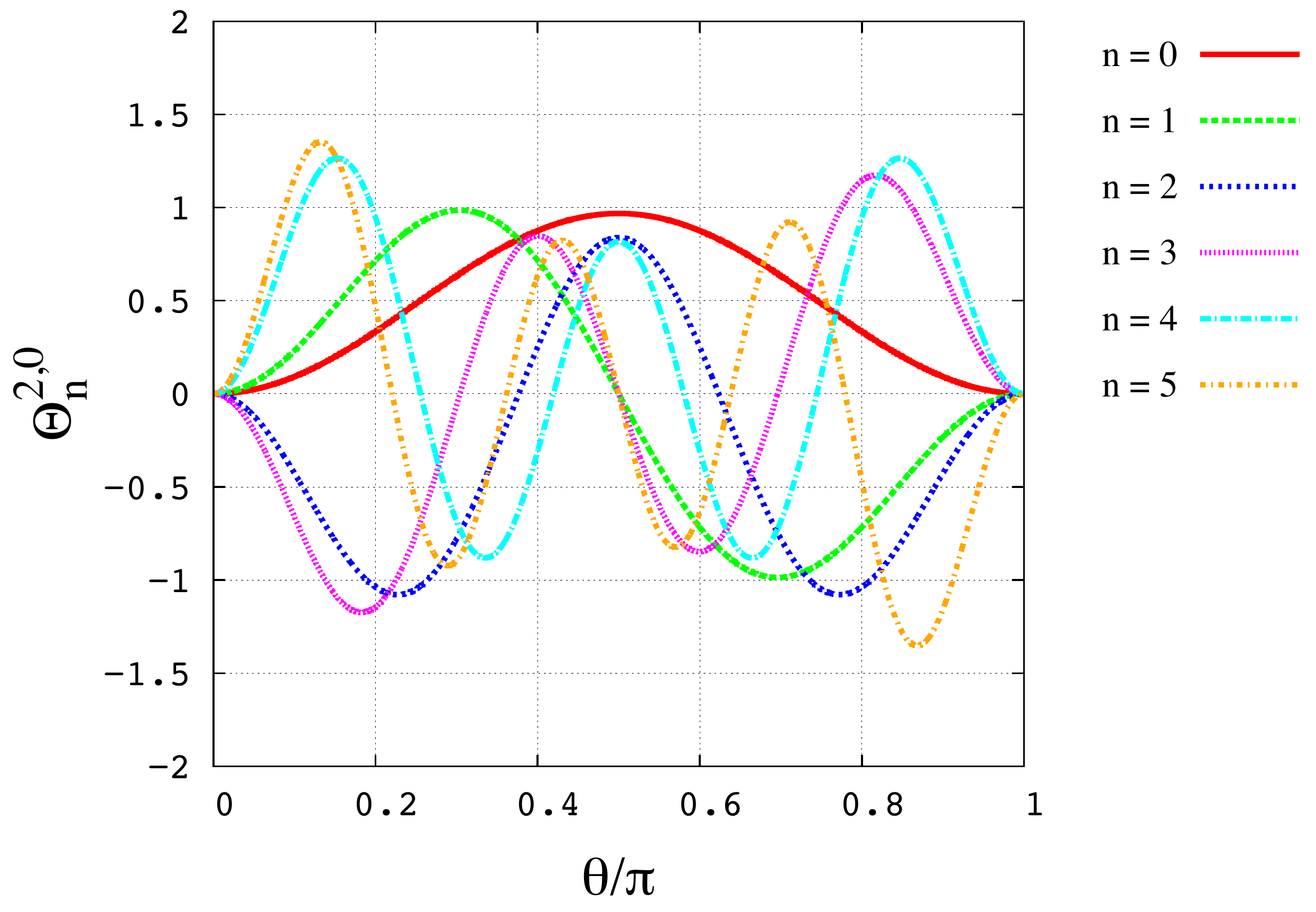} \hspace{1cm}    
 \includegraphics[width=0.42\textwidth,clip]{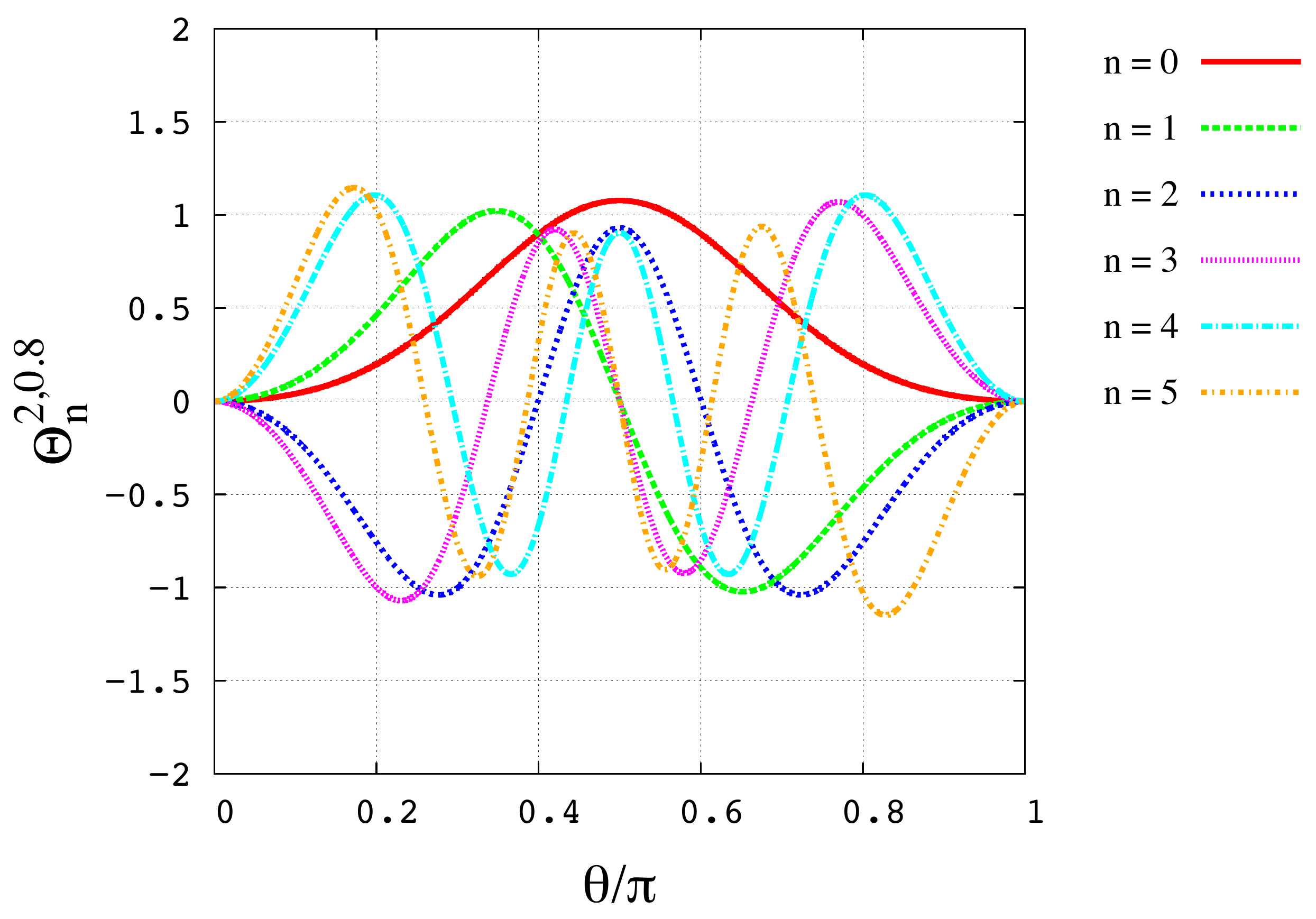} \\
  \includegraphics[width=0.42\textwidth,clip]{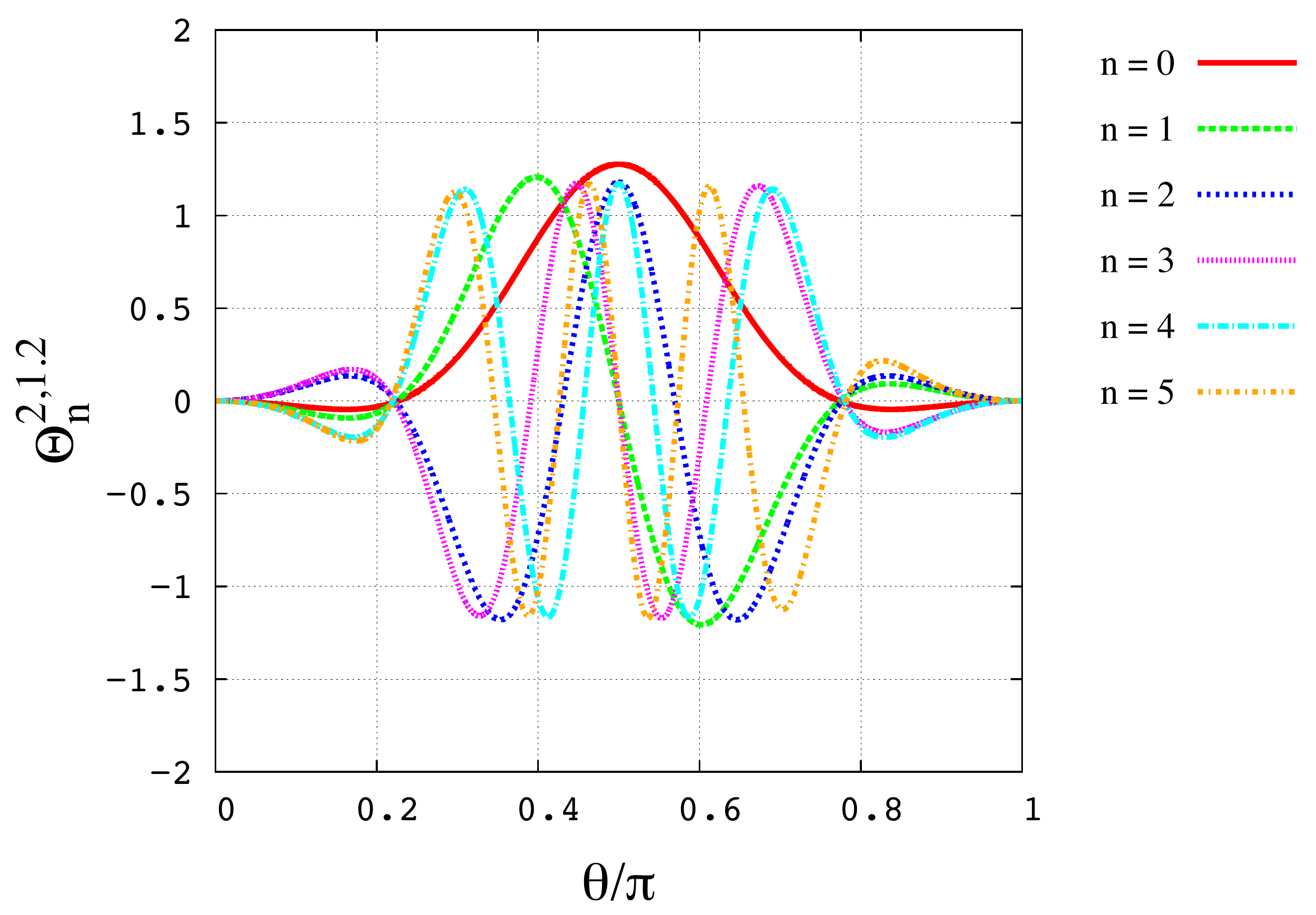} \hspace{1cm}      
 \includegraphics[width=0.42\textwidth,clip]{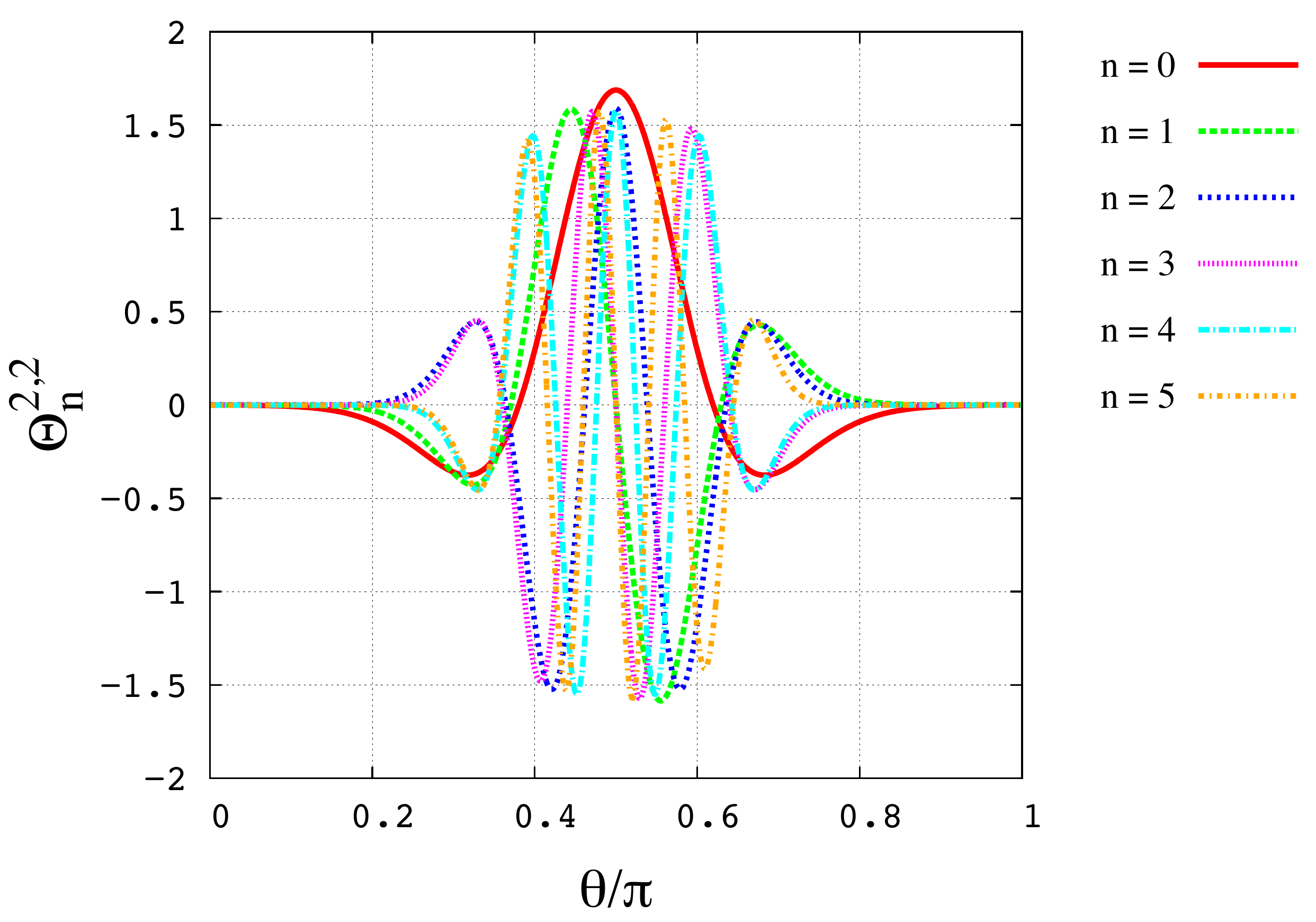} \\
   \includegraphics[width=0.42\textwidth,clip]{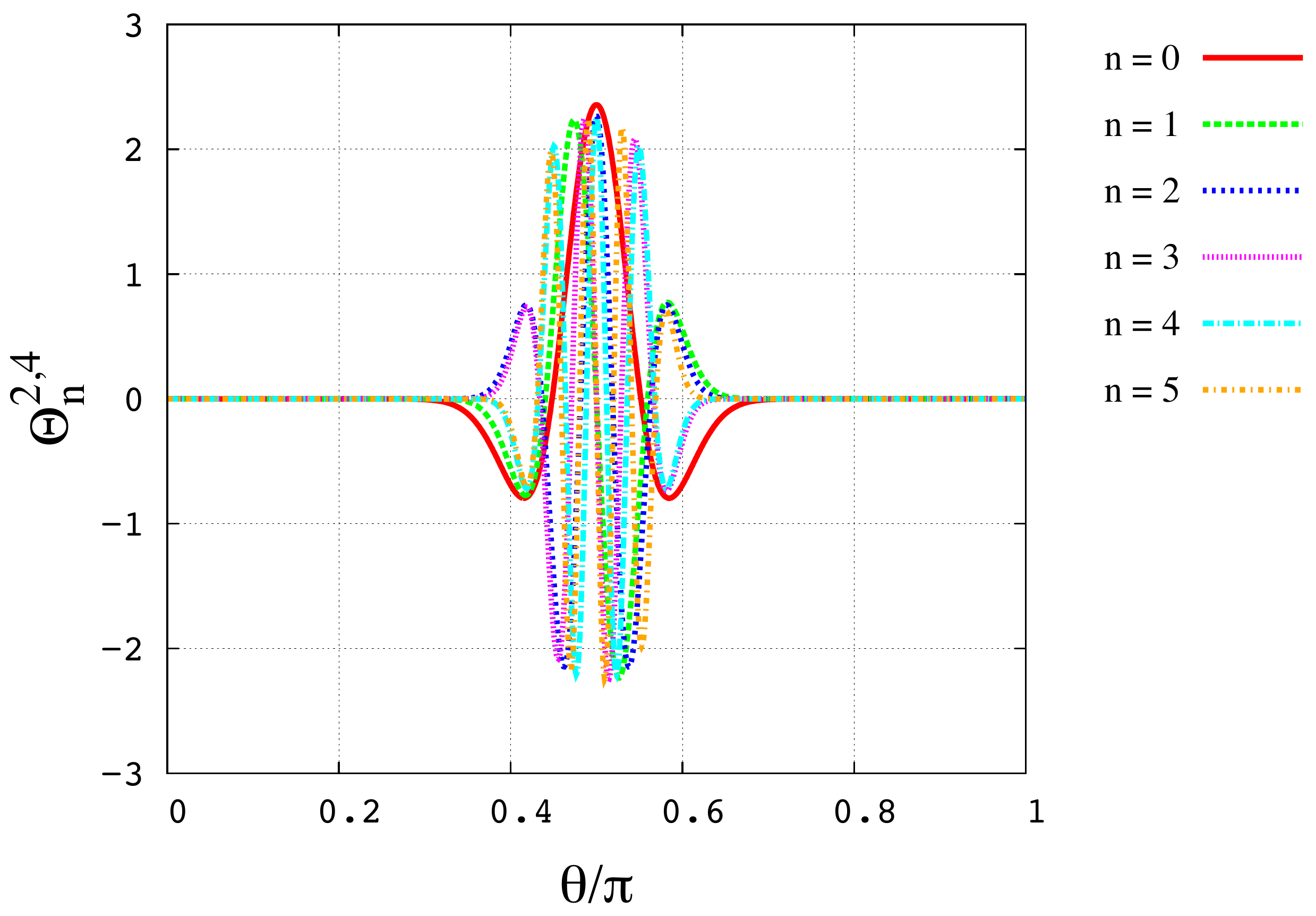} \hspace{1cm}      
 \includegraphics[width=0.42\textwidth,clip]{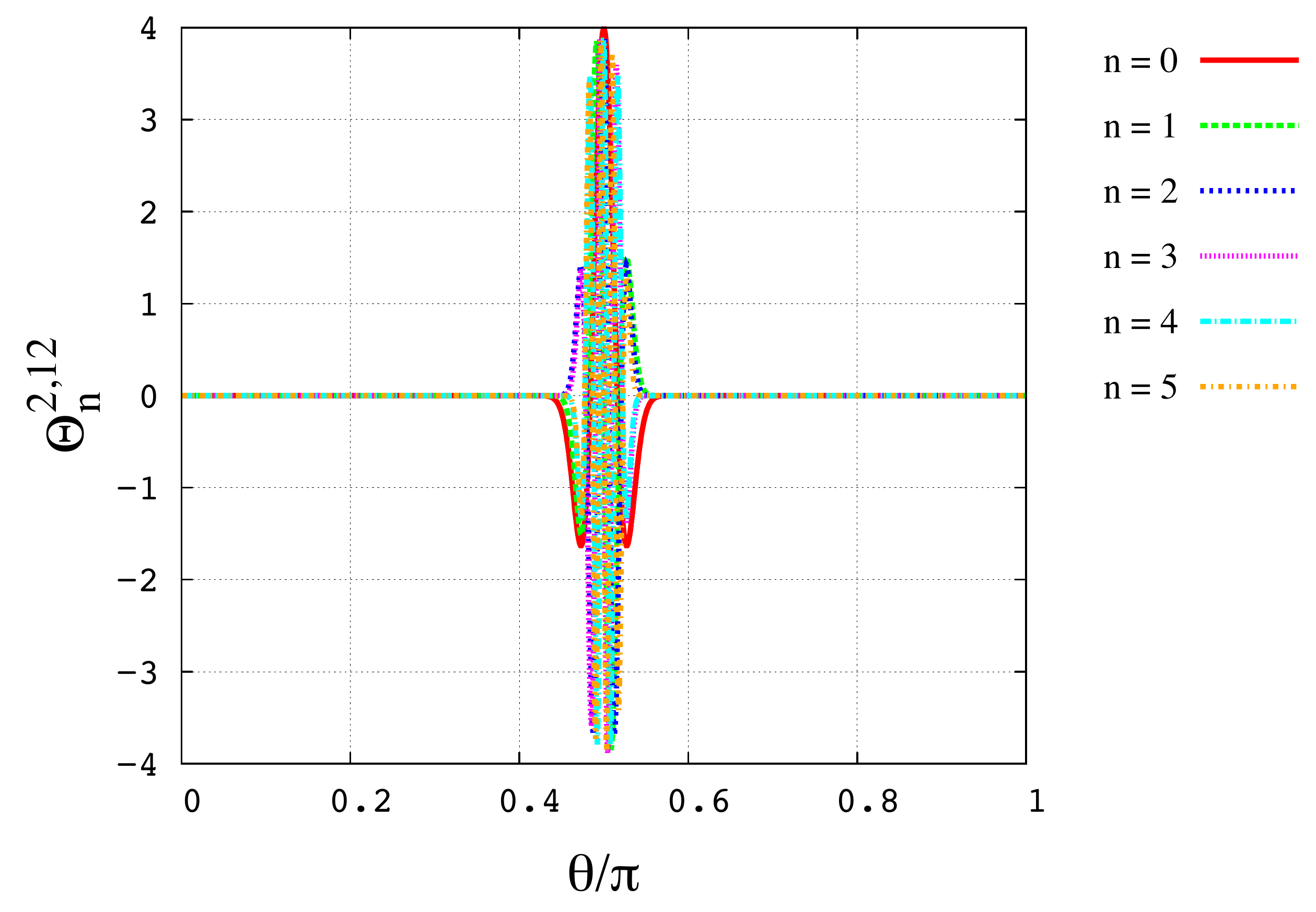}    
  \textsf{ \caption{\label{fig:fonctions_Hough} Normalized Hough functions $ \Theta_n^{2,\nu} $ computed as functions of the colatitude ($ \theta $) for $ 0 \leq n \leq 5 $ (gravity modes) and various values of $ \nu = \left( 2 \Omega \right) / \sigma $. {\it Top-left:} $ \nu = 0 $;  {\it top-right:} $ \nu = 0.8 $;  {\it middle-left:} $ \nu = 1.2 $;  {\it middle-right:} $ \nu = 2 $;  {\it bottom-left:} $ \nu = 4 $; {\it bottom-right:} $ \nu = 12 $. The case $ \nu = 0 $ corresponds to the normalized associated Legendre polynomials $ P_l^2 \left( \cos \theta \right) $. The two plots at the top represent the regime of super-inertial waves ($ \left| \nu \right| \leq 1 $), where Hough functions look like Legendre polynomials. In the regime of sub-inertial waves ($ \left| \nu \right| > 1 $), functions become more and more evanescent when $ \left| \nu \right| $ increases, and oscillations are trapped around the equator in the interval $ \left[ \theta_\nu , \pi - \theta_\nu \right] $, where $ \theta_\nu  =  \arccos \left( \left| \nu \right|^{-1} \right) $. }}
\end{figure*}

\begin{figure*}[htb]
 \centering
  \includegraphics[width=0.42\textwidth,clip]{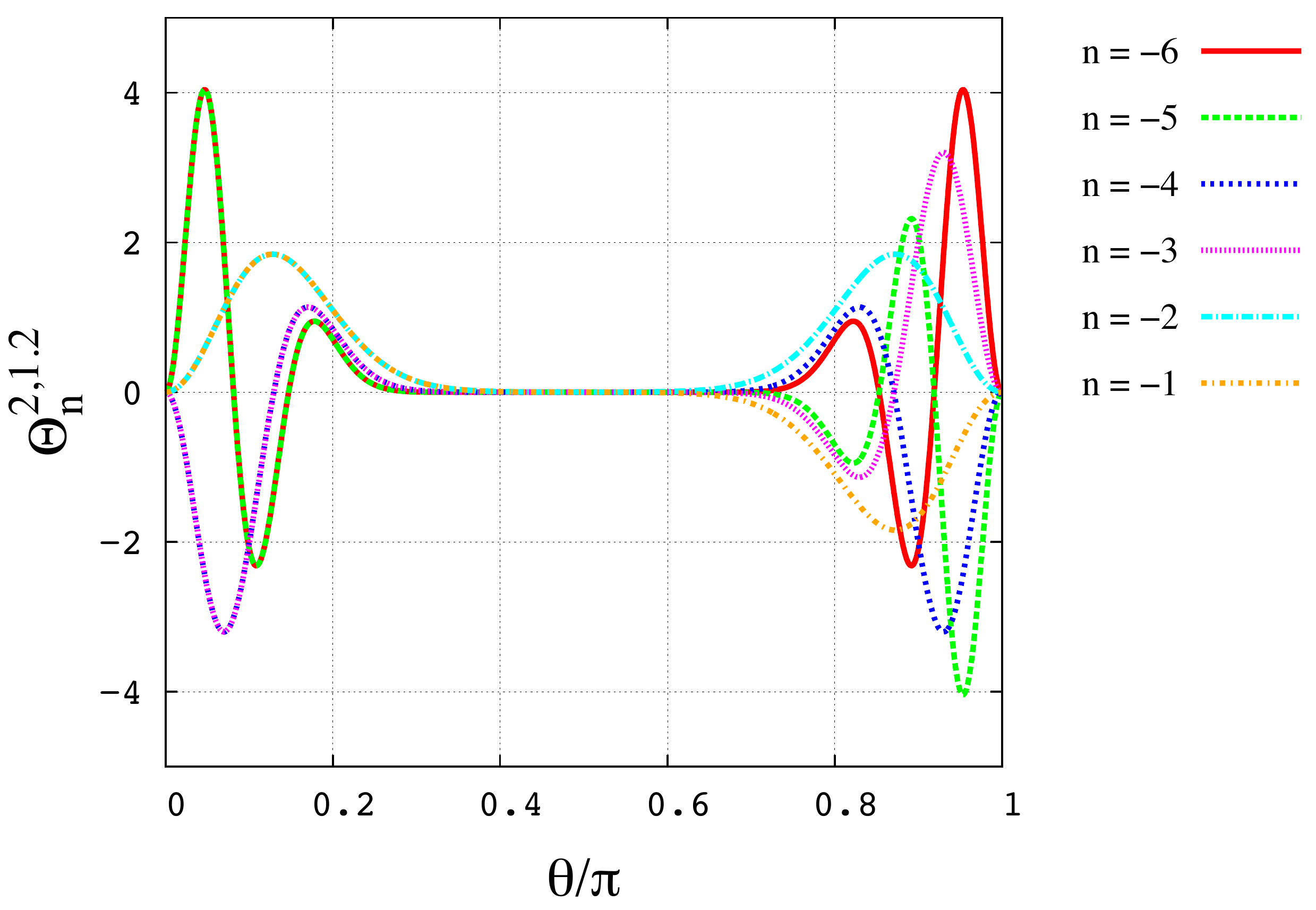} \hspace{1cm}     
 \includegraphics[width=0.42\textwidth,clip]{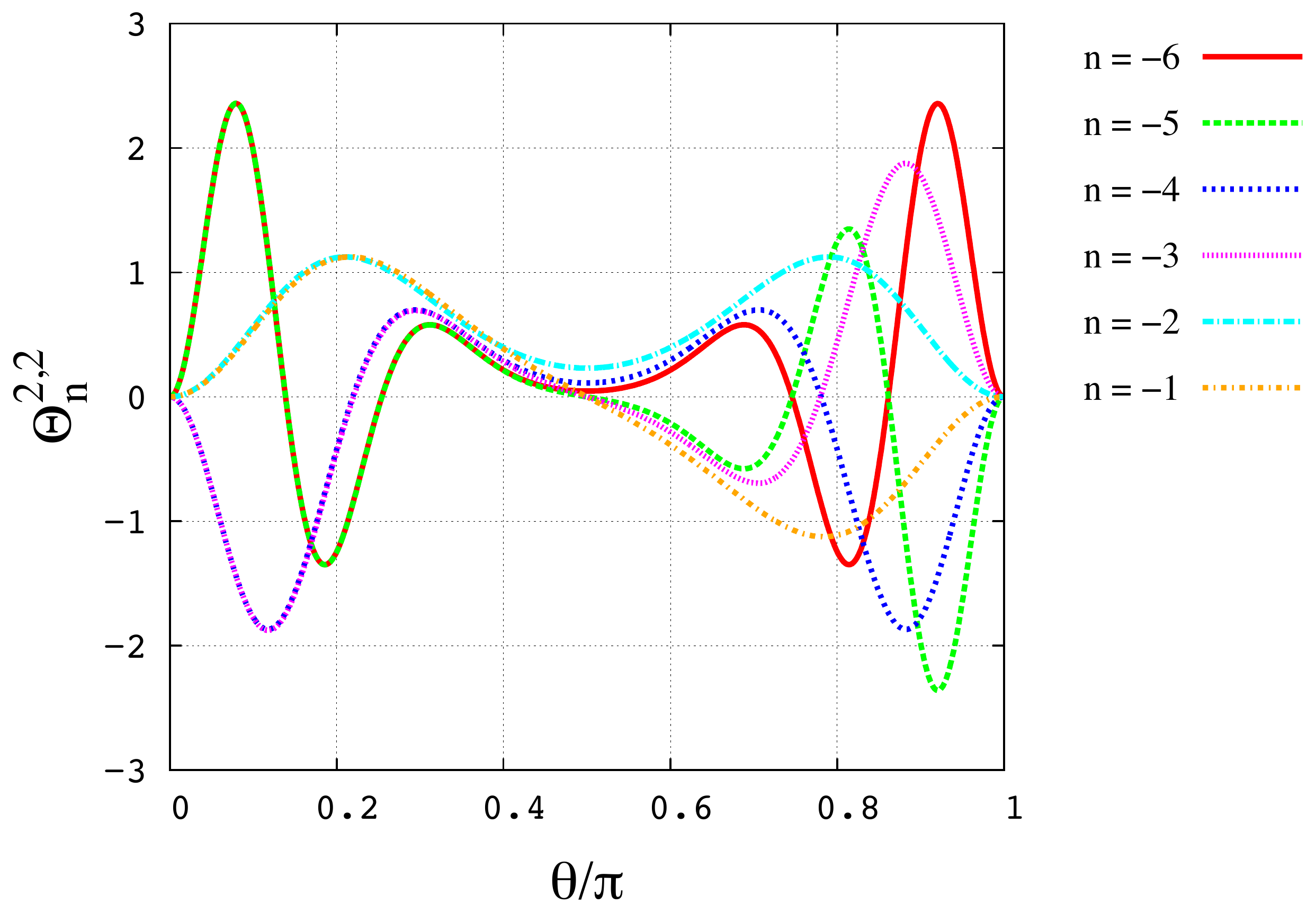} \\
   \includegraphics[width=0.42\textwidth,clip]{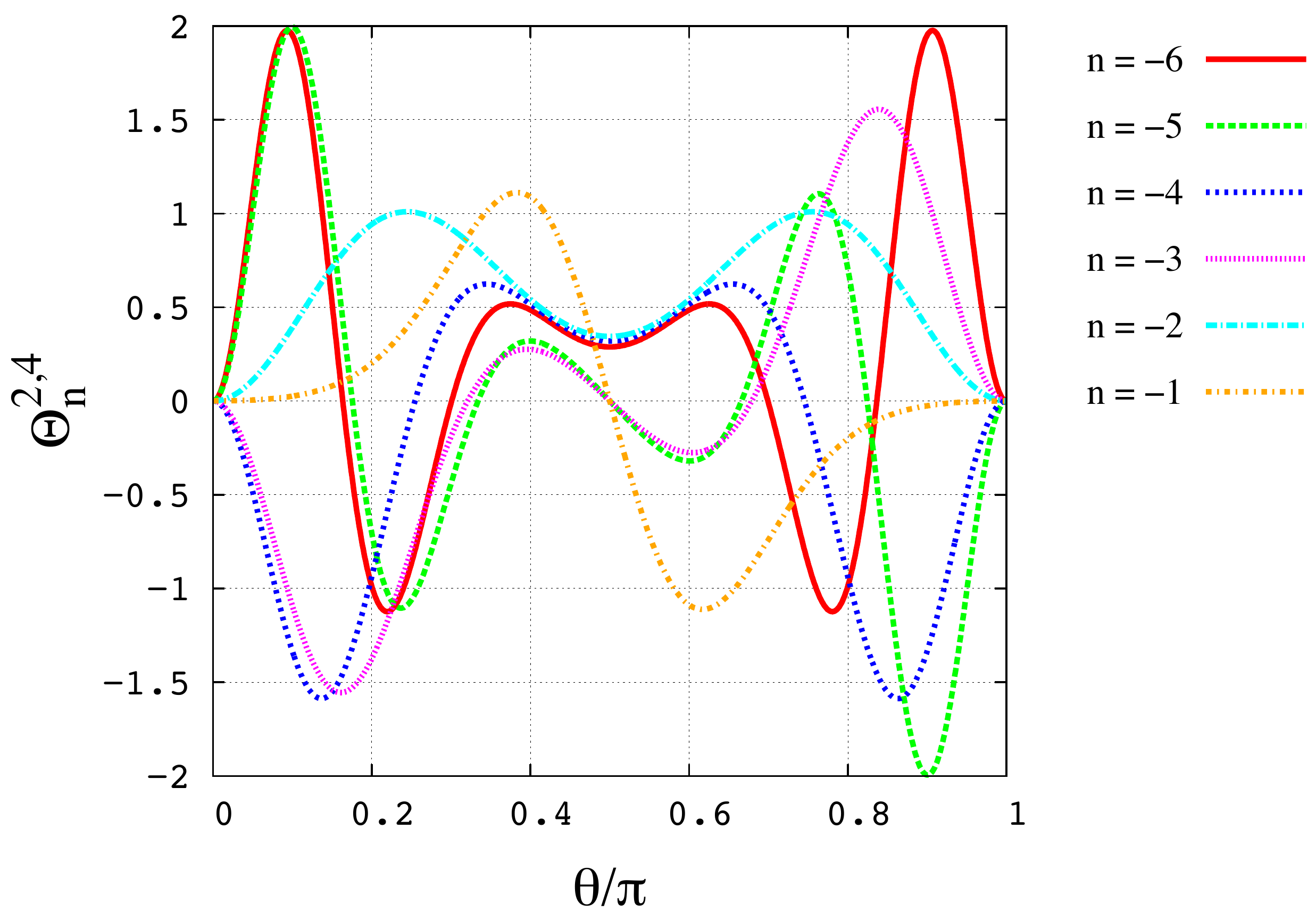} \hspace{1cm}       
 \includegraphics[width=0.42\textwidth,clip]{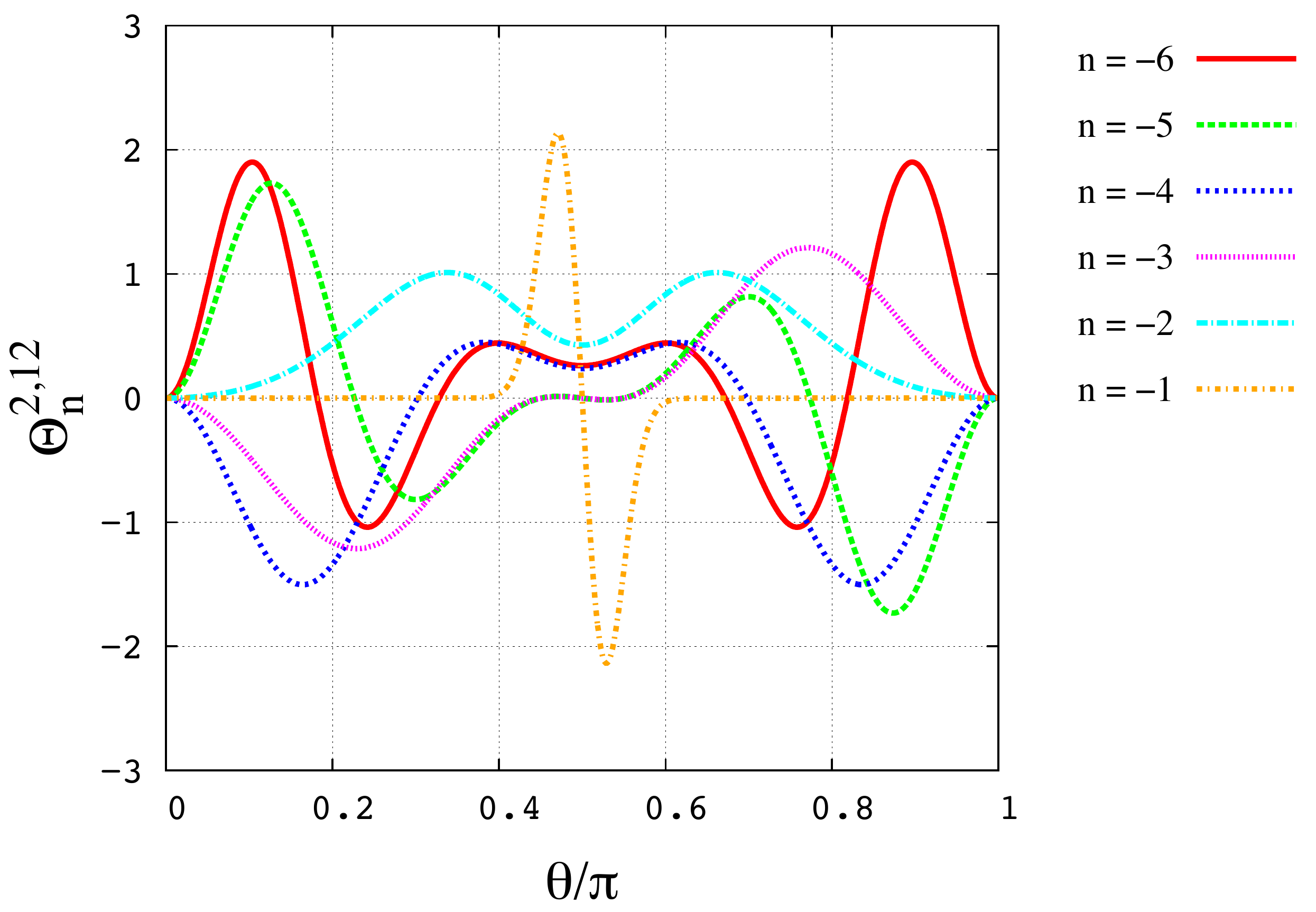}    
  \textsf{ \caption{\label{fig:fonctions_Hough_Rossby} Normalized Hough functions $ \Theta_n^{2,\nu} $ computed as functions of the colatitude ($ \theta $) for $ -6 \leq n \leq -1 $ (Rossby modes) and various values of $ \nu = \left( 2 \Omega \right) / \sigma $. {\it Top-left:} $ \nu = 0 $;  {\it top-right:} $ \nu = 0.8 $;  {\it middle-left:} $ \nu = 1.2 $;  {\it middle-right:} $ \nu = 2 $;  {\it bottom-left:} $ \nu = 4 $; {\it bottom-right:} $ \nu = 12 $. In the regime of sub-inertial waves ($ \left| \nu \right| > 1 $), Rossby modes are trapped around the poles when $ \left| \nu \right| \rightarrow 1 $. }}
\end{figure*}

The solutions and projection matrices given by the spectral method are plotted on Figs.~\ref{fig:fonctions_Hough_Terre}, \ref{fig:fonctions_Hough_Venus}, \ref{fig:fonctions_Hough}, \ref{fig:fonctions_Hough_Rossby}. In the super-inertial regime ($ \left| \nu \right| \leq 1 $), the Hough functions basis is composed of gravity modes only. Therefore, the projection matrix is a band-matrix. In this regime Hough functions are very similar to the associated Legendre polynomials. In the case $ \nu = 0 $ (top left), the bases are exactly the same and the projection matrix is diagonal. \\

The sub-inertial regime is characterized by modes of both types, the transition being identified by Hough functions of the lowest degrees: Rossby modes on the left of the map (lowest eigenvalues) ; gravity modes on the right (highest eigenvalues). The maps reveal the side effect inherent to the spectral method. Hough functions of highest degrees are obviously not well represented by the family of associated Legendre Polynomials; more polynomials are necessary. This is explained by the asymptotial behaviour of Hough functions. At $ \left| \nu \right| \rightarrow 1^+ $, Rossby modes are trapped around the poles and are thus well represented by highly oscillating associated Legendre polynomials (Fig.~\ref{fig:fonctions_Hough_Rossby}). At $ \left| \nu \right| \rightarrow + \infty $, gravity modes behave similarly and are trapped around the equatorial belt (Figs.~\ref{fig:fonctions_Hough_Terre}, \ref{fig:fonctions_Hough}).  \\

Since the spectral method has been proved to be very efficient to compute a large number of eigenvalues and eigenfunctions at once, it is used for all calculations in this work. As it may be noticed, some particular cases cannot be treated using Eq.~(\ref{pb_vp}). These cases have been discussed by CL70 and will not be developed in this work.

\section{Numerical scheme used to integrate spatial differential equations}
\label{app:num_scheme}

We detail here the numerical scheme used to integrate Laplace's tidal equation (with the finite differences method) and the vertical structure equation. This method, introduced by \cite{Bruceetal1953} and described by \cite{ChapmanLindzen1970} for a particular case, is very convenient to solve second-order differential equations with non-constant coefficients. Those equations are written

\begin{equation}
\dfrac{d^2y}{dx^2} + P  \dfrac{dy}{dx} + Q  y = R 
\label{Eqdiff}
\end{equation}

where $ P $, $ Q $ and $ R $ are given functions of $ x $. The domain $ \left[ x_{\rm inf} , x_{\rm sup} \right] $ on which Eq.~(\ref{Eqdiff}) is integrated is divided into $ N $ intervals, i.e. $ N + 1 $ points, identified by the index $ n $. Hence, for a regular mesh with elements of size $ \delta x $, the first and second-order derivative of a solution $ f $ are expressed

\begin{equation}
\begin{array}{lll}
 \displaystyle \dfrac{df}{dx} = \frac{f_{n+1} - f_{n-1}}{2 \delta x} & \mbox{and} & \displaystyle \frac{d^2 f}{dx^2} = \frac{f_{n+1} + f_{n-1} - 2 f_n}{ \delta x^2 }
\end{array}
\label{dfnum}
\end{equation}

with $  n \in \llbracket 1 , N-1 \rrbracket $. Substituting Eq.~(\ref{dfnum}) in Eq.~(\ref{Eqdiff}), we obtain the numerical relations

\begin{equation}
  A_n f_{n+1} + B_{n} f_n + C_n f_{n-1} = D_n,
\end{equation}

with the coefficients 

\begin{equation}
\begin{array}{ll}
     \displaystyle A_n = 1 + \frac{\delta x P_n}{2}, &  \displaystyle B_n = \delta x^2 Q_n - 2, \\[0.5cm]
     \displaystyle C_n = 1 - \frac{\delta x P_n}{2}, &  \displaystyle D_n = \delta x^2 R_n,
\end{array}
\end{equation}

the $ P_n $, $ Q_n $ and $ R_n $ coefficients being the components of the numerical vectors corresponding to $ P $, $ Q $ and $ R $. Then, we introduce the coefficients $ \alpha_n $ and $ \beta_n $, such as

\begin{equation}
 f_n = \alpha_n f_{n+1} + \beta_n.
 \label{fnrec}
\end{equation}

These coefficients are defined straighforwardly by the recursive relations

\begin{equation}
\left\{
\begin{array}{l}
    \displaystyle \alpha_n = - \frac{A_n}{B_n + \alpha_{n-1} C_n} \\[0.5cm]
    \displaystyle \beta_n = \frac{D_n - \beta_{n-1} C_n}{B_n + \alpha_{n-1} C_n}
\end{array}
\right.
.
\end{equation}

The first terms ($ \alpha_0 $ and $ \beta_0 $) are deduced from the boundary condition at $ x = x_{\rm inf} $, which may be written

\begin{equation}
a_{0} \left. \dfrac{d f}{dx} \right|_{x_{\rm inf}} + b_0 f \left( x_{\rm inf} \right) = c_{0}.
\end{equation}

where $ a_0$, $b_0$ and $ c_0 $ are real or complex coefficients. Thus, we obtain 

\begin{equation}
\left\{
\begin{array}{l}
    \displaystyle \alpha_0 = - \frac{a_0}{b_0 \delta x - a_0}\\[0.5cm]
     \displaystyle \beta_0 = \frac{c_0 \delta x}{b_0 \delta x - a_0}
\end{array}
\right.
\end{equation}

and, step by step, all the following terms. At $ x = x_{\rm sup} $, we apply the condition 

\begin{equation}
a_{N} \left. \dfrac{d f}{dx} \right|_{x_{\rm sup}} + b_N f \left( x_{\rm sup} \right) = c_{N}
\end{equation}

with $ \left( a_N , b_N , c_N \right) \in \mathbb{C}^3 $, which gives

\begin{equation}
f_N = \frac{c_N \delta x + \beta_{N-1} a_N }{ b_N \delta x + a_N \left( 1 - \alpha_{N-1} \right)}. 
\end{equation}

The solution is finally obtained by computing the $ f_n $ backward using Eq.~(\ref{fnrec}).

\section{Decomposition of the thermal forcing in Keplerian elements}
\label{app:thermal_forcing}

\begin{figure}[htb]
\centering
{\includegraphics[width=0.95\textwidth]
{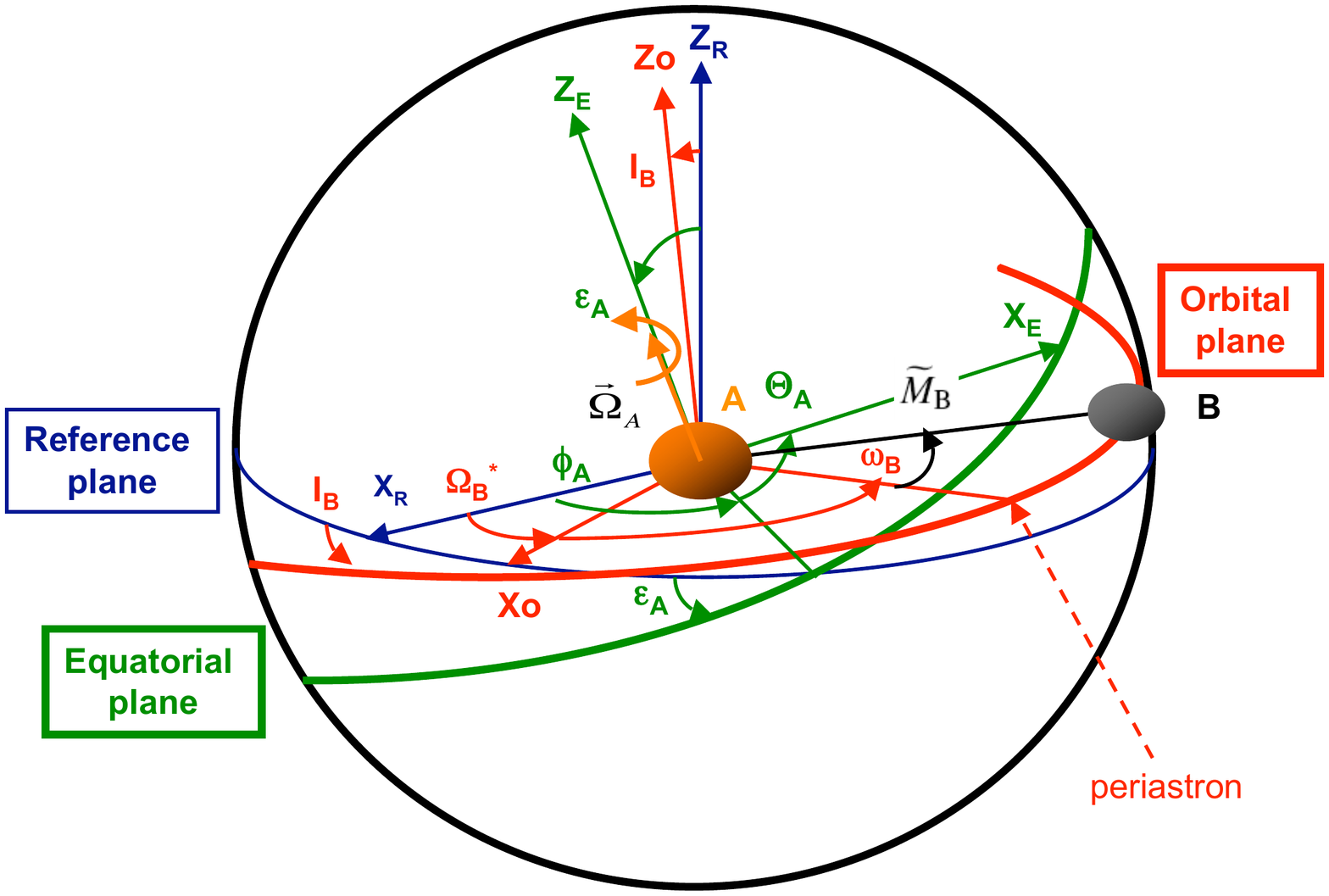}
\textsf{\caption{\label{fig:frames} Inertial reference, orbital, and equatorial rotating frames and associated Euler's angles of orientation. Figure taken from \cite{MLP09}.} }}
\end{figure}

We establish here the general multipole expansion in spherical harmonics of the thermal power given in Eq.~(\ref{UJ_Legendre}) using the notations of \cite{MLP09}. We consider a planet ($ A $) orbiting around its star ($ B $) circularly. Three reference frames, all centered on the center of mass of the planet and represented in Fig.~\ref{fig:frames}, must be defined:
\begin{itemize}
\item[$ \bullet $] $ \mathscr{R}_{\rm R}:\left\{ O_{\rm A} , \textbf{X}_{\rm R} , \textbf{Y}_{\rm R} , \textbf{Z}_{\rm R}  \right\} $, time independent inertial frame, with $ \textbf{Z}_{\rm R} $ the direction of the total angular momentum of the system;
\item[$ \bullet $] $ \mathscr{R}_{\rm O}:\left\{ O_{\rm A} , \textbf{X}_{\rm O} , \textbf{Y}_{\rm O} , \textbf{Z}_{\rm O}  \right\} $, the orbital frame linked to $ \mathscr{R}_{\rm R} $ by the Euler angles:
   \begin{itemize}
    \item $ I_{\rm B} $, the inclination of the orbital frame with respect to $ \left( O_{\rm A} ,  \textbf{X}_{\rm R}  , \textbf{Y}_{\rm R}  \right) $;
    \item $ \omega_{\rm B} $, the argument of the pericenter;
    \item $ \Omega_{\rm B}^* $, the longitude of the ascending node;
   \end{itemize}
\item[$ \bullet $] $ \mathscr{R}_{\rm E;T}:\left\{ O_{\rm A} , \textbf{X}_{\rm E} , \textbf{Y}_{\rm E} , \textbf{Z}_{\rm E}  \right\} $, the spin equatorial frame rotating with the angular velocity $ \Omega_{\rm A} $, and linked to $ \mathscr{R}_{\rm R} $ by three Euler angles:
   \begin{itemize}
    \item $ \varepsilon_{\rm A} $, the obliquity, i.e. the inclination of the equatorial plane with respect to $ \left( O_{\rm A} ,  \textbf{X}_{\rm R}  , \textbf{Y}_{\rm R}  \right) $;
    \item $ \Theta_{\rm A} $, the mean sideral angle such as $ \Omega_{\rm A} = {\rm d} \Theta_{\rm A} / {\rm d}t $, which is the angle between the minimal axis of inertia and the straight line due to the intersection of the planes $ \left( O_{\rm A} , \textbf{X}_{\rm E} , \textbf{Y}_{\rm E} \right) $ and $ \left( O_{\rm A} , \textbf{X}_{\rm R} , \textbf{Y}_{\rm R} \right) $;
    \item $ \phi_{\rm A} $, the general precession angle.
   \end{itemize} 
\end{itemize}

Finally, we denote $ a $ the semi-major axis and $ \tilde{M}_{\rm B} $ the mean anomaly with $ \tilde{M}_{\rm B} \approx n_{\rm B} t $, $ n_{\rm B} $ being the mean motion. Neglecting the action of the centrifugal acceleration, the hydrostatic structure of the atmosphere presents a spherical symmetry, $ p_0 $, $ \rho_0 $, $ T_0 $ being functions of the radial coordinate only. Hence, any heat forcing caused by the star can be written

\begin{equation}
J \left( r , \psi \right) = \frac{1}{a^2} h \left(  r , \psi   \right),
\end{equation}

where $ h $ is a function of $ r $ and $ \psi $, the angle between the points $ M \left( r , \theta , \varphi \right) $ and $ B \left( a , \theta_{\rm AB} , \varphi_{\rm AB} \right) $. We decompose $ J $ in normalized Legendre polynomials

\begin{equation}
J = \frac{1}{a^2} \sum_{l \geq 0} \mathcal{J}_l \left( r \right) P_l \left( \cos \psi \right)
\end{equation}

with $ \mathcal{J}_l \left( r \right) = \langle h , P_l \rangle $. Using the addition theorem

\begin{equation}
P_l \left( \cos \psi \right) = \frac{4 \pi}{ \sqrt{ 2 \left( 2 l + 1 \right)}}    \sum_{m = - l}^{l} Y_{l,m} \left( \theta , \varphi \right) Y_{l,m}^* \left( \theta_{AB} , \varphi_{AB} \right), 
\end{equation}

with 

\begin{equation}
  Y_{l,m} \left(\theta , \varphi \right) = \frac{1}{\sqrt{2\pi}} P_{l}^m \left( \cos \theta \right) e^{i m \varphi},
\end{equation}

we obtain 

\begin{equation}
J = \sum_{l \geq 0}  \sum_{m = - l}^{l} \frac{4 \pi}{ \sqrt{ 2 \left( 2 l + 1 \right)}}  \frac{\mathcal{J}_l \left( r \right)}{a^2}  Y_{l,m} \left( \theta , \varphi \right) Y_{l,m}^* \left( \theta_{AB} , \varphi_{AB} \right).
\label{J_SH1}
\end{equation}

Kaula's transform is finally introduced to express $ Y_{l,m}^* \left( \theta_{AB} , \varphi_{AB} \right) $ as a function of Keplerian elements \citep{Kaula1962,MLP09}. Since the motion of the planet is circular ($ e = 0 $), we get

\begin{equation}
Y_{l,m} \left( \theta_{AB} , \varphi_{AB} \right) = \sum_{j , p,q} \kappa_{l,j} d_{j,m}^l \left( \varepsilon_{\rm A} \right) F_{l,j,p} \left( I_{\rm B} \right) G_{l,p,q} \left( 0 \right) e^{i  \Psi_{l,m,j,p,q} \left( t \right) },
\label{Yls}
\end{equation}

with $ \left( j , p , q \right) \in \llbracket -l ,l \rrbracket \times \llbracket 0 ,l \rrbracket \times \mathbb{Z}  $, where the $ \kappa_{l,j} $ coefficients are given by 

\begin{equation}
\kappa_{l,j} = \sqrt{ \frac{2 l + 1}{4 \pi} \frac{\left( l - \left| j \right| \right)!}{\left( l + \left| j \right| \right) ! } }.
\end{equation}

The functions denoted $ d_{j,m}^l $, $ F_{l,j,p} $ and $ G_{l,p,q} $ are called the obliquity, inclination and eccentricity functions \citep[for more details, see][]{MLP09}. The angle $ \Psi_{l,m,j,p,q} $ can be written

\begin{equation}
\Psi_{l,m,j,p,q} \left( t \right) = - \sigma_{l,m,p,q} t + \psi_{l,m,j,p,q},
\end{equation}

the parameter $ \sigma_{l,m,p,q} = m \Omega_{\rm A} - \left( l - 2 p + q  \right) n_{\rm B}  $ being the tidal frequency of the mode $ \left( l,m,p,q \right) $ and $ \psi_{l,m,j,p,q} = \left( l - 2 p \right) \omega_{\rm B} + j \left( \Omega_{\rm B}^* - \phi_{\rm A} \right) + \left( l - m \right) \pi / 2 $ the phase lag. Substituting Eq.~(\ref{Yls}) in Eq.~(\ref{J_SH1}), we finally get

\begin{equation}
J = J_0 + \sum_{l \geq 1} \sum_{m = - l}^l \sum_{j,p,q} W_{l,m,j,p,q} \frac{\mathcal{J}_l \left( r \right)}{a^2} P_l^m \left( \cos \theta \right) e^{i \left[ \sigma_{l,m,p,q} t + m \varphi \right]},
\end{equation}

where 

\begin{equation}
\begin{array}{ll}
W_{l,m,j,p,q} \left( \varepsilon , I , e \right)  = & \displaystyle  \sqrt{ \frac{  \left( l - \left| j \right| \right)! }{ \left( l + \left| j \right| \right)!} }   \\ 
  & \displaystyle \times \left[ d_{j,m}^l \left( \varepsilon \right) F_{l,j,p} \left( I \right) G_{l,p,q} \left( e \right)  \right] e^{-im \varphi_{l,m,j,p,q}},
\end{array}
\end{equation}

with $ \varphi_{l,m,j,p,q} = \psi_{l,m,j,p,q} / m $. \\

With the same notations, the gravitational potential is written

\begin{equation}
U = U_0 + \sum_{l \geq 1} \sum_{m = - l}^l \sum_{j ,p,q} W_{l,m,j,p,q} \frac{\mathscr{G} M_{\rm B}}{a^{l+1}} r^l P_l^m \left( \cos \theta \right) e^{i \left[ \sigma_{l,m,p,q} t + m \varphi \right]}.
\label{Kaula_potential}
\end{equation}

The tidal frequency, written $ \sigma =   m \Omega - s n_{\rm orb} $, is defined by the doublet $ \left( m,s \right) \in \mathbb{Z}^2 $. \\

The perturbing tidal potential and thermal power finally write

\begin{equation}
\left\{
\begin{array}{l}
     \displaystyle U = \sum_{ \left( s, m \right) \in \mathbb{Z}^2} U^{\sigma_{s,m},m} \left( r , \theta \right) e^{i \left( \sigma_{s,m} t + s \varphi   \right)}, \\[0.5cm]
     \displaystyle J = \sum_{ \left( s, m \right) \in \mathbb{Z}^2} J^{\sigma_{s,m},m} \left( r , \theta \right) e^{i \left( \sigma_{s,m} t + s \varphi   \right)},
     \end{array}
\right.
\end{equation}

where the spatial functions $ U^{m,\sigma} $ and $ J^{m,\sigma} $ are expressed

\begin{equation}
\left\{
\begin{array}{l}
     \displaystyle U^{m,\sigma} \left (r , \theta \right) = \sum_{l \geq \left| m \right|} U_l^{m,\sigma} \left( r \right) P_l^m \left( \cos \theta \right), \\[0.5cm]
     \displaystyle J^{m,\sigma} \left (r , \theta \right) = \sum_{l \geq \left| m \right|} J_l^{m,\sigma} \left( r \right) P_l^m \left( \cos \theta \right),
\end{array}
\right.
\label{UJ_Legendre2}
\end{equation}

with

\begin{equation}
\left\{
\begin{array}{l}
     \displaystyle U_l^{m,\sigma} \left( r \right) =   \sum_{j = -l}^l \sum_{p = 0}^l \sum_{q \in \mathbb{Z}} U_{l,m,j,p,q} \left( r \right) \delta_{l-2p+q,m}, \\[0.5cm]
     \displaystyle J_l^{m,\sigma} \left( r \right) =   \sum_{j = -l}^l \sum_{p = 0}^l \sum_{q \in \mathbb{Z}} J_{l,m,j,p,q} \left( r \right) \delta_{l-2p+q,m},
     \end{array}
\right.
\label{UJls}
\end{equation}

the notation $ \delta $ standing for the Kronecker symbol.

\section{Radiated power as a function of physical parameters}
\label{app:newtonian_cooling}

The power per unit mass emitted by \mybf{an optically thin atmosphere} ($ J_{\rm rad} $) can be expressed as an analytic function of physical parameters in our simplified framework. Indeed, assuming that the flux radiated by a layer of thickness $ dx $, denoted $ \delta I $, propagates along the vertical direction without being absorbed by the other layers, we write

\begin{equation}
\delta I = \frac{1}{2} \phi_{\rm rad} \rho_0  dr,
\label{dI1}
\end{equation}

where $ \phi_{\rm rad} $ is the total power emitted per unit mass. Then, the atmosphere is treated as a grey body of molar emissivity per unit mass $ \epsilon_a $, which, using the molar concentration $ C_0 $ introduced in Sect. 2,  gives

\begin{equation}
\delta I = \epsilon_a C_0 \mathscr{S} T^4 {\rm dr}.
\label{dI2}
\end{equation}

Combining Eqs.~(\ref{dI1}) and (\ref{dI2}) allows us to get

\begin{equation}
\phi_{\rm rad} = \frac{2 \epsilon_a}{M} \mathscr{S} T^4.
\end{equation}

Finally, we derive the perturbed radiative loss $ J_{\rm rad} $ associated to the tidal perturbation as a function of $ \delta T $ by linearizing the previous expression around the equilibrium state:

\begin{equation}
J_{\rm rad} = \frac{8 \epsilon_a}{M} \mathscr{S} T_0^3 \delta T,
\end{equation}

which gives the radiation frequency introduced in Eq.~(\ref{sigma0})


\begin{equation}
\sigma_0 \left( r \right) = \frac{8 \kappa \epsilon_a \mathscr{S} }{\mathscr{R}_{\rm GP} } T_0^3,
\end{equation}

\section{Analytic solution for constant profiles and upper boundary condition} 
\label{app:boundary_conditions}

We show here that the solution of the equation describing the vertical structure strongly depends on the choice of the upper boundary condition. As discussed in Sect. 3, 

\begin{equation}
\Psi_n^{m,\sigma} \left( 0 \right) = 0,
\end{equation}

at the ground level $ x = 0 $. \\

Instead of the energy-bounded condition prescribed in the model, we set at the upper limit ($ x_{\rm atm} $) the commonly used stress-free condition,

\begin{equation}
\delta p_n + \frac{1}{H}  \frac{d p_0}{dx} \xi_{r;n} = 0.
\end{equation}

We obtain, at $ x = x_{\rm atm} $,


\begin{equation}
\dfrac{d \Psi_n^{m,\sigma}}{dx} + \mathcal{P} \Psi_n^{m,\sigma} =  \mathcal{Q},
\end{equation}

with the complex coefficients

\begin{equation}
\left\{
\begin{array}{l}
     \displaystyle \mathcal{P} =  \mathcal{A}_n - \left( 1 - \varepsilon_{s;n} \right) \frac{\sigma_{s;n}^2}{\Gamma_1 \sigma^2}, \\[0.5cm]
     \displaystyle \mathcal{Q} = e^{-  \frac{x_{\rm atm}}{2}} \left[ \frac{\kappa R^2}{g \left( i \sigma + \sigma_0 \right)} J_n^{m,\sigma} + \frac{H \Lambda_n^{m,\nu}}{\sigma^2} U_n^{m,\sigma}  \right]_{x = x_{\rm atm}}.
\end{array}
\right.
\end{equation}

The parameter $ x_{\rm atm} $ must be fixed so that $ \rho_0 \left( x_{\rm atm}  \right) \ll \rho_0 \left( 0 \right) $, which corresponds to $ x_{\rm atm} \gg 1 $. A simple analytic solution can be computed assuming that $ J_n^{m,\sigma} $ and $ U_n^{m,\sigma} $ are constant profiles. The second member of the vertical equation, $ C $, given by Eq.~(\ref{coeffsABC_mince}), becomes

\begin{equation}
C = \frac{\kappa \Lambda_n^{m,\nu}}{H \sigma^2 \left (i \sigma + \sigma_0 \right)} \left\{   \mathcal{K}_n J_n^{m,\sigma}  -  i \sigma U_n^{m,\sigma}  \right\};
\end{equation}

the solution is written

\begin{equation}
    \Psi_n^{m,\sigma} =  \frac{H^2 C}{\hat{k}_n a_n b_n} \left\{  \left( \mathcal{Z} - \frac{1}{2} \right) \sin \left( \hat{k}_n x \right)   + \hat{k}_n \left[ \cos \left( \hat{k}_n x \right) - e^{-\frac{x}{2}}  \right] \right\},
 \label{Psi_ana_sf}
\end{equation}

with $ a_n =  i \hat{k}_n + 1/2 $ and $ b_n =  i \hat{k}_n - 1/2 $, and the complex integration constant

\begin{equation}
\begin{array}{ll}
\mathcal{Z} = & \displaystyle \frac{ \displaystyle \frac{\hat{k}_n a_n b_n}{H^2 C} \mathcal{Q} + \left( \hat{k}_n^2 + \frac{1}{2} \mathcal{P}  \right) \sin \left( \hat{k}_n x_{\rm atm} \right)  }{  \mathcal{P} \sin \left( \hat{k}_n x_{\rm atm} \right) + \hat{k}_n \cos \left( \hat{k}_n x_{\rm atm} \right) } \\[0.5cm]
   & \displaystyle + \frac{ \hat{k}_n \left( \frac{1}{2} - \mathcal{P} \right) \left[  \cos \left( \hat{k}_n x_{\rm atm} \right) - e^{-\frac{x_{\rm atm}}{2}} \right] }{  \mathcal{P} \sin \left( \hat{k}_n x_{\rm atm} \right) + \hat{k}_n \cos \left( \hat{k}_n x_{\rm atm} \right) }.
\end{array}
\end{equation}

Now, the condition of Eq.~(\ref{condition_therm}) characterizing thermal tides, i.e.

\begin{equation}
\left|  \frac{U_n^{m,\sigma}}{J_n^{m,\sigma}} \right| \ll  \left|  \frac{H - h_n}{\sigma H}   \right|,
\label{condition_therm}
\end{equation}

is assumed. The second-order Love number reduced to the main contribution of the thermal semi-diurnal tide, given by Eq.~(\ref{k2_simple}), can be expressed

\begin{equation}
k_2^{2,\sigma_{\rm SD}} \sim \frac{8 \pi}{5} \sqrt{\frac{2}{3}} \frac{a^3 H}{M_S R} \int_0^{x_{\rm atm}}  \delta \rho_{0,2}^{2,\sigma_{\rm SD}}  \left( x \right) dx
\end{equation}


for a thin isothermal atmosphere. As a zero-order approximation, we consider that the forcings are the constant profiles  $ U_2 = U_{2,2,2,0,0} $ and $ J_2 = J_{2,2,2,0,0} $ used in Sect. 2 (with $ J_2 \in \mathbb{R} $). Therefore, introducing the complex impedance $ \Xi_{\rm atm} $, defined by

\begin{equation}
\int_0^{x_{\rm atm}}  \delta \rho_{0,2}^{2,\sigma_{\rm SD}}  \left( x \right) {\rm dx} = \Xi_{\rm atm} J_2,
\end{equation}

and substituting the analytic solution (Eq.~\ref{Psi_ana_sf}) in Eq.~(\ref{deltaq_mince}), one gets

\begin{equation}
k_2^{2,\sigma_{\rm SD}} \sim \frac{8 \pi}{5} \sqrt{\frac{2}{3}} \frac{a^3 H }{M_S R}  \Xi_{\rm atm} J_2 
\end{equation}


where $ \Xi_{\rm atm} $ is explicitly given by 

\begin{equation}
\begin{array}{ll}
    \Xi_{\rm atm} = & \! \! \! \! \displaystyle \int_0^{x_{\rm atm}} \frac{ \kappa \mathcal{K}_n \left( i \sigma + \Gamma_1 \sigma_0  \right)}{ \lambda_0 a_0 b_0 c_s^2 \left( i \sigma + \sigma_0 \right)^2 \left( 1 - \varepsilon_{s;n}  \right)}  \left\{  \mathcal{M} e^{b_0 x}   \right. \\[0.5cm]
    & \! \! \! \! \displaystyle \left.  + \mathcal{N} e^{-a_0 x} + \lambda_0 \left[ \frac{1}{2} - \mathcal{B}_n - \frac{\Gamma_1 a_0 b_0 \left( i \sigma + \sigma_0 \right) }{\mathcal{K}_n \left( i \sigma + \Gamma_1 \sigma_0 \right)}  \right] e^{-x} \right\},
\end{array}
\end{equation}

with

\begin{equation}
\left\{
\begin{array}{l}
    \displaystyle \mathcal{M} = \frac{\lambda_0}{2} \left( \mathcal{Z} - \frac{1}{2} + \mathcal{B}_n  \right)  - \frac{i}{2} \left[  \left( \mathcal{Z} - \frac{1}{2} \right) \mathcal{B}_n - \lambda_0^2  \right],   \\[0.5cm]
    \displaystyle \mathcal{N} = \frac{\lambda_0}{2} \left( \mathcal{Z} - \frac{1}{2} + \mathcal{B}_n  \right)  + \frac{i}{2} \left[  \left( \mathcal{Z} - \frac{1}{2} \right) \mathcal{B}_n - \lambda_0^2  \right].
\end{array}
\right.
\end{equation}

Similarly, the torque of Eq.~(\ref{torque2_simple}) associated with the $ \Theta_0^{2,\nu} $-contribution of the semidiurnal tide can be written

\begin{equation}
\mathcal{T}^{2,\sigma_{\rm SD}} \sim  2 \pi R^2 H  \Im \left\{ \Xi_{\rm atm} \right\}  U_2 J_2.
\end{equation}

\begin{figure}[htb]
\centering
\includegraphics[width=0.950\textwidth]{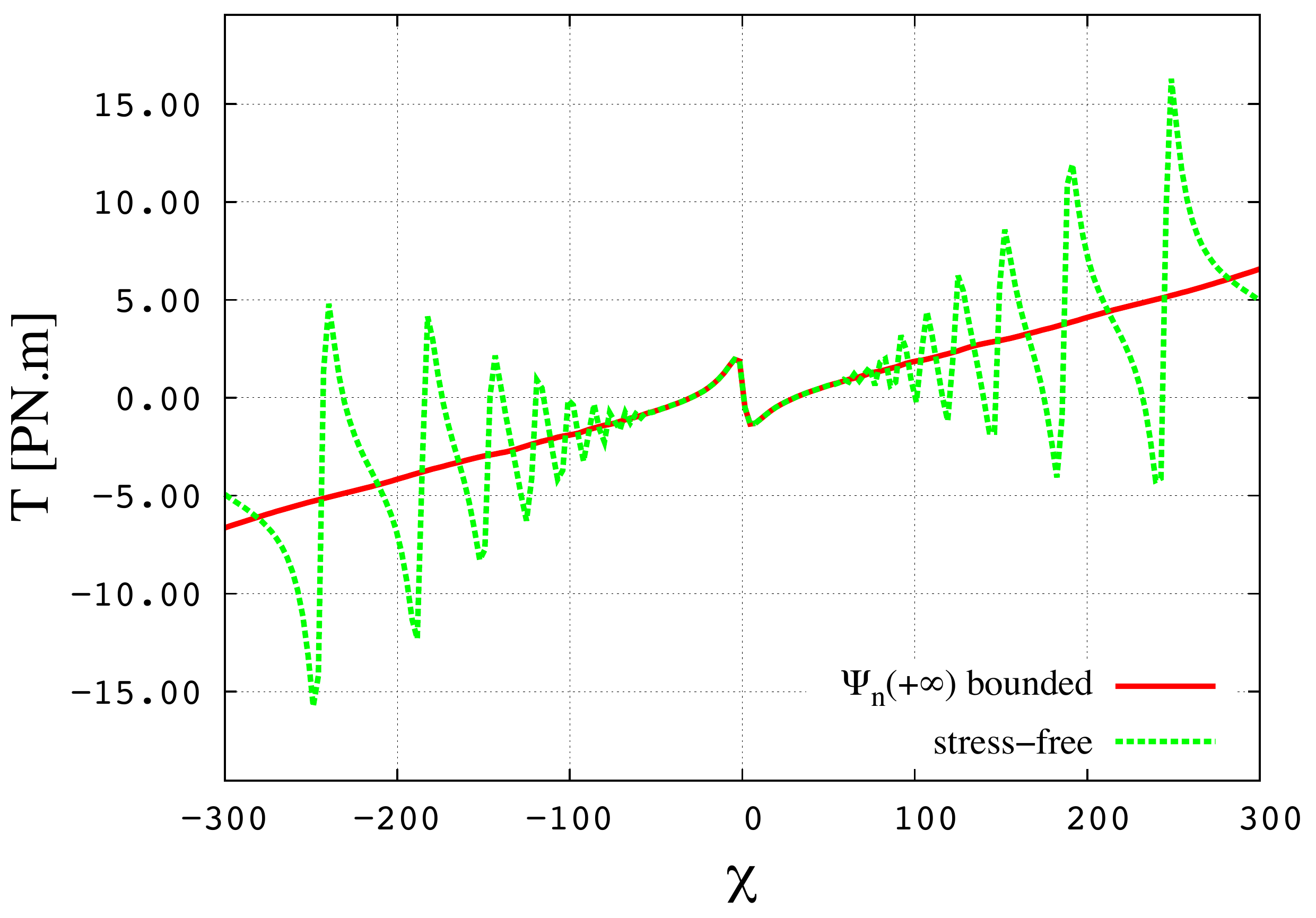}
\textsf{\caption{\label{fig:couple_BC} Tidal torques exerted on the atmosphere (in peta-Newton meters (PN.m)) induced by the first gravity mode of the Earth's semidiurnal tide ($ n = 0 $) as a function of the reduced tidal frequency ($ \chi $) obtained with two different conditions at the upper boundary: the energy-bounded condition chosen for the model (red line) and a stress-free condition (green dashed line). The equation for the structure in the vertical (resp. horizontal) direction has been integrated with 10000 mesh points (resp. 400 associated Legendre polynomials). } }
\end{figure}

To illustrate the difference induced by the choice of the upper boundary condition, we compute the variation of the tidal torque generated by the Earth's semidiurnal tide with the reduced tidal frequency ($ \chi = \left( \Omega - n_{\rm orb} \right) / n_{\rm orb} $)  for two conditions: the energy-bounded condition used in this work and the stress-free condition studied here. To simplify calculations, we focus on the gravity mode of lowest degree ($ n = 0 $). Results are plotted on Fig.~\ref{fig:couple_BC}. In the first case, we obtain the smooth evolution observed on Fig.~\ref{fig:Terre_couple} (with slight differences in the vicinity of the synchronization because of the absence of Rossby modes), but the stress-free condition obviously leads to a highly resonant behaviour. This may be explained by the form of the solution given by Eq.~(\ref{Psi_ana_sf}), where the integration constant $ \mathcal{Z} $ depends on $ \hat{k}_n $ and $ x_{\rm atm} $. By fixing $ x_{\rm atm} $, we implicitly define values of $ \hat{k}_n $ for which $ \mathcal{Z} $ reaches a maximum. On the contrary, the integration constant obtained by applying the energy-bounded condition, given by Eq.~(\ref{Kconst}), does not depend on the upper limit far beyond the critical altitude given by Eq.~(\ref{xcrit}).

\section{Supplementary materials about the dependence of the tidal response on the tidal frequency}
\label{app:material}

\begin{figure*}[htb!]
 \centering
 \includegraphics[width=0.42\textwidth,clip]{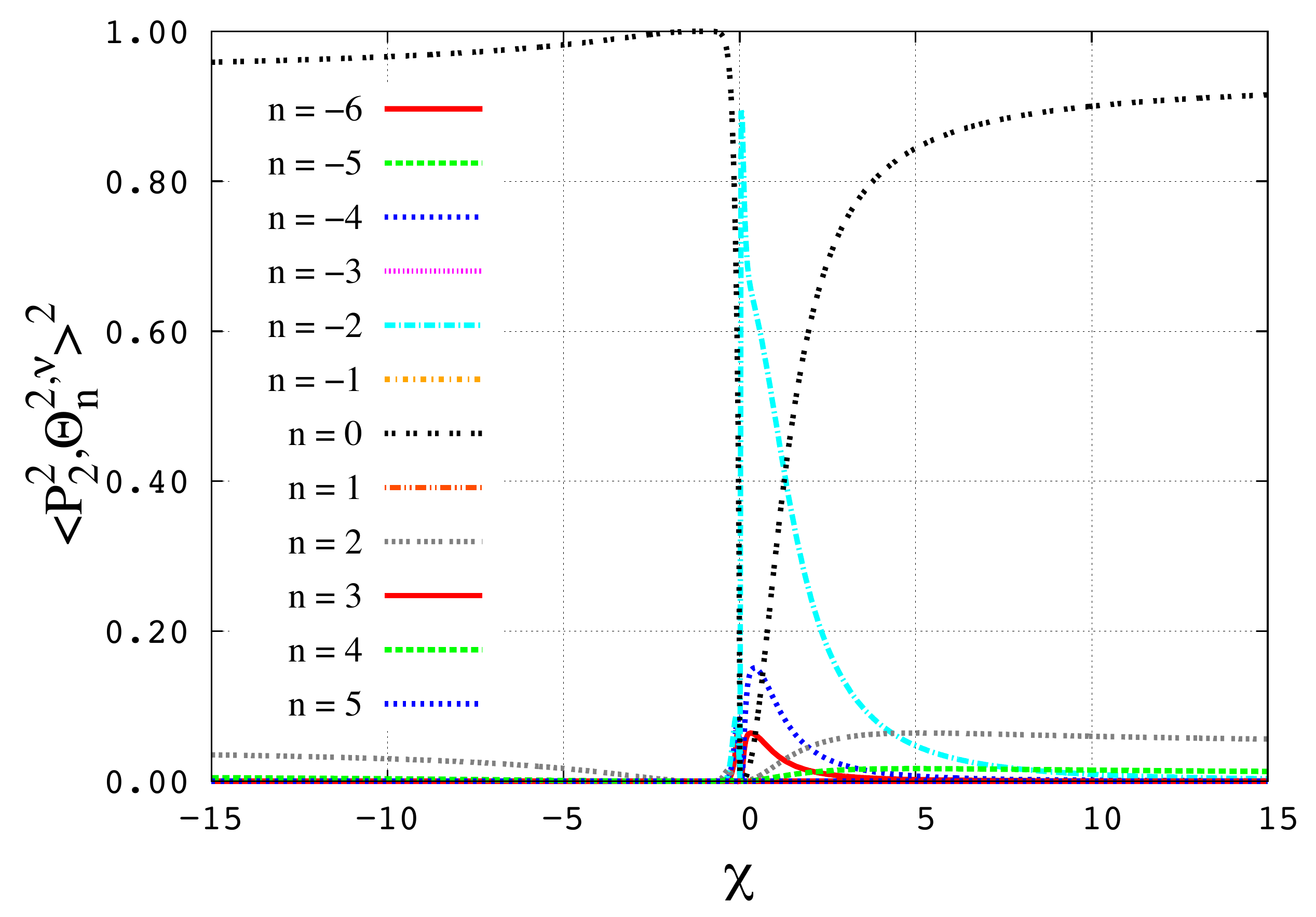} \hspace{1cm}    
 \includegraphics[width=0.42\textwidth,clip]{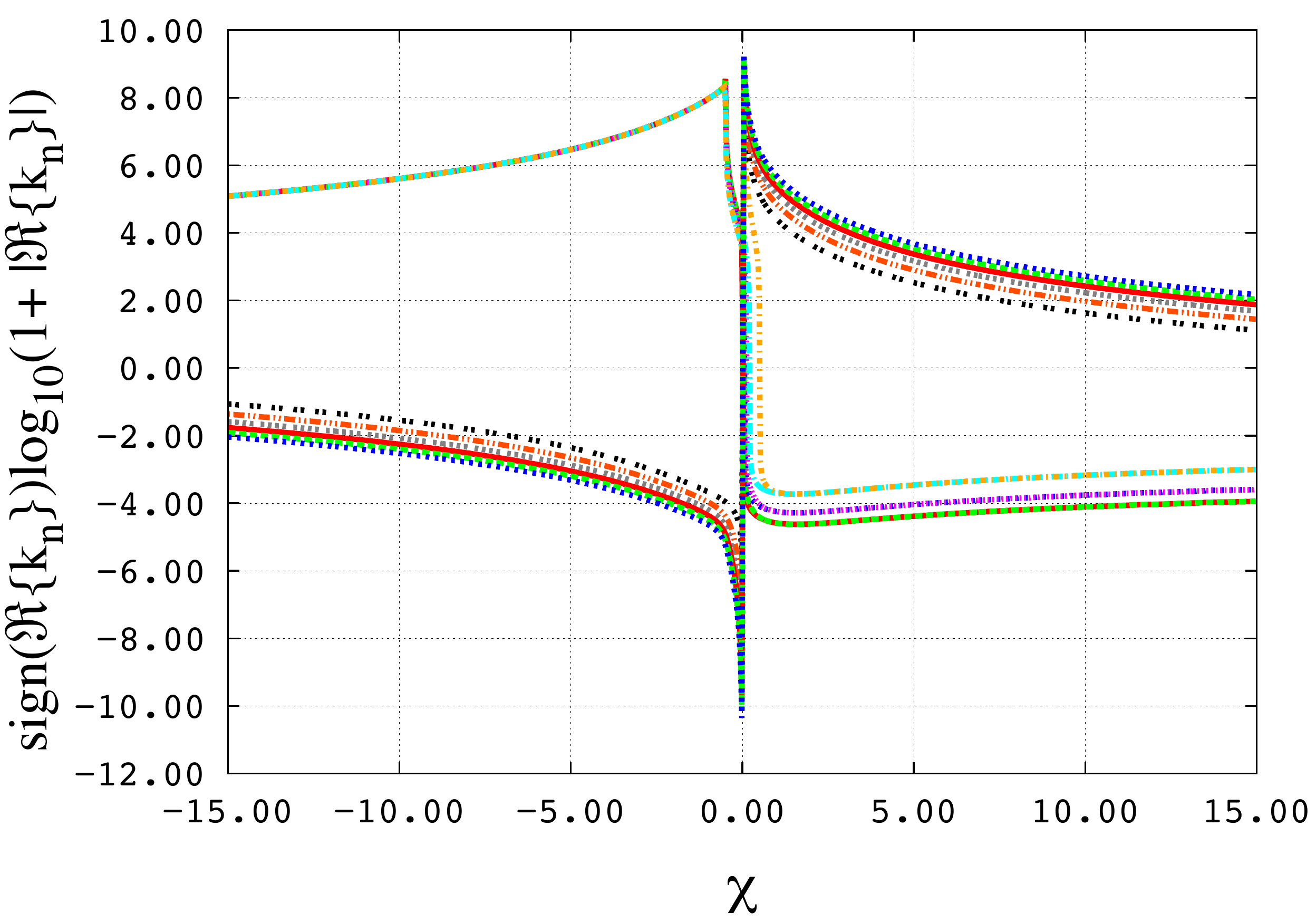} \\
  \includegraphics[width=0.42\textwidth,clip]{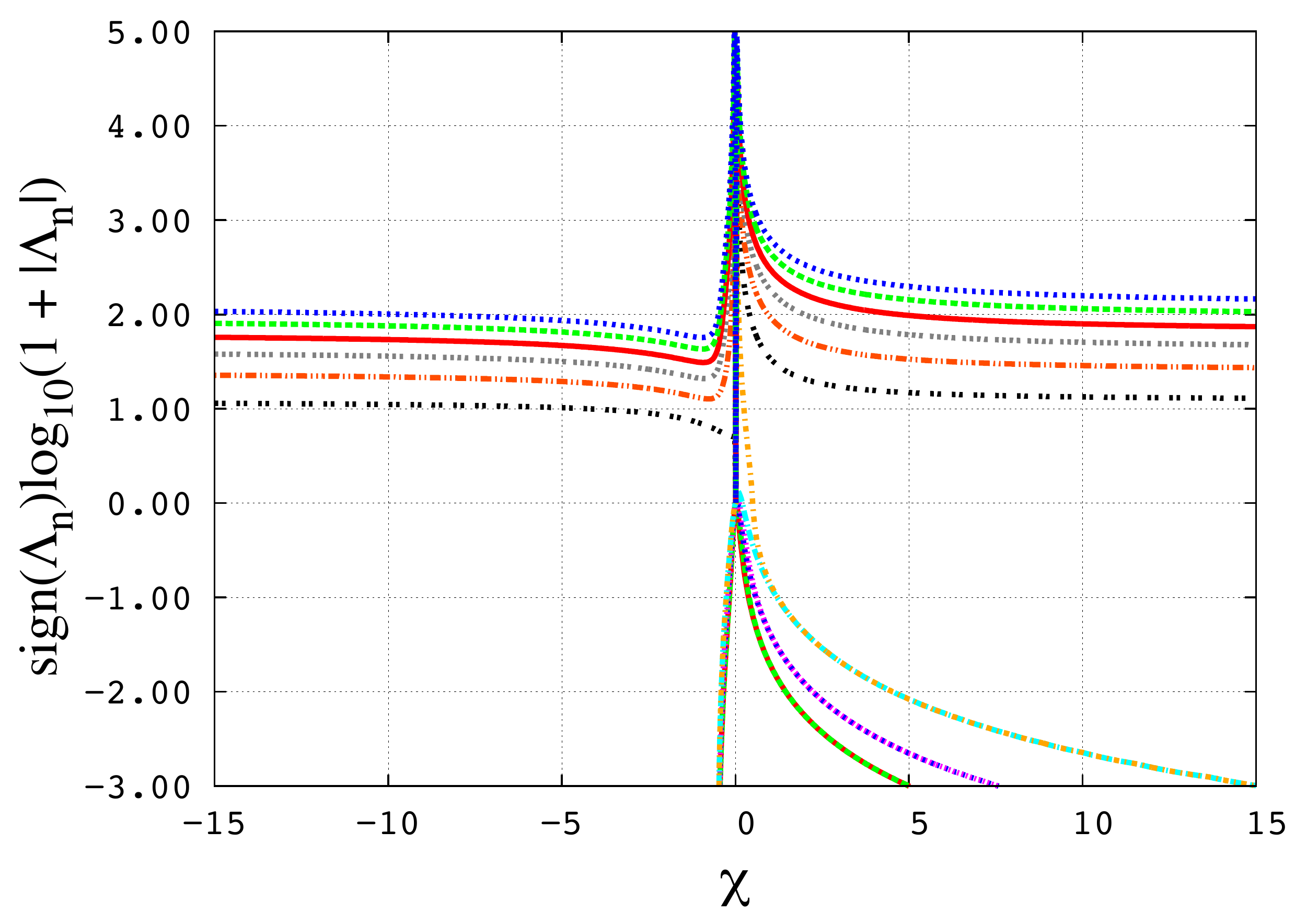} \hspace{1cm}         
 \includegraphics[width=0.42\textwidth,clip]{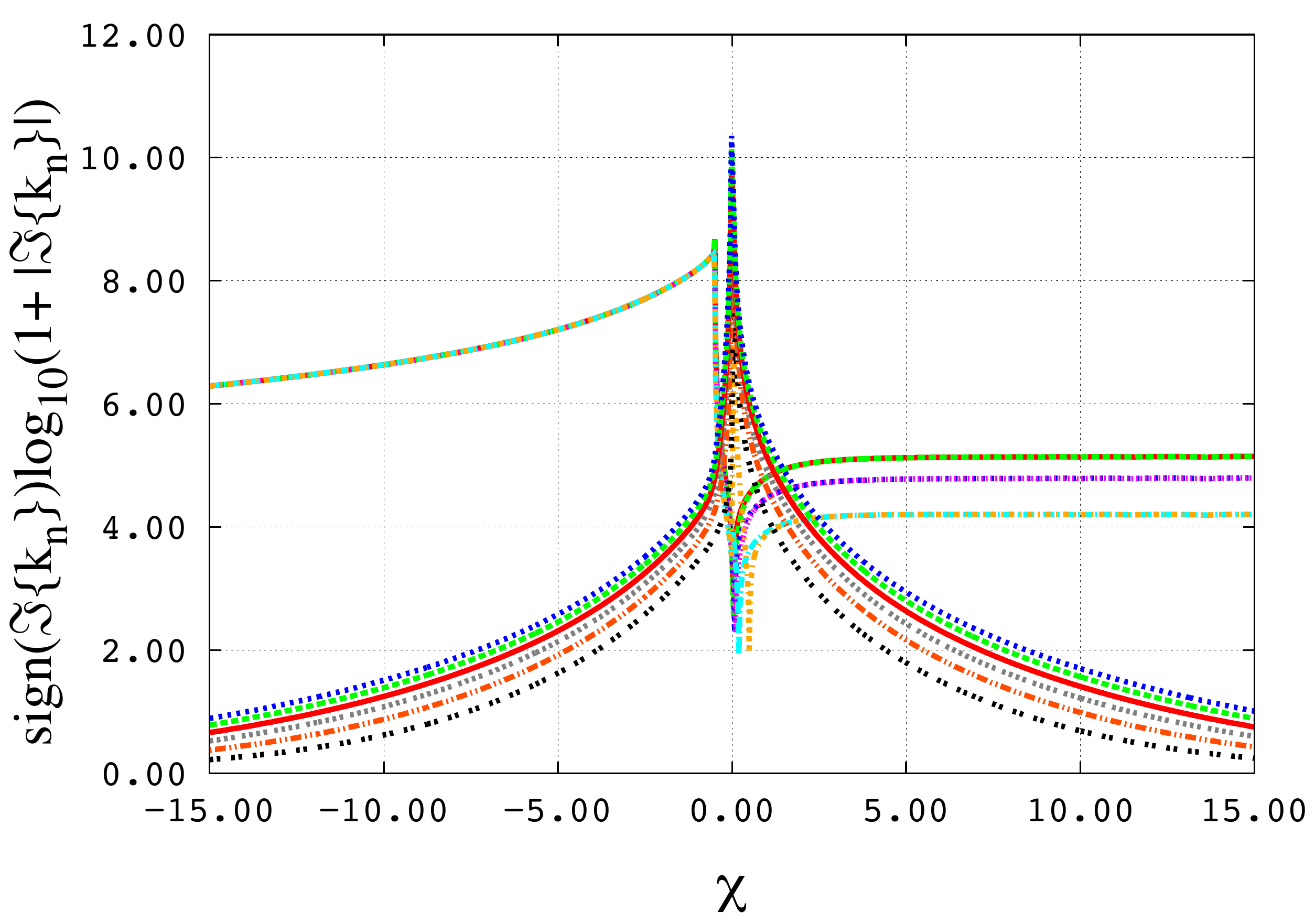} \\
   \includegraphics[width=0.42\textwidth,clip]{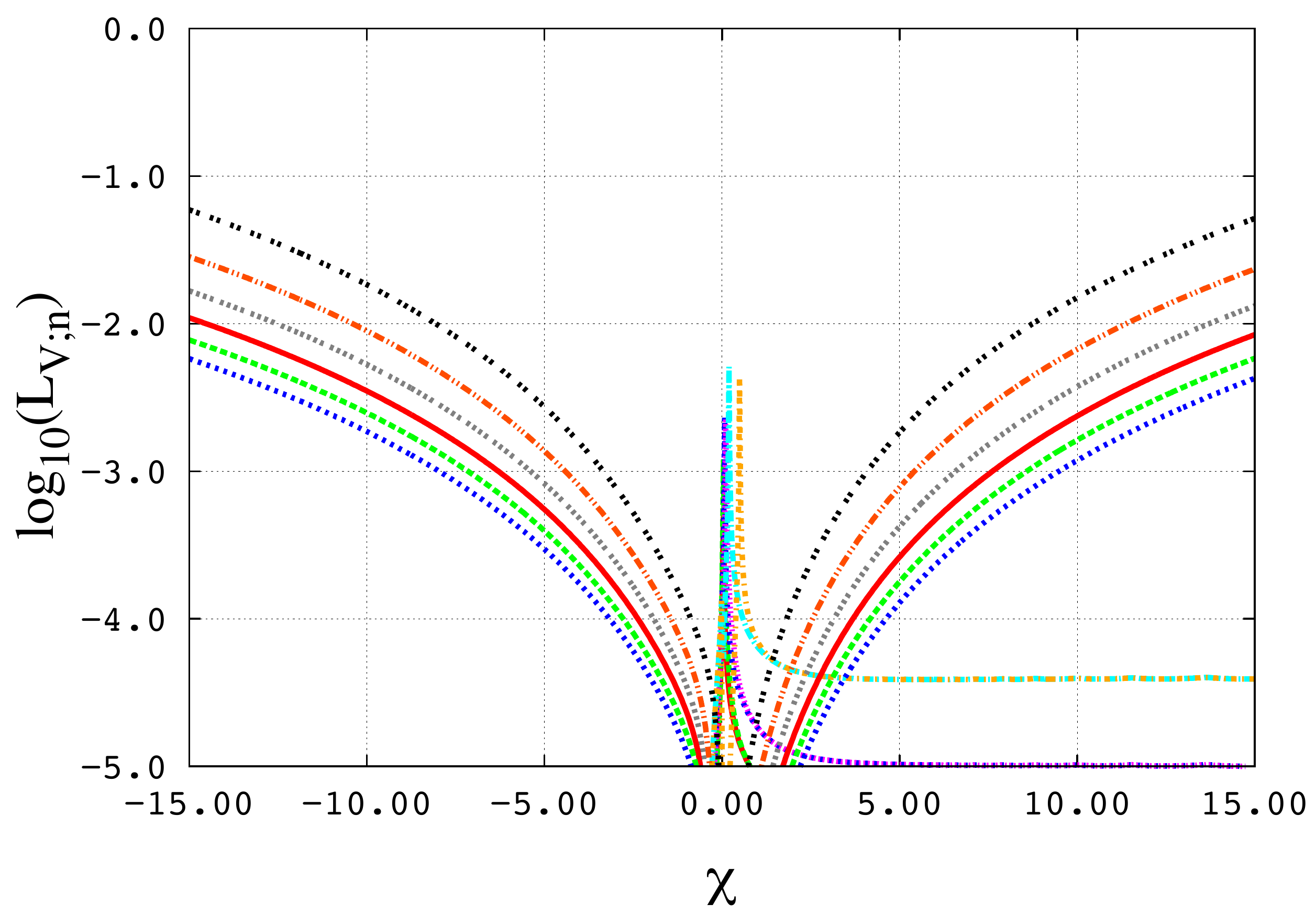} \hspace{1cm}          
 \includegraphics[width=0.42\textwidth,clip]{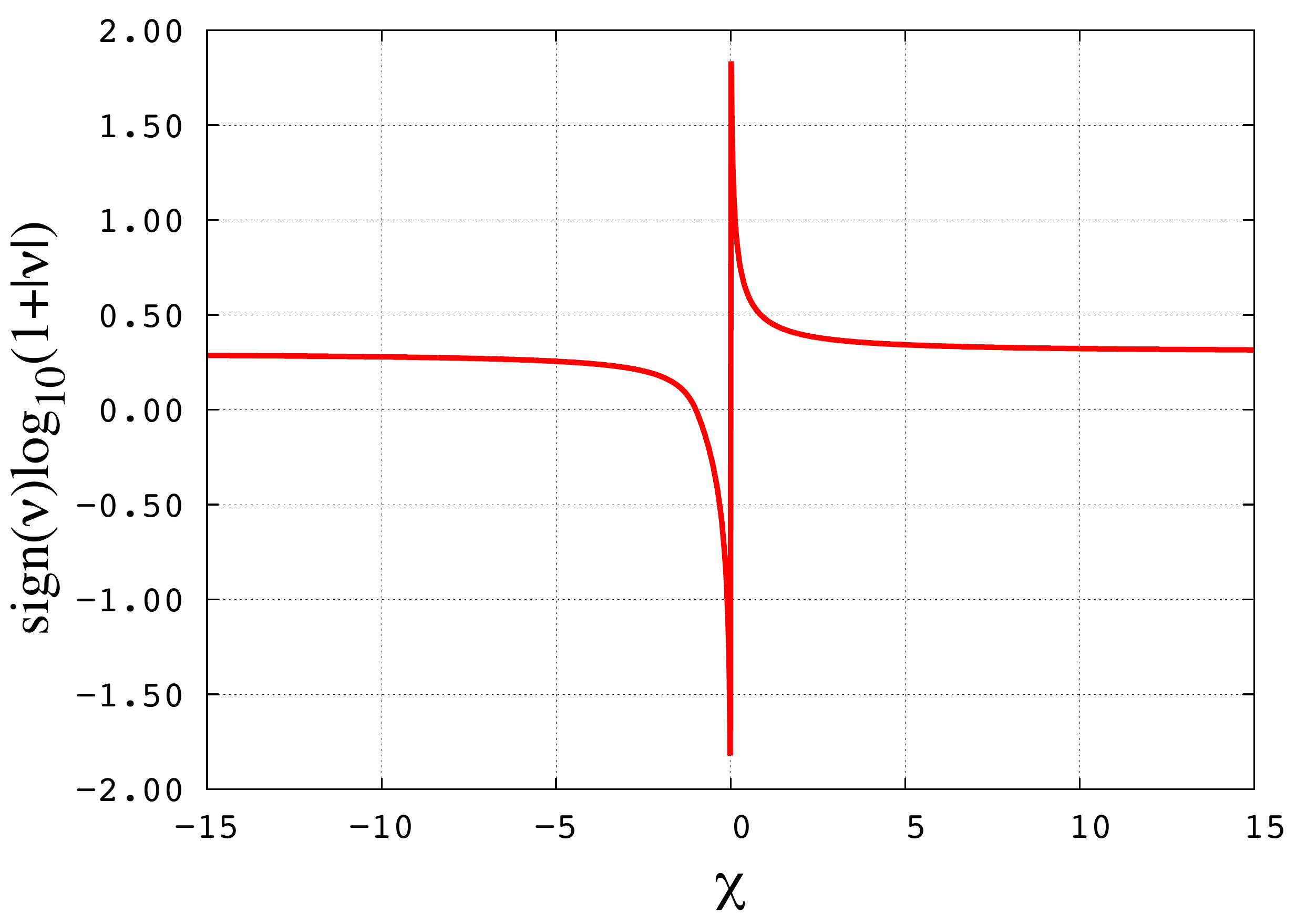}    
  \textsf{ \caption{\label{fig:Terre_kv} Tidal response of the Earth's atmosphere to a quadrupolar forcing ($ U \, \propto \, P_2^2 \left( \cos \theta \right) e^{2i\varphi} $ and $ J \, \propto \, P_2^2 \left( \cos \theta \right) e^{2i\varphi} $) as a function of the reduced tidal frequency $ \chi = \left( \Omega - n_{\rm orb} \right) / n_{\rm orb} $ (horizontal axis). For each quantity $ q_n $ associated with the Hough mode of degree $ n $ except the coefficients $ c_{2,n,2}^{\sigma,2} $ (top left) and the scales of variations $ L_{\rm V;n} $, the function $ f \left( \chi \right) = {\rm sign}\left(q_n \right) \log \left( 1 + \left| q_n \right| \right) $ is plotted (vertical axis) for $ n \in \llbracket -6 ,5 \rrbracket $. {\it Left, from top to bottom:} (a) coefficients $ c_{2,n,2}^{\sigma,2} = \langle P_2^2 , \Theta_n^{\sigma,2} \rangle^2 $ (Eq.~\ref{clnk}); (b) eigenvalues of Laplace's tidal equation ($ \Lambda_n^{m,\nu} $ defined by Eq.~\ref{eq_Laplace}); (c) normalized variation scales of the vertical structure of \mybf{the vertical component}, $ L_{\rm V;n} = 2 \pi / \left( x_{\rm atm} \left| \hat{k}_n  \right| \right) $ (Eq.~\ref{Lvn}), plotted in logarithmic scale. {\it Right, from top to bottom:} (d) real part of the vertical wavelengths ($ \hat{k}_n $, Eq.~\ref{kv2_2}); (e) imaginary part of the vertical wavelengths; (f) spin parameter $ \nu = 2 \Omega / \sigma $ (Eq.~\ref{Ltheta}).  }}
\end{figure*}

As a complement of the tidal torques (Figs.~\ref{fig:Terre_couple}) and perturbed pressure, density and temperature (Figs.~\ref{fig:Terre_variations_sol}) \mybf{resulting from the Earth's Solar semidiurnal tide}, we plot as a function of the tidal frequency several quantities that can be helpful for the diagnostic of results \mybf{(Fig.~\ref{fig:Terre_kv})}. These quantities are the coefficients $ c_{2,n,2} = \langle P_2^2 , \Theta_n^{2,\nu} \rangle^2 $ introduced in Eq.~(\ref{torque_epais_n}), which weight the contribution of Hough modes to the tidal torque, the eigenvalues of the Laplace's tidal equation ($ \Lambda_n^{m,\nu} $), the length scale of vertical variations associated to Hough modes ($ L_{\rm V;n} $), the spin parameter ($ \nu $) and the real and imaginary parts of the complex vertical wavenumber ($ \hat{k}_n $ denoted $ k_n $ in plots), which determines the nature (propagative or evanescent) of tidal waves.\\

These plots highlight the difference between prograde ($ \nu < 0 $) and retrograde ($ \nu > 0 $) modes. Indeed, one notes that the projection of the associated Legendre polynomial $ P_2^2 $ on Hough functions is not the same before and after the synchronization: the weigh of the gravity mode of degree $ 0 $, which is the most representative one, varies and Rossby modes are predominating in a very short interval of positive tidal frequencies. One retrieves this dissymmetry in the variations of the horizontal eigenvalues, where Rossby modes only appear for $ \sigma > 0 $. \\

As predicted by its analytic expression (Eq.~\ref{kv2_1}), the real and imaginary part of the vertical wavenumber diverge at the synchronization. Therefore, Hough modes become both highly oscillating and strongly damped in this region. Their spatial variation scale along the vertical direction decays, which corresponds to what is observed in the variations of $ L_{\rm V;n} $. Actually, below a given typical length scale, these modes are damped by dissipative processes such as thermal diffusion. Finally, the graph of $ \nu $ show that this parameter strongly varies around the synchronization. The transition at $ \sigma = 0 $ is discontinuous, $ \nu $ switching from $ - \infty $ to $ + \infty $.

\section{Consequences of an insufficient spatial resolution on the numerical computation of the vertical structure}
\label{app:spatial_resolution}

As pointed out by CL70 and detailed in Sect.~5, solving numerically the vertical structure equation (Eq.~\ref{Shrodinger_mince}) with the numerical scheme of Appendix~\ref{app:num_scheme} requires to use a sufficiently high spatial resolution. This condition is mathematically expressed by the criterion given by Eq.~(\ref{Ncrit}) for a regular mesh and depends on the vertical wavenumbers that characterize Hough modes. To compute the response of a mode, it is necessary to use a spatial step $ \delta x $ lower than the corresponding variations length scale ($ L_{\rm V;n} $, defined by Eq.~\ref{Lvn}). Otherwise, the signal does not appear. We propose to illustrate that point in this Appendix by considering the simplified forced wave equation

\begin{equation}
\dfrac{d^2 \Psi}{dx^2} + k_{\rm V}^2 \Psi = e^{-x/2},
\label{simple_vertical}
\end{equation}

where $ k_{\rm V} \in \mathbb{C} $ is such that $ \Im \left\{ k_{\rm V} \right\} > 0 $, with the previously chosen boundary conditions,

\begin{equation}
\begin{array}{lcl}
    \Psi \left( 0 \right) = 0 & \mbox{and} &   \left. \dfrac{d \Psi}{dx} \right|_{ x_{\rm sup}} - i k_{\rm V} \Psi \left( x_{\rm sup} \right) = 0.
\end{array}
\end{equation} 

\begin{figure}[htb]
 \centering
 \includegraphics[width=0.96\textwidth,clip]{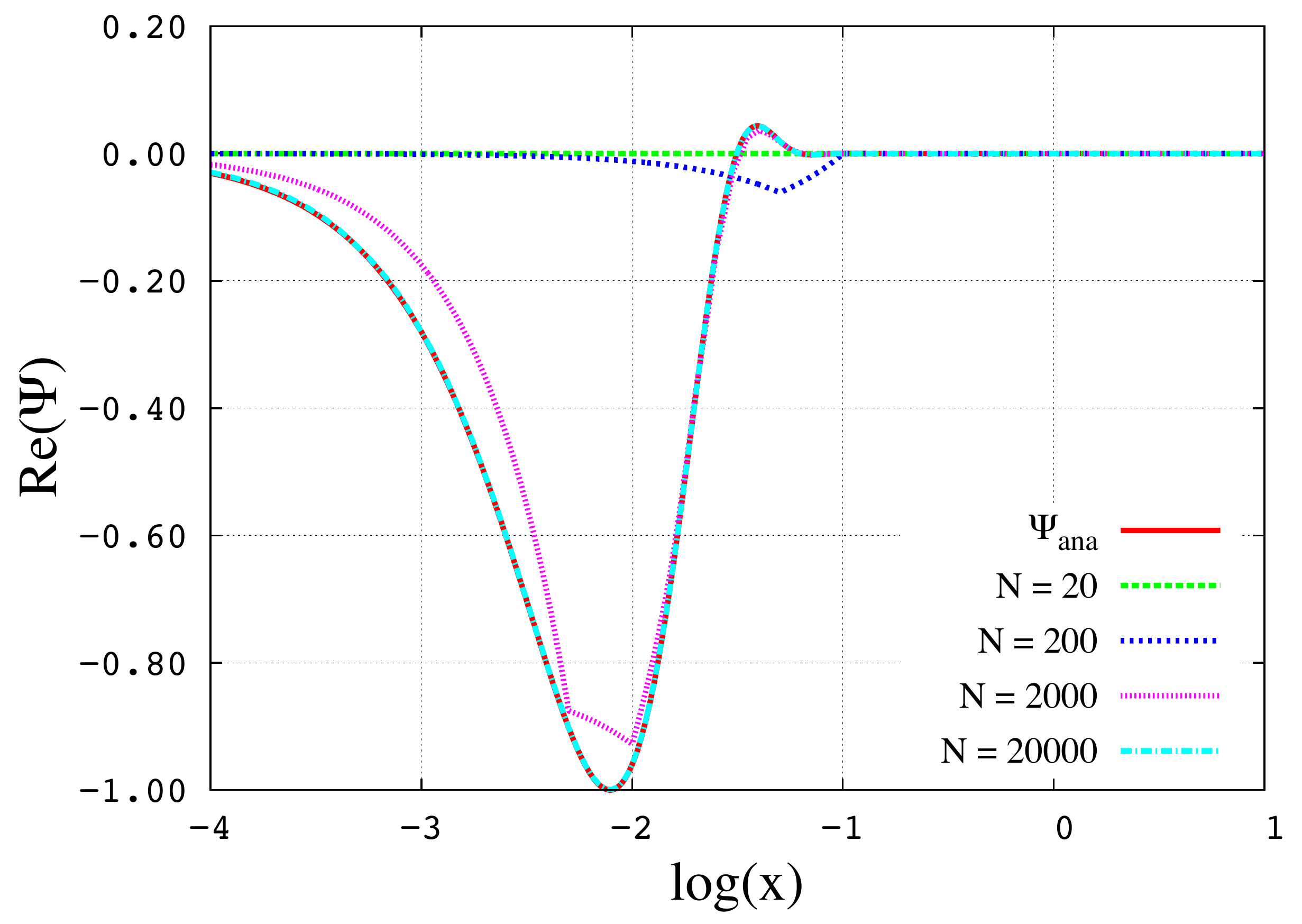}%
  \textsf{ \caption{\label{fig:Re_Psi_num} Real part of the solution of Eq.~(\ref{simple_vertical}) as a function of $ x $ in logarithmic scale computed numerically for various mesh resolutions. The number of points of the spatial discretization is set at $ N = 20 $ (green dashed line), $ N = 200 $ (blue dashed line), $ N = 2000 $ (pink dashed line) and $ N = 20000 $ (cyan dashed line). These numerical solutions are compared to the analytical solution given by Eq.~(\ref{Psi_ana}) (red line).    }}
\end{figure}

We arbitrarily set $ k_{\rm V} = 100 \left( 1 + i \right) $ in order to satisfy the condition $ L_{\rm V} \ll 1 $ (let us recall that $ x_{\rm sup} = 10 $). This corresponds to a strongly damped tidal mode similar to those predicted by the model in a frequency range close to synchronization. The critical number of points necessary to compute the response with our scheme is then $ N_{\rm crit} \sim 400 $. We solve Eq.~(\ref{simple_vertical}) for various resolutions. The corresponding results are plotted on Fig.~\ref{fig:Re_Psi_num}, where they are also compared to the analytical solution, given by Eq.~(\ref{Psi_ana}). The flattening observed for $ N < N_{\rm crit} $ gives an idea of the kind of numerical difficulties that shall be addressed to compute tidal low-frequency waves with a GCM. Particularly, this shows that if we compute the tidal torque numerically by using the scheme of Appendix~\ref{app:num_scheme} with an insufficient resolution, we will only be able to get the large scale contribution associated with the \mybf{horizontal component} (see Figs.~\ref{fig:Terre_couple}).

\end{document}